\documentclass[12pt,a4paper,twoside]{book}

\usepackage{latexsym, amsmath}
\usepackage{amssymb}
\usepackage{amscd}
\usepackage{graphicx}
\usepackage{fancyhdr}
\usepackage{color}

\makeindex
\makeglossary
\newcommand{\helv}{
  \rmfamily\scshape}
\usepackage{amsfonts}
\usepackage{amsmath}
\usepackage{amssymb}
\usepackage{graphicx}
\usepackage[nohug]{diagrams}
\diagramstyle[labelstyle=\scriptstyle]
\usepackage{color}
\def\be{\begin{equation}}
\def\ee{\end{equation}}
\def\bea{\begin{eqnarray}}
\def\eea{\end{eqnarray}}
\def\beann{\begin{eqnarray*}}
\def\eeann{\end{eqnarray*}}
\def\nn{\nonumber}

\newcommand\qed{\begin{flushright} $\Box$ \end{flushright}}
\newtheorem{pr}{Proposition}
\newtheorem{de}{Definition}
\newtheorem{theo}{Theorem}
\newtheorem{rem}{Remark}
\newtheorem{co}{Corollary}
\newtheorem{lem}{Lemma}

\def\circMunit{\bigcirc\!\!\!\!\,\!_{^{\mathbf{1}}}\,}
\def\circMa{\bigcirc\!\!\!\!\,\!_{^{a}}\,}
\def\circMb{\bigcirc\!\!\!\!\,\!_{^{b}}\,}
\def\circMc{\bigcirc\!\!\!\!\,\!_{^{c}}\,}

\def\circMi{\bigcirc\!\!\!\!\,\!_{^{\!\:i}}\,}
\def\circMm{\bigcirc\!\!\!\!\!_{^{\;\!m}}\,}
\def\circMn{\bigcirc\!\!\!\!\,\!_{^{\;\!\!n}}\,}
\def\circMM{\bigcirc\!\!\!\!\,\!\!_{^{M}}\,}
\def\circMr{\bigcirc\!\!\!\!\,\!_{^{\:\!r}}\,}
\def\circMR{\bigcirc\!\!\!\!\!_{^{\;\!R}}\,}
\def\circML{\bigcirc\!\!\!\!\!_{^{\;\!L}}\,}
\def\circMO{\bigcirc\!\!\!\!\!_{^{\;\!\mathcal{O}}}\,}
\def\circMA{\bigcirc\!\!\!\!\!_{^{\;\!\mathcal{A}}}\,}
\def\circMF{\bigcirc\!\!\!\!\!_{^{\;\!F}}\,}

\def\circMI{\bigcirc\!\!\!\!\,\!_{^{\:\!\mathrm{\mathbf{I}}}}\,}

\usepackage{epsfig}

\begin{document}
\frontmatter

\thispagestyle{empty}

\begin{center}

  {\center{\LARGE \bf Physical Determination of the Action  } \\[14mm]}

\end{center}

\begin{center}
  D I S S E R T A T I O N \\[6mm]
  zur Erlangung des akademischen Grades \\[6mm]
  doctor rerum naturalium \\[1mm]
  (Dr.~rer.~nat.)\\[1mm]
  im Fach Physik \\[6mm]
  eingereicht an der \\[1mm]
  Mathematisch-Naturwissenschaftlichen Fakult\"at\\[1mm]
  Humboldt-Universit\"at zu Berlin \\[6mm]
  von \\[1mm]
  \bf Dipl.-Phys. Dipl.-Math. Bruno Hartmann\\[2cm]
\end{center}

\begin{flushleft}
   Pr\"asident der Humboldt-Universit\"at zu Berlin: \\[1mm]
   Prof. Dr. Jan-Hendrik Olbertz\\[6mm]
   Dekan der Mathematisch-Naturwissenschaftlichen Fakult\"at: \\[1mm]
   Prof. Dr. Elmar Kuhlke\\[7mm]

Gutachter:\\[1mm]
  \begin{tabular}{ll}
    1. & Prof. Dr. rer. nat. habil. Thomas Lohse\\[1mm]
    2. & Prof. Dr. rer. nat. habil. Thomas Thiemann\\[1mm]
    3. & Dr. phil. habil. Oliver Schlaudt
  \end{tabular}\\[6mm]

eingereicht am: 29.04.2014\\[2mm]
Tag der m\"undlichen Pr\"ufung: 17.04.2015
\end{flushleft}

\pagebreak
\section*{Acknowledgements}

Thank you to Bruno Hartmann sen. and Peter Ruben for the introduction into the research problem and essential suggestions. I also want to thank Thomas Thiemann for stimulating discussions and support as well as Dieter Suisky, Markus M\"uller, Thomas Lohse and Oliver Schlaudt for orientation. This work was made possible initially by the German National Merit Foundation and finally with support by the Perimeter Institute with special thanks to Lucien Hardy, John Berlinsky, Greg Dick, Michael Duscheness and Howard Burton.

Finally, I would like to express my sincere gratitude to my mother and to Ulrich Hedtke. Without you, this thesis and nothing else would ever have been possible.

\addtocontents{toc}{\vskip -1.2cm}

\chapter{Abstract}
\vskip -0.6cm

Current research on the foundation of physics presupposes mathematical formalisms, which cannot anymore be interpreted and justified by tangible operations. In search for better and better formalisms arises the problem, that one must tie oneself down given an almost infinite number of possible extensions. In this thesis I develop a complementary foundations program, which makes no formal preassumptions at all. My objective is a foundation of physics from the operationalization of its basic observables. We begin with classical and relativistic kinematics. Seizing on a programmatic proposal by Heinrich Hertz we arrive via quantification of energy-momentum at the equations of motion and the action functional. Finally we present a relativistic revision of energy, momentum and inertial mass. The goal is not primarily to change or improve the mathematical structure of mechanics, but to gain a deeper physical understanding of mechanics like Einstein \cite{Einstein-Grundlagen der ART} with his gedanken experiments on relativistic kinematics. Starting from undisputed measurement operations and natural principles the mathematical formalism is only created and thus shown in a new light. This work is in the same context as current efforts to the ''foundation of quantum mechanics'', just with respect to the interpretation of the classical and relativistic calculuses.

We define the basic observables from practical comparison ''longer than'', ''heavier than'', ''more impact than'' etc. Next one wants to specify ''how many times'' more. The procedure for finding these values is the measurement (e.g. of a length by repeated placement of unit sticks one after another or of a weight by successively adding weight units on a balance scale). Hermann von Helmholtz \cite{Helmholtz - Zaehlen und Messen} starts from counting same objects: A basic measurement consists in a reconstruction of the measurement object with a material model of concatenated units. It requires knowledge of the method of comparison (of a particular attribute of both bodies) and of the physical concatenation methods (on objects which represent the unit measure). From actual physics one knows them only as mathematical operation symbols. That is our foundation problem. We give a protophysical foundation \cite{Janich Das Mass der Dinge} of the measurement conventions and norms, which the instruments must obey \cite{Schlaudt}. We specify the test procedures for manufacturing sufficiently constant reference devices (straight rulers, uniform running clocks, equally loaded springs etc.) and for standardizing a reproducible concatenation.

We develop Helmholtz program for measuring relativistic motion. We define a spatiotemporal order "longer than" by the classical comparison, whether one object or process covers the other. Without one word of mathematics one can manufacture uniform running light clocks and place them literally one after the other or side by side; and then count, how many building blocks it takes for assembling a regular grid which covers the measurement object. These basic operations overlap for different observers. We derive the mathematical relation between the physical quantities (number of light clocks), the formal Lorentz transformation. This demonstrates a strictly physical approach to physics where initially mathematics must remain outside - and then every step where mathematics is introduced requires an extra justification.

In this way we determine interactions. Heinrich Hertz \cite{Hertz - Einleitung zur Mechanik} outlined a novel treatment of mechanics based on the notion of energy for directly observable phenomena (independent from Newton's equations of motion and broader). We define energy, momentum and mass from the original principles of Leibniz and Galilei by the elemental comparison ''more capability to work than'' (against same test system) resp. ''more impact than'' (in a collision) and develop Helmholtz \cite{Helmholtz H. v. - Einleitung zur Vorlesung ueber Theoretische Physik} program for direct quantification. Luce, Suppes \cite{Luce Suppes - Theory of Measurement} diagnosed the failure to uncover a suitable empirical concatenation operation for energy and momentum carriers. We construct a solution. We let them coact ''expediently'' in a gedanken-calorimeter.

With a single elementary reference process (inelastic collision of irrelevant inner structure) and symmetry principles we can measure the energy and momentum of all other more complex processes. We construct an instrument, which functions for a basic measurement: Whatever gets absorbed in an external reservoir must generate only standard energy and momentum carriers. We define standard springs and impulse carriers as sufficiently constant and reproducible reference objects for ''capability to work'' and ''impact''. We count how many standard obstacles a moving body can overcome, before it stops and thus determine the magnitude of its energy. We transfer its impact onto a certain number of standard impulse carriers and thus quantify its momentum. From matching the form and layout of the building blocks in the machinery we can count the coinciding numbers of activated standard springs, standard bodies, velocity units. We obtain a geometric proof for the kinetic energy-velocity relation and similarly for the generic momentum-velocity relation, the energy-mass relation etc. We define the basic observables (energy, impulse, mass), quantification and derive their fundamental equations.

By introducing quantity equations from vivid process depictions this approach is also interesting for didactics. Students have a crucial difficulty to ''translate between the physics phenomena and the mathematical formulation. ... The mathematics part is only at the end of a whole process and ... needs interpretation: What is the (physical) meaning of the structure?'' \cite{Pospiech_2006}. ''Students prefer verbal explanations and experiments before calculations and use of formulae'' \cite{Pospiech_2008-Design_curriculum}. We bridge the gap from the natural processes (experiment, basic observables, qualitative relations ''the more - the more'' etc.) to the mathematical language: We reveal the basic actions in experiment and measurement. We generate a mathematical formulation of these tangible operations. All practical steps are understandable from everyday (work) experience and from a colloquial description. In this foundation, which explains the mathematical formalism from the operationalization of the basic observables, one can also understand scope and limitations of the formalisms. The latter have undergone profound revisions. We make the physical and measurement methodical principles (paradigm change) transparent. That helps students to overcome mental barriers and apparent paradoxes.

\chapter{Zusammenfassung}
\vskip -1.1cm

Die aktuelle physikalische Grundlagenforschung setzt mathematische Formalismen voraus, welche nicht mehr durch konkrete physikalische Handlungen interpretiert und gerechtfertigt werden k\"onnen. Auf der Suche nach immer besseren Formalismen entsteht das Problem,  sich angesichts einer schier unendlichen Zahl m\"oglicher Erweiterungen festlegen zu m\"ussen. In dieser Arbeit wird ein komplement\"ares Fundierungsprogramm entwickelt, welches gar keine mathematischen Vorannahmen trifft. Das Ziel ist eine Begr\"undung der Physik aus der Operationalisierung ihrer Grundma\ss{}e. Wir beginnen mit der klassischen und relativistischen Kinematik. Dann greifen wir einen programmatischen Entwurf von Heinrich Hertz auf und gelangen \"uber die Quantifizierung von Energie und Impuls zu den Bewegungsgleichungen und dem Wirkungsfunktional. Den Abschlu\ss{} bildet eine relativistische Revision der Energie, Impuls und Massenvorstellung. Das Ziel ist nicht prim\"ar, den mathematischen Aufbau der Mechanik zu \"andern oder zu verbessern, sondern ein tieferes physikalisches Verst\"andnis zu erlangen, so wie Einstein \cite{Einstein-Grundlagen der ART} mit seinen Gedankenexperimenten zur relativistischen Kinematik. Ausgehend von unbestrittenen Messhandlungen und Naturprinzipien wird der mathematische Formalismus erst erzeugt und somit neu beleuchtet. Die Arbeit steht damit im selben Kontext wie die aktuellen Bem\"uhungen zur ''Fundierung der Quantenmechanik'', nur eben mit Bezug auf die Interpretation der klassischen und relativistischen Kalk\"ule.

Wir definieren die Grundma\ss{}e aus praktischen Vergleichen ''l\"anger als'', ''schwerer als'', ''wuchtiger als''. Man m\"ochte auch bestimmen ''wieviel mal'' mehr. Das Verfahren, diese Werte zu finden, ist die Messung (z.B. von L\"angen durch mehrmaliges Hintereinanderlegen eines Einheitsstabes oder von Gewichten durch hinzupacken von Gewichtseinheiten auf eine Balkenwaage). Hermann von Helmholtz \cite{Helmholtz - Zaehlen und Messen} geht aus vom Z\"ahlen gleicher Objekte: Eine Grundmessung besteht in der Rekonstruktion des Messobjektes durch ein materielles Modell verkn\"upfter Einheiten. Sie erfordert die Kenntnis der Methode zur Vergleichung (eines bestimmten Attributes der beiden Objekte) und der Methode zur physischen Verkn\"upfung (von Objekten, die die Messeinheit repr\"asentieren). Aus der tats\"achlichen Physik kennt man diese nur als mathematische Operationszeichen. Darin liegt das Fundierungsproblem. Wir geben eine protophysikalische Erkl\"arung \cite{Janich Das Mass der Dinge} der Messvorschriften und Normen, welche die Instrumente erf\"ullen m\"ussen \cite{Schlaudt}. Wir spezifizieren die Testprozeduren zur Herstellung von hinreichend konstanten Bezugsger\"aten (gerade Lineale, gleichf\"ormig laufende Uhren, gleich gespannte Federn usw.) und zur Normierung von reproduzierbaren Verkn\"upfungen.

Wir entwickeln Helmholtz Programm f\"ur das Messen von relativistischen Bewegungen. Wir definieren eine raumzeitliche Ordnung ''l\"anger als'' durch den klassischen Vergleich, ob ein Objekt oder Vorgang den anderen \"uberdeckt. Um zu finden ''wieviel mal'' l\"anger, kann man gleichf\"ormig laufende Lichtuhren herstellen und diese wortw\"ortlich nacheinander oder nebeneinander platzieren; und dann z\"ahlen, wieviele Bausteine man braucht, um ein regul\"ares Gitter zusammenzuf\"ugen, welches das Messobjekt \"uberdeckt. Die elementaren Messhandlungen \"uberschneiden sich f\"ur verschiedene Beobachter. Wir leiten die mathematische Beziehung zwischen den physikalischen Gr\"o\ss{}en (Anzahl von Lichtuhreinheiten) her, die formale Lorentz-Transformation. Dies demonstriert einen strikt physikalischer Zugang zur Physik, bei dem die Mathematik zun\"achst au\ss{}en vorbleiben muss – und dann jeder Schritt, bei dem Mathematik eingef\"uhrt wird, einer extra Rechtfertigung bedarf.

In gleicher Weise bestimmen wir Wirkungen. Heinrich Hertz \cite{Hertz - Einleitung zur Mechanik} entwarf ein neuartiges Bild der Mechanik auf der Energievorstellung von direkt beobachtbaren Ph\"anomenen (unabh\"angig von Newton's Bewegungsgleichungen und weiter). Wir halten uns an die urspr\"unglichen Prinzipien von Leibniz und Galilei und definieren Energie, Impuls und Masse durch elementare Vergleichsverfahren ''wirkungsverm\"ogender als'' (gegen dasselbe Testsystem) bzw. ''wuchtiger als'' (im Frontalsto\ss{}-Test) und entwickeln Helmholtz \cite{Helmholtz H. v. - Einleitung zur Vorlesung ueber Theoretische Physik} Programm zur direkten Quantifizierung. Luce, Suppes \cite{Luce Suppes - Theory of Measurement} konstatierten das Scheitern, geeignete empirische Verkn\"upfungshandlung f\"ur Energie und Impulstr\"ager aufzukl\"aren. Wir konstruieren eine L\"osung. Wir lassen sie ''nutzbar'' zusammenwirken in einem Gedanken-Kalorimeter.

Mit einem einzigen elementaren Grundsto\ss{}prozess (inelastische Kollision von irrelevanter Struktur) und Symmetrieprinzipien k\"onnen wir die Energie und Impuls von allen anderen komplexeren Prozessen ausmessen. Wir konstruieren ein Instrument, das zur Grundmessung funktioniert: Alles, was darin absorbiert wird, soll nur Standardenergie- und Impulstr\"ager erzeugen. Wir definieren Standardfedern und Impulstr\"ager als hinreichend konstante Bezugs-objekte f\"ur ''Wirkungsverm\"ogen'' und ''Wucht''. Wir z\"ahlen, wieviele Standardhindernisse ein bewegter K\"orper \"uberwinden kann, bevor er stoppt, und bestimmen somit die Gr\"o\ss{}e seiner Energie. Wir \"ubertragen seine Wucht auf eine bestimmte Anzahl von Standardimpuls-tr\"agern und quantifizieren somit seinen Impuls. Durch Anpassung der Form und Anordnung der Bausteine in der Maschinerie k\"onnen wir die Anzahlen der jeweiligen Bezugseinheiten z\"ahlen. Wir finden einen geometrischen Beweis f\"ur die kinetische Energie-Geschwindigkeit-Beziehung, die Energie-Masse-Beziehung etc. Wir definieren Grundobservablen (Energie, Impuls, Masse), Quantifizierung und leiten deren Grundgleichungen her.

Mit der Einf\"uhrung von Gr\"o\ss{}engleichungen aus anschaulichen Prozessdarstellungen ist dieser Ansatz auch in didaktischer Hinsicht interessant. Sch\"uler haben ein entscheidendes Problem ''zwischen den physikalischen Erscheinungen und mathematischen Formulierungen zu \"ubersetzen. ... Der mathematische Anteil folgt nur am Ende eines ganzen Prozesses und ... bedarf Interpretation: Was ist die Bedeutung dieser Strukturen?'' \cite{Pospiech_2006}. ''Sch\"uler bevorzugen verbale Erkl\"arungen und Experimente vor dem Rechnen und Gebrauch von Formeln'' \cite{Pospiech_2008-Design_curriculum}. Wir \"uberbr\"ucken die L\"ucke von den Naturvorg\"angen (Experiment, Grundobservablen, qualitative Beziehungen ''je mehr - desto mehr'') zur mathematischen Sprache: Wir zeigen die Grundhandlungen in Experiment und Messung und erzeugen dann eine mathematische Formulierung von diesen konkreten Handlungen. Alle praktischen Schritte sind verst\"andlich mit allt\"aglicher (Arbeits-) Erfahrung und umgangssprachlicher Beschreibung. In diesem Zugang, der die mathematischen Formalismen aus der Operationalisierung von Grundma\ss{}en gewinnt, kann man auch Tragweite und Grenzen der Formalismen verstehen.

\tableofcontents

\mainmatter


\chapter{Foreword}\label{Kap - KM Dynamics short Review - Foreword}

Helmholtz and Hertz approach to theoretical physics reflects on aims and methods of physics. They explain the relationship between the theory and the (practical steps in conducting) experiment. Hertz \cite{Hertz - Einleitung zur Mechanik} begins the introduction to mechanics by specifying its purpose
\begin{quote}
    ''Es ist die n\"achste und in gewissem Sinne wichtigste Aufgabe unserer bewu\ss{}ten Naturerkenntnis, da\ss{} sie uns bef\"ahige, zuk\"unftige Erfahrungen vorauszusehen, um nach dieser Voraussicht unser gegenw\"artiges Handeln einrichten zu k\"onnen.''\footnote{''It is the nearest and in a sense main function of our conscious natural knowledge, that it enables us to foresee future experiences so as to be able to set up our present actions according to that foresight.''}
\end{quote}
The know how developed most profitably in everyday work experience.\footnote{Since the ancient Greeks Euclidean geometry was elevated to an exact science (provability of assertions). It took 2000 years until (in the division of labor in industrial revolution) the thinking was influenced by motives of constructing and \emph{engine}ering the tool use \cite{Wolff - Geschichte der Impetustheorie} - until dynamics developed.} One can regard mechanics as a science of the possibility of machines.

The understanding developed historically: Up to the mid of the 19th century ''it appeared as ultimate goal and as ultimate aspired explanation of natural phenomena to trace back the latter to countless distant forces (Fernkr\"afte) between the atoms of matter'' \cite{Hertz - Einleitung zur Mechanik}. ''Starting from considerations which are concerned only with the nearest practical interests of technical work'' (\emph{extractable mechanical work} from dynamical machines, collisions, heat machines and \emph{work equivalent} from electrical, chemical or phase transitions etc.) Helmholtz \cite{Helmholtz - Ueber die Erhaltung der Kraft} was lead to discover the principle of conservation of energy - in practical form the impossibility of perpetuum mobile. Under its overwhelming impression physics begun to favor a new point of view: ''to approach all phenomena falling into its domain as conversion of energy into new forms and to regard tracing back all phenomena to laws of energy conservation as its ultimate goal'' \cite{Hertz - Einleitung zur Mechanik}. The principles of mechanics evolve.

Hertz stood for eliminating the fundamental role of ''force'': The concept of ''force'', as it grew out of Newton's axiomatic system, does not apply the category of causality properly in mechanics \{\ref{Kap - KM Dynamics short Review - Causal misapplication}\}. It is inadmissable for the foundation of the theory. Moreover ''force'' is neither directly measurable nor can it be indirectly determined by Newton's incomplete axiomatic system alone \{\ref{Kap - KM Dynamics short Review - Incompleteness}\}. According to Hertz conception instead ''energy'', independent from Newton's equations of motion and broader, must be taken as a basis for a \emph{comprehensible} foundation of (classical and relativistic) mechanics \{\ref{Kap - KM Dynamics short Review - Hertz program}\}. If one want's to skip the historic and conceptual motivation the development of the new models begins with chapter \{\ref{Kap - Kinematics}\}.

\section{Incomplete formalism}\label{Kap - KM Dynamics short Review - Incompleteness}

Galilei conceived ''dynamics is the science of (movement) actions of forces... and gets to the bottom of the free play of natural forces'' \cite{Szabo Geschichte der Mechanischen Prinzipien}. Also Newton did examine collisions in his study notes, before 20 years later 1687 (with main goal gravitation) he presents an axiomatic system for mechanics \cite{Newton - Principia}. Ultimately he introduced the notions ''force'', ''inertial mass'' indirectly by his basic law of mechanics - which Newton \emph{postulates} (but not proves). Beginning 1734 first Euler \cite{Euler Anleitung} seeks for a direct method for measuring ''inertial mass'', ''external force'' in interactions. Though the basic equation (Newton: $F=m\cdot a$, Euler: $F\cdot \Delta t = m \cdot \Delta v$ etc.) always remained in the status of a postulate. Without derivation its scope and limitations remain a mystery. The origin of basic observables for classical and relativistic mechanics is contested to this day \cite{Okun - concept of mass}. Despite the fact that the assumption is un-natural; from postulating basic equations one generates circular arguments.

There is no quantification of ''force'' as a basic physical observable.\footnote{We distinguish basic and derived measures. In a \emph{basic} measurement (i.e. of length or duration) the result does not require prior knowledge of the quantification of any other observables (provided a rigid meter-stick one simply counts the number of steps or provided a functioning clock \cite{Janich Das Mass der Dinge} one counts the number of ticks). Instead e.g. a measurement of force according to Hooke's empirical law by means of deforming a spring requires knowledge (from guessing?) of the non-linear (!) expansion coefficient of the spring. Likewise for any proposal of a basic measurement for force - please check that implicit assumptions are not circular.} An alternative position regards the notions mass and force inherit to the totality of an axiomatic system. Inl this view one might determine force from known inertial mass by means of measuring the acceleration provided one postulates the basic law
\[
\begin{array}{lccl}
   \mathrm{Newton \;\; II}:   &   F  &  :=  &  m\cdot a \\
   \mathrm{Lorentz}:   &   F_{\mu}^{\nu} v^{\mu} &  :=  &  m \cdot a^{\nu} \;\; .
\end{array}
\]
Despite the ambiguity (if velocity dependence is attributed to mass or force) the quantification rests on reliable basic measurements for inertial mass. Dating back to Euler \cite{Euler Anleitung} and popularized by Mach \cite{Mach - Mechanik in ihrer Entwicklung} one can specify the inertial mass from colliding two objects $\mathcal{A}$, $\mathcal{B}$
\[
   m_{\mathcal{A}} \cdot \Delta v_{\mathcal{A}} \stackrel{\mathrm{(Newton\;II)}}{=} \left( F_{\mathcal{A}} \cdot \Delta t \stackrel{\mathrm{(Newton\;III)}}{=} - F_{\mathcal{B}} \cdot \Delta t \right) \stackrel{\mathrm{(Newton\;II)}}{=}  - m_{\mathcal{B}} \cdot \Delta v_{\mathcal{B}}
\]
by eliminating the undetermined force provided one postulates Newton's third law. With this hypothetical definition one would measure the ratio of the two inertial masses directly from the collision behavior by the inverse ratio of the velocity changes
\be\label{Formel - Mach Hypothese Traegheit}
   \frac{m_{\mathcal{A}}}{m_{\mathcal{B}}} := - \frac{\Delta v_{\mathcal{B}}}{\Delta v_{\mathcal{A}}}   \;\;\; .
\ee

\begin{figure}    
  \begin{center}           
  \includegraphics[height=3.1cm]{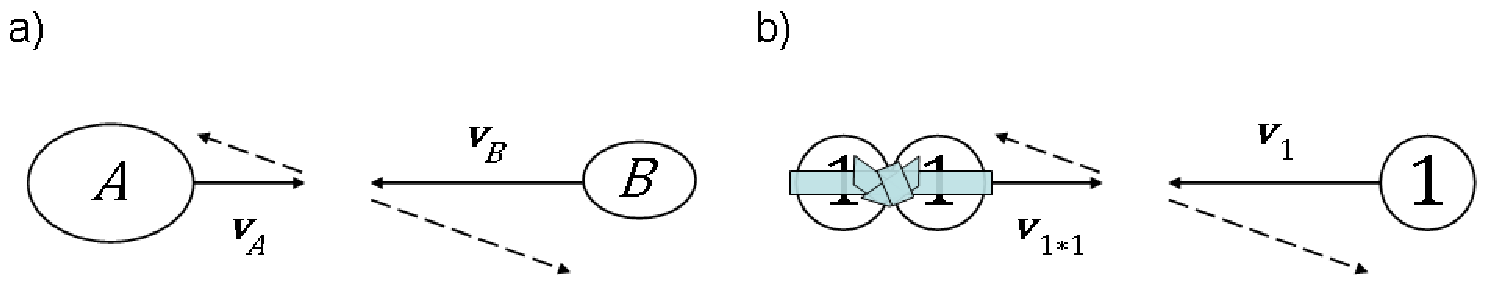}  
  \end{center}
  \vspace{-0cm}
  \caption{\label{pic_BER_composite_collision} a) collision between two generic objects $\mathcal{A}$ and $\mathcal{B}$ b) in a reservoir of identically constituted bodies $\left\{ \circMunit \right\}$ two are bound together (e.g. by a practically massless sling) and collide as a composite $\circMunit\!\ast\!\circMunit$ with another individual object $\circMunit$
    }
  \end{figure}

Consider a simple application where we provide a reservoir with identically constituted objects $\left\{ \circMunit \right\}$. Suppose we can tightly connect two of them such that the composite $\circMunit\!\ast\!\circMunit$ acts like one rigid body (see figure \ref{pic_BER_composite_collision}).
From the head-on collision between one element $\circMunit$ and the composite $\circMunit\!\ast\!\circMunit$ we would determine the ratio of their inertial masses
\[
   m_{\circMunit\!\ast\!\circMunit} \;\; := \; - \frac{\Delta v_{\circMunit}}{\Delta v_{\circMunit\!\ast\!\circMunit}} \; \cdot \; m_{\circMunit}  \;\; .
\]
The axiomatic system does not predict what the ratio of the accelerations is; one has to make the experiment. If one finds $\frac{\Delta v_{\circMunit}}{\Delta v_{\circMunit\!\ast\!\circMunit}}   \stackrel{\mathrm{?}}{=}  -2 $ one deduces by the hypothesis $m_{\circMunit\!\ast\!\circMunit} \stackrel{(\ref{Formel - Mach Hypothese Traegheit})}{:=} 2 \cdot m_{\circMunit}$. In classical mechanics one measures $\Delta v_{\circMunit}/\Delta v_{\circMunit\!\ast\!\circMunit} = -2$ while relativistic measurements give $\Delta v_{\circMunit}/\Delta v_{\circMunit\!\ast\!\circMunit} \neq -2$. Suppose one connects multiple elements $\underbrace{\circMunit\!\ast\!\ldots\ast\! \circMunit}_{N \times}\,$, is the magnitude of the inertial mass in every case an empirical number or can we predict it with certainty?
\\

In classical mechanics one has one more notion of mass. One measures the weight with a functioning beam scale \cite{Schlaudt} and a set of physically identical weight units $\left\{ \circMunit \right\}$. For a generic weight $\mathcal{A}$ on the left arm of the scale one successively adds weight units $\circMunit$ to the right until the scale is in static equilibrium. One quantifies the weight of $\mathcal{A}$ by counting the weight units $\sharp\{\circMunit\}$ on the right
\be
   m_{\mathcal{A}}^{\mathrm{(weight)}} \;\; := \;\; \sharp \{\circMunit\} \; \cdot \;  m_{\circMunit}^{\mathrm{(weight)}}   \;\; .  \nn
\ee
Hence the weight for the composite body $\circMunit\!\ast\!\circMunit$ is unambiguously quantified
\[
   m_{\circMunit\!\ast\!\circMunit}^{\mathrm{(weight)}} \;\; := \;\; 2 \; \cdot \;  m_{\circMunit}^{\mathrm{(weight)}}   \;\;\; .
\]
We justify factor $2$ by the \emph{congruence principle}. Under the conditions of weight measurements all weight units $\circMunit$ are \emph{identically constituted} and, when placed into the static scale, they \emph{behave in the same way}. In a direct weight measurement one counts congruent units.

The simple example illustrates that based on Newton's framework one cannot uniquely determine the dynamics of collision processes. The formalism makes no definite prediction without additional implicit assumptions.\footnote{As a guiding principle in a search for deeper understanding Zeilinger \cite{Zeilinger interpretation and phil foundation of QM} highly recommends ''to follow the guidance of the Copenhagen interpretation, that is, not to make any unnecessary assumptions not supported by a thorough analysis of what it really means to make an experiment.'' Zeilinger regards as first step of a physical interpretation the analysis of ''rules that determine which element of the formalism corresponds to which measurable quantity or to which observable fact in a concrete experimental situation. These rules are a large, mostly not explicated but only implicit, set of instructions. They concern the instructions on how to proceed in experiment in order to demonstrate or test a theoretical prediction.'' As instructive example of a well-founded
theory he refers to Einstein's theory of Special Relativity: ''Almost all relativistic equations... of 1905 were known already before... But only Einstein created the conceptual foundations... \emph{from which the equations} of the theory of relativity \emph{arise}. He did this by introducing the principle of relativity, which asserts that the laws of physics must be the same in all inertial systems... together with the constancy of the velocity of light.''} Newton's axiomatic system is incomplete. It does not account for the congruence principle which underlies every basic measurement.

\section{Misapplication of causality}\label{Kap - KM Dynamics short Review - Causal misapplication}

The content of Newtonian mechanics had proven splendidly in practice. Nevertheless Hertz is concerned about gaps in the form of its presentation. The formalism applies the purely categorial distinction - what is cause and what is effect - improperly. When the cause becomes the effect one can start to listen attentively. Without specifying elements of the interacting system forces against different elements are assigned to the same object.

\begin{figure}    
  \begin{center}           
  \includegraphics[height=3.1cm]{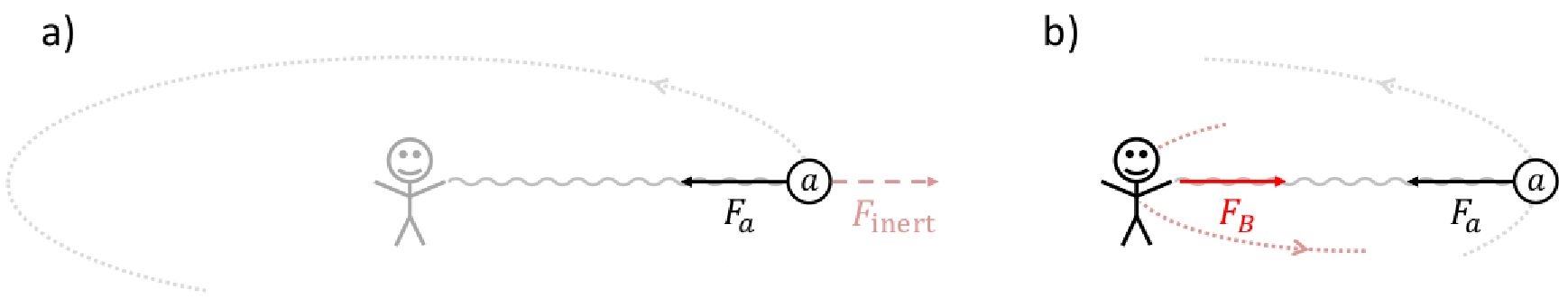}  
  \end{center}
  \vspace{-0cm}
  \caption{\label{pic_Hertz_Sitz_der_Kraefte} a) stone on a string swings around a circle b) balanced forces of the string
    }
  \end{figure}

Hertz \cite{Hertz - Einleitung zur Mechanik} illustrates by the common treatment of a simple example: ''Right from the beginning it must work miracles, how easy it is... to undoubtedly abash clear thinking''. Let Bob swing a stone $\circMa$ by a string around in a circle; Bob exerts a \emph{force} $F_a$ on the stone; that force constantly deflects the stone from a straight path - indeed always with force, mass and radial acceleration in accordance with Newton's second law (see figure \ref{pic_Hertz_Sitz_der_Kraefte}a).
%
But now the third law demands a \emph{counterforce} to the force $F_a$ which Bob's hand exerts on the stone. To the question for the \emph{place} of this counterforce the common answer is: as a result of the ''inertial force'' $F_{\mathrm{inert}}$ (German: Schwungkraft) the stone acts back on the hand, and that inertial force is in effect exactly opposite and equal ''$F_a=-F_{\mathrm{inert}}$'' to the force Bob exerts. Hertz asks ''Is that so-called centrifugal force something other than the inertia of the stone? Can we, without undermining clarity of our conception, account for the effect of inertia twice, namely once as mass and second as a force? In laws of motion we regard force as cause of accelerations. Are we allowed, without confusing our concepts, to suddenly speak of forces which emerge not until the acceleration, which are a consequence of acceleration? To all these questions we must clearly say no! ... But where then remains the claim of the third law, which requires a force, which the dead stone exerts on the hand, and which is to be gratified by a real force, not by a mere name? ... The meaning of the third law is that forces always connect two bodies'' with directions  equally well from Bob to the stone as from the stone to Bob.\footnote{The question of the place of the forces demonstrates the missing association with the elements of the system, the carrier of the forces. Already Euler \cite{Euler Anleitung} abandoned the misconceived expression ''inertial force'' of an accelerating body. He stated more precisely ''external force''. The force (of an interaction) is associated with an interacting \emph{system}. The force \emph{causes} accelerations the its elements, not reversely.

We can anticipate the result of our systematic development, which begins with Hertz question and his approach to use energy-momentum as cornerstone for constructing the theory - and to regard force as a derived quantity (where conditions allow, but not necessary and outright avoidable for quantum mechanics!). We will define \cite{Hartmann-KM_Dyn}: The \emph{force} $\mathbf{F}_a$ (\emph{of interaction} $w$ \emph{in the system} $\circMa \cup \mathcal{S} \cup \circMb $ with initial conditions $\mathbf{x}_I$, $\mathbf{v}_I$) against the element $\circMa$ is a derived physical quantity which specifies how the momentum changes over time
\[
   \mathbf{F}_a \! \left[ w\big|_{\mathbf{x}_I,\mathbf{v}_I} \right] \cdot \Delta t \;\; := \;\; \Delta \mathbf{p}_a  \;\; .
\]
\\
The force of interaction $w$ in the system $\circMa \cup \mathcal{S} \cup \circMb$ requires e.g. a tight string $\mathcal{S}$. The source (Ursache) for an interaction could be a spring, bent bow, charged battery etc. We have the \emph{force of the string} $\mathcal{S}$ against the elements $\circMa$ and $\circMb$; eventually the more massive Bob swings to (see figure \ref{pic_Hertz_Sitz_der_Kraefte}b). The force of the string $\mathcal{S}$ against the stone $\circMa$, symbolized by $\mathbf{F}_{a}$, and the force of the string $\mathcal{S}$ against Bob, symbolized $\mathbf{F}_{B}$, associate the terminus ''force'' with the elements of the system.

The correct interpretation of Newton's third law regards the two forces associated with the source
\[
   \mathbf{F}_{a} \; = \;  - \; \mathbf{F}_{B} \;\; .
\]
The force of the string $\mathcal{S}$ against both attached bodies is balanced. That is the familiar momentum conservation. The expression ''action equals reaction'' falls more precisely under Leibniz equipollence principle (i.e. measure an energy source by its effect \cite{Hartmann-KM_Dyn}) resp. under Helmholtz principle of conservation of energy.}
Without need to show further examples Hertz refers to the experience from teaching ''that it is very difficult, to communicate especially the introduction to mechanics to \emph{thinking audiences} without some embarrassment, without feeling the need to apologize here and there, without the wish to slur fairly quickly over the beginnings towards examples, which speak for themselves. (Hertz) thinks that Newton must have felt this embarrassment, when he defined somewhat violently mass as product of volume and density.''

Hertz is convinced the content of Newton's mechanics is correct. Those indeterminacies, which worry Hertz about the foundations, did certainly not prevent a single of the countless successes of mechanical applications. Rather Hertz criticizes the form of presentation of this content and declares ''this presentation was never penetrated towards scientific completion, it lacks the distinction as to what in the conceptualized picture originated from logical necessity, what from experience and what from our arbitrariness... The dignity and grandeur of the content of mechanics require to render its clarity by a completed presentation.''

\section{Hertz scheme}\label{Kap - KM Dynamics short Review - Hertz program}

Hertz \emph{outlines} a new type of treatment of mechanics based on energy principles and which introduces the concept of energy before the concept of force.
The advantage is simplicity. The concept of force with its difficulties can be avoided.\footnote{On the basis of energy principles the system
of mechanics is completed. ''What then we can add are only mathematical derivations and for example simplifications or auxiliary denominations, which are perhaps useful but certainly not necessary. To the latter belongs the concept of force which in the fundamentals did not appear ... As such the concept of force can not cause difficulties'' \cite{Hertz - Einleitung zur Mechanik}.} ''All assertions can refer (directly) to known characteristics of body systems under consideration, without needing to mask our lack of knowledge of the details by arbitrary and uninfluential hypotheses.''\footnote{Energy theory does not limit itself to repressing the notion ''force'' in substitution for ''energy'' - also their carriers are grasped much broader. Newton's presentation of mechanics is formulated for individual point particles, \emph{atomic elements}. Whereas energy theory addresses the effect potential of \emph{entire systems}. Without problematic extra assumptions energy theory applies in a broader sense.}

Hertz points out that ''in this way we can \emph{avoid} best to talk about things, of which we know very little and which also have no influence on intended essential assertions. ... The reduction of all phenomena onto forces compels us to constantly tie our thinking to considerations of individual atoms and molecules ... but the shape of atoms, their cohesion, their motion, all that is entirely concealed in most cases; their number is unimaginably large. Our notion of atoms is by no means particularly suited to serve as known and secured basis for mathematical theories. ...
Arbitrarily assumed properties of atoms may be of no influence on the final result; the latter may be correct. Nonetheless details of those derivations are presumably in large part wrong - the derivation is an illusory proof.''

''In contrast the theory of energy offers the \emph{advantage} that into the \emph{assumptions} of the problem enter only those features, parameters... of regarded bodies which are immediately accessible to experience; that the consideration \emph{proceeds} by means of these properties in finite and closed form and that also the \emph{final result} can be translated directly into tangible experience. Except for energy in its few forms no auxiliary constructions enter our considerations. ... Not only the final result, but also all steps of its derivation can be acknowledged as correct and meaningful.''

Hertz specifies \emph{fundamental steps} of the program which ''gives priority to the concept of energy and which also uses it as basis and cornerstone of our construction.'' We can consent to his notice that - to this day - ''there is no textbook of mechanics that takes the view of energy theory and introduces the concept of energy before the concept of force.'' One will have to start from ''four mutually independent concepts... space, time, mass, energy. ... The true difficulties occur as soon as we attempt... to \underline{introduce} energy. We must not pursue the usual path, starting from forces... proceeding to energy. Rather it will be necessary to specify, without already presupposing mechanical developments, by means of which \emph{simple direct experiences} we ultimately perceive the \emph{presence} of energy and \emph{determine its quantity}.'' Despite his confidence of feasibility Hertz got into the dilemma that he could not present a tangible form on how to measure energy in a fundamental manner. He ''did only assume but not prove that such a way of measuring (energy, momentum, mass) exists.''

Hertz raises the particular problem ''that energy occurs in two so entirely different forms as kinetic and potential energy.'' Mistakenly he demands that both shall be determined independently.\footnote{As it turns out - in measurement actions - kinetic and potential form of energy are inseparably unified. In our quantification scheme \cite{Hartmann-KM_Dyn} we must first define a carrier with \emph{(standard) potential energy}; to determine the quantity of \emph{kinetic energy} of decelerating body $\circMa_{\,\mathbf{v}}$ by means of counting above (measurement) units; such that we can determine the quantity of \emph{(generic) potential energy} from generic interactions by means of measuring its (extractable) kinetic effect.}
Further ''in order to stipulate relations between all four basic concepts (length, time, energy, mass) and thus temporal evolution of phenomena he resorts to integral principles of ordinary mechanics... he arbitrarily uses Hamilton's principle'' - which is widely used also in contemporary formal approaches to physics. Though Hertz bears in mind that,
''it is unthinkable that in reality Hamilton's principle (or similar) represents a basic law of mechanics and hence a basic law of nature, because from a basic law from the start it is to be expected \emph{simplicity} and \emph{modesty}; though Hamilton's principle expresses an utterly intricate assertion... The integral whose minimum Hamilton's principle requires has no simple physical meaning... in fact for the unprepared mind it is an incomprehensible expression.''

Hertz objection against intricate assertions as basic laws expresses the convincement that:
''If the content of an assertion is correct and comprehensive, then after more expedient choice of basic concepts, the content must let itself pronounce in a simpler form'' and the awakening desire: ''to penetrate from exterior understanding of such law to its deeper and actual meaning - in its presence we are convinced.'' It is evident that conceptual problems of Newton's presentation of mechanics are not resolved by transformation (automorphism) onto one of their completely equivalent formulations. -- Hertz sums up ''Those are all difficulties which ought to be eliminated and circumvented by the intended definition of energy... Such circumvention is presently not achieved.\footnote{Despite initial attempts Hertz concedes defeat and explores a third picture of mechanics of all movement phenomena between course sensual (German: grobsinnlich) demonstrable masses. He substitutes the action of all independent sources by hypothetical assumption of actions of (imaginary) masses and their motion in secrecy. As in kinetic theory of gases where all forces of heat are reduced to the motion of tangible masses - Hertz regards energy and force (of independent sources) as kinetic energy and momentum of hidden masses.} It remains an open question as to whether such a system can be developed altogether.''
\\

We will complete Hertz approach drawing on measurement principles of Galilei, Leibniz \{\ref{Kap - KM Dynamics - Physical Measurement - Pre-theoretical Ordering Relation}\} and Helmholtz \{\ref{Kap - KM Dynamics short Review - Physical method}\} and more recent developments: Einstein's \cite{Einstein '05 - Zur ED bewegter Koerper} relativity principle, light principle and the measurement theoretical foundation of kinematics, Planck's \cite{Planck - Wege zur physikalischen Erkenntnis} discovery of the action as irreducible unit, Wallot \cite{Wallot - Groessengleichungen Einheiten und Dimensionen}, Ruben's \cite{Peter '87 - Dialektische Logik und messende Wissenschaft} conception of a basic measurement as a doubling of the measures and Lorenzen, Janich \cite{Janich Das Mass der Dinge} and Schlaudt's \cite{Schlaudt} protophysical understanding of reference devices and procedures.

We develop a novel strictly physical foundation of basic observables energy, momentum and inertial mass without taking equations of motion etc. as a basis (of unclear origin and status). We reveal the physical and methodical basis for an unambiguous quantification. Our practical method for measuring interactions will entail Newton's equations and also relativistic dynamics. We will understand that ''inertial mass'' (unlike ''weight'' whose connection to collision behavior reveals not before Einstein's general relativistic revision of the background-spacetime) coincides with the amount of matter. We uncover the scope and limitations for ''force'' as a meaningful derived physical quantity.


\chapter{Initial assumptions}\label{Kap - KM Dynamics short Review - Initial assumptions}

We pursue a different approach to the foundation of the theory. Our problem is a physical explanation of physics based on the tangible operations of the human being (steering and measurement interventions) \{\ref{Kap - KM Dynamics short Review - Measurement operations}\}. The subject of our explanation is mathematized physics. Our presupposition for the explanation is non-mathematized physics.

We view basic observables and numbers, the elementary concepts of physics and mathematics, as a result of measurement resp. counting operations \{\ref{Kap - KM Dynamics short Review - On physical objects and mathematical objects}\}. The latter are a product of historic evolution \{\ref{Kap - KM Dynamics short Review - Evolutionary product}\}.\footnote{Our physical-mathematical sign language - Adorno remarks - ''long since cannot be fetched home in intuition, nor any category directly commensurable to human consciousness'' \cite{Peter '79 - Philosophie und Mathematik}. If in addition we regard their work - as Helmholtz, Hertz show \{\ref{Kap - KM Dynamics short Review - Foreword}\} - we can reveal its origin.}
We connect its origin to the historic development of work \{\ref{Kap - KM Dynamics short Review - Work concept}\}. In the practical domain the construction of physical quantities (for the standardization of the conduct of physical experiments) becomes explainable \{\ref{Kap - KM Dynamics short Review - Physical method}\}.

\section{Physical objects and mathematical objects}\label{Kap - KM Dynamics short Review - On physical objects and mathematical objects}

With Ruben \cite{Peter '79 - Philosophie und Mathematik} we can interpret ''mathematics as a particular kind of general work. Since everybody will admit, that counting and measuring are indeed \emph{actions}, not purely passive perceptions (of given) predetermined numbers and magnitudes.'' To imagine such conception ''it is useful to think about the action of a shepherd, who is counting sheep by means of a run-through gate, which occurs in every depiction dedicated to understanding of birth or application of the concept of numbers. In the counting act all sheep are forced into an order (by means of the gate) which they may give up after the counting.'' Helmholtz \cite{Helmholtz H. v. - Einleitung zur Vorlesung ueber Theoretische Physik} demands a \emph{countable} collective of (invariant) objects - coins, chumps, eggs or non-tangible things as words, sounds or light - ... that neither suffer division nor fuse... that each possesses and retains its separate existence during the act of counting.''\footnote{With Wiener's remark on the difference between stars and clouds - ''A star is a distinct object, outstandingly suited for being cataloguized or counted... When we require from a meteorologist to provide a similar sampling for clouds... he would explain indulgently, that in entire language of meteorology there is no thing as a cloud, defined as an object with quasi-permanent identity.'' - Ruben clarifies the relation between natural objects and idealized things in mathematics. ''Clearly we can only talk about an \emph{actual} individuum upon condition of its interaction with environment, whereupon both change... It is that \emph{quasi-permanent identity} under which we have to enforce natural things with more or less effort, to ultimately enable them as unit in a counting act.''} To be countable an object must exist as a multitude of elementary units ''$\mathbf{1}_{\sharp}$''.

Ruben specifies the productive character in counting. Given two countable collectives (irrespective being apples or pears) we determine the equality with regard to number ''$\sim_{\sharp}$'' by a manual counting act: ''In the sense of Cantor's procedure for determining equal cardinality of sets... we successively separate an individuum (provided it admissibly represents a unit) from the collective... and reunite those isolated members elsewhere back to the collective. The process is complete when the original collective (herd of sheep) disappeared and resurrects as a counted collective. Initially there is an uncounted collective, with given but not yet determined amount, and in the end we have the collective back together in a counted state... The separation of members from the collective presents the elementary operation of substraction and division. The reunification of separated members to a new collective represents the elementary operation of addition and multiplication.'' Clearly the concatenation operation ''$\ast_{\,\sharp}$'' in counting has an empirical character. One conducts the addition of length, mass, energy by \emph{different} operations, depending which measure the reference device represents: one adds lengths by aligning rulers side by side at an angle of 180$^{\circ}$; one adds masses by assembling them tightly and we will add energy sources in a calorimeter \{\ref{Kap - KM Dynamics - Basic Dynamical Measures - Energy}\}.

We characterize ''mathematics as a science of the properties and relations between sets. If we are interested in sets 'mathematically' we thereby assume... an abstraction from the empirical determination of presented sets... Empirically determined sets differ in their kind. By \emph{abstraction from their difference in kind} we treat them as specimens for sets of $n$ elements. As such representatives they are... \underline{mathematical objects} in the proper sense, i.e. those objects about which the mathematical cognition makes assertions. Thus we can define mathematics as non-empirical science of the properties and relations between sets'' \cite{Peter '79 - Philosophie und Mathematik}.

Counting as a real work presents a most elementary evolutionary stage of mathematics. Helmholtz \cite{Helmholtz - Zaehlen und Messen} pursues the same productive character in the conduct of basic measurements. The \emph{objects} are physical (not kitchen utensils or medical instruments), that means their \emph{properties} refer to physical behavior (e.g. impact in a collision). Physicists connect their reference devices by physical operations (not farming activities or medical conduct because they are dealing with capability to work and with numbers). Those \emph{physical interventions} are a particular kind of human activity. It is the way how physicists construct experimental apparatuses to make lengths, durations, energies, momenta measurable (to quantify them). We pursue the active role of physicists, their interventions in a basic measurement.

We give the general form of a measurement verdict as ''$m = z \cdot m_{\mathbf{1}} $'' where the attribute (of a phenomenon, body or substance) has a magnitude that can be expressed as a number and a reference \cite{international vocabulary of metrology}. This form should be understood as a historic product, whose genesis has to be explained. Ruben \cite{Peter '79 - Philosophie und Mathematik} clarifies the historic resp. evolution-theoretic problem, ''namely the question of occurrence of measurement expression '$z \cdot m_{\mathbf{1}}$' under the condition, that neither units (reference standards) nor numbers (4-vector calculus etc.) are present ... We comprehend the problematic expression as a verbal representation of the result of a measurement. '$m = z \cdot m_{\mathbf{1}} $' is a completed development product (Evolutions\emph{produkt}). What we need is the evolution prerequisite (Evolutions\emph{voraussetzung}), i.e. the conceivably simplest expression'' for an observable. Obviously we do not intend, to explain the product of a development (quantified observable) from a genetic prerequisite which does not involve observables at all. ''We are not under the suspicion of explaining a certain thing evolutionary under the assumption of its non-existence, but stick to the good old materialist insight: \emph{Nothing comes from nothing}.'' Our problem of the \underline{genesis of physical quantities} is not that interactions exist, but to demonstrate how from pre-theoretic definitions of energy and momentum the quantified observable (and four-vector formulation) arises. We develop the transition from an undetermined interaction to the physical determination of an action.

Once we get to grips with our evolutionary problem \{\ref{Kap - KM Dynamics short Review - Evolutionary product}\} and establish its basis \{\ref{Kap - KM Dynamics short Review - Work concept}\} we can specify Helmholtz physical method more precisely \{\ref{Kap - KM Dynamics short Review - Physical method}\}. With elemental comparison ''$>$'', reference standards ''$\mathcal{S}$'', concatenation ''$\ast$'' we will generate the physical quantities (of energy-momentum) and finally formulate these practical operations mathematically.

\section{Evolutionary product}\label{Kap - KM Dynamics short Review - Evolutionary product}

Following Lorenzen \cite{Lorenzen - Entstehung der exakten Wissenschaften} we can understand our evolution-theoretic problem in a way that ''does not narrow down on stringing together 'facts' ''.
According to the guiding principle of the historic school ''also conceptual schemata of past epoches are to be worked out of these themselves''.
Following Collingwood he demands ''to precisely make sure, how historic problems develop. Ultimately every historic problem arises from real life''.
To think and practice history means to recognize ''how historic problems \emph{originate from practical problems}.''

Thales of Milet -585 marks the beginning of exact science \cite{Lorenzen - Entstehung der exakten Wissenschaften}.\footnote{Thales was able to predict the first solar eclipse, compute the height of pyramides by its 45$^{\circ}$ shadow (geometric problem - a pyramid is not a tree!), prove theorem of Thales by symmetry and seeing rectangles and prove angular sum in a triangle via seeing zigzag patterns and parallels in common decoration of vases. Lorenzen \cite{Lorenzen - Entstehung der exakten Wissenschaften} specifies Thales immortal historic achievement as conveying logical connections, pursuing them and discovering them.} Accumulated bodies of knowledge were generated before. In reception of geometric knowledge from Babylonia and Egypt Thales discovered contradicting assertions - an unprecedented problem! Seeking for certainty Thales developed geometry as a ''theory'', i.e. as system of \emph{logically connected} assertions. Ancient Greeks demand provability. Compared to old Babylonian mathematics that is something radically new. ''Only the Greeks invented logical chains of assertions, by means of which assertions about something 'seen' became undeniable... (A \emph{proof}) is the voluntary and at the same time necessary acknowledgement of an impersonal authority, the 'logos' ''. With Euclid's -300 axiomatization mathematics reached a grand completion.

Why did it take another 2000 years until mechanics? In our view it was not just the time required for analyzing astronomical observations \cite{Barbour - nature of time}. We must distinguish kinematical description and physical explanation in the proper sense. Need for the latter first had to develop historically. From which unprecedented practical problem did it originate?

''The exact natural sciences are a present from the sky'' Lorenzen \cite{Lorenzen - Entstehung der exakten Wissenschaften} evaluates the importance of astronomy.
''From the phenomenon of regular movement of celestial bodies humans conceived the idea of exact regularity in nature. Despite the precise astronomical observations, which the Babylonians made throughout the centuries (and despite their arithmetic rules for projections), one can not accredit to them the thought of a ''natural law'' as we understand it today.'' That Ionian tradition in natural philosophy is the first step to the \emph{de-deification} (German: Entg\"otterung) of nature.

Then the Pythagorean view developed. It treats astronomy as an exact science, i.e. as science with a mathematical theory. They \emph{introduce mathematical methods} but in their intention to ''study reality only to find 'behind' the phenomena mathematical laws as the divine harmony'',
in restoring God they represent a step backwards compared to the Ionian view. Lorenzen judges ''that rationalism of the Pythagoreans - so estranged to modern empiricist thinking - had influenced subsequent history of astronomy. Compared to the (obviously insufficient) astrophysics of the Ionians the subsequent Pythagorean school narrows down to a purely kinematic theory of celestial motion.'' In contrast e.g. ''the astronomy of Anaxogoras is clearly of Ionian spirit, in attempting to provide a physical explanation of phenomena ('explaining' tilted ecliptic... by compression of air masses) as opposed to giving a Pythagorean purely kinematic description of orbits.''
Even the possibility for conversion of the Ptolemaic system (of eccentric epicycles) to the heliocentric system was not implemented ''As long as planetary theory remained purely kinematic there was no particular necessity. The Ptolemaic theory never asserted the \emph{claim}, to be regarded as a \emph{physical explanation}.''

What we understand by physics today, Lorenzen defines, ''is designing of mathematical theories for the interpretation and prognosis of natural or technically effected processes... was not present in ancient times. Ancient science misses precisely what is crucial for modern physics: namely what, in above definition is expressed by the words 'or technically effected processes' (equally natural and steered). Greek enlightenment ('de-deification' of the world) did not unlike contemporary enlightenment let the \emph{ideal} - of technical mastery of nature - \emph{come into effect}.'' Thus we have tracked down the, in our view, crucial access to Helmholtz and Hertz mechanical explanation. In the historic development of work we find an unprecedented problem which lead Leibniz \cite{Hecht - Ruben-Festschrift} and Helmholtz \cite{Helmholtz - Ueber die Erhaltung der Kraft} to the discovery of energy.

\section{Work concept}\label{Kap - KM Dynamics short Review - Work concept}

With Ruben \cite{Peter '79 - Philosophie und Mathematik} we presuppose that humans can think about their work \emph{experience}. It is most profitable! So we can understand, that the historic development of new forms of work and organization entail conceptual revolutions in thinking (about work actions).\footnote{For understanding the first principles (Anfangsgr\"unde) of a science one can look at human history namely at the \emph{development of their work practice} \cite{Peter '79 - Philosophie und Mathematik}: ''Human history is distinguished by generating means of production. The making of history is essentially the making of production means and their inheritance to the following generation. On their production means (tools, machines, buildings, means of transportation etc.) humans win the possibility to override their immediate bonding to given natural environment, hence to become largely independent of the particularities of different geographic milieus. By their production means they create out of a given natural a new type of environment''. In the ''humanization of their surrounding nature, in the subjugation of Earth humans manifest at the same time \emph{their nature as human genus}.''

According to Ruben's work concept ''it is the \emph{development of that nature of human genus}, which provides the primary object of philosophical cognition. ... The existence of that (human) genus becomes concrete as a process of historic development. ... It is the great insight of Hegel that the human genus is not comprehended by one particular historic state, but in the necessity of a genetic succession of those states.''}

We do not need to explain the existence of interactions. Everyone who throws a stone, swings a hammer or has an apple falling onto his head did experience cause and effect. In daily work (architecture, building ships, streets, aqueducts, military technology) one becomes aware of the physical behavior of work objects. Early manufacturers did practice elemental work actions, adjustment and comparison procedures and concatenation methods. Passing on the skills to the next generation involves the demonstration and a colloquial description of e.g. the ''impact from a kick'' and forming its comparative ''more impact than''.

Helmholtz, Hertz introductory examples \{\ref{Kap - KM Dynamics short Review - Foreword}\} demonstrate, during the industrial revolution the thinking of humans was influenced by motives of machine construction and work economy \cite{Wolff - Geschichte der Impetustheorie}. In examination of, partly only recently, published manuscripts from Leibniz time in Paris 1672-1676 Hecht \cite{Hecht - Ruben-Festschrift} reconstructs the methodological principles, by which Leibniz arrived at the observable (kinetic) energy: ''Leibniz knew, that mechanics of his time underwent fundamental change, which placed the elaboration of physical dynamics on the agenda, where at the same time its relation to traditional statics needed to be clarified. This is expressed by mentioning of five ancient machines, whose theory leads to statics.'' Leibniz was searching on machine models (in ever new thought experiments) for a connection between statics and dynamics. ''Statics as a theory of simple machines... is paired with dynamics as a theory of machines, in which accelerations occur. This corresponds with the distinction between the dead and living forces.''

In the industrial revolution 1788-1842 the steam engine and the work machine are coupled together for the first time \cite{Peter '08 - Vom Kondratieff-Zyklus und seinem Erklaerungspotenzial}. Suddenly one can put power sources anywhere, even onto rails (locomotive). This implies a liquidation of the dependence from water mills (or earlier power sources like own muscle strength or slaves which could be \emph{instructed} or exploited until they ''simply'' run dry). That marks the transition from (the personal interplay of) worker and work machine to (the coupling of) engine drive and work machine. While traditional craftsmen learned to handle their tools (hammer, saw, file, needle etc.) intuitively, industrial revolution substitutes the former by an impersonal motor. Steering the latter became an unprecedented problem. Their multiple propulsive potential (coal, oil) allows machines of huge extent. Historically for the first time one builds complex production facilities (weaving factories, steel plants, chemical industry etc.). At the same time they represent enormous investments. In case of uncontrolled fuel supply potential damages are not anymore easy to replace or repair (like a bent nail from an unskilled hammer stroke or a piece of scrap in a carpentry). In order to safe the investments the challenge is to construct and steer the machines \emph{cautiously}, not to overload them. With the industrially revolutionized tool use developed the \emph{practical need} for steering and experimenting with propulsion. At the same time particularly the Germans have established research laboratories in the industry. This means industry pays for (practically relevant) research to a completely new extent.\footnote{In England by contrast only universities had paid and cultivated research. The industry bought it there; though was not as innovative as the German industry (eventually giving rise to ''Made in Germany'').} There the (historically developed need for) reproducible work actions stimulated the standardization of the conduct of (initially work related) physical experiments.\footnote{In view of Ruben's work conception Schlaudt \cite{Schlaudt - Ruben-Festschrift} remarks ''a scientist is not only (passive) observer, but also an experimental and instrumental agent. ... By registering reactions of his object in dependence on the \emph{conditions} of its treatment, a scientist accomplishes the transition from ordinary (work) operations to scientific experimentation.'' That includes \emph{standardization} of reproducible measurement conditions.}

One can \emph{become aware} of the physical processes in work experience and described the pre-theoretic (work) actions in colloquial language. We will demonstrate the \emph{standardization} of the conduct of physical experiments. We assume presuppositions on the (i) thinking behavior (clarity in verbal expressions) and (ii) practical operations (artisan behavior). Schlaudt \cite{Schlaudt} begins with ''intensionally gifted, causal agents (Akteure) in their world''. He illustrates on various examples (weighing scale, straight ruler, uniform running clock etc.) the explication of work norms as measurement norms.

\section{Physical method}\label{Kap - KM Dynamics short Review - Physical method}

With Helmholtz \cite{Helmholtz H. v. - Einleitung zur Vorlesung ueber Theoretische Physik}
we ''depart fairly from common habits'' if in a ''systematic presentation of physics... we first state the general logical and epistemic principles of scientific methodology of empirical sciences. ... Since philosophers... begin their studies mostly only upon knowledge, which can already be phrased in words, and mostly know the underlying processes of \emph{gathering actual experience} not at all or only from hearsay... (Helmholtz) had to develop (the conception of basic measurement practice) all by himself.''

We make serious with presupposing tangible operations and the colloquial description for the foundation of physics. We begin with elemental notions \{\ref{Kap - KM Dynamics - Physical Measurement - Pre-theoretical Ordering Relation}\} which are immediately understandable in everyday language. They are not made up out of thin air; but originate from the historic development of work practice. One describes the demonstrable circumstances by \emph{elementary descriptive sentences} (''This ball is moving.'', ''This motion is quick.'' etc.). The concept of a \emph{property} originates from a sentence analysis.\footnote{''In the sense of our traditional grammar every sentence, as (smallest) unit of meaning in colloquial language, appears in familiar subject-predicate-structure, which we note symbolically short as S/P. The subject S of a sentence S/P denotes, as one says, the \emph{object} of the statement of a sentence; the predicate P provides the language representation of the \emph{statement} about the object'' \cite{Peter '79 - Philosophie und Mathematik}.
} In sentences of the form ''S is P'' the subject S denotes a thing; the predicate ''is P'' a property, whose denomination can be nominalized ''the P-ness''. Quickness (velocity), heaviness (weight), hotness (temperature), massiveness (impulse) etc. are \emph{physical objects} which we specify more precisely. Carnap \cite{Carnap Physikalische Begriffsbildung}
explains ''the most important physical properties... are nothing but manners of reactions of bodies'' to certain external conditions. His maxim is ''All physical statements are conditional statements... they allege: as often as such and such conditions are met, wherever and whenever that might be, such and such will occur''.

With Ruben \cite{Peter '13 - Umgangssprachliche Voraussetzung zur fuer die physikalische Groessenbildung} we notice the gradability of adjectives (long, quick, hot) as a colloquial prerequisite for forming physical quantities: ''just as large as'' and ''larger than'' are the two fundamental examples for all words, which can be used for the denomination of physical measures. ''Just as large as'' would then plausibly introduce the \emph{equality of any kind} and ''larger than'' show the corresponding \emph{ordering relation}. In practice we assess equality, if two objects are neither larger nor smaller than one another (in sufficient precision).

With the denomination ''larger than'' we immediately associate a property or quality as the \emph{type of the ordering relation}. Since all orderings are species-specific. One knows many practical comparison methods ''longer than'', ''heavier than'', ''more impact than'' etc. Helmholtz \cite{Helmholtz H. v. - Einleitung zur Vorlesung ueber Theoretische Physik}
defines ''In all cases the measures, which are declared as equal, are only equal with one another
in a certain relationship.\footnote{''One can never set two bodies as equal in all regards. ... Two weights e.g. can be of the same mass, though they can have very different temperature and very different color.''} To assess equality requires the knowledge of the method by which the comparison is carried out.'' We regard two bodies as equal, if they can substitute one another in a particular relationship (like two weights on a weighing scale). The procedure for determining substitutability must be admissible. Assessing the equality must be independent from the order of the objects (commutativity), ruling out e.g. non equal-arm weighing scales. When two measures equal a third, then they must also be equal among themselves (transitivity). Helmholtz measurement-methodical norms guarantee reproducibility! ''Only those examination methods can be used for definition of equality, which satisfy the arithmetic axiom. Whether it is fulfilled is decided by the experiment.''

Helmholtz \cite{Helmholtz H. v. - Einleitung zur Vorlesung ueber Theoretische Physik}
regards two bodies as of the same kind ''which may be compared among one another by the same method. But bodies of the same kind can still be unequal in their magnitude.'' With the ordering relation ''larger than'' we induce the question about the measurement unit: ''how much'' larger? And only once a reference device is defined, one can count how many copies of that standard have been expended in measuring; only then therefore comes the counting! This is the beginning of physical (technical) language without presupposing one word of mathematics. Mathematics comes into being at the moment we introduce units.

With reference devices (ruler $\mathcal{R}$, light-clock $\mathcal{L}$, standard process of energy source $\mathcal{S}$ against impulse carriers $\circMunit_{\:\mathbf{v}}$ etc.) we enter the domain of \emph{physical operations} on objects, which represent the unit measures. From physics one knows them only as mathematical operation symbols. We want to reveal the underlying vivid operations (Handlungen) \{\ref{Kap - KM Dynamics short Review - Measurement operations}\}. For the \emph{direct} measurement Helmholtz seeks for
\begin{quote}
''a method, to conduct the comparison, so that one treats the one body as equal to an aggregate of two or more bodies. For this purpose... two bodies must be connected or, as Grassmann says, concatenated with regard to one property or one effect. By concatenation Grassmann understands any kind of... natural connection, as it may occur in \emph{all sorts of coaction of different bodies}.''
\end{quote}
We will concatenate energy sources ''$\ast_{E}$'', impulse carriers ''$\ast_{\mathbf{p}}$'' and masses ''$\ast_{m}$'' in a calorimeter model \{\ref{Kap - KM Dynamics - Basic Dynamical Measures Quantification - Concatenation}\}. For reproducibility Helmholtz requires ''that the result of the concatenation is independent... from the combination in groups and from the order in which we put together individual quantities'' (distributivity). By recourse to the impossibility of a perpetuum mobile, relativity principle and others our measurement operations will fulfill the axioms of addition.\footnote{Carnap \cite{Carnap Physikalische Begriffsbildung}
in contrast takes the view that ''physical measurement means assignment of numbers to any physical objects (things, properties, phases of a process etc.) of a particular domain.'' His principle is: ''Every measurement essentially goes back to counting.'' Such representation theoretical conception views every measure as numerical assignment. The form of scale of e.g. a ruler (conformal factor) seems arbitrary. ''Any two line segments which are 'measured' as equal according to one scale, will also appear as equal in any new scale; is one path longer than another according to one scale, so also in any other scale.'' The ordering relation (i.e. with regard to length ''$>_l$'') is \emph{scale-invariant}. But Carnap admits: they can not specify ''\emph{which} length a path has and by how many times longer the length of one path is than the other.''

In the case of ''known, very old procedure of length measurements by repeated placement of unit sticks one after the other'' Carnap realizes that ''the common (linear) scale form of length measures follows by fixing that: two line elements must be considered as equal, when they have same length'' (congruence). Here such linear scale form can be given because ''measures of type length are of such a nature that differences of paths can be considered and measured again as a path... This does not hold for other types of measures, e.g. a temperature difference can not be regard as temperature itself.'' Principles of (reproducible) measurement practice will imply that also energy-momentum, mass are additive; their quantities have precise meaning!} We begin from a colloquially represented theory of work experience and demonstrate the passage to a technically speaking and mathematical terminology using representation of mechanics.

\subsection{Measurement operations}\label{Kap - KM Dynamics short Review - Measurement operations}

We are searching for a strictly physical foundation of physics where initially mathematics must remain outside - and then every step where mathematics is introduced requires extra justification (by abstraction). Our problem is the measurement and thus the construction of physical measures. That is a question of \emph{quantification and qualification} since a measure is the unity of quantity and quality (which physicists also name dimension). The English term ''quantification'' constantly misleads to speak about numbers\footnote{Like Carnap \cite{Carnap Physikalische Begriffsbildung} - Luce, Suppes so called ''Theory of Measurement'' \cite{Suppes - Theory of Measurement}; a \emph{representation theoretical conception} grasps measurements as ''the assignment of numbers to objects and phenomena'' and focusses on the (formal) analysis of ''their invariance under appropriate transformations''. We instead look into what actually happens in measurement practice itself.} where in truth the talk is about measures.\footnote{The international vocabulary of metrology \cite{international vocabulary of metrology} calls the result of a measurement ''quantity'' and defines: \textit{A ''quantity'' is a property of a phenomenon, body, or substance, where the property has a magnitude that can be expressed as a number and a reference.} commonly symbolized: The ''quantity'' equals $Q = \left\{ Q \right\} \cdot \left[ Q \right] + \Delta Q$ with unit measure or dimension $\left[ Q \right]$, numerical value $\left\{ Q \right\} := \sharp \left[ Q \right]$ determined in the measurement method and sufficiently small measurement uncertainty $\Delta Q$. The methodical question with regard to basic dynamical observables focusses on the origin of both the ''reference'' and the operation for determining that ''number''.} Our problem is called physical measurement (as unity of quantification and qualification) or ''On the construction of physical measures''.

Therein we encounter the following set of problems: What are the tangible operations which we denote in the theory simply by mathematical operation symbols. Is the ''$+$'' in $1\mathrm{kg} + 1\mathrm{kg}$ the same symbol as in $1\mathrm{sec} + 1\mathrm{sec}$? Asking this question means to immediately recognize: No! The concatenation of two weights is not the same as the connection of two durations etc., the concatenation of two lengths requires Euclidian geometry as a physical theory, because the very operation requires to place rulers along a straight line. Such determination of a measurement operation makes of course no sense with regard to seconds. The next question is: What do physicists actually mean when they formulate $\mathbf{F}=\mathbf{p}/t$? One can not divide an impulse by a duration. There must be clarity what kind of operation the formation of that proportion is. Physicists combine quantities by ''multiplication'' $\mathbf{p}=m\cdot \frac{\mathrm{d}\mathbf{s}}{\mathrm{d}\tau}$ which however is only explained if respective variables are meant to be numbers. How can we physically understand what a physicist does when he combines a mass $m$ with a four-velocity $\frac{\mathrm{d}\mathbf{s}}{\mathrm{d}\tau}$? We can raise these questions for all mathematical operation symbols in physical equations; and only therewith we actually pose the foundation of physics from solely physical grounds as a general problem \cite{Peter '13 - Programm der physikalischen Selbstbegruendung}. This work concerns \emph{physical objects} and the problem of determining \emph{physical operations} really in a strictly physical way.

Helmholtz \cite{Helmholtz - Zaehlen und Messen} distinguishes the act of counting and measuring. Mathematical addition and physical ''addition'' are different operations. In a direct measurement physicists conduct tangible operations; not formal mathematical. To assess the ''equality'' and ''multiple'' of two observables one conducts a pair comparison: e.g. one measures the length of Otto (measurement object) by placing copies of a ruler (reference unit ''$\mathcal{R}$'') side by side in a straight line (concatenation ''$\ast$'') until the layout covers Otto (comparison ''$>_l$'') sufficiently precise. In this way Hartmann sen. \cite{Vati - Logik und Arbeit} grasps the logical core of measuring as a continued comparison of a measurement object and a constructed material model.

One can become aware of physically interesting attributes on demonstrable objects in work experience. Galilei and Leibniz define the basic observables independently from \emph{comparison methods} ''longer than'', ''more impact than'' (in a collision), ''more capability to work than'' (against same test system). Ruben \cite{Peter '69 - Dissertation} describes the program for direct quantification. In the family of carriers (of the same kind of measure) one introduces an \emph{order} and addition resp. \emph{concatenation operation} as instructions on how to proceed in practice and represents the \emph{reference unit} as a sufficiently constant and arbitrarily reproducible material object \cite{Schlaudt}. One operates with material objects according to the practical norms of the measurement instructions. One counts how many reference units the \emph{material model} contains, to reproduce the observable of the measurement object; e.g. a stack of rulers that covers Otto. The measurement object and the material model can \emph{substitute} one another with regard to the interesting property (length, impact, capability to work etc.).

\begin{de}
A \underline{basic physical observable} is an attribute of an object, a property which in a practical comparison allows the difference of larger, equal or smaller \cite{Helmholtz - Zaehlen und Messen}. The quality is determined by the kind of comparison method and the quantity by the extent.
\end{de}
\begin{de}
One wants to express the value also numerically (''how many times'' more). The procedure for finding these values is the \underline{measurement}.
\end{de}
\begin{rem}
In a basic measurement one \underline{quantifies} the direct comparison in a standardized (reproducible) way: By concatenating sufficiently constant reference devices one constructs a material model which can substitute the measurement object in the respective comparison. By counting the congruent building blocks one finds ''how many times'' larger the observable of the object is than the reference unit.
\end{rem}

Our goal is to develop the tangible steps for dynamics. A basic measurement requires knowledge of the method of comparison (of a particular attribute of both bodies) and of the method of their physical concatenation \cite{Helmholtz - Zaehlen und Messen}. We search for pre-theoretic comparisons (more capability to work $>_E$, more impact $>_{\mathbf{p}}$), reference standards (energy sources $S_E$, impulse carriers $S_{\mathbf{p}}$) and concatenation operations ($\ast_E$, $\ast_{\mathbf{p}}$) and underlying measurement-methodical principles. We want to grasp all operations which bring together measures (in the theory) as tangible operations on the carriers. While one measures length, duration in rather vivid (anschaulich) way, Einstein's similar measurements of relativistic motion require a considerably refined procedure \cite{Einstein-Grundlagen der ART}. Until now energy and momentum have not been grasped as basic observables, neither in classical nor relativistic or quantum mechanics.

For the operationalization of Euclidean geometry and Galilei kinematics (without presupposing equations of motion etc. \cite{Brown - Physical Relativity}) we can refer to the Erlangen school protophysics \cite{Janich Das Mass der Dinge} as known and given. In chapter \{\ref{Kap - Kinematics}\} we develop Helmholtz program of basic measurements for relativistic motion. By tangible operations with light clocks and GPS-procedures we measure length and duration and define the mathematical structure of special relativity (4-vector $v^{\mu}$, Lorentz transformation $\Lambda_{\mu\nu}$, Minkowski metrik $g$ etc.). Luce, Suppes diagnose: A major hindrance to understanding of fundamental measurements in dynamics has been the failure to uncover suitable empirical concatenation operations ''$\ast$'' for attributes ''capability to work'' and ''impact''. By means of which measurement devices and procedures can they be measured directly?\footnote{Instead Luce, Suppes et al define ''properties kinetic energy and momentum having two independent components: mass and velocity... In conjoint-measurement theory, the way in which each component affects
the kinetic energy resp. momentum is studied by discovering which changes must be made in one component to compensate changes
in the other'' \cite{Suppes - Theory of Measurement}. They can \emph{verify} quantity equations $E_{\mathrm{kin}} = \frac{m}{2}\cdot \mathbf{v}^{2}\:$, $\mathbf{p}=m\cdot \mathbf{v}$ in individual cases. But they \emph{cannot judge scope and limitation} without taking underlying measurement-methodical principles into consideration.

Luce, Suppes conception of ''Theory of Measurement'' \emph{primarily focusses on numerical representations} of pre-theoretic ordering relations (i.e. of empirical relational system onto numbers) \emph{and uniqueness theorems}. They presuppose abstract mathematics (representation spaces etc.) as given and characterize ''measurement statements as empirically meaningful only if its truth value is invariant under the appropriate transformations of the numerical quantities involved'' \cite{Luce Suppes - Theory of Measurement}. Whether associated numbers have meaning, remains uncertain; one solely asks if such assigning is unique. Their representation theoretical conception is \emph{complementary} to our \emph{physical perspective}: (i) we can perceive meaningful comparison methods in practice (ii) Physics turns out as mother of its mathematics in empirical practice - without presupposing it.} In chapter \{\ref{Kap - Mechanics}\} we demonstrate the operationalization of energy-momentum for classical mechanics and in chapter \{\ref{Kap - Relativistic energy-momentum}\} we revise the same method for relativistic dynamics.


\chapter{Kinematics}\label{Kap - Kinematics}

We demonstrate the definition of basic observables from physical operations, the key to overcome hidden stumbling blocks and apparent paradoxes from unscrutinized (classical) formalisms. We develop Helmholtz program of basic measurements for relativistic motion. We define the basic observables by direct comparison: ''longer than'' if one object or process covers the other. To express the spatiotemporal order also numerically (how many times longer) we cover them by a locally regular grid of light clocks. These are the basic physical operations. From their interrelation we derive mathematical relations, e.g. for different observers the formal Lorentz transformation; for accelerating observers we reveal a measurement-methodical view on the apparent Twin paradox.
\\

One usually explains kinematics axiomatically. So one can trace back the whole mathematical formalism to a manageable system of initial propositions which are logically independent from one another. Though this formal bookkeeping of physics already begins in the abstract. The axiomatic formulation assumes scalars, four-vectors, metric etc. as known objects of its description. Without further implicit assumptions it lacks interpretation and physical meaning. Origin, scope and limitations of the variables and algebra remain unclear. The lack of alternative approaches seems to justify the formal path for developing novel theories. For the foundation of elementary kinematics (next also for dynamics \{\ref{Kap - Mechanics}\}) we develop a complementary program; we begin from the primary measurement operations \{\ref{Kap - SRT Massbestimmung}\}.

Our objective is a foundation of relativistic kinematics from the operationalization of its basic observables. The goal is not to change or improve the mathematical structure, but to gain a deeper physical understanding of kinematics. Like Einstein \cite{Einstein-Grundlagen der ART} for the concept of simultaneity we reveal the underlying physical operations. We begin from undisputed natural and measurement principles. We stress the active role of physicists, the interventions to define basic observables, quantification and then derive the four-vector formulation second. With Helmholtz' method \{\ref{Kap - KM Dynamics short Review - Measurement operations}\} we will show, how physics generates its own mathematics in empirical practice.

Theodor H\"ansch - inventor of the optical frequency comb generator which facilitates the construction of most precise clocks - defines: ''Time is what one measures with a clock''. In his case a light clock. The origin of basic \emph{reference devices} and measurement \emph{procedures} is not in the domain of non-empirical mathematics. We require them (like Einstein's clock postulate and laser ranging technique) \emph{before} having a theory of matter and as a basis for developing the latter \cite{Hartmann-KM_Dyn}. We develop all protophysical prerequisites from everyday work experience. We make a digression into watchmaking and understand by what actions one provides these reference devices if one did not have them before \{\ref{Kap - SRT Massbestimmung - classical metric - Watchmaking}\}. With classical rulers and clocks one determines the universal motion of light \{\ref{Kap - SRT Massbestimmung - light principle}\}, it propagates locally uniform. For basic measurements we introduce (light) clocks as an unstructured unit \{\ref{Kap - SRT Massbestimmung - light clock}\}.

Let every observer place them side by side and one after another until the measurement object is covered; for the technical description we introduce \emph{measurement termini} (simultaneity lines, projections etc.). In the mathematical formulation of the procedures all corresponding \emph{terms} have physical meaning. From the underlying operations we derive the Lorentz symmetry \{\ref{Kap - Masszsh}\}. Every formal calculation, e.g. in the configuration of the apparent Twin paradox \{\ref{Kap - Twinparadox}\}, assumes connected basic measurement operations. For an accelerating observer they become impracticable. From vivid pre-theoretic principles we develop all mathematical variables and operations and finally the relativistic equations.


\section{Intrinsic measurement practice}\label{Kap - SRT Massbestimmung}

For the origin of colloquial notions - \emph{motion}, \emph{space} and \emph{rigid body} - from common sensual experience we refer to Poincare \cite{Poincare - Wissenschaft und Hypothese} and Mach \cite{Mach - Raum und Geometrie}. According to Poincare \emph{geometric properties} essentially characterize the relative \emph{motion} between neighboring \emph{objects}. Leibniz characterizes ''space'' and ''time'' as relations between the observable things. \emph{Space} brings order into things which happen simultaneously. \emph{Time} brings order into things which happen sequentially. From everyday practice one knows the direct comparison
\begin{itemize}
\item   $>_{l}$ \;\;\; if two extended objects lie on top of each other - one will \emph{cover} the other
\item   $>_{t}$ \;\;\; if two processes begin simultaneously - one will \emph{outlast} the other.
\end{itemize}
The \emph{ordering relation} is reproducible in an observer independent way. Next one wants to find ''how many times'' longer.

For reproducible measurements one provides sufficiently constant reference devices and standardized procedures. The \emph{construction} and the \emph{conventions} historically developed from daily work experience; we sketch the transition to physical experimentation. One works with natural objects in a natural environment. Their behavior depends on external conditions (some known and others undiscovered). One wants to control the interrelation of work conditions; it pays off. We regard the origin of basic measurements as a standardization in the conduct of reproducible experiments (to specify known work conditions \cite{Ruben - Arbeitskonzept}). With empirical knowledge on feasibility and outcomes of work \emph{actions} one can probe the objects and rehearse expedient ways of handling. We define all basic observables (length, duration; in \{\ref{Kap - Mechanics}\} also impulse, inertial mass, capability to work/energy) and the associated comparison and concatenation operations in the practical domain. The development has a social dimension: a master inherits the demonstrable practice to a student, first simply by pointing a finger and then defines a colloquial and technical language. We presuppose usage of common denominations (''$\mathcal{A}$lice, $\mathcal{B}$ob and $\mathcal{O}$tto move relative to one another.'', ''They signal with light.'' etc.) with their common meaning as a known part of work experience.\footnote{We explain the meaning of colloquial expressions by exemplary demonstration (of sufficiently constant phenomena). We can neither demonstrate pure matter in isolation from its behavior nor pure behavior detached from matter. In colloquial speech we express a demonstrable fact by a simple descriptive sentence like ''Otto is long''. These represent the smallest unit of meaning. The subject Otto $\mathcal{O}$ and its attribute length $l$ are distinguishable but inseparably unified \cite{Peter '76 - Praedikationstheorie} \cite{Vati - Logik und Arbeit}. The subject terminus ''long Otto'' emphasizes the subject $\mathcal{O}$ which embodies the property long $l$; we symbolize the \emph{long object} by $\mathcal{O}_l$. The predicate terminus ''Otto's length'' emphasizes the property which Otto represents; we symbolize the \emph{object's length} $l_{\mathcal{O}}$. From elemental operations on tangible things $\mathcal{O}_l$ we develop basic measurements for the attribute $l_\mathcal{O}$.} With Lorenzen, Janich \cite{Janich Das Mass der Dinge} we can presuppose (circularity free and without mathematical presuppositions) that every observer can manufacture ''straight'' ''rigid'' rulers and ''uniform running'' clocks \{\ref{Kap - SRT Massbestimmung - classical metric - Watchmaking}\}. In a direct measurement one \emph{concatenates} ''$\ast_{s}$'' the rulers ''$\mathcal{R}$'' side by side in a straight way until the layout, symbolized $ \mathcal{R} \ast_{s} \ldots \ast_{s} \mathcal{R} \sim_{s} \mathcal{O}$, covers the object. The ordering relation ''longer than'' becomes measurable $s_{\overline{O}} = \sharp\left\{\mathcal{R}\right\} \cdot s_{\mathcal{R}}$ by the number of connected rulers and their standard length $s_{\mathcal{R}}$; similarly for durations.

In starting figure \ref{figure-1}a we illustrate objects and observers in motion. Consider a (hidden) railway track along which $\mathcal{A}$lice, $\mathcal{B}$ob and $\mathcal{O}$tto specify their relative motion and the light. After including the historical depth \{\ref{Kap - KM Dynamics short Review - Evolutionary product}\} of work experience \{\ref{Kap - KM Dynamics short Review - Work concept}\} they are equipped with the local Euclidean metric. We will demonstrate the transition to relativistic kinematics. Each observer measures $\mathcal{O}$tto's relative motion with classically constructed light clocks. They place them one after the other or side by side; connected by coinciding rays of light \{\ref{Kap - SRT Massbestimmung - direct connecting products}\}. A regular grid of light clocks covers their relative distances. Each building block is congruent with the next; by counting them they measure the magnitude of the length. They measure their relative motion with (the motion of light in) their reference device.


\subsection{Watchmaking}\label{Kap - SRT Massbestimmung - classical metric - Watchmaking}

The protophysical foundation of Euclidean geometry \cite{Janich Das Mass der Dinge} explains the standardization of length comparisons circularity free in the categories of purpose and expedient means of everyday work. One works on raw materials and reshapes them for practical needs. One builds rulers and clocks as sufficiently constant representatives of ''length'' and ''duration''. The success of (tentative) manufacturing methods is secured by \emph{test procedures} for the \emph{straight form} of a constructed ruler and the \emph{uniform running} of a clock \cite{Janich Das Mass der Dinge}.\footnote{Before $\mathcal{B}$ob can specify the form of $\mathcal{O}$tto's relative motion he has to find out if his own clock provides a uniform ticking reference. Before $\mathcal{A}$lice can determine whether $\mathcal{O}$tto's nose is crooked she needs to know if her ruler is straight. The test rules for admissible reference devices do \emph{not} presuppose an already existent \emph{prototype} for a straight line and an ur-clock which one can simply copy or transport.} For this reason \emph{measurement instruments} are to be understood - not simply as arbitrary designations of natural objects but - as \emph{artifacts}; our manufacturing actions must realize test norms \cite{Schlaudt}.

The testing method for geometrical shapes originates from grinding practice: A body has a ''flat'' surface if one can produce two moldings of that body and then fittingly (!) shift the two imprint surfaces against one another. If there is a gap continue grinding them against one another; make two new moldings and check again. Similarly if one has manufactured a body with two flat surfaces which intersect one another, then the intersecting edge represents a ''straight'' line. The test norms originate from intuitively controlled actions in everyday (technical) work, which is governed by the rationality of purpose and expedient means. Ultimately one explicates expedient work norms as measurement norms. Practicable rules of pre-scientific technical behavior develop into norms for measurement operations \cite{Schlaudt}.

A watchmaker evaluates by test procedures whether a tentative device runs uniform. In the empirical interplay of analyzing manufacturing conditions and examining the respective products the \emph{manufacturing method} is continually refined until the device \emph{realizes} the aspired \emph{ideal} of uniform motion sufficiently precise. In this process we make the practical experience that the ideal is never completely realizable. The closer one wants to approach the more effort and workload is required in the production and also for the conservation of the product (shielding fragile clock). He takes guidance by a test procedure: Take two structurally identical copies of the clock and align them such that their clock hands are running straight into fixed (e.g. perpendicular) directions. Then one can couple the motion of the two clock hands e.g. by a mechanical transmission; draw down the stretch of way of their superposed motion and \emph{check} its geometric form. The clocks run uniform if - independently from when each was started and coupled together - their superposed stretch of way always has the form of a straight line. As before the path in question represents the ideal of a straight line if any two segments can be fittingly (!) shifted against one another.\footnote{The protophysical norm for \emph{uniform} motion originates from a test of the \emph{straight} shape of rigid objects.}

A clock is manufactured and tested as a representative for a uniform motion. By metricizing the length of the traversed stretch of way (of the moving clock hand) one obtains a metrical measurement instrument for ''durations''. In practice (accumulated friction etc.) clocks will run (approximately) uniform for only finite durations. Such ''finite duration'' measurement standards can be aligned synchronously one after the other to cover longer processes. By this \emph{substitution} we measure the magnitude of ''durations''. We arrive at classical Galilei kinematics for space and time.
Despite the uniform motion of the clock hand - the clock (housing) can be under acceleration, sitting still or free falling.

\subsection{Principle of Inertia}\label{Kap - SRT Massbestimmung - classical metric - principle of inertia}

We link uniform motion to the behavior of natural (work) objects by the principle of inertia. Bodies move (without external agent) on their own. ''Every body with no (external) forces acting on it remains - as judged from the (inertial) lab - in a state of rest or of uniform rectilinear motion'' \cite{Sexl Urbantke Relativity}. In isolation their motion is preserved. We identify the presence of interactions by changing state of motion and the absence from practical reasoning. We postulate an inertial reference as a reproducible experimental prerequisite which we must shield from all external disturbances. We can provide it after probing all empirical conditions of an interaction (e.g. set up a billiard table \emph{horizontally} and \emph{test} that a prepared arrangement of object balls does not roll off to any side before the actual experiment). As Galilei and Huygens we develop via principle of inertia and relativity principle elementary dynamical concepts \emph{before} the latter were transferred by Newton onto gravitational systems. There the (initially practically solved) problem of selecting inertial references is revised; for building and steering machines (for the arising need to mechanize tool use based on the division of labor in industrial revolution) the latter had no practical significance.

Newton could draw on (in \emph{top}-heavy circles proscribed) ''literature of practical (\emph{hand}craft) mechanics on problems of machine construction and work economy, which is considered to little by historiography of science.'' For the context of origin of dynamical concepts Wolff's genetic reconstruction of Impetus theory - mechanics in epoch from 6. to 17. century - provides ''plausible arguments for the proposition that the inner conceptual content of mechanics was influenced by motives, which developed during that economic and technical revolution'' \cite{Wolff - Geschichte der Impetustheorie}.


\subsection{Light principle}\label{Kap - SRT Massbestimmung - light principle}

In order to give physical meaning to the concept of time Einstein \cite{Einstein-Grundlagen der ART} requires the use of some process which establishes relations between distant locations. In principle one could use any type of process. Most favorable for the theory one chooses a process about which we know something certain. For the free propagation of light this holds much more than for any other process.

\begin{figure}         
  \begin{center}         
  \includegraphics[height=7.3cm]{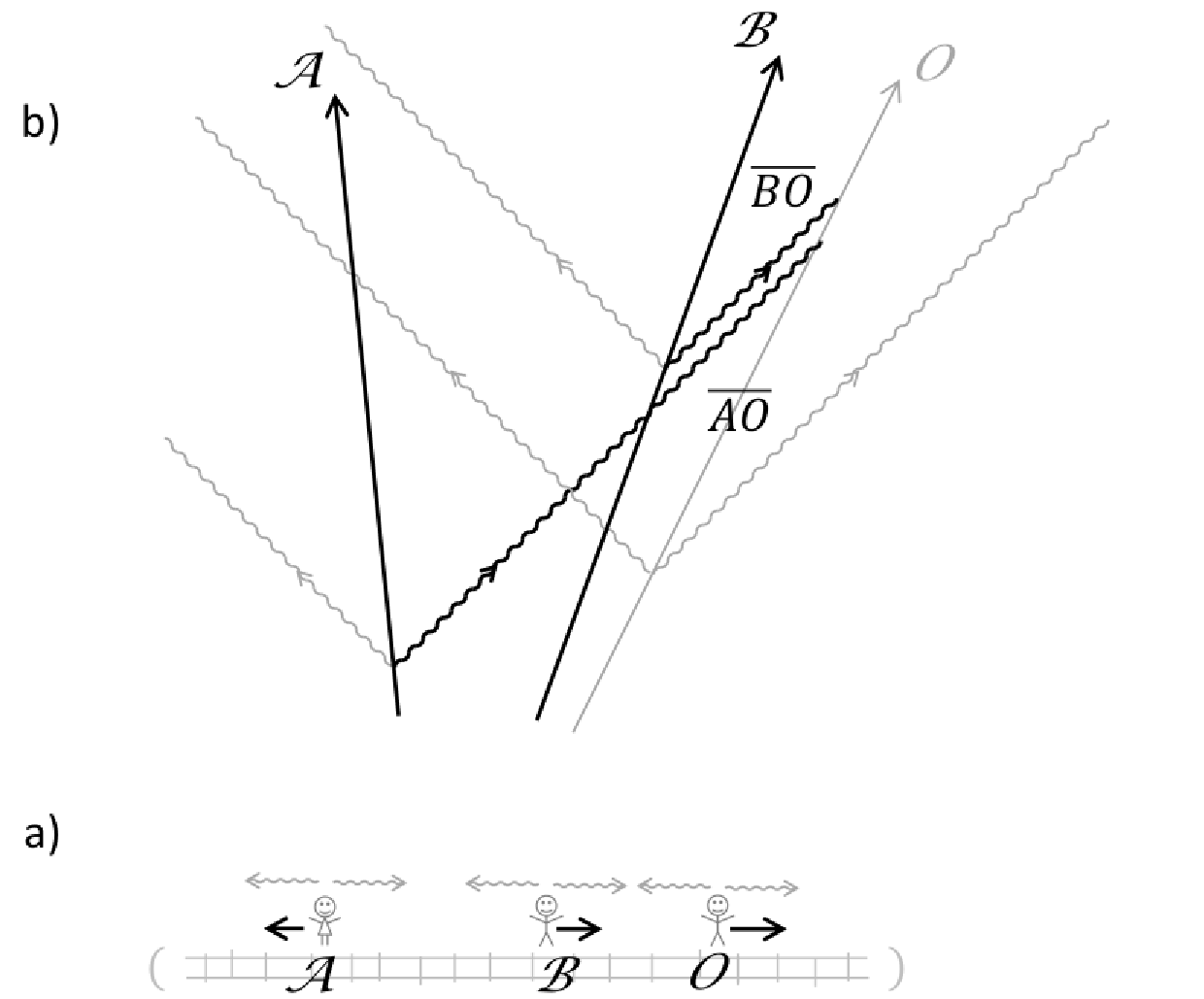}  
  \end{center}
  \vspace{-0.0cm}
  \caption{\label{figure-1} a) moving objects and moving light b) interrelation of corresponding processes
    }
  \end{figure}

One measures the motion of light with rulers and clocks. We depict the relative motion between all objects in a \emph{spacetime diagram} (see figure \ref{figure-1}b).
$\mathcal{A}$lice, $\mathcal{B}$ob or $\mathcal{O}$tto may move equally or not, but no object can overtake free light. Let $\mathcal{A}$lice and $\mathcal{B}$ob emit light towards $\mathcal{O}$tto. It propagates independently no matter how they move the source. If $\mathcal{A}$lice sends her light to $\mathcal{O}$tto and shortly after it passes $\mathcal{B}$ob he sends his own light to $\mathcal{O}$tto as well, then both rays $\overline{\mathcal{AO}}$ and $\overline{\mathcal{BO}}$ coincide. One measures the ''magnitude'' of distances and durations and the ''form'' of motion with classical rulers and clocks. They approximate a straight line and uniform motion. By local comparison with these reference instruments free light propagates in a uniform and straight way. In a spacetime diagram we represent it by a \emph{straight line}. Locally the light $\mathcal{A}$lice or $\mathcal{B}$ob send to $\mathcal{O}$tto $\overline{\mathcal{AO}}$ and $\overline{\mathcal{BO}}$ remains parallel.

With the classical metric (in the domain of classical measurements of length and duration) we discover: \emph{Locally} free light represents a \emph{uniform, isotropic and straight form of motion}. It provides a universal reference for any intrinsic observer. Based on the light principle we conduct laser ranging measurements. We presuppose this hypothesis also along \emph{global paths of light} which can be thought of as a connected covering of multiple local segments.

\subsection{Light clock}\label{Kap - SRT Massbestimmung - light clock}

\begin{figure}         
  \begin{center}         
  \includegraphics[height=9.3cm]{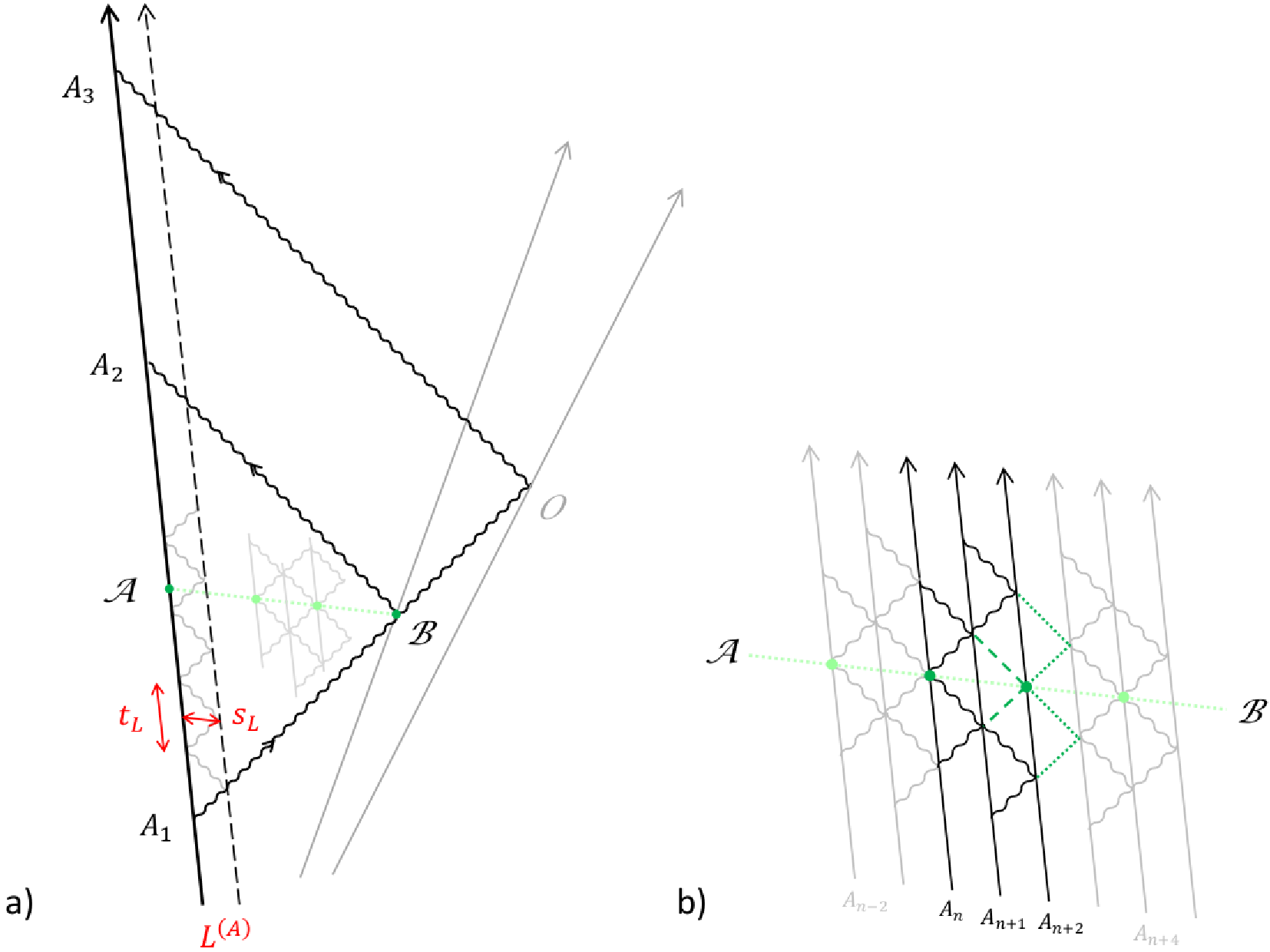}  
  \end{center}
  \vspace{-0.0cm}
  \caption{\label{figure-2} a) laser ranging b) consecutive and adjacent connection of light clocks
    }
  \end{figure}

Because of the universal light principle ''laser ranging'' is a reliable \emph{practice of navigation}.\footnote{The procedure developed naturally. Throughout millennia of evolution bats, coordinating their living at night, or dolphins, hunting under invisible conditions, discovered and rehearsed the practice of (i) \emph{producing} sonar waves and (ii) exploiting that \emph{tool} to master given living conditions.

Upon developing the classical metric one understands why it works so reliably in practice. With rulers and clocks one can measure the prerequisites. For durations of each sonar ranging act the emitting organism represents a sufficiently rigid body at constant motion. Sound propagates much faster, sufficiently straight and uniform. Thus by successive echoing animals can maneuver within an environment of comparably small relative motions. Based on common navigation actions Einstein demonstrates standardization of the conduct of spatiotemporal measurements. He discovered the Light principle as extra condition for physical measurements. Its theoretical conception led engineers into a revolution of technical applications (GPS, Lunar-Laser-Ranging, synchronization and coordination of partition of work on a global scale etc.).}
Let $\mathcal{A}$lice send out light towards $\mathcal{O}$tto\; $\mathcal{A}_1\!\rightsquigarrow\mathcal{O}\rightsquigarrow\mathcal{A}_2$ and towards $\mathcal{B}$ob\; $\mathcal{A}_1\!\rightsquigarrow\mathcal{B}\rightsquigarrow\mathcal{A}_3$ and wait until their reflection returns (see figure
\ref{figure-2}a). In radar round trips we focus on the distance covered and $\mathcal{A}$lice waiting time. For two ranging cycles $\mathcal{A}_1\!\rightsquigarrow\mathcal{O}\rightsquigarrow\mathcal{A}_2$ and $\mathcal{A}_1\!\rightsquigarrow\mathcal{B}\rightsquigarrow\mathcal{A}_3$ $\mathcal{A}$lice notices the order in which the light returns. By the light principle more waiting time $t_{\overline{\mathcal{A}_1\mathcal{A}_2}}>t_{\overline{\mathcal{A}_1\mathcal{A}_3}}$ corresponds to a larger distance covered $s_{\overline{\mathcal{AO}}}>s_{\overline{\mathcal{AB}}}$ from $\mathcal{A}$lice to turning point $\mathcal{O}$tto resp. $\mathcal{B}$ob and back.

For quantification $\mathcal{A}$lice constructs a reference device,
a \emph{light clock} $\mathcal{L} : \mathcal{L}_I\rightsquigarrow \mathcal{L}_{II} \rightsquigarrow \mathcal{L}_I \ldots$ with two nearby mirrors $\mathcal{L}_{I}$ and $\mathcal{L}_{II}$ in a rigid frame. The light constantly oscillates back and forth. Each tick of her \emph{measurement unit} $\mathcal{L}$ covers the same \emph{standard distance} $s_{\mathcal{L}}$ and takes the same \emph{standard time} $t_{\mathcal{L}}$ \{\ref{Kap - SRT Massbestimmung - light principle}\}.

$\mathcal{A}$lice light clock substitutes the traditional rulers and clocks. The protophysical test norms - for manufacturing both, traditional clocks and light clocks - remain the same. In practice one replaces the former by new light clocks because they realize the aspired ideal of uniform running more precisely. The motion of light is not anymore measured with traditional rulers and clocks; instead we determine all other motions with respect to the motion of light in (classical protophysically manufactured) light clocks. The light principle implies a \emph{paradigm shift}. One abandons the former priority of classical measurement devices in favor for the universal propagation of light. The motion of light becomes a measurement standard itself.\footnote{The definition of standard length $s_{\mathcal{L}}$ is based on a \emph{given} standard duration $t_{\mathcal{L}}$ and the universal speed of light $c$. Contemporary metrology regards speed of light $c$ as invariant natural constant and introduces optical clocks as frequency standards. Our world's current time standard (a laser-cooled cesium fountain known as NIST-F1 based on resonant transitions between quantized energy levels in atoms) is accurate to within $\Delta f/f \sim 10^{-16}$. It represents the ultimate reference for time intervals $t_{\mathrm{Cs}}$ with accuracy $~10^{-16}\mathrm{sec}$.

Bureau of Standards defines the atomic second $1\mathrm{sec}^{(\mathrm{SI})}:= 9192631770 \cdot t_{\mathrm{Cs}}$ - on paper - as a multiple of that standard duration. ''Cesium provides a 'physical' second that can be realized in laboratories and used for other measurements. ... The basic principle of the atomic oscillator is simple: Since all atoms of a specific element are identical, they should produce the exact same frequency...'' \cite{NIST - Frequency Standards and Realization of SI Second}. They refer to an \emph{intrinsic} property of an atom under standardized conditions: ''cesium atom at rest at a thermodynamic temperature of $0\mathrm{K}$''. An unperturbed atomic transition is identical from atom to atom (reproducibility).} We use the classical light clock as a new \emph{measurement unit}.

\subsection{Direct connections}\label{Kap - SRT Massbestimmung - direct connecting products}

We essentially refer to the oscillating light inside. In practice the two dimensions ''length'' and ''duration'' of a light clock $\mathcal{L}$ are always addressed unified. Depending on the concatenation:
\begin{itemize}
   \item   adjacently connected $\ast_s$ (ticking) light clocks represent a distance unit $\mathcal{L}_s$ and
   \item   consecutively connected $\ast_t$ (light clock) ticks represent a duration unit $\mathcal{L}_t$.
\end{itemize}
The width of the light clock $\mathcal{L}$ becomes our \emph{unit length} $s_{\mathcal{L}}$ and each tick lasts \emph{unit time} $t_{\mathcal{L}}$.


\subsubsection{Time-like concatenation}

Let $\mathcal{A}$lice join together light clock ticks $\mathcal{L}$ \emph{one after another} until the sequence - symbolized by $\mathcal{L}\ast_{t}\ldots\ast_{t}\mathcal{L}$ - covers the waiting interval of her laser ranging cycle $\mathcal{A}_1\!\rightsquigarrow\mathcal{B}\rightsquigarrow\mathcal{A}_2$
\be\label{Formel - radar duration konstruierbare Ersetzung}
   \overline{\mathcal{A}_1\mathcal{A}_2} \; \sim_{t} \;
   \mathcal{L}\ast_{t}\ldots\ast_{t}\mathcal{L}  \;\; .
\ee
In her material model $\mathcal{A}$lice can count the number of ticks, symbolized $\sharp \left\{ \mathcal{L}_t \right\} =: t^{(\mathcal{A})}_{\overline{\mathcal{A}_1\mathcal{A}_2}}\,$. The ordering relation ''longer than'' becomes quantified. $\mathcal{A}$lice measures the duration of her laser ranging interval
\be\label{Formel - radar duration physical measure}
   t_{\overline{\mathcal{A}_1\mathcal{A}_2}} \;\;
   \stackrel{(\ref{Formel - radar duration konstruierbare Ersetzung})}{=} \;\;
   t_{\mathcal{L}\ast_t\ldots\ast_t\mathcal{L}} \;\; \stackrel{(\mathrm{Congr.})}{=:} \;\;
   t^{(\mathcal{A})}_{\overline{\mathcal{A}_1\mathcal{A}_2}} \; \cdot \; t_{\mathcal{L}}
\ee
by the sequence of ticks and the latter according to the congruence principle by the number of congruent (light clock) ticks and its reference duration  $t_{\mathcal{L}}$.

\subsubsection{Space-like concatenation}

Furthermore $\mathcal{A}$lice can place ticking light clocks $\mathcal{L}$ literally \emph{side by side}. She utilizes the same \emph{units} $\mathcal{L}$ to produce an adjacent layout of comoving light clocks.
\begin{lem}\label{Lem - SRT Kin - construct straight measurement path}
It represents $\mathcal{A}$lice \underline{simultaneous straight measurement path} towards $\mathcal{B}$ob $\overline{\mathcal{AB}}$.
\end{lem}
\textbf{Proof:}
Imagine a swarm of identically constituted light clocks $\mathcal{L}^{(\mathcal{A}_i)}$. Beginning with her own in moment $\mathcal{A}$ $\mathcal{A}$lice successively places pairs of light clocks $\mathcal{L}^{(\mathcal{A}_i)} \ast_s \mathcal{L}^{(\mathcal{A}_{i+1})}$ next to one another by letting their inner light rays overlap. She builds a locally regular grid of light clocks in an intrinsically \emph{simultaneous and straight way} (see figure
\ref{figure-2}b):
\begin{enumerate}
\item[(a)]   Suppose we have successively laid out light clocks from $\mathcal{L}^{(\mathcal{A}_1)}$ all the way to $\mathcal{L}^{(\mathcal{A}_n)}$. Consider the two ticks of light clock $\mathcal{L}^{(\mathcal{A}_n)} \: \ast\!\mid_{\mathcal{A}_n}\mathcal{L}^{(\mathcal{A}_n)}$ around the moment $\mathcal{A}_n$.
\item[(b)]   The next comoving light clock $\mathcal{L}^{(\mathcal{A}_{n+1})}$ is placed so that the extended (dashed) light ray from $\mathcal{L}^{(\mathcal{A}_n)} \: \ast\!\mid_{\mathcal{A}_n}\mathcal{L}^{(\mathcal{A}_n)}$ coincides with the light ray from $\mathcal{L}^{(\mathcal{A}_{n+1})}$.
\item[(c)]   Starting from $\mathcal{A}_{n+1}$ - by isotropy - light travels in identical \emph{round trip duration} $t_{\mathcal{L}^{(\mathcal{A}_{n+1})}}$ the same distance back to left light clock $\mathcal{L}^{(\mathcal{A}_{n})}$ as to the right light clock $\mathcal{L}^{(\mathcal{A}_{n+1})}$.\footnote{Let two synchronously ticking light clocks $\mathcal{L} \ast_s \mathcal{L}$ sit side by side. We assume that \emph{two-way} light cycle on the left covers same standard distance $s_{\mathcal{L}}$ to its turning point as the other \emph{two-way} light cycle on the right.}
\item[(d)]   Consider a series of three (preceding and following) ticks of light clock $\mathcal{L}^{(\mathcal{A}_{n+1})}$.
\item[(e)]   Place light clock $\mathcal{L}^{(\mathcal{A}_{n+2})}$ so that the extended (dotted) light rays from $\mathcal{L}^{(\mathcal{A}_{n+1})}$ coincide with the two ticks of light clock $\mathcal{L}^{(\mathcal{A}_{n+2})} \: \ast\!\mid_{\mathcal{A}_{n+2}}\mathcal{L}^{(\mathcal{A}_{n+2})}$ around the moment $\mathcal{A}_{n+2}$.
\item[(f)]   By analogous induction steps $\;\mathcal{L}^{(\mathcal{A}_n)} \ast\!\mid_{\mathcal{A}_n}\mathcal{L}^{(\mathcal{A}_n)} \: \Rightarrow \: \mathcal{L}^{(\mathcal{A}_{n+1})} \: \Rightarrow \: \mathcal{L}^{(\mathcal{A}_{n+2})} \: \ast\!\mid_{\mathcal{A}_{n+2}}\mathcal{L}^{(\mathcal{A}_{n+2})}  \; \forall n$
    $\mathcal{A}$lice proceeds towards $\mathcal{B}$ob.

    In every step the extended (dashed resp. dotted) light rays coincide. In her straight comoving \emph{connection} $\mathcal{L}^{(\mathcal{A})}\!\!\mid_{\mathcal{A}} \ast \; \mathcal{L}^{(\mathcal{A}_2)} \ast \; \mathcal{L}^{(\mathcal{A}_3)}\!\!\mid_{\mathcal{A}_3}  \ldots \mathcal{L}^{(\mathcal{A}_n)}\!\!\mid_{\mathcal{A}_n} \ast \; \mathcal{L}^{(\mathcal{A}_{n+1})} \ast \; \mathcal{L}^{(\mathcal{A}_{n+2})}\!\!\mid_{\mathcal{A}_{n+2}}$ all light clocks tick synchronized
    along connecting moments $\mathcal{A}\:, \mathcal{A}_3 \ldots \mathcal{A}_n, \mathcal{A}_{n+2} \ldots \mathcal{B}$.
\end{enumerate}
The construction steps do not depend on the scale of light clock $\mathcal{L}$ (e.g. refining the locally regular layout with twice the light clocks of half the size coincides with the original pattern). They are locally well-defined; the global measurement path $\mathcal{L}\ast_s\ldots\ast_s\mathcal{L}$ is universal.
\qed
$\mathcal{A}$lice covers the laser ranging path to $\mathcal{B}$ob by an adjacent layout of light clocks
\be\label{Formel - radar distance konstruierbare Ersetzung}
   \overline{\mathcal{A}\mathcal{B}} \; \sim_{s} \;   \mathcal{L}\ast_s\ldots\ast_s\mathcal{L} \;\; .
\ee
Each represents a length unit $\mathcal{L}_s$. $\mathcal{A}$lice measures the length along her laser ranging path
\be\label{Formel - radar distance - direct physical measure}
   s_{\overline{\mathcal{AB}}} \;\;
   \stackrel{(\ref{Formel - radar distance konstruierbare Ersetzung})}{=} \;\;
   s_{\mathcal{L}\ast_s\ldots\ast_s\mathcal{L}}
   \;\; \stackrel{(\mathrm{Congr.})}{=:} \;\; s^{(\mathcal{A})}_{\overline{\mathcal{AB}}} \cdot s_{\mathcal{L}}
\ee
by the adjacent layout  $\mathcal{L}\ast_s\ldots\ast_s\mathcal{L}$  and the latter according to the congruence principle by the number $\sharp \left\{ \mathcal{L}_s \right\} =:  s^{(\mathcal{A})}_{\overline{\mathcal{AB}}}\,$ of congruent clocks $\mathcal{L}_s$ and its standard length $s_{\mathcal{L}}$.

\subsubsection{Spacetime-like concatenation}

With every laser ranging ping $\mathcal{A}_1\!\rightsquigarrow\mathcal{B}\rightsquigarrow\mathcal{A}_2$ $\mathcal{A}$lice measures the position of $\mathcal{B}$ob at the moment $\mathcal{B}$ when her signal reflects (see figure \ref{figure-2}a). $\mathcal{A}$lice covers the outgoing light ray $\overline{\mathcal{A}_1\mathcal{B}}$ by a swarm of light clocks in both space-like and time-like way: She connects a consecutive sequence until ''half-time'' moment $\mathcal{A}$ (after waiting half of her laser ranging interval)
\[
   \overline{\mathcal{A}_1\mathcal{A}} \; \sim_t \;
   \mathcal{L}\!\!\mid_{\mathcal{A}_1} \ast_t \ldots \ast_t \mathcal{L}\!\!\mid_{\mathcal{A}}
\]
in light clock $\mathcal{L}\!\!\mid_{\mathcal{A}}$ to an adjacent layout of (ticking) light clocks that reaches to moment $\mathcal{B}$
\[
   \overline{\mathcal{A}\mathcal{B}} \; \sim_s \;
   \mathcal{L}\!\!\mid_{\mathcal{A}} \ast_s \ldots \ast_s \mathcal{L}\!\!\mid_{\mathcal{B}} \;\; .
\]
The collective motion of light inside the composite of ticking light clocks - symbolized by $\mathcal{L}\!\!\mid_{\mathcal{A}_1} \ast_t \ldots \ast_t \mathcal{L}\!\!\mid_{\mathcal{A}} \ast_s \ldots \ast_s \mathcal{L}\!\!\mid_{\mathcal{B}}$ - covers the light ray from $\mathcal{A}$lice towards $\mathcal{B}$ob
\be\label{Formel - radar spatiotemporal distance konstruierbare Ersetzung}
   \overline{\mathcal{A}_1\mathcal{B}} \; \sim_{t,s} \; \mathcal{L} \ast_t \ldots \ast_t \mathcal{L} \ast_s \ldots \ast_s \mathcal{L} \;\; .
\ee
$\mathcal{A}$lice utilizes copies of the same light clock $\mathcal{L}$ as spatiotemporal units. Along a consecutive segment $\mathcal{L} \ast_t \ldots \ast_t \mathcal{L}$ each represents a time unit $\mathcal{L}_t$ and along an adjacent segment $\mathcal{L} \ast_s \ldots \ast_s \mathcal{L}$ a distance unit $\mathcal{L}_s$. In both segments she counts the congruent ticks $\sharp \left\{ \mathcal{L}_t \right\} =: t^{(\mathcal{A})}_{\overline{\mathcal{A}_1\mathcal{B}}}$ and the congruent clocks $\sharp \left\{ \mathcal{L}_s \right\} =: s^{(\mathcal{A})}_{\overline{\mathcal{A}_1\mathcal{B}}}\,$. $\mathcal{A}$lice measures the spatiotemporal distance towards $\mathcal{B}$ob
\be \label{Formel - radar direkt spatiotemporal physical measure}
   (t,s)_{\overline{\mathcal{A}_1\mathcal{B}}}
   \;\; \stackrel{(\ref{Formel - radar spatiotemporal distance konstruierbare Ersetzung})}{=} \;\;
   (t,s)_{\mathcal{L} \ast_t \ldots \ast_t \mathcal{L} \ast_s \ldots \ast_s \mathcal{L}}
   \;\; \stackrel{(\ref{Formel - radar duration physical measure})(\ref{Formel - radar distance - direct physical measure})}{=} \;\; \left(\; t^{(\mathcal{A})}_{\overline{\mathcal{A}_1\mathcal{A}}} \cdot t_{\mathcal{L}} \; , \; s^{(\mathcal{A})}_{\overline{\mathcal{A}\mathcal{B}}} \cdot s_{\mathcal{L}} \; \right)
\ee
by her composite layout. It is reproducible from the number of congruent light clocks, the consecutive or adjacent way of their connection and their standard length $s_{\mathcal{L}}$ and duration $t_{\mathcal{L}}$.
\\

\begin{figure}         
  \begin{center}         
  \includegraphics[scale=0.48]{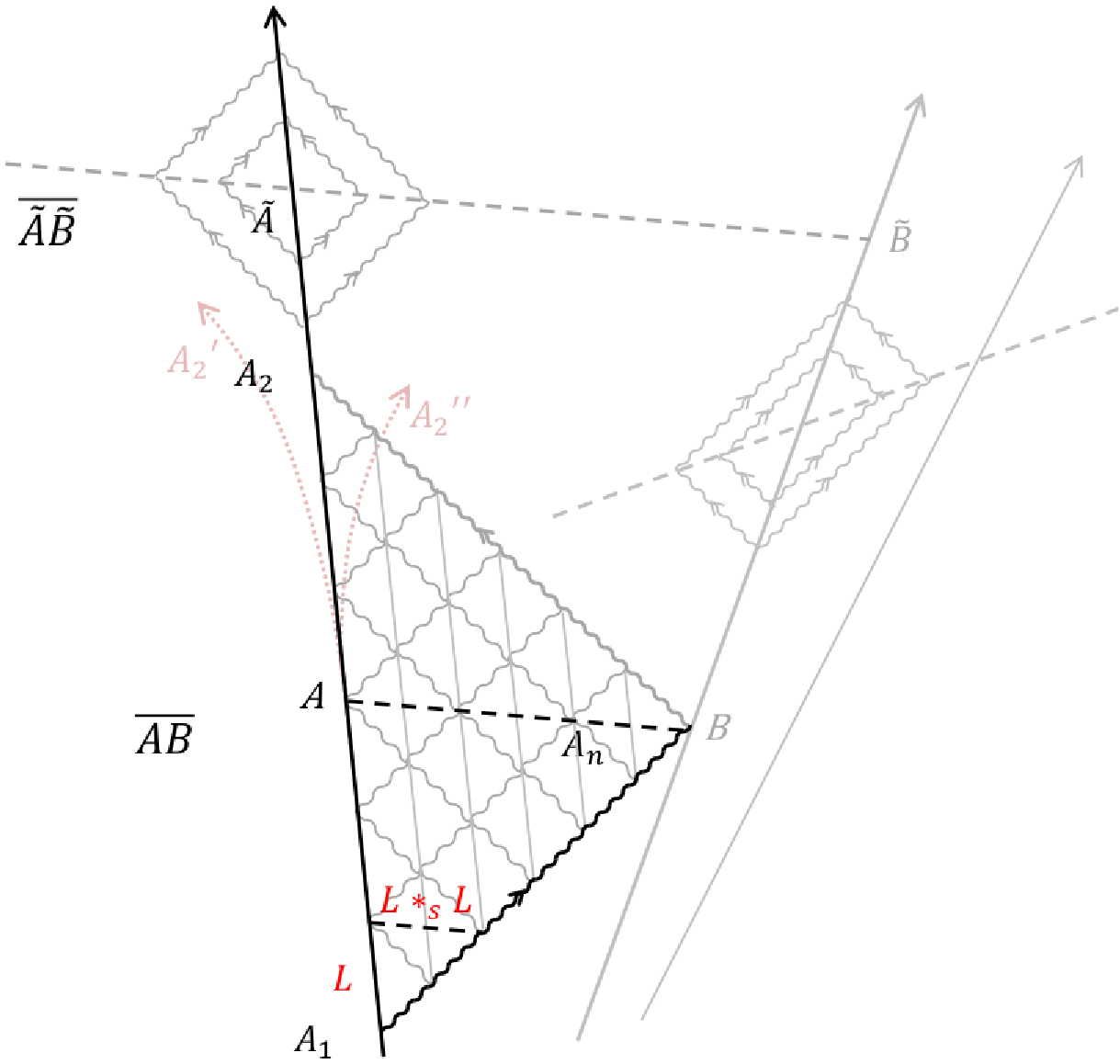}  
  \end{center}
  \vspace{-0.0cm}
  \caption{\label{figure-3} (local) indirect characterization of simultaneous straight measurement paths
    }
  \end{figure}

In the \emph{direct} measurement we cover the object or process by a grid of light clocks. Now consider the measurement of a ray of light. In figure \ref{figure-3}
$\mathcal{A}$lice covers the smallest segment $\overline{\mathcal{A}_1\mathcal{B}'} \sim_{t,s} \mathcal{L}_t \ast (\mathcal{L} \ast_s \mathcal{L})$ with one light clock tick and two adjacent light clocks. Step by step she covers the uniform motion of the outgoing light ray $\overline{\mathcal{A}_1\mathcal{B}}$ by a (locally) regular pattern of light clocks (and the same for the returning light ray $\overline{\mathcal{B}\mathcal{A}_2}$). No matter to what extent $\mathcal{A}$lice covers the departing light ray $\overline{\mathcal{A}_1\mathcal{B}'}\subset\overline{\mathcal{A}_1\mathcal{B}}$ by similarity any pair of durations scales proportional with the corresponding pair of distances\footnote{The division in $s_1/s_2$ resp. $t_1/t_2$ symbolizes $\mathcal{A}$lice dividing \emph{operation}. The formulation $s_{\overline{\mathcal{A}\mathcal{B}}}/s_{\mathcal{L}}:=n$ means: by connecting $n$ congruent reference paths $\mathcal{L} \ast_s \ldots \ast_s \mathcal{L} \sim_{s} \overline{\mathcal{A}\mathcal{B}}$ she will cover the path $\overline{\mathcal{A}\mathcal{B}}$.}
\be \label{Formel - radar direkt measurement c - proportionality}
   \frac{s_{\overline{\mathcal{A}_1\mathcal{B}}}}{s_{\overline{\mathcal{A}_1\mathcal{B}'}}} \;\; = \;\; \frac{t_{\overline{\mathcal{A}_1\mathcal{B}}}}{t_{\overline{\mathcal{A}_1\mathcal{B}'}}}  \;\; .
\ee
For a generic segment $(t_c,s_c)$ of the uniform light ray we express the proportionality relation $\frac{s_c}{2s_{\mathcal{L}}} = \frac{t_c}{t_{\mathcal{L}}}$ between pairs of \emph{same} basic observables by a proportionality constant
\be \label{Formel - radar direkt measurement c - proportionality constant}
   \left\{ \frac{s_c}{s_{\mathcal{L}}} \right\} \;\; \stackrel{(\ref{Formel - radar direkt measurement c - proportionality})}{=} \;\; \underbrace{2}_{\equiv \: c^{(\mathcal{L})}} \; \cdot \; \left\{ \frac{t_c}{t_{\mathcal{L}}} \right\}   \;\; .
\ee
We define the \emph{velocity} of light $c^{(\mathcal{L})} := s_c^{(\mathcal{L})}\! / t_c^{(\mathcal{L})} = 2$ (in standard light clock dimensions $s_{\mathcal{L}}, t_{\mathcal{L}}$) as a derived physical quantity. From known ''distance \emph{for each} time \emph{unit}'' $2\cdot s_{\mathcal{L}}$ and the ''number of time units'' $\frac{t_c}{t_{\mathcal{L}}}$ along the way one gets the total distance as a product of velocity\footnote{The formal expression $\frac{s_{\mathcal{L}}}{t_{\mathcal{L}}}$ has no \emph{physical} meaning. One cannot divide a path by a time \cite{Wallot - Groessengleichungen Einheiten und Dimensionen}. The formal reduction of fractions, that ''same dimensions (unit length, unit mass etc.) cancel one another'', gives back the relation (\ref{Formel - radar direkt measurement c - proportionality constant}) between quantities (ratios) which can all be measured directly by concatenation operations.} and time of flight
\be\label{Formel - radar direkt measurement c}
   s_c \;\; \stackrel{(\ref{Formel - radar direkt measurement c - proportionality constant})}{=} \;\; 2\cdot s_{\mathcal{L}} \cdot \frac{t_c}{t_{\mathcal{L}}}
   \;\; =: \;\; \underbrace{\left( 2\cdot \frac{s_{\mathcal{L}}}{t_{\mathcal{L}}} \right)}_{\equiv \: c} \;\;\cdot\;\; t_c  \;\; .\footnote{For measuring $c = c^{(\mathcal{L})} \cdot \frac{s_{\mathcal{L}}}{t_{\mathcal{L}}}$ we utilize a light clock with dimensions: width $s_{\mathcal{L}}$ and cycle length $t_{\mathcal{L}}$. Basic dimensions (unit length, unit time etc.) are ''arbitrarily chosen constant reference measures'' \cite{Wallot - Groessengleichungen Einheiten und Dimensionen}. Contemporary metrology refers to units based on the standard duration $t_{\mathrm{Cs}}$ of an intrinsic Cesium period and the invariant speed of light $c$ \{\ref{Kap - SRT Massbestimmung - light clock}\}. The atomic second $\mathrm{sec}_{\mathrm{SI}} := 9192631770 \cdot t_{\mathrm{Cs}}$ is a multiple of that standard duration (factor chosen to match traditional calendrical second). The (multiple of the) standard meter $299792458\cdot \mathrm{m}_{\mathrm{SI}} := (c \cdot  \mathrm{sec}_{\mathrm{SI}})$ is the distance of free light in one atomic second of flight. The numerical factor is fixed by convention (to cover $1 \mathrm{m}_{\mathrm{SI}} \simeq 1 \mathrm{m}_{\mathrm{bar}}$ the traditional platinum-iridium standard in the Bureau of Weights and Measures). One refers to the traditional units (meter bar and fraction of a tropical year) one last time, to match the conversion factors. Now one defines the international unit measures $\mathrm{sec}_{\mathrm{SI}}$, $\mathrm{m}_{\mathrm{SI}}$ independently and more precise from invariant natural processes (intrinsic $\mathrm{Cs}$-period and speed of light) and the fixed numerical factors; the old prototypes stay in the museum.

A light clock with SI-unit period $t_{\mathcal{L}}:=t_{\mathrm{SI}}$ and corresponding width $2 \cdot s_{\mathcal{L}} = (c \cdot 1 \mathrm{sec}_{\mathrm{SI}}) =299792458\cdot \mathrm{m}_{\mathrm{SI}}$ has the proportionality constant $\left\{ \frac{s_c}{\mathrm{m}} \right\}  \stackrel{(\ref{Formel - radar direkt measurement c - proportionality constant})}{=}  299792458 \; \cdot \; \left\{ \frac{t_c}{\mathrm{sec}} \right\}$; thus measures speed of light $c = 299792458 \cdot \frac{\mathrm{m}}{\mathrm{sec}}$.}
\ee

\subsection{Indirect laser ranging}\label{Kap - SRT Massbestimmung - indirect laser ranging}

In direct laser ranging $\mathcal{A}_1\!\rightsquigarrow\mathcal{B}\rightsquigarrow\mathcal{A}_2$ $\mathcal{A}$lice measures the distance $s_{\overline{\mathcal{AB}}}$ to $\mathcal{B}$ob (\ref{Formel - radar distance - direct physical measure}) by counting units along the (potentially global) simultaneous measurement path $\overline{\mathcal{AB}}$. From a direct measurement the departing and returning ray of light\footnote{Locally regular pattern of light clocks (see figure \ref{figure-3}) covers laser ranging waiting interval $\overline{\mathcal{A}_1\mathcal{A}_2}$ by same number of consecutive (light clock) ticks as there are adjacent (ticking) clocks along laser ranging route $\overline{\mathcal{AB}}$.} cover in the same duration $t_{\overline{\mathcal{A}_1\mathcal{B}}} = t_{\overline{\mathcal{B}\mathcal{A}_2}} = \frac{1}{2} \cdot t_{\overline{\mathcal{A}_1\mathcal{A}_2}}$ the proportional distance $s_{\overline{\mathcal{A}\mathcal{B}}} \stackrel{(\ref{Formel - radar direkt measurement c})}{=} c \cdot t_{\overline{\mathcal{A}_1\mathcal{B}}} = \frac{c}{2} \cdot t_{\overline{\mathcal{A}_1\mathcal{A}_2}}$. Thus (locally) $\mathcal{A}$lice can also indirectly compute that length
\be \label{Formel - radar indirekt spatiotemporal physical measure}
   (t,s)_{\overline{\mathcal{A}_1\mathcal{B}}} \;\; \stackrel{(\ref{Formel - radar direkt spatiotemporal physical measure})(\ref{Formel - radar direkt measurement c})}{=} \;\; \left( \;\; \frac{1}{2} \cdot t_{\overline{\mathcal{A}_1\mathcal{A}_2}} \;\; , \;\; \frac{c}{2} \cdot t_{\overline{\mathcal{A}_1\mathcal{A}_2}} \;\; \right)
\ee
from measuring the round trip time $t_{\overline{\mathcal{A}_1\mathcal{A}_2}}$; the familiar principle of indirect laser ranging.

For the resulting equation $\mathcal{A}$lice must obey a \emph{measurement condition} (underlying the direct measurement of light rays in figure \ref{figure-3}), that during the radar waiting interval $\overline{\mathcal{A}_1\mathcal{A}_2}$ her motion is preserved. After emitting the light pulse $\overline{\mathcal{A}_1\mathcal{B}}$ she neither accelerates away $\mathcal{A}_2'$ nor towards $\mathcal{A}_2''$ the returning light pulse $\overline{\mathcal{B}\mathcal{A}_1}$. In local laser ranging practice accelerations are negligible; for larger configurations the effects accumulate. Then $\mathcal{A}$lice can characterize all elements of her \emph{simultaneity line} $\mathcal{A}_n \in \overline{\mathcal{AB}}$ by moments $\mathcal{A'}, \mathcal{A''} \in \overline{\mathcal{A}_1\mathcal{A}_2}$ along the waiting interval (see figure \ref{figure-3}). In local laser ranging $\mathcal{A'}\!\rightsquigarrow\mathcal{A}_n\rightsquigarrow\mathcal{A''}$ the preceding emission and subsequent reception are symmetric $t_{\overline{\mathcal{A'A}}} = t_{\overline{\mathcal{AA''}}}$ \emph{with respect to $\mathcal{A}$lice moment $\mathcal{A}$}.

\begin{figure}         
  \begin{center}         
  \includegraphics[scale=0.30]{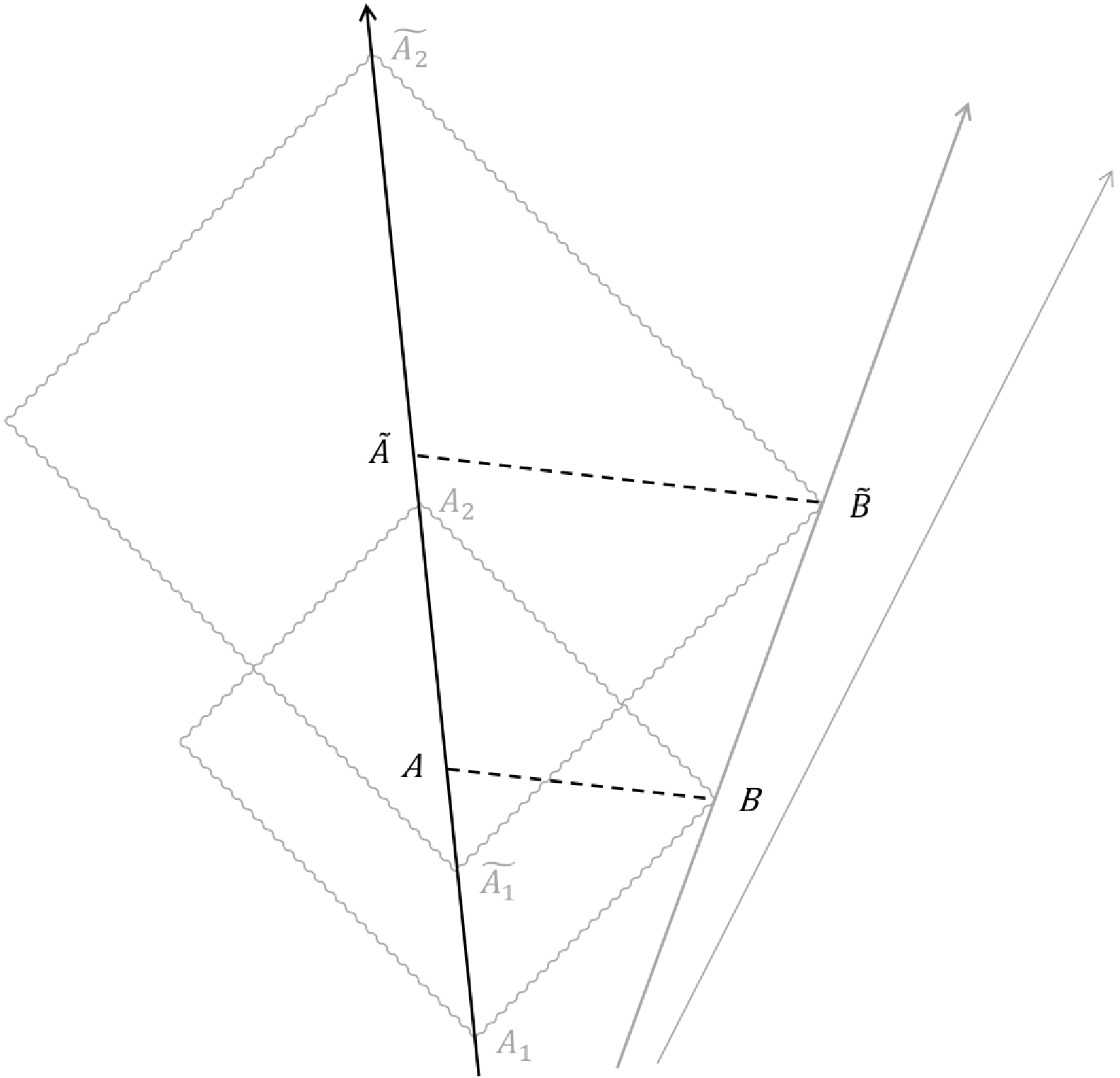}  
  \end{center}
  \vspace{-0.0cm}
  \caption{\label{figure-4} combination of two elementary laser ranging measurements
    }
  \end{figure}
With two elementary laser rangings $\mathcal{A}_1\!\rightsquigarrow\mathcal{B}\rightsquigarrow\mathcal{A}_2$ and $\tilde{\mathcal{A}_1}\!\rightsquigarrow\tilde{\mathcal{B}}\rightsquigarrow\tilde{\mathcal{A}_2}$ towards the two consecutive moments $\mathcal{B}$ and $\tilde{\mathcal{B}}$ (see figure \ref{figure-4}) $\mathcal{A}$lice measures the relative motion of $\mathcal{B}$ob $\overline{\mathcal{B}\tilde{\mathcal{B}}}$.\footnote{In a direct measurement her layout of light clocks $\mathcal{L}\!\!\mid_{\mathcal{B}} \ast_s \ldots \ast_s \mathcal{L}\!\!\mid_{\mathcal{A}} \ast_t \ldots \ast_t  \mathcal{L}\!\!\mid_{\tilde{\mathcal{A}}} \ast_s \ldots \ast_s \mathcal{L}\!\!\mid_{\tilde{\mathcal{B}}} \;\;\equiv\;\;  \overline{\mathcal{B}\mathcal{A}} \ast \overline{\mathcal{A}\tilde{\mathcal{A}}} \ast \overline{\tilde{\mathcal{A}}\tilde{\mathcal{B}}} \;\; \sim_{t,s} \;\; \overline{\mathcal{B}\tilde{\mathcal{B}}}$ covers a segment of his motion.} Her indirect laser ranging involves three steps:
\begin{enumerate}
\item   \emph{construct} her straight simultaneous measurement paths towards $\mathcal{B}$ob
\item   \emph{enclose} the measurement object $\overline{\mathcal{B}\tilde{\mathcal{B}}}$ by her simultaneity lines $\overline{\mathcal{AB}}$ and $\overline{\mathcal{\tilde{A}\tilde{B}}}$
    \be
       \overline{\mathcal{B}\tilde{\mathcal{B}}} \; \sim_{t,s} \; \overline{\mathcal{B}\mathcal{A}} \; \ast \; \overline{\mathcal{A}\tilde{\mathcal{A}}} \; \ast \; \overline{\tilde{\mathcal{A}}\tilde{\mathcal{B}}} \label{Formel - radar direkt motion Eingrenzung}
    \ee
\item   \emph{project} between and along the simultaneity lines for temporal and spatial components
\bea
   (t,s)_{\overline{\mathcal{B}\tilde{\mathcal{B}}}} & \stackrel{(\ref{Formel - radar direkt motion Eingrenzung})}{=} & (t,s)_{\overline{\mathcal{B}\mathcal{A}} \; \ast \; \overline{\mathcal{A}\tilde{\mathcal{A}}} \; \ast \; \overline{\tilde{\mathcal{A}}\tilde{\mathcal{B}}}} \nn \\
    & \stackrel{(\ref{Formel - radar direkt spatiotemporal physical measure})}{=} & \!\! \underbrace{(t,s)_{\overline{\mathcal{B}\mathcal{A}}}}_{ \left( 0 \; , \; - s_{\overline{\mathcal{A}\mathcal{B}}}\right)} \; + \; \underbrace{(t,s)_{\overline{\mathcal{A}\tilde{\mathcal{A}}}}}_{ \left( t_{\overline{\mathcal{A}\tilde{\mathcal{A}}}}  \; , \; 0 \right)} \; + \; \underbrace{(t,s)_{\overline{\tilde{\mathcal{A}}\tilde{\mathcal{B}}}}}_{ \left( 0 \; , \; s_{\overline{\tilde{\mathcal{A}}\tilde{\mathcal{B}}}}\right)}
    \;\; = \; \left( \; t_{\overline{\mathcal{A}\tilde{\mathcal{A}}}} \;\: , \; s_{\overline{\tilde{\mathcal{A}}\tilde{\mathcal{B}}}} - s_{\overline{\mathcal{A}\mathcal{B}}}  \; \right) \label{Formel - radar direkt motion physical measure}
\eea
\end{enumerate}
The vectorial addition of components $(t,s)_{\overline{\mathcal{A}\tilde{\mathcal{A}}}} + (t,s)_{\overline{\tilde{\mathcal{A}}\tilde{\mathcal{B}}}} = (t,s)_{\overline{\mathcal{A}\tilde{\mathcal{B}}}}\,$ corresponds with direct measurement operations. In the underlying material model we concatenate a number of light clocks $\mathcal{L} \ast_t \ldots \ast_t \mathcal{L}$ and $\mathcal{L} \ast_s \ldots \ast_s \mathcal{L}$ to create a composite layout $\mathcal{L} \ast_t \ldots \ast_t \mathcal{L} \ast_s \ldots \ast_s \mathcal{L}$.

Let $\mathcal{A}$lice and $\mathcal{B}$ob coincide (without loss of generality) in the initial moment $\mathcal{P}$. Now the first laser ranging configuration becomes trivial and we are left with $\mathcal{P}\rightarrow \mathcal{A}_1\!\rightsquigarrow\mathcal{B}\rightsquigarrow\mathcal{A}_2$ (see figure \ref{figure-5}). $\mathcal{A}$lice measures the spatiotemporal interval of $\mathcal{B}$ob's \emph{motion}
\be
   (t,s)_{\overline{\mathcal{PB}}} \;\; \stackrel{(\ref{Formel - radar direkt motion physical measure})(\ref{Formel - radar indirekt spatiotemporal physical measure})}{=} \;\; \left( \; t_{\overline{\mathcal{P}\mathcal{A}_1}} \; + \; \frac{1}{2} \cdot t_{\overline{\mathcal{A}_1\mathcal{A}_2}}   \;\; , \;\;
      \frac{c}{2} \cdot t_{\overline{\mathcal{A}_1\mathcal{A}_2}}
   \; \right) \label{Formel - radar indirekt motion physical measure} \;\; .
\ee


\section{Lorentz transformation}\label{Kap - Masszsh}

\begin{figure}         
  \begin{center}         
  \includegraphics[height=19cm]{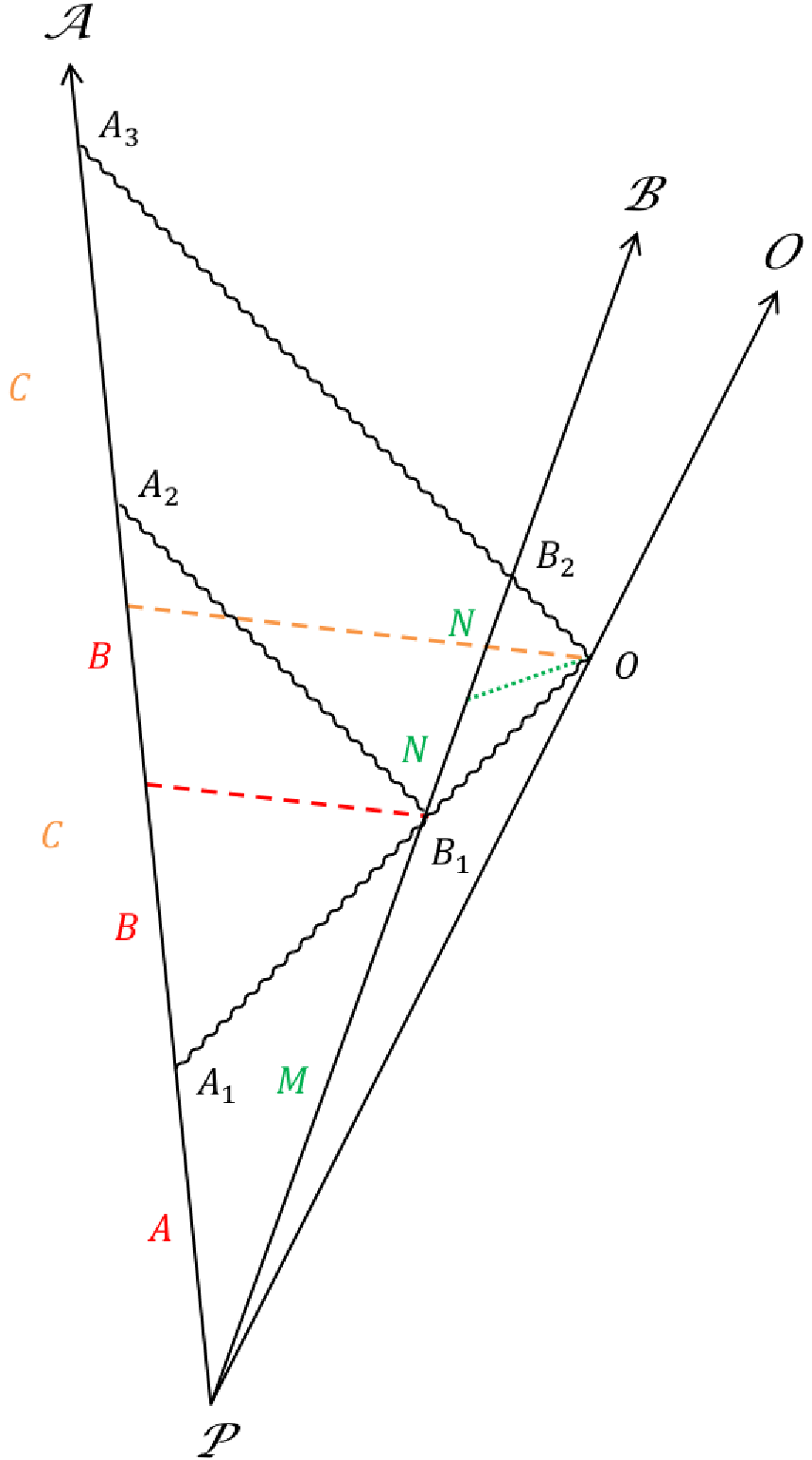}  
  \end{center}
  \vspace{-0.0cm}
  \caption{\label{figure-5} coinciding light rays in intrinsic measurements by $\mathcal{A}$lice and $\mathcal{B}$ob
    }
  \end{figure}

We have defined the termini of $\mathcal{A}$lice laser ranging measurements towards $\mathcal{O}$tto $\mathcal{P}\rightarrow\mathcal{A}_1\rightsquigarrow\mathcal{O}\rightsquigarrow\mathcal{A}_3$. Let another observer $\mathcal{B}$ob measure the same segment $\overline{\mathcal{PO}}$ of $\mathcal{O}$tto's motion (see figure \ref{figure-5}). $\mathcal{B}$ob conducts laser ranging $\mathcal{P}\rightarrow\mathcal{B}_1\rightsquigarrow\mathcal{O}\rightsquigarrow\mathcal{B}_2$ in the same way as $\mathcal{A}$lice. Following protophysical principles he manufactures his own light clock $\mathcal{L}^{(\mathcal{B})}$ and uses it in a standardized way. Step by step $\mathcal{B}$ob develops analogous measurement termini \{\ref{Kap - SRT Massbestimmung}\}.

$\mathcal{B}$ob \emph{constructs} his (dotted) simultaneity lines towards $\mathcal{O}$tto $\overline{\mathcal{BO}}$ (or back to $\mathcal{A}$lice $\overline{\mathcal{BA}}$). Directly, by adjacent connection of comoving light clocks $\mathcal{L}^{(\mathcal{B})} \ast_s \ldots \ast_s \mathcal{L}^{(\mathcal{B})}$, or indirectly, from round trip signaling times. Though, the \emph{same measurement principle} and intrinsic operations (independent propagation of light and intrinsic construction and connection of their respective light clocks) lead not to the same results. $\mathcal{A}$lice constructs simultaneity lines $\overline{\mathcal{AB}}$, $\overline{\mathcal{AO}}$ with different orientation than $\mathcal{B}$ob's simultaneity lines $\overline{\mathcal{BA}}$, $\overline{\mathcal{BO}}$ (see figure \ref{figure-3}).

Next $\mathcal{B}$ob \emph{encloses} measurement object $\mathcal{O}$tto $\overline{\mathcal{P}\mathcal{O}}$ in between his simultaneity lines $\overline{\mathcal{BO}}$
\[
   \overline{\mathcal{PO}} \;\; \stackrel{(\ref{Formel - radar direkt motion Eingrenzung})}{\sim_{t,s}}\;\; \overline{\mathcal{P}\mathcal{B}} \; \ast \; \overline{\mathcal{B}\mathcal{O}}
\]
and projects $\mathcal{O}$tto's relative motion onto the spatial and temporal components
\bea
   (t,s)_{\overline{\mathcal{PO}}} & = & (t,s)_{\overline{\mathcal{P}\mathcal{B}_1} \; \ast \; \overline{\mathcal{B}_1\mathcal{B}} \; \ast \; \overline{\mathcal{B}\mathcal{O}}} \nn \\
    & \stackrel{(\ref{Formel - radar direkt motion physical measure})}{=} & \left( \; t_{\overline{\mathcal{P}\mathcal{B}_1}} +  t_{\overline{\mathcal{B}_1\mathcal{B}}} \;\; , \;\; s_{\overline{\mathcal{B}\mathcal{O}}}  \; \right) \label{Formel - radar Bobs direkt motion physical measure of Otto} \;\;\; .
\eea

The intrinsic procedure is the same. By consecutive and adjacent connection of her light clocks $\mathcal{L}^{(\mathcal{A})}$ $\mathcal{A}$lice covers same segment of $\mathcal{O}$tto's motion as $\mathcal{B}$ob with his light clocks $\mathcal{L}^{(\mathcal{B})}$
\bea
   \overline{\mathcal{PO}} & \sim_{t,s} &  \;\; \left( \; t^{(\mathcal{A})}_{\overline{\mathcal{P}\mathcal{O}}} \;\: , \;\; s^{(\mathcal{A})}_{\overline{\mathcal{P}\mathcal{O}}} \; \right) \cdot \mathcal{L}^{(\mathcal{A})} \nn \\
    & \sim_{t,s} &  \;\; \left( \; t^{(\mathcal{B})}_{\overline{\mathcal{P}\mathcal{O}}} \;\: , \;\; s^{(\mathcal{B})}_{\overline{\mathcal{P}\mathcal{O}}} \; \right) \cdot \mathcal{L}^{(\mathcal{B})} \;\;\; . \nn
\eea
Let both also measure the same segment of $\mathcal{B}$ob's and of  $\mathcal{A}$lice' motion
\bea
   \overline{\mathcal{P}\mathcal{B}_1} & \sim_{t,s} &  \;\; \left( \; t^{(\mathcal{A})}_{\overline{\mathcal{P}\mathcal{B}_1}} \;\: , \;\; s^{(\mathcal{A})}_{\overline{\mathcal{P}\mathcal{B}_1}} \; \right) \cdot \mathcal{L}^{(\mathcal{A})} \nn \\
    & \sim_{t,s} &  \;\; \left( \; t^{(\mathcal{B})}_{\overline{\mathcal{P}\mathcal{B}_1}} \;\: , \;\; 0 \; \right) \cdot \mathcal{L}^{(\mathcal{B})}  \nn \\
    & & \nn \\
   \overline{\mathcal{P}\mathcal{A}_1} & \sim_{t,s} &  \;\; \left( \; t^{(\mathcal{A})}_{\overline{\mathcal{P}\mathcal{A}_1}} \;\: , \;\; 0 \; \right) \cdot \mathcal{L}^{(\mathcal{A})} \nn \\
    & \sim_{t,s} &  \;\; \left( \; t^{(\mathcal{B})}_{\overline{\mathcal{P}\mathcal{A}_1}} \;\: , \;\; s^{(\mathcal{B})}_{\overline{\mathcal{P}\mathcal{A}_1}} \; \right) \cdot \mathcal{L}^{(\mathcal{B})} \;\;\; . \nn
\eea
The coinciding light rays in their laser ranging processes are depicted in figure \ref{figure-5}.

From the interrelation of their physical prerequisites and the same measurement principle we derive the transformation $\left( \; t^{(\mathcal{A})}_{\overline{\mathcal{P}\mathcal{O}}} \; , \: s^{(\mathcal{A})}_{\overline{\mathcal{P}\mathcal{O}}} \; \right) \: \leftrightarrow \: \left( \; t^{(\mathcal{B})}_{\overline{\mathcal{P}\mathcal{O}}} \; , \: s^{(\mathcal{B})}_{\overline{\mathcal{P}\mathcal{O}}} \; \right)$ between $\mathcal{A}$lice and $\mathcal{B}$ob's measured values of the same measurement object: $\mathcal{O}$tto's motion $\overline{\mathcal{PO}}$. Provided measurements of their own motion $\overline{\mathcal{P}\mathcal{A}_1}$, $\overline{\mathcal{P}\mathcal{B}_1}$ it follows by successive substitution in three steps:
\[
\begin{array}{ccccc}
   \mathrm{\mathbf{I}} & \overbrace{\left( \; t^{(\mathcal{B})}_{\overline{\mathcal{P}\mathcal{O}}} \; , \; s^{(\mathcal{B})}_{\overline{\mathcal{P}\mathcal{O}}} \; \right)}^{\overline{\mathcal{PO}} \; \mathrm{measured \; by \;} \mathcal{B}\mathrm{ob}}
      &
    \left( \; M \; , \; N \; \right)  &  &
    \nn \\
   \mathrm{\mathbf{II}} &  & \left( \; M \; , \; N \; \right) & \left( \; A \; , \; B \; , \; C \; \right) &
    \nn \\
   \mathrm{\mathbf{III}} &  &  & \left( \; A \; , \; B \; , \; C \; \right) & \underbrace{\left(
    \left( \; t^{(\mathcal{A})}_{\overline{\mathcal{P}\mathcal{O}}} \; , \; s^{(\mathcal{A})}_{\overline{\mathcal{P}\mathcal{O}}} \; \right) \&
    \left( \; t^{(\mathcal{A})}_{\overline{\mathcal{P}\mathcal{B}_1}} \; , \; s^{(\mathcal{A})}_{\overline{\mathcal{P}\mathcal{B}_1}} \; \right) \right)}_{\overline{\mathcal{PO}} \; \& \; \overline{\mathcal{P}\mathcal{B}_1} \; \mathrm{measured \; by \;} \mathcal{A}\mathrm{lice}}
      \nn
\end{array}
\]
where we use the following abbreviations for indirect laser ranging measurements by $\mathcal{A}$lice
\[
   A := t^{(\mathcal{A})}_{\overline{\mathcal{P}\mathcal{A}_1}}  \;\;\;\;\;\;\;\;\;\;\;
   B := \frac{1}{2} \cdot t^{(\mathcal{A})}_{\overline{\mathcal{A}_1\mathcal{A}_2}} \;\;\;\;\;\;\;\;\;\;\;
   C := \frac{1}{2} \cdot t^{(\mathcal{A})}_{\overline{\mathcal{A}_1\mathcal{A}_3}}
\]
and by $\mathcal{B}$ob
\[
   M := t^{(\mathcal{B})}_{\overline{\mathcal{P}\mathcal{B}_1}}  \;\;\;\;\;\;\;\;\;\;\;
   N := \frac{1}{2} \cdot t^{(\mathcal{B})}_{\overline{\mathcal{B}_1\mathcal{B}_2}} \;\;\; .
\]

In \textbf{step I} we express $\mathcal{B}$ob's indirectly determined values of $\mathcal{O}$tto's motion $(t,s)^{(\mathcal{B})}_{\overline{\mathcal{PO}}}$ in terms of $\mathcal{B}$ob's direct measurements of round-trip signaling durations $M$, $N$
\bea
   t^{(\mathcal{B})}_{\overline{\mathcal{P}\mathcal{O}}}
   & \stackrel{(\ref{Formel - radar indirekt motion physical measure})}{=} & M + N \label{Formel - Masszsh - Substitution st MN i}\\
   s^{(\mathcal{B})}_{\overline{\mathcal{P}\mathcal{O}}}
   & \stackrel{(\ref{Formel - radar indirekt motion physical measure})}{=} & c \cdot N \label{Formel - Masszsh - Substitution st MN ii} \;\;\; .
\eea

In \textbf{step II} we express $\mathcal{B}$ob's measured durations $M$, $N$ in terms of $\mathcal{A}$lice' duration measurements $A$, $B$, $C$. In order to substitute the two physical quantities $M$, $N$ in terms of physical quantities $A$, $B$, $C$ we need two relations between corresponding measurements.

$\mathcal{A}$lice' and $\mathcal{B}$ob's laser ranging processes overlap in figure \ref{figure-5}. The outgoing light rays $\overline{\mathcal{A}_1\mathcal{B}_1}$ and $\overline{\mathcal{A}_1\mathcal{O}}$ partially coincide and the reflected light rays $\overline{\mathcal{B}_1\mathcal{A}_2}$ and $\overline{\mathcal{B}_2\mathcal{A}_3}$ are parallel \{\ref{Kap - SRT Massbestimmung - light principle}\}. From similar triangles $\mathcal{P}\mathcal{B}_1\mathcal{A}_2$ and $\mathcal{P}\mathcal{B}_2\mathcal{A}_3$ (all sides are pairwise parallel) we get one relation
\be\label{Formel - Masszsh - Substitution MN ABC i}
   \frac{M}{A+B+B} \; = \; \frac{M+N+N}{A+C+C} \;\;\; .
\ee

For a second relation we analyze the two triangles $\mathcal{P}\mathcal{A}_1\mathcal{B}_1$ and $\mathcal{P}\mathcal{B}_1\mathcal{A}_2$. We can regard each as ''degenerate trapezoid'', as a \emph{calibration} procedure by means of which $\mathcal{A}$lice and $\mathcal{B}$ob can compare their light clocks $\mathcal{L}^{(\mathcal{A})}$ and $\mathcal{L}^{(\mathcal{B})}$:
\begin{itemize}
\item   In $\mathcal{P}\mathcal{A}_1\mathcal{B}_1$  $\mathcal{A}$lice sends out two light signals along $\overline{\mathcal{P}\mathcal{A}_1} = A \cdot \mathcal{L}^{(\mathcal{A})}_t$ - the first at moment $P$ and the second  at moment $\mathcal{A}_1$ after $\sharp A$ ticks of her light clock - which $\mathcal{B}$ob receives along $\overline{\mathcal{P}\mathcal{B}_1} = M \cdot \mathcal{L}^{(\mathcal{B})}_t$ - at moments $P$ and $\mathcal{B}_1$ after $\sharp M$ ticks of his light clock.
\item   In $\mathcal{P}\mathcal{B}_1\mathcal{A}_2$  $\mathcal{B}$ob sends out two light signals along $\overline{\mathcal{P}\mathcal{B}_1} = M \cdot \mathcal{L}^{(\mathcal{B})}_t$ - at moments $P$ and $\mathcal{B}_1$ after $\sharp M$ ticks of his light clock - which $\mathcal{A}$lice receives along $\overline{\mathcal{P}\mathcal{A}_2} = (A+B+B) \cdot \mathcal{L}^{(\mathcal{A})}_t$ - the first in $P$ and the second in $\mathcal{A}_2$ after $\sharp (A+B+B)$ ticks of her light clock.
\end{itemize}
If $\mathcal{A}$lice and $\mathcal{B}$ob use identically constituted reference devices (light clocks made from same material) then both encounter the same dilation effect for each others relative motion. According to the \emph{relativity principle} both configurations are intrinsically similar. $\mathcal{A}$lice and $\mathcal{B}$ob have no way to specify absolute motion. By means of intrinsic measurements both determine the same ratio between the two durations for receiving both signals (heard from the other) and the duration of the sending interval (measured by themselves)
\be\label{Formel - Masszsh - Substitution MN ABC ii}
   \frac{M}{A} \; \stackrel{!}{=} \; \frac{A+B+B}{M} \;\;\; .
\ee

In \textbf{step III} we express $\mathcal{A}$lice laser ranging durations $A$, $B$, $C$ in terms of $\mathcal{A}$lice indirect determined values (\ref{Formel - radar indirekt motion physical measure}) for $\mathcal{O}$tto's motion $(t,s)^{(\mathcal{A})}_{\overline{\mathcal{PO}}}$ and for $\mathcal{B}$ob's motion $(t,s)^{(\mathcal{A})}_{\overline{\mathcal{PB}_1}}$
\bea
   A & = &  t^{(\mathcal{A})}_{\overline{\mathcal{P}\mathcal{O}}} \; - \; \frac{1}{c} \cdot s^{(\mathcal{A})}_{\overline{\mathcal{P}\mathcal{O}}} \label{Formel - Masszsh - Substitution ABC st i} \\
   & = &  t^{(\mathcal{A})}_{\overline{\mathcal{P}\mathcal{B}}} \; - \; \frac{1}{c} \cdot s^{(\mathcal{A})}_{\overline{\mathcal{P}\mathcal{B}}} \label{Formel - Masszsh - Substitution ABC st ii}\\
   B & = & \frac{1}{c} \cdot s^{(\mathcal{A})}_{\overline{\mathcal{P}\mathcal{B}}} \label{Formel - Masszsh - Substitution ABC st iii}\\
   C & = & \frac{1}{c} \cdot s^{(\mathcal{A})}_{\overline{\mathcal{P}\mathcal{O}}} \label{Formel - Masszsh - Substitution ABC st iv} \;\;\; .
\eea

After successive insertion of these three steps (see appendix A) we can express $\mathcal{B}$ob's physical quantities of $\mathcal{O}$tto's motion $(t,s)^{(\mathcal{B})}_{\overline{\mathcal{PO}}}$ in terms of $\mathcal{A}$lice measurements of $\mathcal{O}$tto's motion $(t,s)^{(\mathcal{A})}_{\overline{\mathcal{PO}}}$ and of the relative motion of $\mathcal{B}$ob $(t,s)^{(\mathcal{A})}_{\overline{\mathcal{PB}_1}}$
\bea\label{Formel - Masszsh - Lorentz Trafo st matrix}
   t^{(\mathcal{B})}_{\overline{\mathcal{P}\mathcal{O}}} & \stackrel{(\ref{Formel - Masszsh - Lorentz Trafo t})}{=} &
   \;\;\;\;\;\: \frac{1}{\sqrt{1-\frac{v_{\mathcal{B}}^2}{c^2}}} \cdot \;\; t^{(\mathcal{A})}_{\overline{\mathcal{P}\mathcal{O}}} \;\;\;\;\: - \;\; \frac{1}{\sqrt{1-\frac{v_{\mathcal{B}}^2}{c^2}}} \cdot \frac{v_{\mathcal{B}}}{c^2} \cdot \: s^{(\mathcal{A})}_{\overline{\mathcal{P}\mathcal{O}}} \\
   s^{(\mathcal{B})}_{\overline{\mathcal{P}\mathcal{O}}} & \stackrel{(\ref{Formel - Masszsh - Lorentz Trafo s})}{=} & \!\!
   - \; \frac{1}{\sqrt{1-\frac{v_{\mathcal{B}}^2}{c^2}}} \cdot v_{\mathcal{B}} \cdot \: t^{(\mathcal{A})}_{\overline{\mathcal{P}\mathcal{O}}}  \;\;\; + \;\;\;\;\;\; \frac{1}{\sqrt{1-\frac{v_{\mathcal{B}}^2}{c^2}}} \cdot \;\; s^{(\mathcal{A})}_{\overline{\mathcal{P}\mathcal{O}}}  \nn
\eea
where $\mathcal{A}$lice determines $\mathcal{B}$ob's relative velocity from $v_{\mathcal{B}} := s^{(\mathcal{A})}_{\overline{\mathcal{P}\mathcal{B}}} \!\left/\! t^{(\mathcal{A})}_{\overline{\mathcal{P}\mathcal{B}}}\right.$.

\begin{figure}[t!]         
  \begin{center}         
  \includegraphics[height=8.5cm]{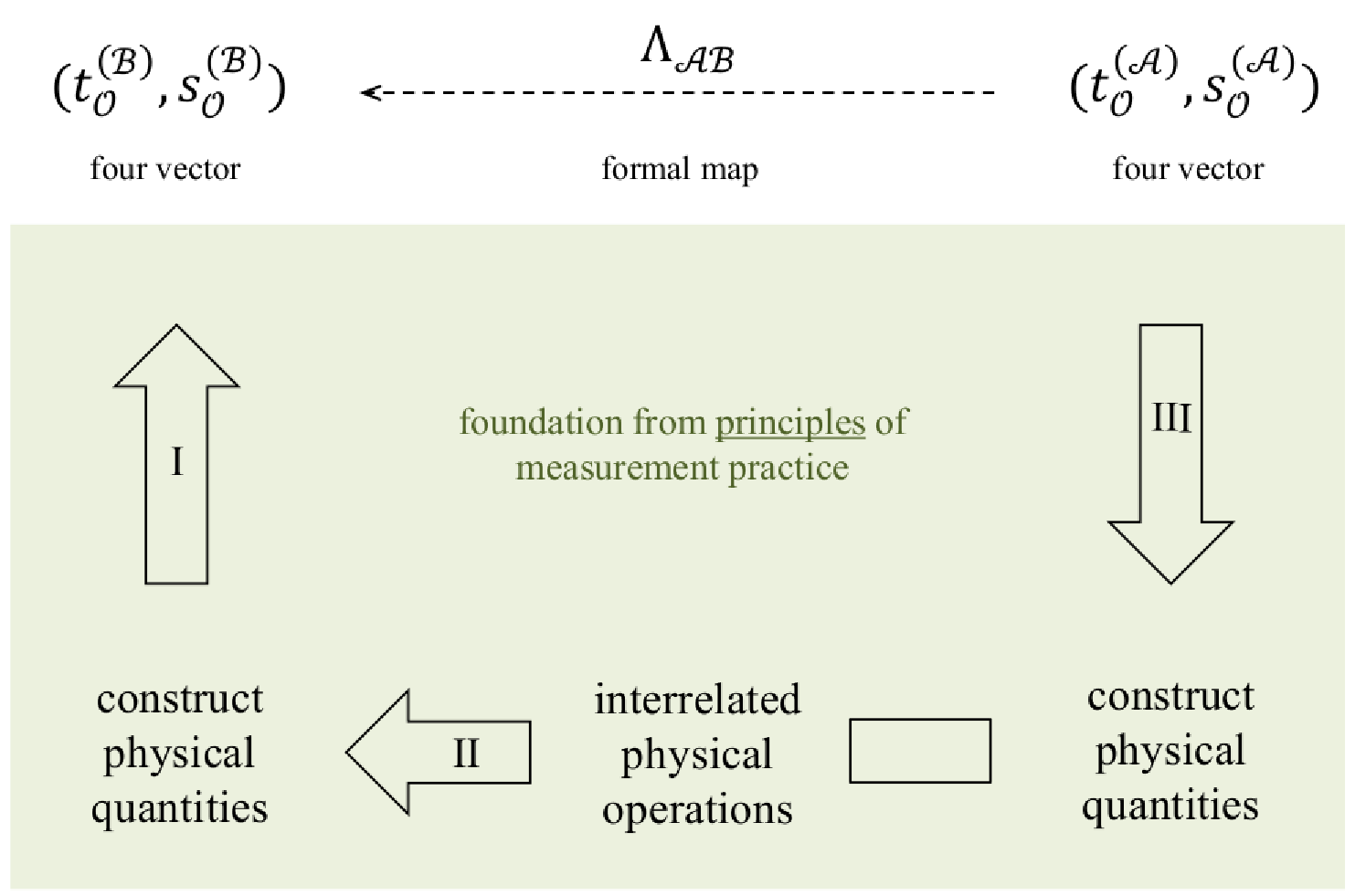}  
  \end{center}
  \vspace{-0.0cm}
  \caption{\label{figure-KERN-Bild} physical basis of the mathematical formulation
    }
  \end{figure}

We derive the Lorentz transformation $\Lambda_{\mathcal{A}\mathcal{B}} : (t,s)^{(\mathcal{A})} \mapsto (t,s)^{(\mathcal{B})}$ in a commutative diagram by three successive pull backs through - the \emph{principles of tangible measurement} in - in the natural world (see figure \ref{figure-KERN-Bild}). We derive the mathematical relation from $\mathcal{A}$lice and $\mathcal{B}$ob's intrinsic construction of physical quantities of $\mathcal{O}$tto's motion (step I and III) and from their overlapping laser ranging operations (step II) (see figure \ref{figure-KERN-Bild}).
While the formal approach assumes vectors, Lorentz symmetry for its description; we form the mathematical framework. From simple measurement-methodical principles - without mathematical presuppositions - we generate physical quantities  $\left(t^{(\mathcal{A})}_{\overline{\mathcal{P}\mathcal{O}}} \,,\: s^{(\mathcal{A})}_{\overline{\mathcal{P}\mathcal{O}}}\right)$. They specify the layout and number of building blocks in the material model $\mathcal{L}\ast\ldots\ast\mathcal{L}$ which $\mathcal{A}$lice assembles to cover the ''duration'' and ''length'' of measurement object $\overline{\mathcal{P}\mathcal{O}}$. From the interrelation of the underlying practical operations we derive the ''local Lorentz symmetry''.

From classical measurement practice we get Galilei kinematics. Einstein analyzed mutual measurements of moving objects and recognized the need to establish a physical connection between clocks at different locations and speeds. Intrinsic operations with light clocks represent the classical metric locally (Euclidean geometry). For their connection Einstein chose the universal motion of light. By including the (local) light principle and the relativity principle we derive Poincare kinematics. Our locally regular composable grid of light clocks can potentially grow into every direction. It is our \emph{metric connection} between distant measurements. Then the formerly isolated and local notions of the classical metric (absolute time, space, local flatness etc.) will reveal new intricate interrelations.


\section{Twin paradox}\label{Kap - Twinparadox}

\begin{figure}         
  \begin{center}         
  \includegraphics[height=18cm]{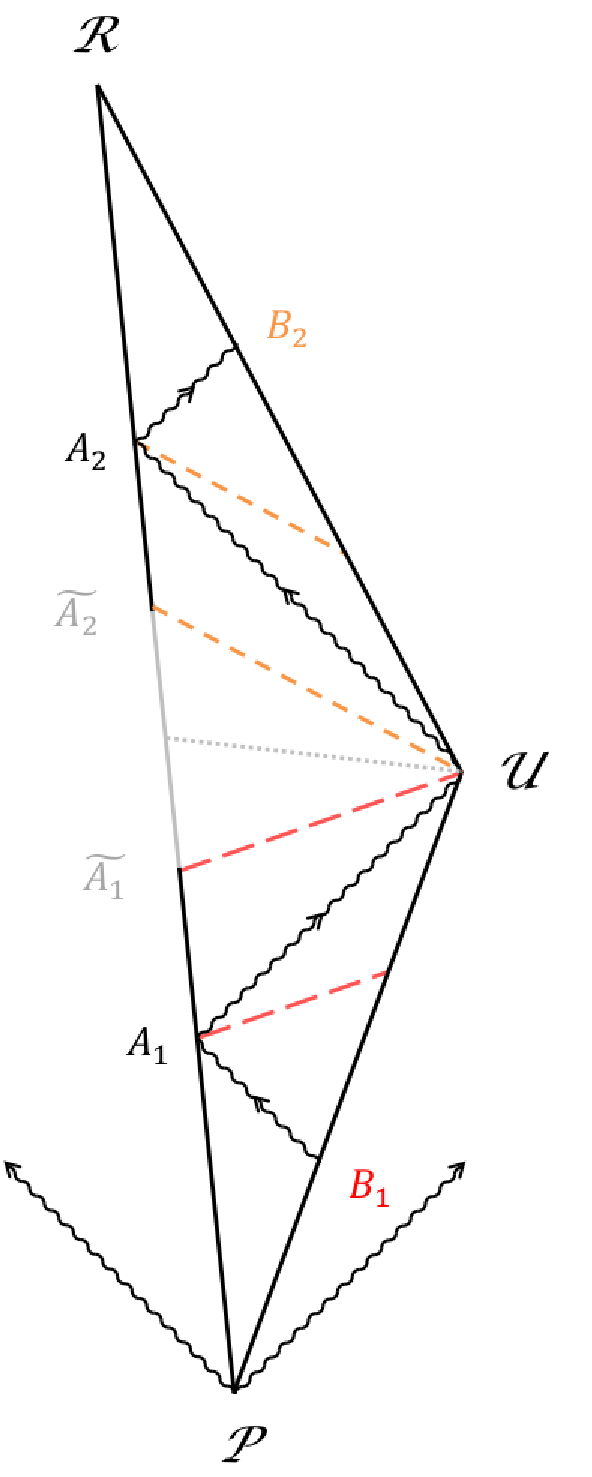}  
  \end{center}
  \vspace{-0.0cm}
  \caption{\label{figure-6} connected laser rangings in the Twin configuration of $\mathcal{A}$lice and $\mathcal{B}$ob$_{1,2}$
    }
  \end{figure}

In the Twin configuration $\mathcal{A}$lice and $\mathcal{B}$ob explore their mutual time dilation in a round trip experiment. They depart at moment $\mathcal{P}$. While $\mathcal{A}$lice remains at rest $\mathcal{B}$ob rides with uniform motion $v^{(\mathcal{A})}_{\mathcal{B}}$ to a distant turning point $\mathcal{U}$ and returns with same velocity $-v^{(\mathcal{A})}_{\mathcal{B}}$ to reunite with $\mathcal{A}$lice in future moment $\mathcal{R}$ (see figure \ref{figure-6}).

Throughout the whole round trip $\mathcal{A}$lice can observe $\mathcal{B}$ob. Vice versa $\mathcal{B}$ob will receive all light signals which $\mathcal{A}$lice sends from departure until return. The time during which $\mathcal{A}$lice measures $\mathcal{B}$ob's journey $t^{(\mathcal{A})}_{\mathcal{B}} \equiv t^{(\mathcal{A})}_{\mathcal{A}}$ coincides with her (proper) waiting time; then $\mathcal{B}$ob spends less time on tour $t^{(\mathcal{B})}_{\mathcal{B}} = \underbrace{\sqrt{1-\frac{v_{\mathcal{B}}^2}{c^2}}}_{<1} \; \cdot \; t^{(\mathcal{A})}_{\mathcal{A}}$
than $\mathcal{A}$lice waiting and watching (the previous measurement object $\overline{\mathcal{P}\mathcal{O}}\equiv\mathcal{B}$ coincides with $\mathcal{B}$ob's ride $(t,s)^{(\mathcal{B})}_{\mathcal{B}} \stackrel{(\ref{Formel - Masszsh - Lorentz Trafo st matrix})}{=} \left(\sqrt{1-\frac{v_{\mathcal{B}}^2}{c^2}} \cdot t^{(\mathcal{A})}_{\mathcal{B}}\, , \: 0 \: \right)$). She observes her own clock ticks faster than the moving clock of $\mathcal{B}$ob and vice versa by the symmetry of their relative motion $v^{(\mathcal{B})}_{\mathcal{A}} = - v^{(\mathcal{A})}_{\mathcal{B}}$. During his trip $\mathcal{B}$ob can see all of $\mathcal{A}$lice; if we assume his \emph{observation} time $t^{(\mathcal{B})}_{\mathcal{B}}$
\be
   t^{(\mathcal{B})}_{\mathcal{B}} \stackrel{?}{=} t^{(\mathcal{B})}_{\mathcal{A}} \label{Formel - twins - Bob Beobachtungszeit Wartezeit}
\ee
is the time to \emph{measure} all of $\mathcal{A}$lice $t^{(\mathcal{B})}_{\mathcal{A}}$, then she spends less time waiting  $t^{(\mathcal{A})}_{\mathcal{A}} \! \stackrel{(\ref{Formel - twins - Bob Beobachtungszeit Wartezeit})}{=} \! \underbrace{\sqrt{1-\frac{v_{\mathcal{B}}^2}{c^2}}}_{<1} \; \cdot \; t^{(\mathcal{B})}_{\mathcal{B}}$ than the ride takes for $\mathcal{B}$ob himself (using the reverse Lorentz transformation of $\mathcal{A}$lice waiting interval
$(t,s)^{(\mathcal{A})}_{\mathcal{A}} \stackrel{(\ref{Formel - Masszsh - Lorentz Trafo st matrix})}{=} \left( \sqrt{1-\frac{v_{\mathcal{B}}^2}{c^2}} \; \cdot \; t^{(\mathcal{B})}_{\mathcal{A}} \, , \: 0 \: \right)$). The combined \emph{conclusion} is a contradiction $t^{(\mathcal{B})}_{\mathcal{B}} < t^{(\mathcal{A})}_{\mathcal{A}} < t^{(\mathcal{B})}_{\mathcal{B}}$; the so called \emph{Twin paradox}.

The assumption (\ref{Formel - twins - Bob Beobachtungszeit Wartezeit}) was incorrect; $\mathcal{B}$ob's observation period $t^{(\mathcal{B})}_{\mathcal{B}} \neq t^{(\mathcal{B})}_{\mathcal{A}}$ is not the time for measuring $\mathcal{A}$lice. The Lorentz transformation (\ref{Formel - Masszsh - Lorentz Trafo st matrix}) between physical measures does not refer to durations of observations; it refers to durations of their measurements. For a physically meaningful calculation we remember that basic physical quantities $(s,t)$ originate from tangible operations. The implicit conditions (for constructing the underlying grid of light clock units) must be fulfilled \emph{before} one can apply the Lorentz transformation between results of supposed measurements. Our completed view prevents unreflective calculations in the Lorentz formalism and provides an explanation of the apparent Twin paradox from a measurement-methodical perspective.

$\mathcal{A}$lice can observe and measure $\mathcal{B}$ob throughout the whole trip, unlike $\mathcal{B}$ob. He can receive all light signals from $\mathcal{A}$lice. Though the middle segment of $\mathcal{A}$lice motion $\overline{\widetilde{\mathcal{A}}_1\widetilde{\mathcal{A}}_2}$ is \emph{observable but not measurable} for him. $\mathcal{B}$ob's light clocks $\mathcal{L}^{(\mathcal{B}_1)}$ and $\mathcal{L}^{(\mathcal{B}_2)}$ function properly on the way out and back. Though on each leg $\mathcal{B}$ob cannot cover measurement objects beyond the $\mathcal{U}$-turn point. $\mathcal{B}$ob$_1$'s last indirect laser ranging $\mathcal{B}_1\rightsquigarrow\mathcal{A}_1\rightsquigarrow\widetilde{\mathcal{B}}_1$ reaches $\mathcal{A}$lice at moment $\mathcal{A}_1$ so the reflection (to analyze round trip times) returns before he changes his motion at moment $\mathcal{U}$ (violating measurement condition \{\ref{Kap - SRT Massbestimmung - indirect laser ranging}\}). His direct laser ranging by consecutive and adjacent connection of his light clocks covers $\mathcal{A}$lice up to moment $\widetilde{\mathcal{A}}_1$. Later $\mathcal{B}$ob$_2$ assembles his light clocks $\mathcal{L}^{(\mathcal{B}_2)}$ next to one another forming simultaneity lines with respect to his new state of motion. His direct and indirect laser ranging measurements cover $\mathcal{A}$lice from moment $\widetilde{\mathcal{A}}_2$ resp. $\mathcal{A}_2$ up to the point of $\mathcal{R}$eturn. $\mathcal{B}$ob$_1$ and $\mathcal{B}$ob$_2$ cannot measure segment $\overline{\widetilde{\mathcal{A}}_1\widetilde{\mathcal{A}}_2}$ of $\mathcal{A}$lice motion by intrinsic use of their light clock units.

During his entire round trip $\mathcal{B}$ob measures two segments $\overline{\mathcal{P}\widetilde{\mathcal{A}}_1}$ and $\overline{\widetilde{\mathcal{A}}_2\mathcal{R}}$ of $\mathcal{A}$lice motion. He sees all processes for moving $\mathcal{A}$lice run slower by the same factor $t^{(\mathcal{A})}_{\mathcal{A}} = \sqrt{1-\frac{v_{\mathcal{B}}^2}{c^2}} \; \cdot \; t^{(\mathcal{B})}_{\mathcal{A}}$ as in $\mathcal{A}$lice reverse perspective. Though $\mathcal{B}$ob$_1$ covers a shorter segment of $\mathcal{A}$lice relative motion
\be
   \overline{\mathcal{P}\widetilde{\mathcal{A}}_1} \;\; \sim_{t,s} \;\; \left( \;\; t^{(\mathcal{B}_1)}_{\mathcal{B}_1} \; , \;\; v_{\mathcal{A}} \cdot t^{(\mathcal{B}_1)}_{\mathcal{B}_1} \; \right) \: \cdot \mathcal{L}^{(\mathcal{B}_1)} \nn
\ee
with proper duration $t^{(\mathcal{A})}_{\overline{\mathcal{P}\widetilde{\mathcal{A}}_1}} \stackrel{(\ref{Formel - Masszsh - Lorentz Trafo st matrix})}{=} \sqrt{1-\frac{v_{\mathcal{B}}^2}{c^2}} \cdot t^{(\mathcal{B}_1)}_{\mathcal{B}_1} \stackrel{(\ref{Formel - Masszsh - Lorentz Trafo st matrix})}{=} \left( 1-\frac{v_{\mathcal{B}}^2}{c^2} \right) \cdot \frac{t^{(\mathcal{A})}_{\mathcal{A}}}{2}$. Similarly during his return $\mathcal{B}$ob$_2$ measures a segment of $\mathcal{A}$lice with same proper duration $t^{(\mathcal{A})}_{\overline{\widetilde{\mathcal{A}}_2\mathcal{R}}}$. With regard to measurability by $\mathcal{B}$ob the waiting process of $\mathcal{A}$lice $\overline{\mathcal{P}\mathcal{R}} \equiv \overline{\mathcal{P}\widetilde{\mathcal{A}}_1} \ast \overline{\widetilde{\mathcal{A}}_1\widetilde{\mathcal{A}}_2} \ast \overline{\widetilde{\mathcal{A}}_2\mathcal{R}}$ splits into measurable segments $\overline{\mathcal{P}\widetilde{\mathcal{A}}_1}$ and $\overline{\widetilde{\mathcal{A}}_2\mathcal{R}}$ and non-measurable segment $\overline{\widetilde{\mathcal{A}}_1\widetilde{\mathcal{A}}_2}$. Her waiting time divides accordingly
\bea
   t_{\mathcal{A}} \; & = & \; t_{\overline{\mathcal{P}\widetilde{\mathcal{A}}_1}} \; + \; t_{\overline{\widetilde{\mathcal{A}}_2\mathcal{R}}} \;\;\;\; + \;\;\;\; t_{\overline{\widetilde{\mathcal{A}}_1\widetilde{\mathcal{A}}_2}} \nn \\
   & = & \left( 1-\frac{v_{\mathcal{B}}^2}{c^2} \right) \cdot t_{\mathcal{A}} \;\;\; + \;\;\; \frac{v_{\mathcal{B}}^2}{c^2} \cdot t_{\mathcal{A}} \;\; . \nn
\eea
In the limit where $\mathcal{B}$ob approaches speed of light $v_{\mathcal{B}_i} \rightarrow c$ his total trip duration vanishes
\[
   t_{\mathcal{B}_1} + t_{\mathcal{B}_2} \stackrel{(\ref{Formel - Masszsh - Lorentz Trafo st matrix})}{=} \sqrt{1-\frac{v_{\mathcal{B}}^2}{c^2}}  \cdot t_{\mathcal{A}} \;\; \rightarrow \;\; 0
\]
while during (observable but) unmeasurable part of her waiting process $\mathcal{A}$lice grows older
\[
   t_{\overline{\widetilde{\mathcal{A}}_1\widetilde{\mathcal{A}}_2}} = \frac{v_{\mathcal{B}}^2}{c^2} \cdot t_{\mathcal{A}} \;\; \rightarrow \;\; t_{\mathcal{A}} \;\;\; .
\]

While the formal explanation of the Twin paradox assumes four-vectors $x^{\mu}$, Minkowski metric $\eta_{\mu\nu}$ and ''integrates proper time along a curved worldline'' our measurement-methodical foundation reveals the physical reason and it derives the rules of the calculus as well.

\section{Mathematical formulation of physical operations}

We have applied Helmholtz program of direct measurements to relativistic kinematics. We can compare the spatiotemporal order of two objects by the classical probe, whether one of them covers the other. Without one word of mathematics one can manufacture identically constituted light clocks $\mathcal{L}$ and place them literally side by side or one after the other. $\mathcal{A}$lice concatenates these measurement units by a physical process, by letting their inner light rays overlap. Her measurements are based on light principle, classical construction of light clocks $\mathcal{L}$ and their direct and indirect connection by the independent motion of light.

Mathematics comes into being at the moment we introduce units and count, how many congruent building blocks it takes for assembling a regular grid $\mathcal{L} \ast \dots \ast \mathcal{L} \sim_{s} \mathcal{O}$ which covers the measurement object. If both are interchangeable in the comparison they have same length (up to a non-vanishing but practically admissible \emph{measurement error})
\[
   s_{\mathcal{O}} \;\; = \;\;  s_{\mathcal{L} \ast \dots \ast \mathcal{L}} \; + \; \Delta s  \;\; .
\]
\begin{rem}\label{Rem - SRT Kin - doubling of physical measures}
Helmholtz way of basic measurement involves a pair comparison. Measurement object and material model are natural objects. $\mathcal{A}$lice covers the spatiotemporal interval of e.g. relative motion of $\mathcal{B}$ob by a regular grid of ticking light clocks. It is built of solely congruent building blocks $\mathcal{L}$ and it is (locally) invariant under permuting their order; by counting them $\mathcal{A}$lice can measure ''how many times'' further and longer $\mathcal{B}$ob's relative motion spreads than the (universal) light in her reference device.
\end{rem}
$\mathcal{A}$lice builds wide layouts of (ticking) light clocks $\mathcal{L} \ast_s \ldots \ast_s \mathcal{L}$ and enduring sequences of (light clock) ticks $\mathcal{L} \ast_t \ldots \ast_t \mathcal{L}$. All steps of her procedure (assembling the material model and conducting length comparison $\sim_s$) are reproducible by any other physicist -- the practical purpose of standardizing measurements (of the magnitude of durations and lengths).\footnote{Her technique is preserved until in empirical practice one oversteps unforeseen conditions, physically specifies them further and thus evolves - in a continual historic process - measurement practice and its (physically determined) mathematical formulation. -- Remembering Feynman's motto: \emph{Yesterday's sensation is today's calibration and tomorrow's background}.}

In starting figure \ref{figure-1} we illustrate only observed objects and observers in motion. Provided the construction of light clocks (measurement unit $\mathcal{L}$) and their connection in consecutive and adjacent ways (concatenation $\ast_t$, $\ast_s$) we introduce increasingly complex material models which where not yet assembled in the uncultivated beginning. As a colloquial expression we introduce \emph{measurement termini}. Along figures \ref{figure-2}, \ref{figure-3}, \ref{figure-4} we define \emph{physical notions} based on measurement operations with light clocks (simultaneous straight measurement path, spatial and temporal projection etc.). Step by step we introduce operational denominations which specify aspects of $\mathcal{A}$lice measurement practice precisely.

From the underlying operational definitions their interrelation becomes transparent. Their common origin inherits a genetic interrelation between measurement termini and the corresponding terms in the mathematical formulation. Based on measurement-methodical principles of their formation we can avoid apparent paradoxes in blind calculations. In retrospect of practical operations with light clocks we emphasize, that all our assertions on the one-way propagation of light strictly come from closed two-way cycles. In laser ranging configuration $\mathcal{A}_1\!\rightsquigarrow\mathcal{B}\rightsquigarrow\mathcal{A}_2$ $\mathcal{A}$lice has no way to measure whether the departing light ray $\overline{\mathcal{A}_1\mathcal{B}}$ towards $\mathcal{B}$ob takes more time than the returning ray $\overline{\mathcal{B}\mathcal{A}_2}$. Similarly we cannot figure what happens \emph{inside} the measurement unit $\mathcal{L} : \mathcal{L}_I\rightsquigarrow \mathcal{L}_{II} \rightsquigarrow \mathcal{L}_I \ldots$ (when light travels between both mirrors to the right $\mathcal{L}_I\rightsquigarrow \mathcal{L}_{II}$ vs. to the left $\mathcal{L}_{II}\rightsquigarrow \mathcal{L}_{I}$).

Direct and indirect laser ranging with light clocks $\mathcal{L}$ always involves both, outgoing and returning pulse. In practice we solely deal with two-way light cycles: (i) inside individual light clocks $\mathcal{L}$ and (ii) in suitably connected configurations of light clocks. Our measurement unit $\mathcal{L}$ comprises both dimensions, width $s_{\mathcal{L}}$ and duration $t_{\mathcal{L}}$ of an elementary two-way light cycle. For the measurement we generate complex configurations of two-way light cycles in suitable layouts of ticking light clocks $\mathcal{L} \ast_t \ldots \ast_t \mathcal{L} \ast_s \ldots \ast_s \mathcal{L}$. Thus $\mathcal{A}$lice covers one-way light rays $\overline{\mathcal{A}_1\mathcal{B}}$ and $\overline{\mathcal{B}\mathcal{A}_2}$ by a layout of two-way light cycles (in light clock grid figure \ref{figure-3}). By counting ticks along her waiting interval $\overline{\mathcal{A}_1\mathcal{A}_2}$ and the light clocks sitting side by side to cover all of laser ranging path $\overline{\mathcal{A}\mathcal{B}}$ $\mathcal{A}$lice measures the magnitude of their length and duration
$ (t,s)_{\overline{\mathcal{A}_1\mathcal{B}}} \stackrel{(\ref{Formel - radar indirekt spatiotemporal physical measure})}{=} \frac{1}{2} \cdot t_{\overline{\mathcal{A}_1\mathcal{A}_2}} + \frac{c}{2} \cdot t_{\overline{\mathcal{A}_1\mathcal{A}_2}}$ resp. $ (t,s)_{\overline{\mathcal{B}\mathcal{A}_2}} = \frac{1}{2} \cdot t_{\overline{\mathcal{A}_1\mathcal{A}_2}} - \frac{c}{2} \cdot t_{\overline{\mathcal{A}_1\mathcal{A}_2}}$.
\begin{rem}\label{Rem - SRT Kin - inseparable unit}
The meaning of basic physical quantities arises - not by chopping measurement units $\mathcal{L}$ into pieces but instead - by concatenating many congruent measurement units $\mathcal{L}$ (each taken as inseparable unity) to construct material models $\mathcal{L} \ast \dots \ast \mathcal{L}$.
\end{rem}
We introduce all arithmetic operations between measures ''$+$'', ''$-$'', ''$\frac{1}{2} \; \cdot$'' etc. via underlying connection of congruent light clocks. Basic physical quantities specify a reproducible layout of reference devices which covers the measurement objects sufficiently precise.
\\

Our objective is a definition of basic observables from physical operations (what one does in measurement practice). \emph{In absence} of interactions we have developed Helmholtz method for basic measurements of relativistic motion. In this approach, which derives the mathematical formalism of kinematics from this operationalization of length and duration, one can address scope and limitations of the formalism. It can be taken as a basis for our next step, basic measurements of interactions. Next we develop Helmholtz method for the foundation of classical \{\ref{Kap - Mechanics}\} and relativistic dynamics \{\ref{Kap - Relativistic energy-momentum}\}.

\pagebreak

\section*{Appendix A: Successive substitution}\label{Appendix - Rechnungsdetails}

In step I of our series of substitutions we express the space and time component of $\mathcal{B}$ob's physical measure
$\left( \; t^{(\mathcal{B})}_{\overline{\mathcal{P}\mathcal{O}}} \; , \; s^{(\mathcal{B})}_{\overline{\mathcal{P}\mathcal{O}}} \; \right)$
in terms of his laser ranging duration measurements $M$, $N$
\bea
   t^{(\mathcal{B})}_{\overline{\mathcal{P}\mathcal{O}}}
   & \stackrel{(\ref{Formel - Masszsh - Substitution st MN i})}{=} & M + N \nn \\
   s^{(\mathcal{B})}_{\overline{\mathcal{P}\mathcal{O}}}
   & \stackrel{(\ref{Formel - Masszsh - Substitution st MN ii})}{=} & c \cdot N \nn \;\;\; .
\eea
In step II we substitute $\mathcal{B}$ob's laser ranging durations $M$, $N$ - due to the interrelation of their measurement conditions - with $\mathcal{A}$lice laser ranging durations $A$, $B$, $C$
\bea
   M & \stackrel{(\ref{Formel - Masszsh - Substitution MN ABC ii})}{=} & \sqrt{A\cdot(A+B+B)} \label{Formel - Masszsh - Substitution MN ABC - N explizit i} \\
   N & \stackrel{(\ref{Formel - Masszsh - Substitution MN ABC i})(\ref{Formel - Masszsh - Substitution MN ABC ii})}{=} & \frac{1}{2} \cdot \sqrt{A\cdot(A+B+B)} \cdot \frac{A+C+C}{A+B+B} \;\; - \;\; \frac{1}{2} \cdot \sqrt{A\cdot(A+B+B)}
    \label{Formel - Masszsh - Substitution MN ABC - N explizit ii} \;\;\; .
\eea
In step III finally we reformulate $\mathcal{A}$lice laser ranging durations $A$, $B$, $C$ in terms of the space and time components of $\mathcal{A}$lice's physical measures $\left( \; t^{(\mathcal{A})}_{\overline{\mathcal{P}\mathcal{O}}} \; , \; s^{(\mathcal{A})}_{\overline{\mathcal{P}\mathcal{O}}} \; \right)$ and $\left( \; t^{(\mathcal{A})}_{\overline{\mathcal{P}\mathcal{B}}} \; , \; s^{(\mathcal{A})}_{\overline{\mathcal{P}\mathcal{B}}} \; \right)$. We successively insert all substitutions for the space and time component separately
\bea
   s^{(\mathcal{B})}_{\overline{\mathcal{P}\mathcal{O}}} \!\!
   & \stackrel{(\ref{Formel - Masszsh - Substitution st MN ii})}{=} & c \cdot N \nn \\
   & \stackrel{(\ref{Formel - Masszsh - Substitution MN ABC - N explizit ii})}{=} & c \cdot \left[ \frac{1}{2} \cdot \sqrt{A\cdot(A+B+B)} \cdot \frac{A+C+C}{A+B+B} \; - \; \frac{1}{2} \cdot \sqrt{A\cdot(A+B+B)} \cdot \underbrace{\frac{\sqrt{A+B+B}}{\sqrt{A+B+B}}}_{=1} \; \right]
   \nn \\
   & = & c \cdot \frac{1}{2} \cdot \frac{\sqrt{A}}{\sqrt{A+B+B}} \cdot (A+C+C) \; - \; c \cdot \frac{1}{2} \cdot \frac{\sqrt{A}}{\sqrt{A+B+B}} \cdot (A+B+B) \nn \\
   & = & c \cdot \underbrace{\frac{\sqrt{A}}{\sqrt{A}}}_{=1} \cdot \frac{\sqrt{A}}{\sqrt{A+B+B}} \cdot (C-B) \nn \\
   & = & \frac{1}{\sqrt{A\cdot(A+B+B)}} \; \cdot \; \left(\; -\; c \cdot A \cdot B \; + \; c \cdot A \cdot C\right) \nn \\
   & = & \frac{1}{\sqrt{A\cdot(A+B+B)}} \; \cdot \left( \; -\; c \cdot A \cdot B \; \underbrace{ -\; c \cdot B \cdot C \; + \; c \cdot B \cdot C}_{=0} \; + \; c \cdot A \cdot C \right) \nn \\
   & = & \frac{1}{\sqrt{A\cdot(A+B+B)}} \; \cdot \; \left( \; -\; c \cdot B \cdot ( A + C ) \;\; + \;\; c \cdot C \cdot ( A + B ) \right) \nn
\eea
\bea
   & \stackrel{(\ref{Formel - Masszsh - Substitution ABC st i})-(\ref{Formel - Masszsh - Substitution ABC st iv})}{=} & \frac{1}{\sqrt{ \left( t^{(\mathcal{A})}_{\overline{\mathcal{P}\mathcal{B}}} \; - \; \frac{1}{c} \cdot s^{(\mathcal{A})}_{\overline{\mathcal{P}\mathcal{B}}} \right) \cdot \left( t^{(\mathcal{A})}_{\overline{\mathcal{P}\mathcal{B}}} \; + \; \frac{1}{c} \cdot s^{(\mathcal{A})}_{\overline{\mathcal{P}\mathcal{B}}} \right) }} \cdot  \left( - s^{(\mathcal{A})}_{\overline{\mathcal{P}\mathcal{B}}} \cdot t^{(\mathcal{A})}_{\overline{\mathcal{P}\mathcal{O}}} \; + \; s^{(\mathcal{A})}_{\overline{\mathcal{P}\mathcal{O}}} \cdot t^{(\mathcal{A})}_{\overline{\mathcal{P}\mathcal{B}}}  \right) \nn \\
   & = & \frac{t_{\overline{\mathcal{P}\mathcal{B}}}}{\sqrt{  {t_{\overline{\mathcal{P}\mathcal{B}}}}^2 \; - \; \frac{1}{c^2} \cdot {s_{\overline{\mathcal{P}\mathcal{B}}}}^2  }} \; \cdot \; \left( - \frac{s_{\overline{\mathcal{P}\mathcal{B}}}}{t_{\overline{\mathcal{P}\mathcal{B}}}} \cdot t_{\overline{\mathcal{P}\mathcal{O}}} \; + \; s_{\overline{\mathcal{P}\mathcal{O}}}  \right) \nn \\
   & = & \frac{1}{\sqrt{ \frac{{t_{\overline{\mathcal{P}\mathcal{B}}}}^2}{{t_{\overline{\mathcal{P}\mathcal{B}}}}^2}  \; - \; \frac{1}{c^2} \cdot \frac{{s_{\overline{\mathcal{P}\mathcal{B}}}}^2}{{t_{\overline{\mathcal{P}\mathcal{B}}}}^2}  }} \; \cdot \; \left( - v_{\mathcal{B}} \cdot t_{\overline{\mathcal{P}\mathcal{O}}} \; + \; s_{\overline{\mathcal{P}\mathcal{O}}}  \right) \nn \\
s^{(\mathcal{B})}_{\overline{\mathcal{P}\mathcal{O}}}
   & = &   - \; \frac{1}{\sqrt{ 1  \; - \; \frac{ {v_{\mathcal{B}}}^2 }{c^2} }} \; \cdot \; v_{\mathcal{B}} \cdot \;\; t^{(\mathcal{A})}_{\overline{\mathcal{P}\mathcal{O}}} \;\;\; + \;\;\; \frac{1}{\sqrt{ 1  \; - \; \frac{ {v_{\mathcal{B}}}^2 }{c^2} }} \; \cdot \;\; s^{(\mathcal{A})}_{\overline{\mathcal{P}\mathcal{O}}} \label{Formel - Masszsh - Lorentz Trafo s}
\eea
where $\mathcal{A}$lice has determined the velocity of the relative motion of Bob $v_{\mathcal{B}} := s^{(\mathcal{A})}_{\overline{\mathcal{P}\mathcal{B}}} \!\left/\! t^{(\mathcal{A})}_{\overline{\mathcal{P}\mathcal{B}}}\right.$ and with notation simplified in last steps on $\mathcal{A}$lice right hand side by suppressing her indices$^{(\mathcal{A})}$.
\bea
   \;\;\; t^{(\mathcal{B})}_{\overline{\mathcal{P}\mathcal{O}}} \!\!
   & \stackrel{(\ref{Formel - Masszsh - Substitution st MN i})}{=} & N + M \nn \\
   & \!\!\! \stackrel{(\ref{Formel - Masszsh - Substitution MN ABC - N explizit i})(\ref{Formel - Masszsh - Substitution MN ABC - N explizit ii})}{=} \!\!\! & \left[ \frac{1}{2} \cdot \sqrt{A\cdot(A+B+B)} \cdot \frac{A+C+C}{A+B+B} \; + \; \frac{1}{2} \cdot \sqrt{A\cdot(A+B+B)} \cdot \underbrace{\frac{\sqrt{A+B+B}}{\sqrt{A+B+B}}}_{=1} \; \right]
   \nn \\
   & = & \frac{1}{2} \cdot \frac{\sqrt{A}}{\sqrt{A+B+B}} \cdot (A+C+C) \; + \; \frac{1}{2} \cdot \frac{\sqrt{A}}{\sqrt{A+B+B}} \cdot (A+B+B) \nn \\
   & = & \underbrace{\frac{\sqrt{A}}{\sqrt{A}}}_{=1} \cdot \frac{\sqrt{A}}{\sqrt{A+B+B}} \cdot (A+B+C) \nn \\
   & = & \frac{1}{\sqrt{A\cdot(A+B+B)}} \; \cdot \; \left(\; A \cdot A \; + \; A \cdot B \: + \; A \cdot C \; \underbrace{+ \; B \cdot C \: - \; B \cdot C}_{=0} \right) \nn \\
   & = &  \frac{1}{\sqrt{A\cdot(A+B+B)}} \; \cdot \; \left(\; (A + C) \cdot (A+B) \: - \; B \cdot C \right) \nn
\eea
\bea
   & \!\!\!\!\! \stackrel{(\ref{Formel - Masszsh - Substitution ABC st i})-(\ref{Formel - Masszsh - Substitution ABC st iv})}{=} \!\!\!\!\! & \frac{1}{\sqrt{ \left( t^{(\mathcal{A})}_{\overline{\mathcal{P}\mathcal{B}}} \; - \; \frac{1}{c} \cdot s^{(\mathcal{A})}_{\overline{\mathcal{P}\mathcal{B}}} \right) \cdot \left( t^{(\mathcal{A})}_{\overline{\mathcal{P}\mathcal{B}}} \; + \; \frac{1}{c} \cdot s^{(\mathcal{A})}_{\overline{\mathcal{P}\mathcal{B}}} \right) }} \cdot  \left(  t^{(\mathcal{A})}_{\overline{\mathcal{P}\mathcal{O}}} \cdot t^{(\mathcal{A})}_{\overline{\mathcal{P}\mathcal{B}}} \; - \; \frac{1}{c} \cdot s^{(\mathcal{A})}_{\overline{\mathcal{P}\mathcal{B}}} \cdot  \frac{1}{c} \cdot s^{(\mathcal{A})}_{\overline{\mathcal{P}\mathcal{O}}}  \right) \;\;\;\;\;\;\;\: \nn \\
   & = & \frac{t_{\overline{\mathcal{P}\mathcal{B}}}}{\sqrt{  {t_{\overline{\mathcal{P}\mathcal{B}}}}^2 \; - \; \frac{1}{c^2} \cdot {s_{\overline{\mathcal{P}\mathcal{B}}}}^2  }} \; \cdot \; \left( t_{\overline{\mathcal{P}\mathcal{O}}} \; - \; \frac{1}{c^2} \cdot \frac{s_{\overline{\mathcal{P}\mathcal{B}}}}{t_{\overline{\mathcal{P}\mathcal{B}}}} \cdot
   s_{\overline{\mathcal{P}\mathcal{O}}}  \right) \nn \\
   t^{(\mathcal{B})}_{\overline{\mathcal{P}\mathcal{O}}}
   & = &  \frac{1}{\sqrt{ 1  \; - \; \frac{ {v_{\mathcal{B}}}^2 }{c^2} }} \; \cdot \;\; t^{(\mathcal{A})}_{\overline{\mathcal{P}\mathcal{O}}} \;\;\;\; - \;\;\; \frac{1}{\sqrt{ 1  \; - \; \frac{ {v_{\mathcal{B}}}^2 }{c^2} }} \cdot \frac{v_{\mathcal{B}}}{c^2} \; \cdot
   \;\; s^{(\mathcal{A})}_{\overline{\mathcal{P}\mathcal{O}}} \label{Formel - Masszsh - Lorentz Trafo t}
\eea


\chapter{Mechanics}\label{Kap - Mechanics}

Our objective is a foundation of mechanics from the operationalization of its basic observables. The problem is to first determine the physical operations really in a strictly physical way and then formulate a mathematical notation. From daily work experience and in play one can become familiar with interactions of motion \{\ref{Kap - KM Dynamics - Physical Measurement - Interaction of Motion}\}. According to Leibniz resp. Galilei we define energy, momentum and mass from the elemental comparison ''more capability to work than'' (against same test system) and ''more impact than'' (in a collision) \{\ref{Kap - KM Dynamics - Physical Measurement - Pre-theoretical Ordering Relation}\} and develop Helmholtz program for direct quantification \{\ref{Kap - KM Dynamics - Basic Dynamical Measures}\}. From vivid pre-theoretic (principle of inertia, impossibility of perpetuum mobile, relativity principle) and measurement methodical principles we derive the fundamental equations of mechanics \{\ref{Kap - KM Dynamics - Basic Dynamical Measures - Metrization - physical quantity}\}.

\section{Interaction of motion}\label{Kap - KM Dynamics - Physical Measurement - Interaction of Motion}

In theoretical physics the term ''action'' is occupied by the action functional (which physicists simply name action). Only the expression ''physical interaction'' makes unambiguous reference to the actual process between natural objects.\footnote{The \emph{action functional} is a formal map (from trajectories onto numbers); the attribute ''physical'' plays no role. In an \emph{interaction} we observe the physical behavior of the elements
in a system. Not from calculus but direct measurement of their actual behavior we begin the strictly physical foundation of mechanics.} One knows many types of interactions: In chemical interactions (called ''reaction'') the chemical bond between atoms is changing. Quantum mechanical interactions involve changes of internal degrees of freedom as spin or particle generation. The billiard collision is an example for an ''interaction of motion'' (Bewegungswirkung) according to Euler \cite{Euler Anleitung}. The identity of the interacting objects is preserved, only their state of motion changes. They include contact interactions like billiard, gravitational interactions in the solar system or electromagnetic interactions between charged particles. We use the term action strictly in this empirical sense.

We are referring to domains of everyday work experience where conditions for meaningful colloquial denominations ''material body'' and ''motion'' are practically sufficiently satisfied \cite{Peter '69 - Dissertation} \cite{Suppes - Aristotle's concept of matter}. We grasp an action impartially as the collective behavior of an interacting system. According to principle of inertia free objects move on their own. Before an interaction each state of motion is preserved until it is affected by an external cause \cite{Euler Anleitung}. The interaction involves the entire system $G_1\cup . . . \cup G_N$ (all passive elements and e.g. binding ''sources''). After an ''interaction of motion'' the elements preserve their character of a physical body (billiard balls keep the color and number) and solely changed their velocity $\mathbf{v}_i$.

\begin{figure}    
  \begin{center}           
  \includegraphics[height=8.7cm]{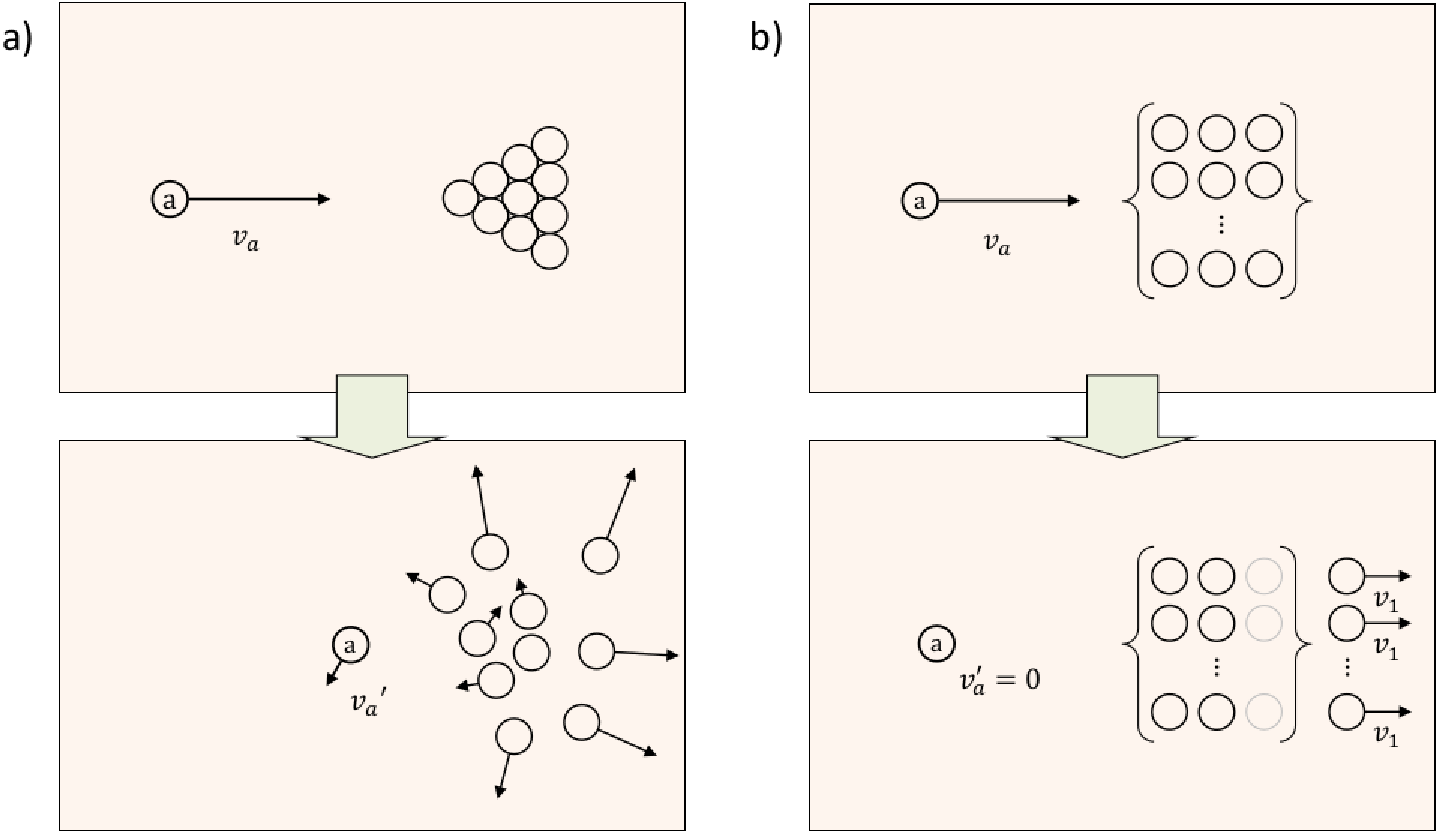}  
  \end{center}
  \vspace{-0cm}
  \caption{\label{pic_Billiard_Illustration} initial and final state of motion in a) generic billiard collision and b) in a steered process with a calorimeter reservoir $\left\{ \circMunit _{\,\mathbf{v}=\mathbf{0}} \right\}$
    }
  \end{figure}

When we step into a billiard room we deal with physical bodies and physical behavior. Free billiard balls move on their own. Motion is preserved until they hit the cushions or another ball. During the collision they form an interacting system. The elastic deformation (balls are impenetrable) causes a change of their respective state of motion. Dynamics explains that change due to external causes (rest of the system). Let Bob execute the first break shot with his cue ball against object balls which are racked tightly and initially at rest (see figure \ref{pic_Billiard_Illustration}a).
In a generic collision the cue ball eventually stops while the object balls fly off in an uncontrolled way. We make the experience that a heavier and faster cue ball has more impact than a slower. With a moving cue ball we associate a capability to execute work. We can compare the absorption effect on the same standard pile of object balls. We can also compare the impact directly in a head-on collision, when one cue ball overruns the other. We start with a pre-theoretic manner of speaking before we bring the demonstrable circumstances under the determination of the theory. We specify standardized methods for the impact comparison \{\ref{Kap - KM Dynamics - Physical Measurement - Pre-theoretical Ordering Relation}\}. For the ball one wants to find ''how many times'' more absorption effect or impact it carries than a reference device.

\begin{figure}    
  \begin{center}           
  \includegraphics[height=3.5cm]{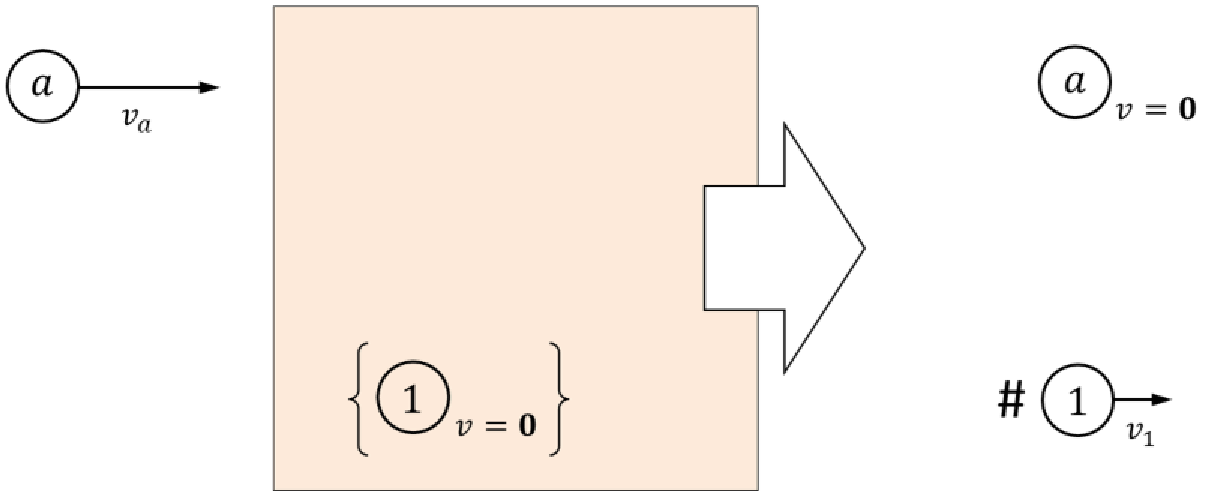}  
  \end{center}
  \vspace{-0cm}
  \caption{\label{pic_calorimeter} whatever comes in, the calorimeter generates only standard impulse carriers
    }
  \end{figure}

Who would think that at the billiard table one acts as a physicist? When Bob strikes he is dealing with interactions of motion. He begins to behave as a physicist if he steers and sorts these interactions in a controlled way. We illustrate the construction of an experimental instrument for a basic measurement of ''capability to work'' (energy) and ''impact'' (momentum) \{\ref{Kap - KM Dynamics - Basic Dynamical Measures - Metrization}\}. We construct a Gedanken-model on the operation of a particle detector: a calorimeter (see figure \ref{pic_calorimeter} with internal process as black box). It contains a reservoir with identically constituted elements $\left\{ \circMunit_{\,\mathbf{v}=\mathbf{0}} \right\}$ which are initially at rest. An incoming particle $\circMa_{\:\mathbf{v}}$ with velocity $\mathbf{v}$ will be slowed down $\circMa_{\:\mathbf{v}=0}$ by successive collisions with the resting elements from the reservoir (see figure \ref{pic_Billiard_Illustration}b). While in a generic billiard collision all object balls fly off with arbitrary impetus, now the process is set up and controlled from outside. A team of assistants will steer the process. As a result of stopping the incident cue ball a certain number of reservoir elements $\sharp\left\{\circMunit_{\,\mathbf{v}_\mathbf{1}}\right\}$ (with standardized velocity $\mathbf{v}_{\mathbf{1}}$) will be knocked out of the calorimeter. Each ball has exactly the same impact behavior and represents a unit of momentum. \emph{In this model we can count} the number of extracted reference units. The elemental ordering relations ''more capability to work than'' (energy) and ''more impact than'' (momentum) become measurable.

We introduce the following notation for tangible objects
\begin{itemize}
\item $\circMa\!\!\!\;\!\longrightarrow \!\!\!\!\!\!\!\!_{\mathbf{v}}\;\;\;$ for a moving particle with individual name $a$
\item $\circMa_{\:\mathbf{v}=0}$ or simply $\circMa_{\:\mathbf{0}}$ when the object rests
\end{itemize}
and in formal mathematical expressions:
\begin{itemize}
\item $\circMa _{\:\mathbf{x},\mathbf{v}}$  for a particle $\circMa$ at the place $\mathbf{x}$ with velocity $\mathbf{v}$
\item $\circMa _{\:\mathbf{v}}$ when the place does not matter, similarly for a moving standard spring $\mathcal{S}\big|_{\mathbf{v}}$ etc.
\item $\circMa _{\:\mathbf{v}=0}$ or $\circMa _{\:\mathbf{0}}$ and $\mathcal{S}\big|_{\mathbf{0}}$ when the respective object rests.
\end{itemize}

\section{Basic observable}\label{Kap - KM Dynamics - Physical Measurement - Pre-theoretical Ordering Relation}

From daily work experience and in play (where valid prognoses about natural processes pay off most) one can \emph{notice} the ''impact'' of decelerating bodies and the associated ''capability to work''. One develops \emph{pre-theoretic comparison} methods by examples and in words. Physicists fix the \emph{conventions} for observer independent and reproducible procedures. Without viewing motion as a mere mathematical map $\gamma: \tau \mapsto \left(t,\mathbf{x}\right)\,$; we specify comparison methods so that
\begin{itemize}
\item   the procedure is universally available and
\item   the result does not depend on the individual observer; all physicists find the same \cite{Janich Das Mass der Dinge}.
\end{itemize}

For collisions of irrelevant inner structure Galilei defines an elementary \emph{ordering criterion}. Let two generic bodies $\circMa$ and $\circMb$ run into each other with initial velocities $v_a$ resp. $v_b\,$, collide and stick together.
\begin{de}\label{Def - vortheor Ordnungsrelastion - impulse}
\underline{Momentum} $\mathbf{p}[{\circMa_{\:\mathbf{v}}}]$ is the striking power, impact (Wucht) of the moving body $\circMa_{\:\mathbf{v}}$ \cite{Wolff - Geschichte der Impetustheorie}. Object $\circMa_{\:\mathbf{v}_{\!a}}$ has more impact
than object $\circMb_{\:\mathbf{v}_{\!b}}$
\be\label{Abschnitt -- vortheor Ordnungsrelastion - impulse verhalten}
   \circMa_{\:\mathbf{v}_{\!a}} \;\; >_{\mathbf{P}} \;\; \circMb_{\:\mathbf{v}_{\!b}}
\ee
if in a head-on collision test one body \underline{overruns} the other.
\end{de}
If the bound aggregate $\circMa_{\:\mathbf{v}_a} , \circMb_{\:\mathbf{v}_b} \Rightarrow \circMa\ast\circMb_{\;\mathbf{v}=\mathbf{0}}$ moves neither with object $\circMa_{\:\mathbf{v}_{a}}$ to the right nor with object $\circMb_{\:\mathbf{v}_{b}}$ to the left, then their impact is practically the same $\mathbf{p}[\circMa_{\:\mathbf{v}_{\!a}}] = \mathbf{p}[\circMb_{\:\mathbf{v}_{\!b}}]$.
\begin{de}\label{Def - inertia}\label{Def - vortheor Ordnungsrelastion - inertial mass}
\underline{Inertia} $m\left[\circMa\right]$ is the - passive - resistance against changes of the state of motion of object $\circMa$ \cite{Euler Anleitung} \cite{Peter '69 - Dissertation}. According to Galilei object $\circMa$ is more massive
than object $\circMb$
\be\label{Abschnitt -- vortheor Ordnungsrelastion - inertial verhalten}
  \circMa \;\; >_m \;\; \circMb
\ee
if after an inelastic head-on collision test $\circMa_{\:\mathbf{v}} , \circMb_{-\mathbf{v}} \Rightarrow \circMa\ast\circMb_{\;\mathbf{v}'}$ with same initial velocity the bound composite moves in the direction $\mathbf{v}'\propto \mathbf{v}$ of the heavier object \cite{Weyl - Philosophie der Mathematik und Naturwissenschaft}.
\end{de}
\begin{rem}
One conducts a special case of impulse comparison
\[
   >_m \;\; := \;\; >_{\mathbf{P}}\big|_{\:\mathbf{v}_{\!a}=-\mathbf{v}_{\!b}}
\]
with an \underline{extra condition} that both objects initially move head-on with same velocity $\mathbf{v}_{\!a} \stackrel{!}{=} - \mathbf{v}_{\!b}$.
\end{rem}

\begin{figure}    
  \begin{center}           
  \includegraphics[height=16.8cm]{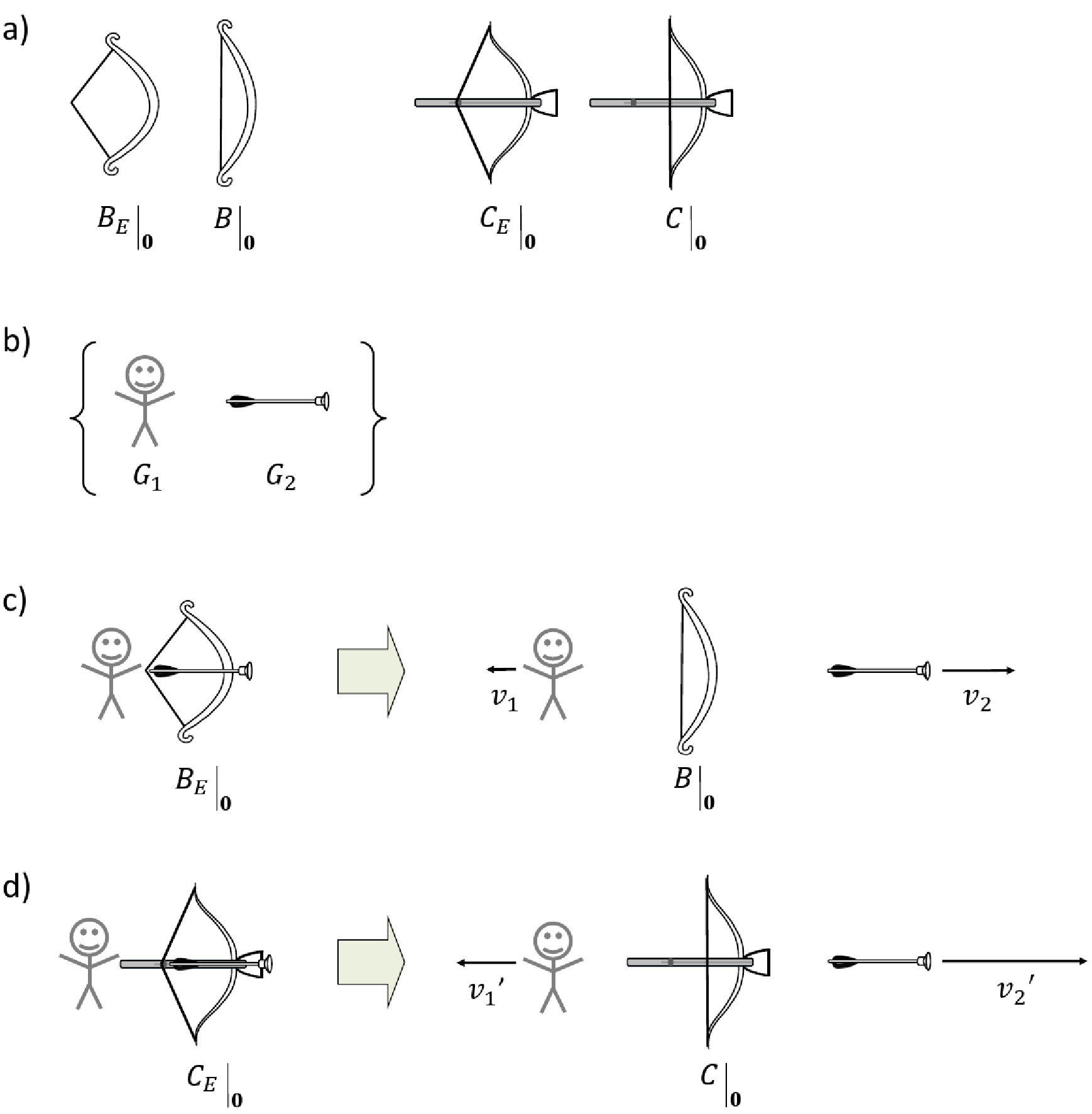}  
  \end{center}
  \vspace{-0cm}
  \caption{\label{pic_Aequipollenz} a) charged bow $\mathcal{B}_E$ or crossbow $\mathcal{C}_E$ represent energy sources which can be coupled into b) a two-body system of initially resting archer and arrow $\mathcal{G}_1 \cup \mathcal{G}_2$ c) charged bow $\mathcal{B}_E$ expends its energy on test system $\mathcal{G}_1 \!\cup\!\, \mathcal{G}_2$ (discharged bow $\mathcal{B}\big|_{\mathbf{0}}$ remains at rest) d) charged crossbow $\mathcal{C}_E$ causes a larger kinetic effect $\Delta v'_i > \Delta v_i$ against the same test system $\mathcal{G}_1 \cup \mathcal{G}_2$
  }
  \end{figure}

Consider a bow $\mathcal{B}$ and a crossbow $\mathcal{C}$. When the string is tightened (and mechanically locked) the charged devices $\mathcal{B}_E$, $\mathcal{C}_E$ become energy sources. With the charged state we associate a capability to work (Wirkungsverm\"ogen). According to Leibniz equipollence principle (detailed discussion in \{\ref{Kap - KM Dynamics - Basic Dynamical Measures - kinetic Energy}\}) we compare it indirectly by measuring their (kinetic) effect against the same test system; whether the same test particles (e.g. archer $\mathcal{G}_1$ and arrow $\mathcal{G}_2$ in figure \ref{pic_Aequipollenz})
repulse with larger velocity $\Delta v'_{1,2} > \Delta v_{1,2}$ than from a shot with the weaker bow. Without restricting generality the charged sources (crossbow $\mathcal{C}_E\big|_{\mathbf{0}}$, bow $\mathcal{B}_E\big|_{\mathbf{0}}$) may initially be at rest and after pulling the trigger, after expending the associated capability to work both \emph{discharged sources} $\mathcal{C}$, $\mathcal{B}$ remain at rest. This allows a clear separation. In return the test particles (archer $\mathcal{G}_1$ and arrow $\mathcal{G}_2$) begin to fly apart. With their motion we associate another form of capability to work (kinetic energy), which they can expend against third parties etc. According to Helmholtz measurement principle: the total capability to work is conserved. In a measurement we consume one specific form entirely (e.g. potential energy until a spring is entirely relaxed; kinetic energy until all projectiles come to rest etc.) in a transformation into other forms (preferably carried by separate elements of the system).
\begin{de}\label{Def - vortheor Ordnungsrelastion - energie}
\underline{Energy} is the capability of a separate source or system to work against an external system $\mathcal{G}$. The kinetic, potential, binding etc. form of energy is associated with exhausting a particular condition of the source (motion, configuration size, chemical bound). According to Leibniz one source
\be\label{Abschnitt -- vortheor Ordnungsrelastion - energetisches verhalten}
   \mathcal{S}_E \; >_{E} \; \tilde{\mathcal{S}}_{\tilde{E}}
\ee
has more potential than another source $\tilde{\mathcal{S}}_{\tilde{E}}$ if the effect of source $\mathcal{S}_E$ on the same test system $\mathcal{G}$ \underline{exceeds} the effect of source $\tilde{\mathcal{S}}_{\tilde{E}}$.
\end{de}
In our calorimeter model we will measure the kinetic energy of a moving body $\circMa_{\:\mathbf{v}_{a}}$ by the number of obstacles it overcomes. We count how many standard springs can be compressed (repeat elementary processes) before the body $\circMa_{\:\mathbf{0}}$ stops. The kinetic energy (of projectiles) transforms into potential energy (of the absorber material). If the latter comes in standard portions, which are all congruent with one another, our quantification is complete.

We define the practical comparison circularity free, without presupposing numerical values (on a ratio scale) of unclear origin and status. Physics goes beyond the formal equations: Mathematics postulates its abstracta as known and given; instead we scrutinize the formation from tangible operations. Under the abstraction ''energy'' and ''momentum'' one compares two interactions with regards to an elemental ordering criterion.
\begin{de}\label{Def - vortheor Ordnungsrelastion - abstraction}
In an \underline{abstraction} we regard the common quality of both objects for itself without needing to  consider the dissimilarity (of both objects in other regards).\footnote{Helmholtz \cite{Helmholtz - Zaehlen und Messen} explains ''bodies whose weight we are comparing can be made from most different materials, different shape and volume. The weight - which we set equal - is only one of their properties and obtained by abstraction. We are only justified - to call those bodies themselves weights and designate these weights as quantities - in circumstances where we can disregard all other properties of these bodies''.

Ruben \cite{Peter '67 - zum Streit um das wahre Mass der Kraft} thinks about: ''a tree e.g. is in general a subject of biology. If the tree is cut down then it is - for the worker who has to get out of the way of the falling tree - a mechanical object. In this context the tree is essentially important as carrier of weight; it is unimportant whether the tree is a linden or an oak. All natural things are always also carrier of mass. Insofar as they are they are a subject of mechanics.''}
\end{de}
The comparison procedure distinguishes one singular aspect (kinetic effect) for observation. We regard all objects (moving billiard ball, compressed spring, battery etc.) solely as substitutable carriers of their common quality ''capability to work'' and ''impact''.\footnote{In the theory one makes propositions about abstract ''energy'' and ''momentum''. The transition is implemented by a \emph{limitation in the manner of speaking} onto invariant assertions \cite{Lorenzen - Konstruktive Wissenschaftstheorie}. One restricts from simple descriptive sentences (about various colloquial aspects of an interaction) onto assertions, which remain unchangedly valid under substituting equivalent energy-momentum carriers; e.g. an equally charged crossbow $\mathcal{C}_E$ and bow $\mathcal{B}_E$ rebound all test particles in the same way (despite different inner structure or materials).}


\section{Quantification}\label{Kap - KM Dynamics - Basic Dynamical Measures} \label{Kap - KM Dynamics - Basic Dynamical Measures - Metrization}

We define the basic observables from direct comparisons. For standardization of experiment and measurement one wants to express their value also numerically (''how many times'' more absorption effect or impetus).

\subsection{Reference standards}\label{Kap - KM Dynamics - Basic Dynamical Measures - Quantification - Dynamical Unit}

With Ruben \cite{Peter '69 - Dissertation} we acknowledge the important methodical distinction between measurement object and reference device. Both are natural objects. In a measurement, which is always a \emph{pair comparison} between measurement object and material model (see Remark \ref{Rem - SRT Kin - doubling of physical measures}), they have different functions. Physicists compare the ''capability to work'' and ''impact'' of the measurement object (e.g. after a scattering process) while they have to provide a reference device in a suitable way \cite{Janich Das Mass der Dinge}.

We specify a reference process as a sufficiently constant representative of ''capability to work'' and ''impact''. We may pick out a \emph{standard} which is reproducible and available anywhere and anytime and in any number. Leibniz presents various candidates including the compression of equivalent springs up to a fixed mark.\footnote{Also D'Alembert utilizes in his \emph{Trait\'{e} de dynamique} congruent actions of a spring. This is a very instructive approach, Schlaudt remarks \cite{Schlaudt}. The action is quantified - not by the depth of compression (in one spring) but instead - by the number of springs which are compressed by a fixed distance. In this way one can \emph{disregard} completely from the \emph{inner dynamics} of the compression process.} We provide a reservoir $\left\{\circMunit\,, \mathcal{S}\right\}$ with standard bodies ''$\circMunit$'' with same inertia. According to Galilei we can test pre-theoretically if in a head-on collision no one overruns the other (see Definition \ref{Def - inertia}). We can charge standard springs ''$\mathcal{S}$'' with same capability to work. Following Leibniz each must catapult our standard objects in the same way (see Definition \ref{Def - vortheor Ordnungsrelastion - energie}).
\begin{de}\label{Def -- basic dynamical measures - Einheitswirkung}
For measurements in entire mechanics we refer to an elementary standard process (of irrelevant inner structure). Let the compressed spring
\be\label{Abschnitt -- basic dynamical measures - Einheitswirkung}
        \mathcal{S}_{E=\mathbf{1}}\big|_{\mathbf{v}=\mathbf{0}} \,,\: \circMunit_{\:\mathbf{v}=\mathbf{0}} \,,\: \circMunit_{\:\mathbf{v}=\mathbf{0}} \;\;\; \stackrel{w_{\mathbf{1}}}{\Rightarrow} \;\;\; \circMunit_{\:\mathbf{v}_{\mathbf{1}}} \,,\: \circMunit_{-\mathbf{v}_{\mathbf{1}}}
\ee
catapult two resting standard objects into diametrically opposed directions (see figure \ref{pic_Wirkungseinheit_Feder}b)
\begin{figure}    
  \begin{center}           
  \includegraphics[height=8.0cm]{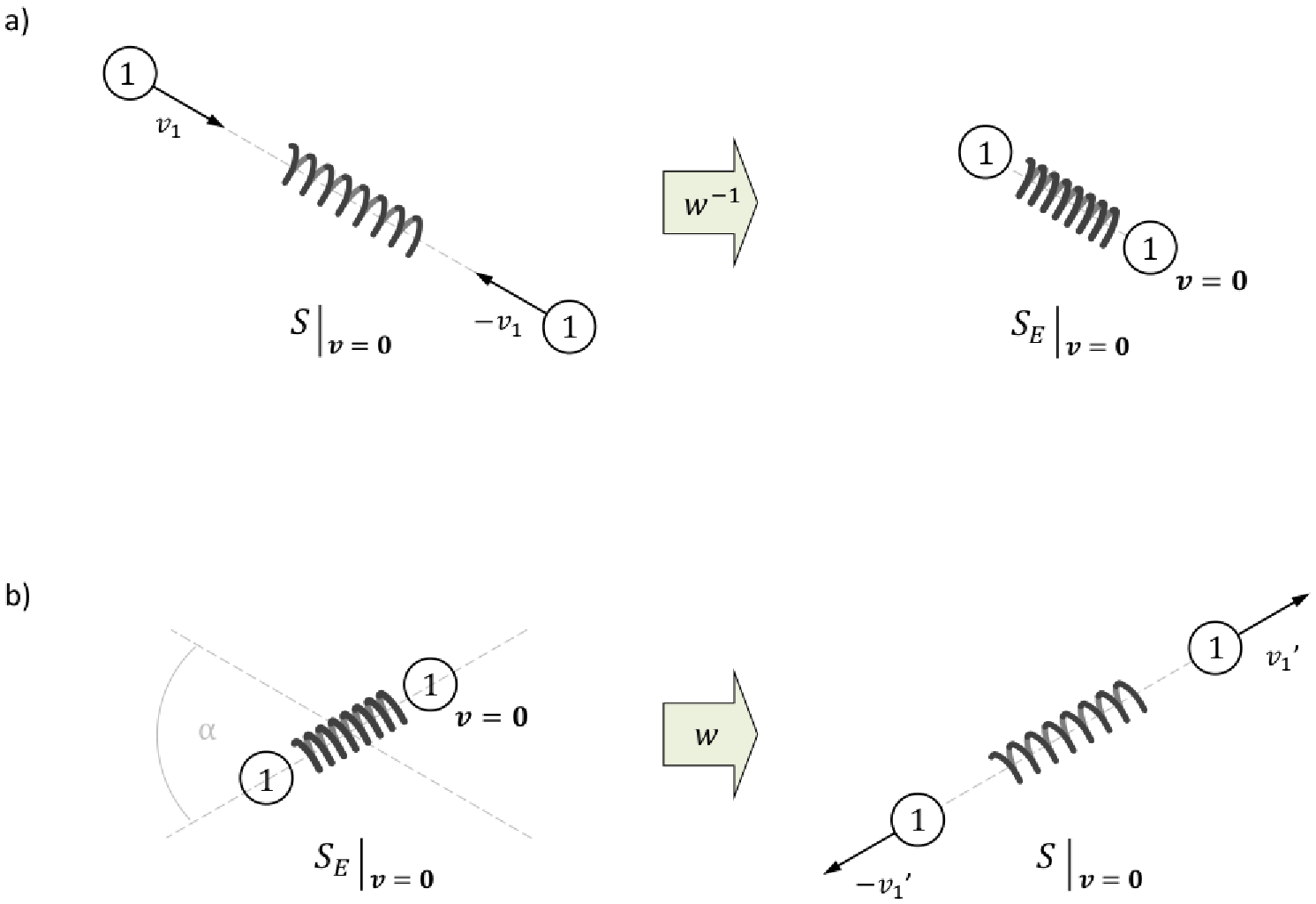}  
  \end{center}
  \vspace{-0cm}
  \caption{\label{pic_Wirkungseinheit_Feder} a) compression, rotation and b) decompression $w$ of charged spring $\mathcal{S}_E\big|_{\mathbf{0}}$ kicks a particle pair into unit velocity and vice versa (neutral spring $\mathcal{S}_{E=0}\big|_{\mathbf{v}=\mathbf{0}}$ remains at rest)
      }
  \end{figure}
or reversely let the particle pair (with standard velocity $\pm\mathbf{v}_{\mathbf{1}}$) compress a neutral spring (see figure \ref{pic_Wirkungseinheit_Feder}a). We suppress empty springs in the notation; they stay in the reservoir.
\end{de}
The standard spring $\mathcal{S}_{\mathbf{1}}$ turns standard particles $\circMunit$ into standard impulse carriers $\circMunit_{\:\mathbf{v}_{\mathbf{1}}}$ and vice versa. We call this reference process ''unit action'' ($\mathcal{W}irkung$ $w_{\mathbf{1}}$). With regards to the basic observable ''capability to work'' and ''impact'' the inelastic collision is well-defined by symmetry and relativity principle.

Huygens did study symmetrical collisions between equivalent objects together with the relativity principle to derive the collision laws for billiard balls. For the same reason Einstein \cite{Einstein '35 - mass energy equivalence} and Feynman \cite{Feynman lectures I} examine interactions between objects which collide and stick together. We build a calorimeter model from elementary building blocks: the standard process $w_{\mathbf{1}}$.

\subsection{Concatenation}\label{Kap - KM Dynamics - Basic Dynamical Measures Quantification - Concatenation}

The model building involves a steering task. Physicists couple a series of standard kicks against the respective particles.
\begin{de}
They \underline{concatenate} ''$\ast$'' two consecutive collisions $w\ast \tilde{w}$
\bea
   & & \mathcal{S}_{E}\big|_{\mathbf{v}_{\mathcal{S}}} \,,\: \circMa_{\:\mathbf{v}_a} \,,\: \circMb_{\:\mathbf{v}_b} \;\stackrel{w}{\Rightarrow}\; \circMa_{\:\mathbf{v}_a'} \,,\: \circMb_{\:\mathbf{v}_b'} \nn \\
   & & \tilde{\mathcal{S}}_{\tilde{E}}\big|_{\mathbf{v}_{\tilde{\mathcal{S}}}} \,,\: \circMb_{\:\mathbf{v}_b'} \,,\: \circMc_{\:\mathbf{v}_c} \;\stackrel{\tilde{w}}{\Rightarrow}\; \circMb_{\:\mathbf{v}_b''} \,,\: \circMc_{\:\mathbf{v}_c'} \nn
\eea
in a coinciding element $\circMb_{\:\mathbf{v}_b'}$ which between every two interventions moves freely.
\end{de}

\begin{figure}    
  \begin{center}           
  \includegraphics[height=8.7cm]{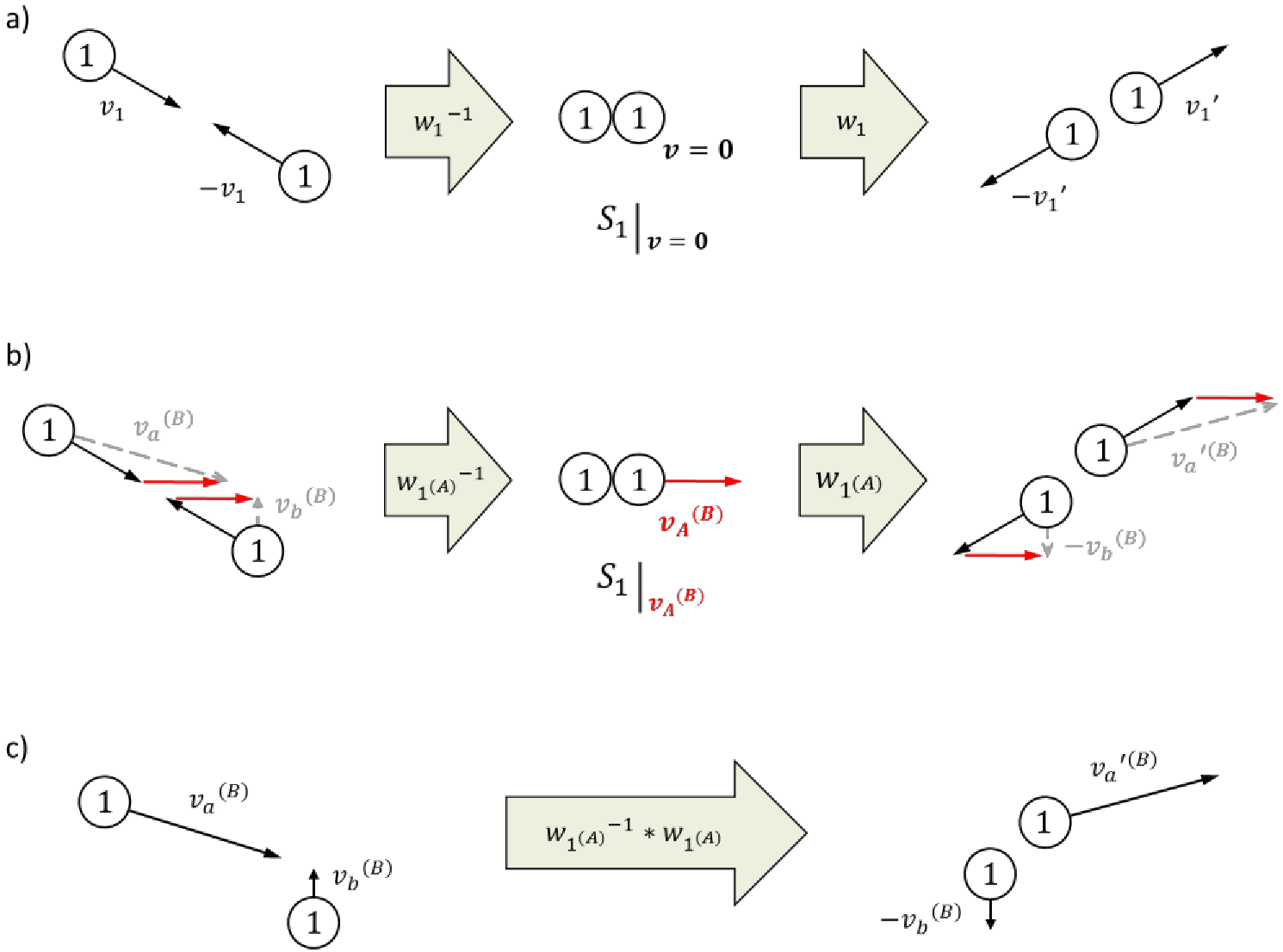}  
  \end{center}
  \vspace{-0cm}
  \caption{\label{pic_dynamical_unit} a) elastic connection of reversible unit actions $w_{\mathbf{1}}^{-1} \ast w_{\mathbf{1}}$ for $\mathcal{A}$lice
  b) covariant transformation to perspective of $\mathcal{B}$ob
  c) appears as ''elastic transversal collision''
    }
  \end{figure}

In figure \ref{pic_dynamical_unit}a we connect two unit actions $w_{\mathbf{1}}^{-1} \ast w_{\mathbf{1}}$ in the temporarily resting unit objects $\circMunit_{\:\mathbf{v}=0}$. In figure \ref{pic_impulse_inversion_means} we connect a sequence of transversal kicks $w_{T} \ast w_{T} \ast w_{T} \ast \ldots$ against the same object $\circMunit_{\:v_{(1)}}$ which in between each intervention moves freely with same velocity $v_{(1)}$.

\begin{rem}\label{Rem - covariant kinematical transformation}
Let $\mathcal{A}$lice and $\mathcal{B}$ob observe the same process $\mathcal{S}_{E}\big|_{\mathbf{v}_{\mathcal{S}}} , \circMa_{\:\mathbf{v}_a} , \circMb_{\:\mathbf{v}_b} \Rightarrow \circMa_{\:\mathbf{v}_a'} , \circMb_{\:\mathbf{v}_b'}$. Let $\mathcal{A}$lice move relative to $\mathcal{B}$ob with velocity $\mathbf{v}_{\mathcal{A}} = v_{\mathcal{A}}^{(\mathcal{B})} \cdot \mathbf{v}_{\mathcal{L}^{(\mathcal{B})}}$. For the same objects $i=\mathcal{S}_{E}$, $\circMa$, $\circMb$ they measure the velocities $\mathbf{v}_{i} = v_{i}^{(\mathcal{A})} \cdot \mathbf{v}_{\mathcal{L}^{(\mathcal{A})}}  = v_{i}^{(\mathcal{B})} \cdot \mathbf{v}_{\mathcal{L}^{(\mathcal{B})}}$
in a covariant way \cite{Hartmann-SRT-Kin}. In Galilei kinematics the numerical values transform by vectorial addition $v_{i}^{(\mathcal{B})} = v_{i}^{(\mathcal{A})} + v_{\mathcal{A}}^{(\mathcal{B})} $.
\end{rem}
$\mathcal{A}$lice standard process $\mathcal{S}_{\mathbf{1}}\big|_{\mathbf{0}} ,  \circMunit_{\:\mathbf{0}} ,  \circMunit_{\:\mathbf{0}} \Rightarrow \circMunit_{\:\mathbf{v}_1} , \circMunit_{-\mathbf{v}_1}$ of the resting spring against reservoir elements appears to $\mathcal{B}$ob as an interaction
$\mathcal{S}_{\mathbf{1}}\big|_{\mathbf{v}_{\mathcal{A}}} ,  \circMunit_{\:\mathbf{v}_{\mathcal{A}}} ,  \circMunit_{\:\mathbf{v}_{\mathcal{A}}} \stackrel{w_{\mathbf{1}^{(\mathcal{A})}}}{\Rightarrow} \circMunit_{\:\mathbf{v}_1 + \mathbf{v}_{\mathcal{A}}} , \circMunit_{-\mathbf{v}_1 + \mathbf{v}_{\mathcal{A}}}$ between the initially comoving energy source $\mathcal{S}_{\mathbf{1}}\big|_{\mathbf{v}_{\mathcal{A}}}$ and bodies $\circMunit_{\:\mathbf{v}_{\mathcal{A}}}$, $\circMunit_{\:\mathbf{v}_{\mathcal{A}}}$ (see figure \ref{pic_dynamical_unit}b).

\subsection{Physical model}\label{Kap - KM Dynamics - Basic Dynamical Measures - Quantification - Physical Model}

\begin{figure}    
  \begin{center}           
  \includegraphics[height=8cm]{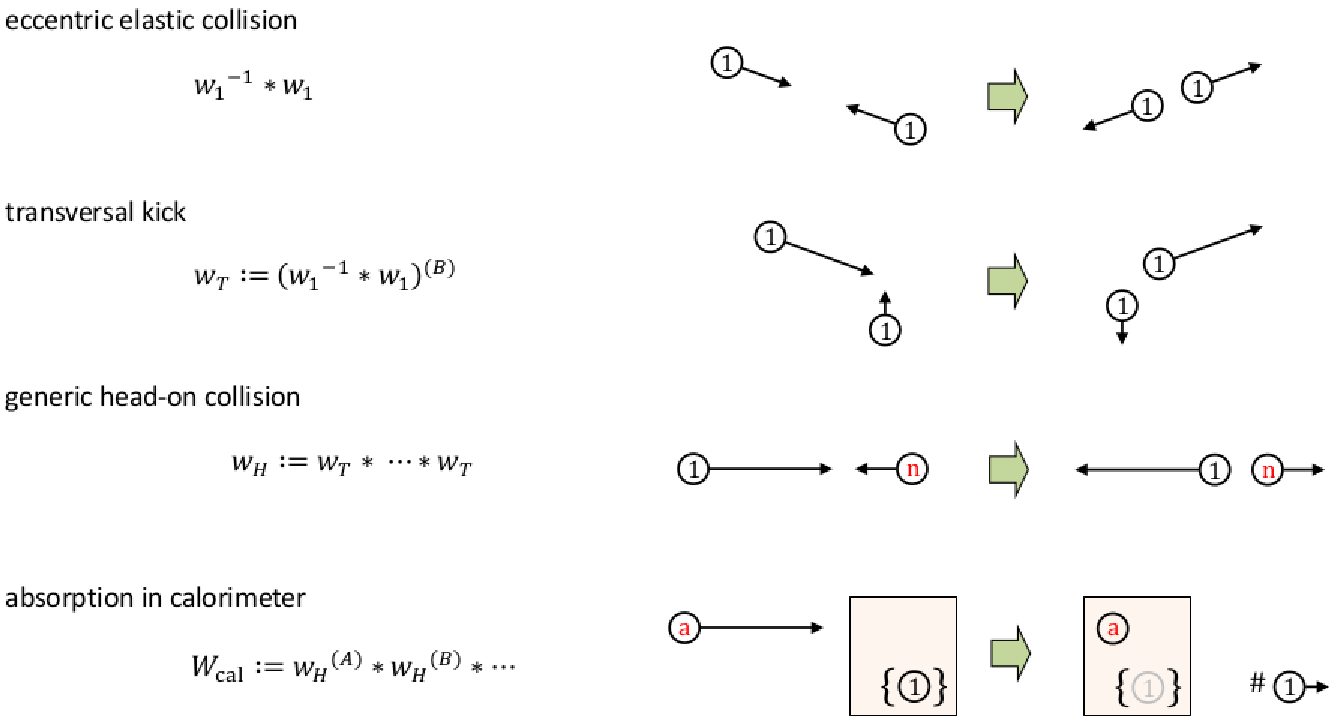}  
  \end{center}
  \vspace{-0cm}
  \caption{\label{pic_Zusammensetzung_Kalorimeter} assemble absorption process $W_{\mathrm{cal}}$ in calorimeter reservoir $\{\circMunit_{\:\mathbf{0}}\}$
    }
  \end{figure}

Such energy sources can be charged, reoriented in space and discharged in a reversible way. Let Alice successively couple compression and decompression of her spring; then the two standard particles, which initially flew towards each other, will instantly be catapulted apart. Consecutive compression and decompression of her spring $w_{\mathbf{1}}^{-1} \ast w_{\mathbf{1}}$ gives an eccentric elastic collision between bodies of same mass (from the initial state in figure \ref{pic_Wirkungseinheit_Feder}a to the final state in figure \ref{pic_Wirkungseinheit_Feder}b). A drive-by observer will see the process as a transversal kick $w_T$ (see figure \ref{pic_dynamical_unit}c). The kinematics is well-defined by symmetry and covariance. From those standard kicks we assemble increasingly complex collision models (outlined in figure \ref{pic_Zusammensetzung_Kalorimeter}).
By controlled linkage of those building blocks and by relativity principle (view from the moving observer) we model an \emph{elastic head-on collision} $w_H$ between generic (non equivalent!) bodies \{\ref{Kap - KM Dynamics - Basic Dynamical Measures - Measurement Means - Kinematic Determination}\} and an \emph{absorption process} $W_{\mathrm{cal}}$ for a generic particle in a calorimeter \{\ref{Kap - KM Dynamics - Basic Dynamical Measures - Measurement Means - Kinematic Quantification Calorimeter Action}\}.

We do not presuppose how velocities of two generic objects change in an elastic collision. The \emph{trick} is to mediate their direct interaction by a steered replacement process with an external reservoir. Our model solely consists of elastic collisions between standard elements which must behave in a symmetrical way. From their layout we derive the generic collision law; and similarly for the absorption process in a calorimeter \{\ref{Kap - KM Dynamics - Basic Dynamical Measures - Measurement Means - Kinematic Quantification Calorimeter Action}\}.

One can conduct calorimetric measurements for individual elements before and after an interaction and for the entire system \{\ref{Kap - Analytical mechanics}\}. We measure the energy-momentum of a generic interaction $w$ with reference interactions $w_{\mathbf{1}}$. We treat each action as an irreducible atom. We presuppose the ''measurement object'' $w$ and the ''reference device'' $w_{\mathbf{1}}$ solely as completed processes with known final state of motion. Thus also the reference interaction $w_{\mathbf{1}}$ is \emph{unquantified}; but the units are all \emph{congruent} with one another (like light clocks in basic measurements for kinematics; see Remark \ref{Rem - SRT Kin - inseparable unit}). It is the task of the physicist to couple these congruent units $w_{\mathbf{1}}\,$! A team of assistants steer the initiation timely and at suitable position, such that the desired effect is achieved (see figure \ref{pic_impulse_inversion_means}). In every individual action only the final changes in motion matter without needing to know the internal structure of the kicks. The coupling of standard kicks is an elementary operation in a measurement. Like placing rulers, light clocks etc. along a \emph{straight} line these practical interventions are well-defined.

We introduce a measurement method where a sequence of standard interactions is steered and coupled. Our material model generates the same kinetic effect $w_{\mathbf{1}} \ast \ldots \ast w_{\mathbf{1}} \: \sim_{E,\,\mathbf{p}} \: w$ (element by element the same change of motion $\Delta\mathbf{v}_i$ and associated energy-momentum gain) like from the generic process $w$. We model the elastic collision of two generic particles \{\ref{Kap - KM Dynamics - Basic Dynamical Measures - Measurement Means - Kinematic Determination}\}, the absorption process in a calorimeter \{\ref{Kap - KM Dynamics - Basic Dynamical Measures - Measurement Means - Kinematic Quantification Calorimeter Action}\} and finally the concatenation of impulses from multiple carriers \{\ref{Kap - KM Dynamics - Basic Dynamical Measures - Momentum}\}.

The models are made of standard energy sources $\sharp \left\{\mathcal{S}_{\mathbf{1}}\right\}$ and impulse carriers $\sharp \left\{\circMunit_{\:\mathbf{v}_{\mathbf{1}}}\right\}$; by counting them we find ''how many times'' more energy the generic interaction $w$ transforms than one standard spring $\mathcal{S}_{\mathbf{1}}$ in reference process $w_{\mathbf{1}}$. The number of congruent units is a reproducible \emph{physical quantity}. We construct the basic measurement instrument from simple practical principles (without using a single formal presupposition). We count the respective elements in the model to derive the fundamental equations between these quantities \{\ref{Kap - KM Dynamics - Basic Dynamical Measures - Metrization - physical quantity}\}.\footnote{Basic measurements do not presuppose equations of motion or formal expressions for conserved quantities. We define basic observables and quantification from physical operations. A basic measurement quantifies the basic observable. First we construct physical quantities; then we slip into a mathematical formulation.}


\subsection{Elastic collision model}\label{Kap - KM Dynamics - Basic Dynamical Measures - Measurement Means - Kinematic Determination}

\begin{figure}    
  \begin{center}           
  \includegraphics[height=3.4cm]{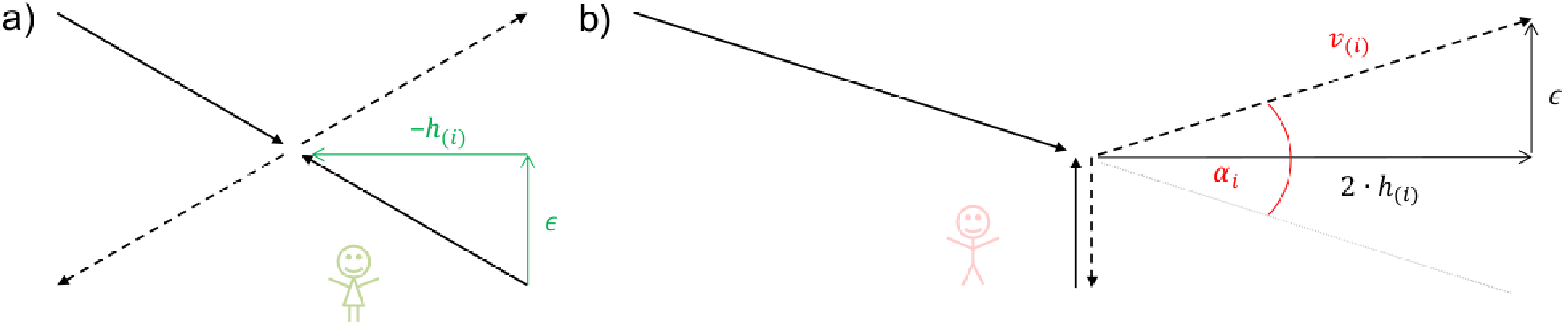}  
  \end{center}
  \vspace{-0cm}
  \caption{\label{pic_elastic_transversal_collision} a) symmetric elastic collision with scattering angle set up by $\mathcal{A}$lice b) appears as elastic transversal collision $w_T$ when $\mathcal{B}$ob drives by with same horizontal velocity to left
    }
  \end{figure}

\begin{lem}\label{Lem - kin quant elast coll - step I Winkel und Geschwindigkeit}
Let in elastic transversal collision between equivalent objects (see figure \ref{pic_elastic_transversal_collision}b)
\be\label{Formel - transversal standard kick}
     \circMunit_{\:v_{(i)}} \,,\: \circMunit_{\:\epsilon\cdot v_{\mathbf{1}}} \;\stackrel{w_{T}}{\Rightarrow}\; \circMunit_{\:v_{(i)}'} \,,\: \circMunit_{-\epsilon\cdot v_{\mathbf{1}}}
\ee
reservoir particle $\circMunit_{\:\epsilon\cdot v_{\mathbf{1}}}$ kick in from below with fixed velocity $\epsilon\cdot v_{\mathbf{1}}$ and rebound antiparallel. Then incident object $\circMunit_{\:v_{(i)}}$ moves on with same velocity $v_{(i)}'=v_{(i)}$ deflected by angle $\alpha_{i}$
\be\label{Formel - v-alpha-elastic transversal collision}
   \sin\left( \frac{\alpha_{i}}{2} \right) = \frac{\epsilon}{v_{(i)}} \;\; .
\ee
\end{lem}
\textbf{Proof:} The elastic collision of identically constituted bodies $\circMunit$ is well-defined by symmetry and Galilei covariance. Let $\mathcal{A}$lice prepare the initial velocities for an eccentric collision\footnote{She can freely adjust the deflection $\tan (\frac{\tilde{\alpha}_i}{2}) = \frac{\epsilon}{h_{(i)}}$ by rotating the spring between two standard processes ${w_{\mathbf{1}}}^{-1}\ast w_{\mathbf{1}}$ (in figure \ref{pic_Wirkungseinheit_Feder}b) or with a suitable impact parameter in an eccentric collision of rigid balls.}
\[
   \mathbf{v}_{(i)} = \left(
               \begin{array}{c}
                 h_{(i)}  \\
                 - \epsilon \\
               \end{array}
             \right) \cdot \mathbf{v}_{\mathbf{1}^{(\mathcal{A})}} \;\;\;\;\;\;\;\;\;\;\;
   \mathbf{v}_{\mathcal{R}} = - \left(
               \begin{array}{c}
                 h_{(i)}  \\
                 - \epsilon \\
               \end{array}
             \right) \cdot \mathbf{v}_{\mathbf{1}^{(\mathcal{A})}}
\]
with fixed horizontal and vertical components (see figure \ref{pic_elastic_transversal_collision}a).

Let $\mathcal{A}$lice move relative to $\mathcal{B}$ob at constant velocity $
   \mathbf{v}_{\mathcal{A}} = \left(
               \begin{array}{c}
                 h_{(i)}  \\
                 0 \\
               \end{array}
             \right) \cdot \mathbf{v}_{\mathbf{1}^{(\mathcal{B})}}
$
in the horizontal direction. Velocities transform by vectorial addition (see Remark \ref{Rem - covariant kinematical transformation}). For $\mathcal{B}$ob incident body $\circMunit _{\: \mathbf{v}_{(i)}}$ has twice the horizontal velocity $2\cdot h_{(i)} \cdot v_{\mathbf{1}^{(\mathcal{B})}}$ with same vertical component $\epsilon\cdot v_{\mathbf{1}^{(\mathcal{B})}}$
\[
   \mathbf{v}_{(i)} = \left(
               \begin{array}{c}
                 2\cdot h_{(i)}  \\
                 \mp\epsilon \\
               \end{array}
             \right) \cdot \mathbf{v}_{\mathbf{1}^{(\mathcal{B})}} \;\;\;\;\;\;\;\;\;\;\;
   \mathbf{v}_{\mathcal{R}} = \left(
               \begin{array}{c}
                 0 \\
                 \pm\epsilon \\
               \end{array}
             \right) \cdot \mathbf{v}_{\mathbf{1}^{(\mathcal{B})}}
\]
while $\mathcal{R}$eservoir particle $\circMunit _{\: \epsilon \cdot \mathbf{v}_{\mathbf{1}}}$ moves up and down vertically with same velocity $\epsilon \cdot v_{\mathbf{1}^{(\mathcal{B})}}$. For the same elastic collision $\mathcal{B}$ob determines a scattering angle $\alpha_{i}$ according to figure \ref{pic_elastic_transversal_collision}b.
\qed
For given initial velocity $v_{(i)}$ and fixed transversal impact velocity $\epsilon\cdot v_{\mathbf{1}}$ we can determine the deflection angle $\alpha_{i}$ - and vice versa provided the latter we find the \emph{admissible velocity} $v_{(i)}$.

\begin{figure}    
  \begin{center}           
  \includegraphics[height=20cm]{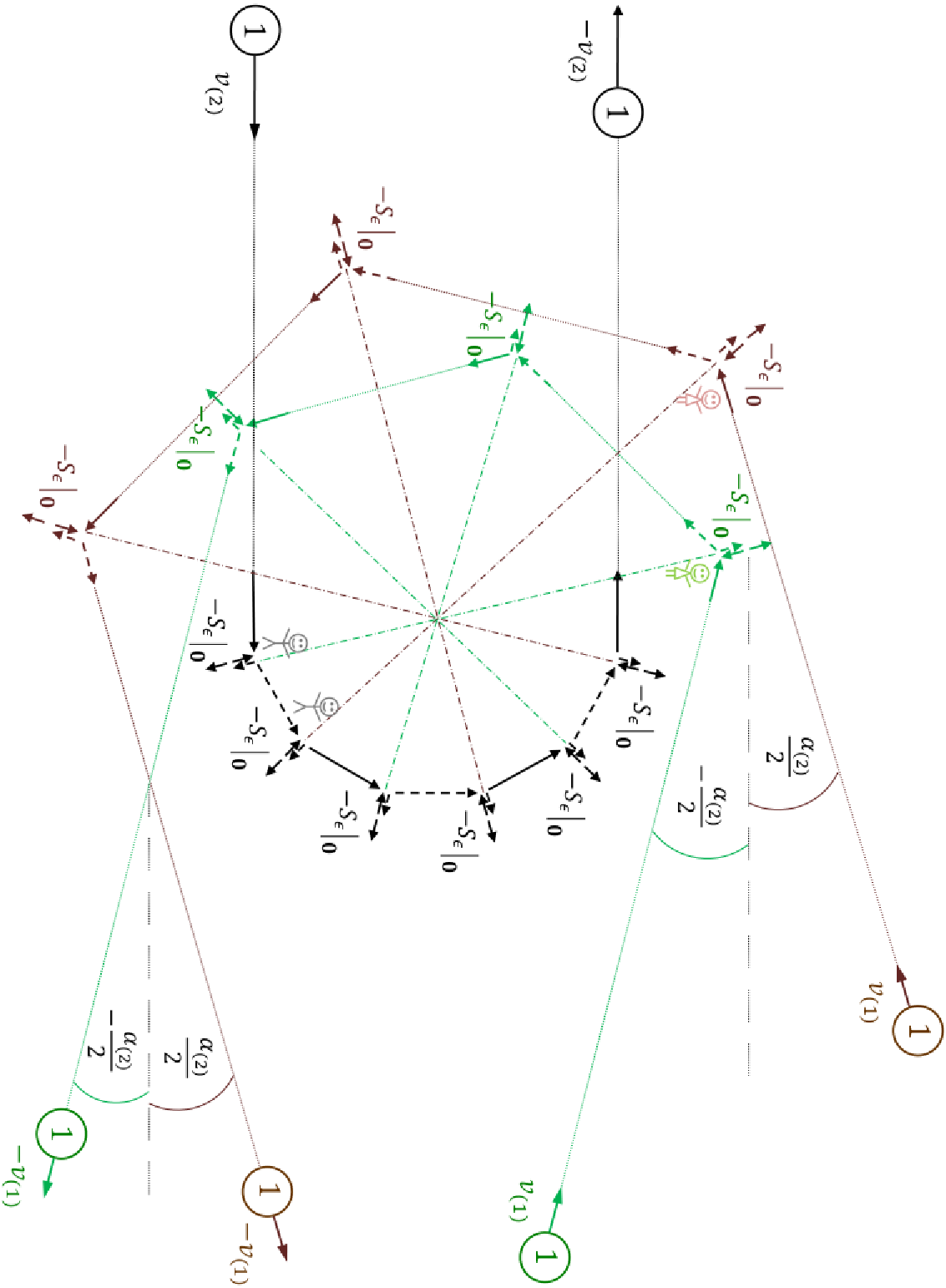}  
  \end{center}
  \vspace{-0cm}
  \caption{\label{pic_Composition1} align $2\times\!$ impulse reversion process $W_{(1)}$ and $1\times\!$ impulse reversion process $W_{(2)}$
    }
  \end{figure}

\begin{figure}    
  \begin{center}           
  \includegraphics[height=8.8cm]{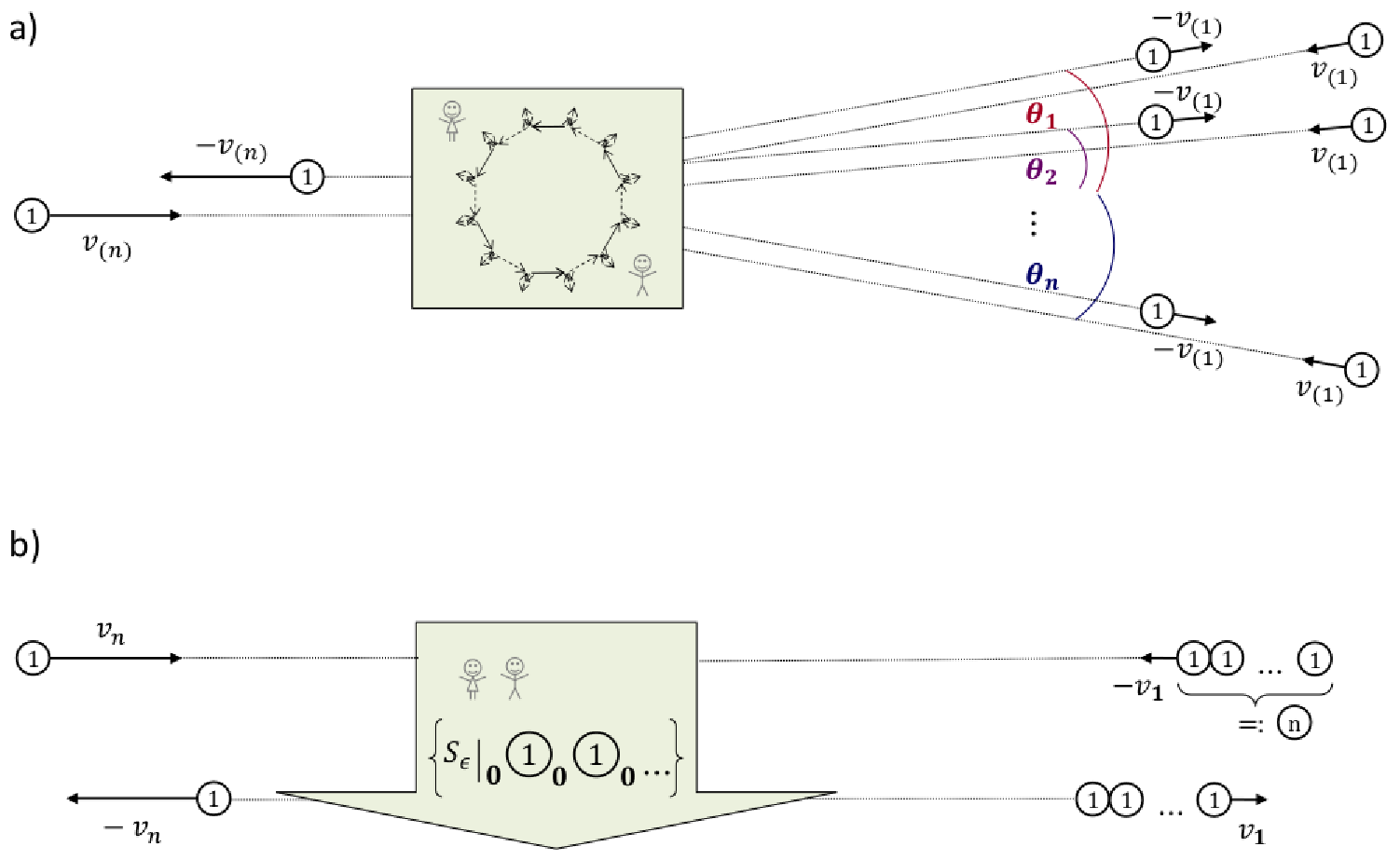}  
  \end{center}
  \vspace{-0cm}
  \caption{\label{pic_composition_coarse_graind} a) coarse grained perspective and b) in refinement limit bundle becomes a ray
    }
  \end{figure}

By a series of transversal standard kicks $w_T$ from reservoir particles we steer a \emph{reversion process} for an incident particle $\circMunit_{\:\mathbf{v}_{(1)}}$ with velocity $\mathbf{v}_{(1)}$ (see figure \ref{pic_impulse_inversion_means}); and similar for a faster particle $\circMunit_{\:\mathbf{v}_{(2)}}$ which requires twice the standard reservoir kicks, until its motion is exactly reversed. We align them in the depicted way (see figure \ref{pic_Composition1}), so that all temporarily mobilized steering elements from the center can be captured again and recycled. In the total balance the reservoir particles do not appear. In the net result \emph{only} the motion of the three incident particles (one from left and two from right side) is exactly reversed. We determine the relation between their \emph{admissible velocities} $v_{(i)}$ from \emph{matching} the building blocks $w_T \left[ \circMunit_{\:\mathbf{v}_{(1)}} \right]$ and $w_T \left[ \circMunit_{\:\mathbf{v}_{(2)}} \right]$ (so that the total configuration functions). By refinement of building blocks we construct similar models for the elastic collision of $n+1$ equivalent particles (see figure \ref{pic_composition_coarse_graind}a) and in the refinement limit (where the spreading bundle narrows to a ray) for rigid composites of $n+1$ equivalent elements (see figure \ref{pic_composition_coarse_graind}b).

\begin{theo}\label{Theorem - kin quant elast coll}
Consider a reservoir with identically constituted elements $\left\{\circMunit \right\}$. Suppose we can tightly connect $n$ of them $\underbrace{\circMunit\ast\ldots\ast\circMunit}_{n\times} =: \circMn$ such that the composite acts like one rigid unit. Let in an elastic head-on collision two different composites of standard objects
\be\label{Abschnitt -- kin quant elast coll - elast head-on collision}
   \circMunit_{\:\mathbf{v}} \,,\: \circMn_{\:\mathbf{w}} \;\;\stackrel{w_H}{\Rightarrow}\;\;\circMunit_{-\mathbf{v}} \,,\: \circMn_{-\mathbf{w}}
\ee
repulse from one another with reversed velocities. In Galilei kinematics the velocities satisfy
\be\label{Abschnitt -- kin quant elast coll - kinemtical relations elast collision two generic objects}
   \mathbf{v} \;\; = \;\; - n\cdot \mathbf{w}  \;\; .
\ee
\end{theo}
\textbf{Proof:} We approximate the collision between two generic objects. Without restricting generality we assume they are composites of unit objects $\circMunit$. When we unlock its inner binding the composite $\circMunit\ast\ldots\ast\circMunit$ becomes a swarm of $n$ unit objects $\circMunit_{\:\mathbf{w}},\ldots,\circMunit_{\:\mathbf{w}}$ with initial velocity $\mathbf{w}$; it runs head-on into one unit object $\circMunit_{\:\mathbf{v}}$ with velocity $\mathbf{v}$. A team of physicists reverse every element $\circMunit$ from both sides separately by standard kicks from an external reservoir, which in the end must stay unaffected. We model the process in three auxiliary steps.\footnote{The detailed construction can be thought of as an appendix; the end is marked by the ''$\Box$'' symbol.} We know the collision law for $1+1$ equivalent objects by symmetry and relativity principle (Lemma \ref{Lem - kin quant elast coll - step I Winkel und Geschwindigkeit}). Based on it we construct the collision model for $2+1$ equivalent objects (step I) and for $n+1$ equivalent objects (step II) and ultimately for composites of $n+1$ equivalent objects (step III) to derive the amount of matter-velocity relation for two generic objects (\ref{Abschnitt -- kin quant elast coll - kinemtical relations elast collision two generic objects}). The model is based on pre-theoretic elements \{\ref{Kap - KM Dynamics - Basic Dynamical Measures - Quantification - Dynamical Unit}\} and principles summarized in \{\ref{Kap - KM Dynamics - Discussion - Principles}\}.

In \textbf{step I} we examine the elastic head-on collision between one unit object $\circMunit_{\:v_{(2)}}$ from left with initial velocity $v_{(2)}$ and two unit objects $\circMunit_{\:\mathrm{R}_{15^{\circ}} v_{(1)}}$ and $\circMunit_{\:\mathrm{R}_{-15^{\circ}}v_{(1)}}$ from right with velocity $v_{(1)}$ under suitable orientation $15^{\circ}$ resp. $-15^{\circ}$ (see figure \ref{pic_Composition1}). We model the process by a series of transversal standard kicks and derive the admissible velocities.

\begin{figure}    
  \begin{center}           
  \includegraphics[height=18.5cm]{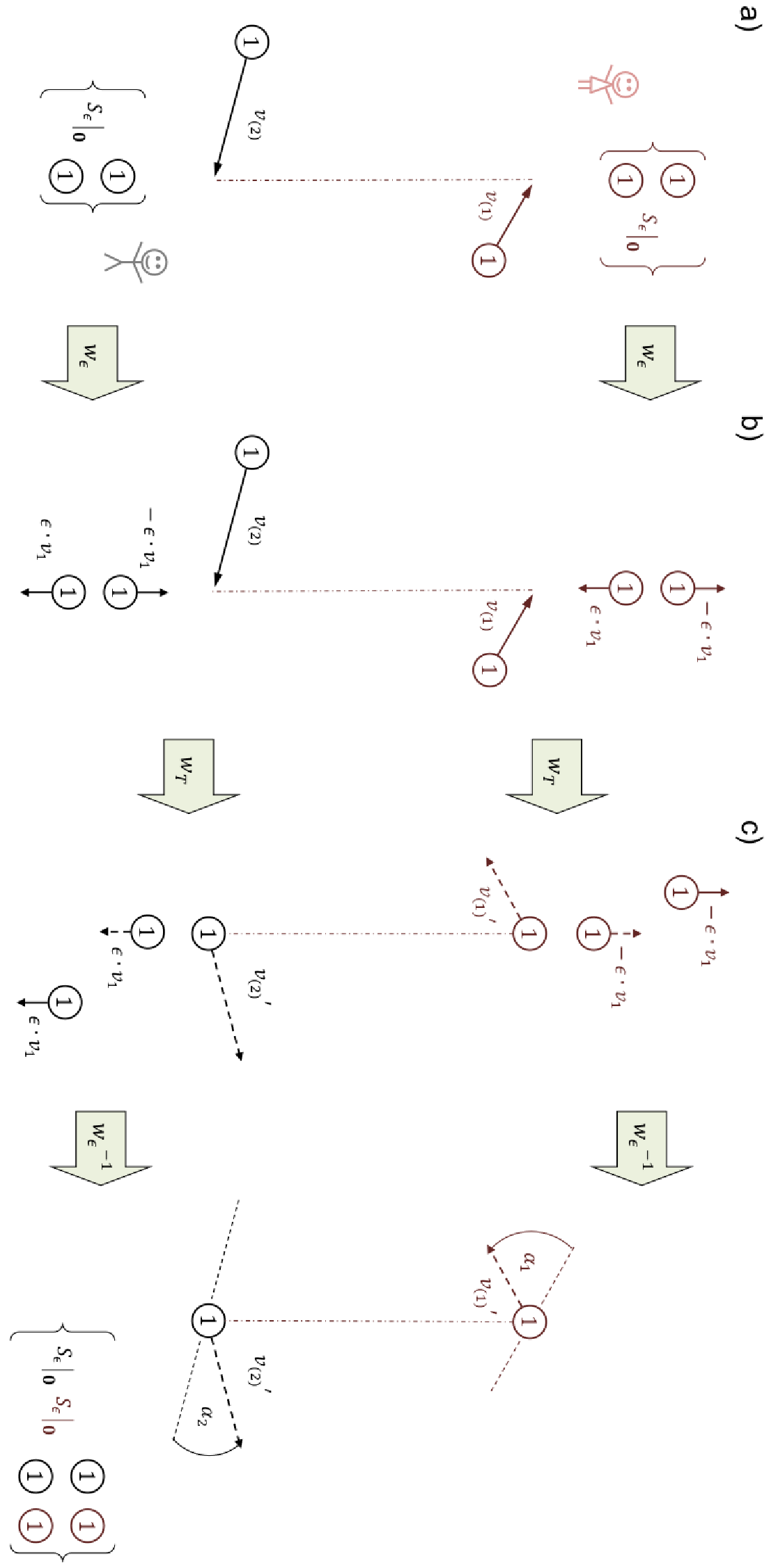}  
  \end{center}
  \vspace{-0.5cm}
  \caption{\label{pic_exact_annihilation1} at diametrically opposed positions Alice and Bob steer a) standard springs $\mathcal{S}_{\epsilon}\:\big|_{\mathbf{0}}$ against resting reservoir elements $\circMunit_{\:\mathbf{0}}$ and $\circMunit_{\:\mathbf{0}}$ to generate a pair of antiparallel impulse carriers for an b) elastic transversal collision with the incident particles $\circMunit\!\!\!\;\!\longrightarrow \!\!\!\!\!\!\!\!\!\!_{v_{\!(1)}}\;\;\;$ resp. $\circMunit\!\!\!\;\!\longrightarrow \!\!\!\!\!\!\!\!\!\!_{v_{\!(2)}}\;\;\;$ c) the antiparallel recoil particles recharge the two temporarily expended springs $\mathcal{S}_{\epsilon}\big|_{\mathbf{0}}$ and return as resting particles back into the reservoir $\left\{\circMunit_{\:\mathbf{0}} \right\}$
    }
  \end{figure}

Let Alice and Bob share an external reservoir $\left\{ \mathcal{S}_{\epsilon}\big|_{\mathbf{v}=\mathbf{0}} , \circMunit_{\:\mathbf{v}=\mathbf{0}} \right\}$. They temporarily expend standard energy sources $\mathcal{S}_{\epsilon}\big|_{\mathbf{v}=\mathbf{0}}$ (of strength $\epsilon$) against initially resting reservoir elements
\be\label{Abschnitt -- kin quant elast coll - prepare congruent transversal impulse}
   \mathcal{S}_{\epsilon}\big|_{\mathbf{0}} \,,\: \circMunit_{\:\mathbf{0}} \,,\: \circMunit_{\:\mathbf{0}} \;\;\stackrel{w_{\epsilon}}{\Rightarrow}\;\; \circMunit_{\,\epsilon\cdot\mathbf{v}_{\mathbf{1}}} \,,\: \circMunit_{-\epsilon\cdot\mathbf{v}_{\mathbf{1}}}
\ee
to \emph{prepare} transversal impulse carriers $\circMunit_{\:\epsilon\cdot \mathbf{v}_{\mathbf{1}}}$ with velocity $\epsilon\cdot v_{\mathbf{1}}$  into suitable direction $\theta$ (see figure \ref{pic_exact_annihilation1}a). They fire them into the momentary way of incident particle $\circMunit_{\:v_{(1)}}$ resp. $\circMunit_{\:v_{(2)}}$ such that the former repulse antiparallel. Each transversal kick $w_T$ successively deflects incident particle $\circMunit_{\:v_{(i)}}$ $i=1,2$ by corresponding angle $\alpha_{i}$ (see figure \ref{pic_exact_annihilation1}b). For \emph{fixed} impact velocity $\epsilon\cdot v_{\mathbf{1}}$ of the reservoir element $\circMunit_{\:\epsilon\cdot v_{\mathbf{1}}}$ and \emph{matching} deflection $\alpha_{1}=60^{\circ}$ and $\alpha_{2}=30^{\circ}$ the admissible velocities are given by $v_{(i)}\stackrel{(\ref{Formel - v-alpha-elastic transversal collision})}{:=}\sin^{-1}\!\left( \frac{\alpha_{i}}{2} \right) \cdot \epsilon\cdot v_{\mathbf{1}}$.

Alice steers the reversion process for incident object $\circMunit_{\:v_{(1)}}$ from the left (see figure \ref{pic_impulse_inversion_means}):
\begin{figure}    
  \begin{center}           
  \includegraphics[height=16.4cm]{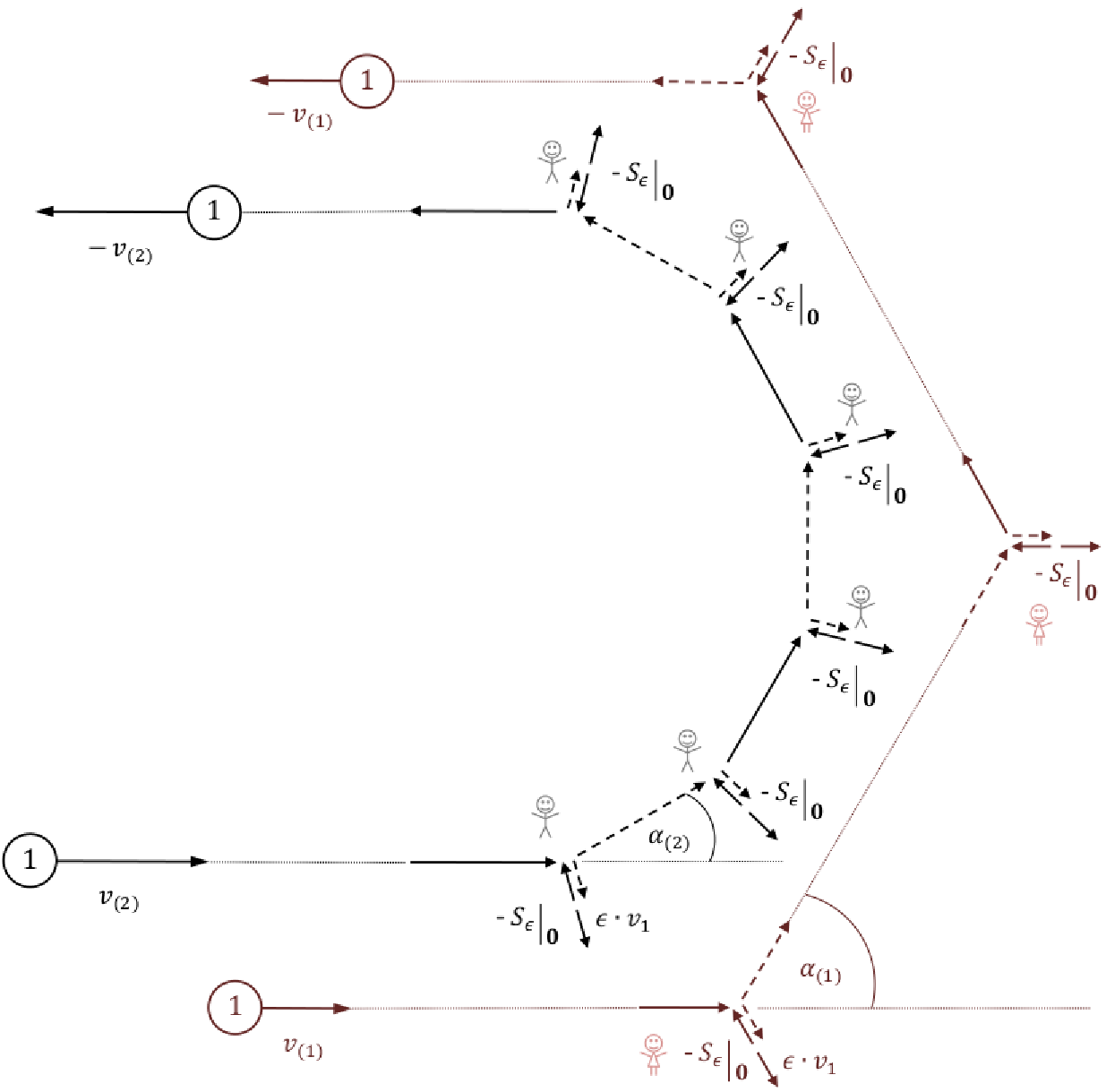}  
  \end{center}
  \vspace{-0cm}
  \caption{\label{pic_impulse_inversion_means} In a coordinated effort Alice and Bob's team of physicists steer a series of transversal standard kicks $w_{T}$ to reverse the impulse of particle $\circMunit_{\:v_{(1)}}$ resp. $\circMunit_{\:v_{(2)}}$
    }
  \end{figure}
Three assistants have to line up at the corners and know \emph{when} and \emph{where} to pick the next initially resting element $\circMunit$ from the reservoir and \emph{how} to fire it into the way of the incident particle $\circMunit_{\:v_{(1)}}$. Alice directs the initiation of each steering kick timely and at suitable position. After a series of three transversal kicks
\be
   W_{(1)} \; := \;  w_{T}^{(30^{\circ})} \ast w_{T}^{(90^{\circ})} \ast w_{T}^{(150^{\circ})} \nn
\ee
against the same object $\circMunit_{\:v_{(1)}}$, which between the kicks moves freely with same velocity $v_{(1)}$, its motion gets reversed. Bob's team steers a separate reversion process for the faster particle $\circMunit_{\:v_{(2)}}$ with velocity $v_{(2)}>v_{(1)}$ which requires twice the standard kicks (\ref{Abschnitt -- kin quant elast coll - prepare congruent transversal impulse}). Six men line up at the corners and know how to fire reservoir elements $\circMunit$ into its way
\be
   W_{(2)} \; := \; w_{T}^{(15^{\circ})} \ast w_{T}^{(45^{\circ})} \ast \ldots \ast
   w_{T}^{(165^{\circ})}  \;\; . \nn
\ee
After six successive kicks of the same strength its direction of motion is reversed too.

Alice and Bob align the reversion processes $W_{(1)}$ and $W_{(2)}$ for the three incident objects. Alice rotates her reversion process for the first incident particle $\circMunit_{\:v_{(1)}}$ by $\beta=195^{\circ}$
\[
   \mathrm{R}_{\beta}\!\left[ w_{T}^{(30^{\circ})} \ast w_{T}^{(90^{\circ})} \ast w_{T}^{(150^{\circ})} \right] \; = \; w_{T}^{(30^{\circ}+\beta)} \ast w_{T}^{(90^{\circ}+\beta)} \ast w_{T}^{(150^{\circ}+\beta)} \;\; \mathrm{;}
\]
for the second incident particle $\circMunit_{\:v_{(1)}}$ she rotates the entire configuration $\mathrm{R}_{165^{\circ}}\!\left[W_{(1)}\right]$ by an angle $\beta=165^{\circ}$. Her assistants rebuild the same model from the same building blocks in a modified orientation (symbolized by operator $\mathrm{R}_{\beta}[\;\cdot\;]$). For every transversal steering kick $w_{T}^{(\theta)} \; := \;  w_{\epsilon}^{(\theta)} \ast \; w_T$ they pick two resting unit objects $\circMunit_{\:\mathbf{0}}$ from the reservoir and generate two recoil particles $\circMunit_{-\epsilon\cdot\mathbf{v}_{\mathbf{1}}}$ with same velocity $-\epsilon\cdot\mathbf{v}_{\mathbf{1}}$: one in the preparation $w_{\epsilon}^{(\theta)}$ (see figure \ref{pic_exact_annihilation1}a) and the other after the elastic kick $w_T$ (see figure \ref{pic_exact_annihilation1}b). In order to retrieve those resources Alice and Bob align their reversion processes
\be\label{Abschnitt -- kin quant elast coll - associate three impulse reversion processes}
   W_{(2)} \; \ast \; \mathrm{R}_{165^{\circ}}\!\left[W_{(1)}\right] \; \ast \; \mathrm{R}_{195^{\circ}}\!\left[W_{(1)}\right]
\ee
such that all transversal standard kicks
\[
\begin{array}{l}
   \left\{w_{T}^{(15^{\circ})} \ast w_{T}^{(45^{\circ})} \ast \ldots \ast w_{T}^{(165^{\circ})}\right\}
    \;\ast\; \left\{ w_{T}^{(30^{\circ}+165^{\circ})} \ast w_{T}^{(90^{\circ}+165^{\circ})} \ast w_{T}^{(150^{\circ}+165^{\circ})} \right\} \nn \\
   \;\;\;\;\;\;\;\;\;\;\;\;\;\;\;\;\;\;\;\;\;\;\;\;\;\;\;\;\;\;\;\;\;\;\;\;\;\;\;\;\;\;\;\;\;\;\;\;\;\;\;\;\;
   \;\ast\; \left\{ w_{T}^{(30^{\circ}+195^{\circ})} \ast w_{T}^{(90^{\circ}+195^{\circ})} \ast w_{T}^{(150^{\circ}+195^{\circ})} \right\} \nn
\end{array}
\]
pair up along the dashed lines in diametrically opposed locations (see figure \ref{pic_Composition1})
\[
   \left(w_{T}^{(15^{\circ})} \ast w_{T}^{(30^{\circ}+165^{\circ})}\right) \ast
   \left(w_{T}^{(45^{\circ})} \ast w_{T}^{(30^{\circ}+195^{\circ})}\right) \ast \ldots \ast
   \left(w_{T}^{(165^{\circ})} \ast w_{T}^{(150^{\circ}+195^{\circ})}\right) \;\; .
\]
The four antiparallel recoil particles $\circMunit_{\:\epsilon\cdot\mathbf{v}_{\mathbf{1}}}$, $\circMunit_{\:\epsilon\cdot\mathbf{v}_{\mathbf{1}}}$, $\circMunit_{-\epsilon\cdot\mathbf{v}_{\mathbf{1}}}$ and $\circMunit_{-\epsilon\cdot\mathbf{v}_{\mathbf{1}}}$ recharge the two - temporarily expended - standard springs $\mathcal{S}_{\epsilon}\big|_{\mathbf{v}=\mathbf{0}}$ and return four resting particles back into the external reservoir $\left\{ \mathcal{S}_{\epsilon}\big|_{\mathbf{0}} , \circMunit_{\:\mathbf{0}} \right\}$ (see figure \ref{pic_exact_annihilation1}c). In the end the reservoir remains unaltered.
The net process (\ref{Abschnitt -- kin quant elast coll - associate three impulse reversion processes}) provides an elastic collision between three equivalent objects:
\be
   \circMunit_{\:\mathbf{v}_{(2)}} \,,\: \circMunit_{\:\mathrm{R}_{15^{\circ}} \mathbf{v}_{(1)}} \,,\: \circMunit_{\:\mathrm{R}_{-15^{\circ}}\mathbf{v}_{(1)}} \;\Rightarrow\; \circMunit_{-\mathbf{v}_{(2)}} \,,\: \circMunit_{-\mathrm{R}_{15^{\circ}}\mathbf{v}_{(1)}} \,,\: \circMunit_{-\mathrm{R}_{-15^{\circ}}\mathbf{v}_{(1)}}  \;\; .  \nn
\ee
In the final state their motion is exactly reversed (see figure \ref{pic_Composition1}). Alice and Bob mediate their elastic repulsion by well-defined resources from an external reservoir. Those were temporarily expended but finally all recycled back. Every act of their procedure is reversible.
\\

For \textbf{step II} we refine the building blocks. We model the elastic head-on collision between one unit object $\circMunit_{\:v_{(n)}}$ from left and a spreading bundle of $n$ unit objects $\circMunit_{\,\mathrm{R}_{\theta_1} v_{(1)}}, \ldots ,\circMunit_{\,\mathrm{R}_{\theta_n} v_{(1)}}$ from right. From the layout of our standard building blocks we determine the admissible velocities $v_{(n)}$ resp. $v_{(1)}$ and the suitable orientations $\theta_k$ for $k=1,\ldots,n$ (see figure \ref{pic_composition_coarse_graind}a).

Alice and Bob refine the strength $\epsilon$ of their radial standard kicks $w_{T}$ (\ref{Formel - transversal standard kick}). Each reservoir element $\circMunit_{\:\epsilon\cdot v_{\mathbf{1}}}$ deflects incident particle $\circMunit_{\:v_{(i)}}$ with admissible velocity $v_{(i)}$ by corresponding angle $\alpha_{i}$ $i=1,n$. Let Alice concatenate $N_{(1)}:=\frac{\pi}{\alpha_{1}}$ radial standard kicks
\be\label{Abschnitt -- kin quant elast coll - associate N_(1) transversal collisions}
   W_{(1)} \; := \;  w_{T}^{( -\frac{\alpha_1}{2} + \alpha_1 )} \ast w_{T}^{( -\frac{\alpha_1}{2} + 2\cdot \alpha_1 )} \ast \ldots \ast w_{T}^{( -\frac{\alpha_1}{2} + N_{(1)}\cdot\alpha_1 )}
\ee
to reverse the motion for each element $\circMunit_{\:v_{(1)}}$ of the right incident bundle. Similarly Bob steers $N_{(n)}:=\frac{\pi}{\alpha_n}$ radial kicks of \emph{same} strength $\epsilon$ like Alice
\be
   W_{(n)} \; := \; w_{T}^{( -\frac{\alpha_n}{2} + \alpha_n )} \ast w_{T}^{( -\frac{\alpha_n}{2} + 2\cdot \alpha_n )} \ast \ldots \ast w_{T}^{( -\frac{\alpha_n}{2} + N_{(n)}\cdot\alpha_n )}  \nn
\ee
until the direction of motion for the left particle $\circMunit_{\:v_{(n)}}$ with velocity $v_{(n)}$ is reversed too.\footnote{We operate with irreducible units and their physical concatenation. In relativistic kinematics \{\ref{Kap - SRT Massbestimmung}\} we operate with light clocks. We construct kinematical models e.g. $\mathcal{L} \ast_t \ldots \ast_t \mathcal{L}^{(1)} \ast_s \ldots \ast_s \mathcal{L}^{(n)} \; \sim_{t,s} \; \overline{\mathcal{A}_1\mathcal{O}}$ (\ref{Formel - radar spatiotemporal distance konstruierbare Ersetzung}) by connecting those congruent kinematical units $\mathcal{L}$ in a consecutive $\ast_t$ resp. adjacent $\ast_s$ way. The way of concatenation of dynamical units is more subtle. Here we construct an impulse reversion process $w_T \ast^{(i)}w_T\ast^{(j)} \ldots \ast^{(k)} w_T \sim_{E,\mathbf{p}} w$ which reproduces the same effects of a direct elastic collision $w$ and the associated energy and momentum. The model solely consists of standard kicks $w_T$ from an
external reservoir. We symbolize the way of concatenation with the index $\ast^{(i)}$. In our superscript $^{(i)}$ we suppress a whole list of specifications (in which particle, timing, position etc.). From figure \ref{pic_impulse_inversion_means} they are obvious. For the layout in (\ref{Abschnitt -- kin quant elast coll - associate N_(1) transversal collisions}) it is sufficient to specify the spatial orientation $\theta$ in the coupling of consecutive reservoir kicks $w_{T}^{( \theta )}$.}

For alignment both reversion processes $W_{(1)}$ and $W_{(n)}$ must match with one another. We impose \emph{matching} deflection angles
\be\label{Abschnitt -- kin quant elast coll - matching condition}
   \alpha_1 \stackrel{!}{=} n\cdot \alpha_n \;\; .
\ee
Then for fixed radial impact velocity $\epsilon\cdot\mathbf{v}_{\mathbf{1}}$ of the reservoir element $\circMunit_{\:\epsilon\cdot\mathbf{v}_{\mathbf{1}}}$ and deflection angle $\alpha_i$ the admissible velocities $v_{(i)}$ of the incident particle $\circMunit_{\:v_{(i)}}$ $i=1,n$ are known (\ref{Formel - v-alpha-elastic transversal collision}).

Let Alice align the $n$ bundle elements $\circMunit_{\,\mathrm{R}_{\theta_1} v_{(1)}}, \ldots ,\circMunit_{\,\mathrm{R}_{\theta_n} v_{(1)}}$ from right with velocity $v_{(1)}$
\begin{itemize}
\item   under orientations $\theta_k:= \frac{n+1}{2}\cdot\alpha_n - k \cdot \alpha_n$ for $k=1,\ldots,n$ \footnote{For step I with $n=2$, $\alpha_1=60^{\circ}$, $\alpha_2=30^{\circ}$ we verify $\theta_1:=\frac{3}{2}\cdot \alpha_2 - \alpha_2 = \frac{1}{2}\cdot \alpha_2$ and $\theta_2:=\frac{3}{2}\cdot \alpha_2 - 2\cdot\alpha_2 = -\frac{1}{2}\cdot \alpha_2$ in accordance with figure \ref{pic_Composition1}.}
\item   with equal spacing $\Delta\theta = \alpha_n$ ranging between $\theta_1=+\frac{\alpha_1}{2}-\frac{\alpha_n}{2}$ ,  $\theta_n=-\frac{\alpha_1}{2}+\frac{\alpha_n}{2}$
\end{itemize}
with the orientation of Bob's reversion process $W_{(n)}$ for incident object $\circMunit_{\:v_{(n)}}$ from left. Then Alice turns the complete reversion process (\ref{Abschnitt -- kin quant elast coll - associate N_(1) transversal collisions}) for the first bundle element $\circMunit_{\,\mathrm{R}_{\theta_1}v_{(1)}}$
\[
   \mathrm{R}_{\beta_1}\!\left[ w_{T}^{(\vartheta_1)} \ast \ldots \ast w_{T}^{(\vartheta_{N_{(1)}})} \right] \; = \; w_{T}^{(\vartheta_1+\beta_1)} \ast \ldots \ast w_{T}^{(\vartheta_{N_{(1)}}+\beta_1)}
\]
with $\vartheta_j:= -\frac{\alpha_1}{2}+j\cdot \alpha_1$ for $j=1,\ldots,N_{(1)}$ around angle $\beta_1:=\pi+\underbrace{\frac{\alpha_1}{2}-\frac{\alpha_n}{2}}_{=:\theta_1}$ and similarly she turns the reversion process $\mathrm{R}_{\beta_k}\!\left[W_{(1)}\right]$ for every other element $\circMunit_{\,\mathrm{R}_{\theta_k}v_{(1)}}$ of the incident bundle $k=1,\ldots,n$ around angle $\beta_k:=\pi+\theta_k$. Like in figure \ref{pic_Composition1} Alice and Bob connect the reversion processes
\be\label{Abschnitt -- kin quant elast coll - associate n+1 impulse reversion processes}
   W_{(n)} \; \ast \; \mathrm{R}_{\beta_1}\!\left[W_{(1)}\right] \; \ast \; \ldots \; \ast \; \mathrm{R}_{\beta_n}\!\left[W_{(1)}\right]
\ee
such that all radial steering kicks with $\gamma_l:= -\frac{\alpha_n}{2}+l\cdot \alpha_n$ for $l=1,\ldots,N_{(n)}$
\[
\begin{array}{l}
   \left\{w_{T}^{(\gamma_1)} \ast \ldots \ast w_{T}^{(\gamma_{N_{(n)}})}\right\}
    \;\ast\; \left\{ w_{T}^{(\vartheta_1+\beta_1)} \ast \ldots \ast w_{T}^{(\vartheta_{N_{(1)}}+\beta_1)} \right\} \nn \\
   \;\;\;\;\;\;\;\;\;\;\;\;\;\;\;\;\;\;\;\;\;\;\;\;\;\;\;\;\;\;\;\;\;\;\;\;\;\;\;\;\;\;\;\; \ast \; \ldots
   \;\ast\; \left\{ w_{T}^{(\vartheta_1+\beta_n)} \ast \ldots \ast w_{T}^{(\vartheta_{N_{(1)}}+\beta_n)}  \right\} \nn
\end{array}
\]
divide into antiparallel pairs\footnote{Straight forward insertion confirms that first pair is aligned antiparallel and analogous for all the rest
\bea
   \gamma_1 - (\delta_1+\beta_n) & := &  -\frac{\alpha_n}{2}+1\cdot \alpha_n \;\; - \;\; \left( -\frac{\alpha_1}{2}+1\cdot \alpha_1 \;\; + \;\; \pi + \frac{n+1}{2}\cdot\alpha_n - n \cdot \alpha_n \right) \nn \\
   & = &  \frac{\alpha_n}{2} \; - \; \frac{\alpha_1}{2} \; - \; \pi + \frac{n\cdot\alpha_n}{2} - \frac{\alpha_n}{2} \;\; \stackrel{(\ref{Abschnitt -- kin quant elast coll - matching condition})}{=} \;\; -\pi \;\; .\nn
\eea
}
\[
   \left(w_{T}^{(\gamma_1)} \ast w_{T}^{(\delta_1+\beta_n)}\right) \ast
   \left(w_{T}^{(\gamma_{2})} \ast w_{T}^{(\delta_1+\beta_{n-1})}\right) \ast \ldots \ast
   \left(w_{T}^{(\gamma_{N_{(n)}})} \ast w_{T}^{(\delta_{N_{(1)}}+\beta_1)}\right)
\]
where again all byproducts from the preparation can be recycled completely as before in figure \ref{pic_exact_annihilation1}a-c. The net process (\ref{Abschnitt -- kin quant elast coll - associate n+1 impulse reversion processes}) mediates an elastic collision of $n+1$ equivalent objects
\be
   \circMunit_{\:v_{(n)}} \,,\: \circMunit_{\:\mathrm{R}_{\theta_1} v_{(1)}} \,, \ldots ,\: \circMunit_{\:\mathrm{R}_{\theta_n} v_{(1)}}
   \;\Rightarrow\;
   \circMunit_{-v_{(n)}} \,,\: \circMunit_{-\mathrm{R}_{\theta_1} v_{(1)}} \,, \ldots ,\: \circMunit_{-\mathrm{R}_{\theta_n} v_{(1)}}  \;\; ; \nn
\ee
their motion is exactly reversed.
\\

In \textbf{step III} we refine the building blocks in model (\ref{Abschnitt -- kin quant elast coll - associate n+1 impulse reversion processes}) to the limit $\epsilon\rightarrow 0$ where the impact of individual reservoir elements $\circMunit_{\:\epsilon\cdot\mathbf{v}_{\mathbf{1}}}$ diminishes. Each radial standard kick $w_T$ (\ref{Formel - transversal standard kick}) deflects the right bundle element $\circMunit_{\:v_{\mathbf{1}}}$ with \emph{fixed} velocity $v_{(1)}\stackrel{!}{=}v_{\mathbf{1}}$ by angle $\sin \frac{\alpha_{1}}{2} \stackrel{(\ref{Formel - v-alpha-elastic transversal collision})}{:=} \frac{\epsilon}{v_{\mathbf{1}}}$ and the left particle $\circMunit_{\:v_{(n)}}$ by matching angle $\alpha_n\stackrel{(\ref{Abschnitt -- kin quant elast coll - matching condition})}{:=}\frac{1}{n}\cdot\alpha_1$. To compensate the diminishing deflection angle we integrate an increasing number $N_{(1)}:=\frac{\pi}{\alpha_{1}}$ resp. $N_{(n)}:=n\cdot N_{(1)}$ of refined standard kicks into the model until the motion of every particle from the right bundle and from the left is reversed. In return the spreading of the bundle $\theta_1-\theta_n := \lim_{\epsilon\rightarrow0} \; (n-1) \cdot \alpha_{1}(v_{\mathbf{1}},\epsilon) = 0$ narrows.

We rewrite the matching condition $\alpha_{1}(v_{\mathbf{1}},\epsilon) \stackrel{!}{=} n\cdot\alpha_{n}(v_{n},\epsilon)$ between the deflection angles of Alice and Bob's radial steering kicks
\bea
   \sin\left(\frac{\alpha_1}{2}\right) & \stackrel{!}{=} & \sin\left(n\cdot \frac{\alpha_n}{2}\right) \nn \\
   & = & \sum^{n-1}_{k=0} \left(
                            \begin{array}{c}
                              n \\
                              k \\
                            \end{array}
                          \right) \cdot
     \cos^{k}\left(\frac{\alpha_n}{2}\right) \cdot \sin^{n-k}\left(\frac{\alpha_n}{2}\right) \cdot \sin\left(\frac{1}{2}(n-k)\cdot \pi\right) \nn
\eea
with trigonometric identity of multiple angles. With substitution $\sin\frac{\alpha_i}{2}\stackrel{(\ref{Formel - v-alpha-elastic transversal collision})}{=}\frac{\epsilon}{v_i}$ we find
\be
   \frac{\epsilon}{v_{\mathbf{1}}} \;\; \stackrel{!}{=} \;\; \sum^{n-1}_{k=0} \left(
                            \begin{array}{c}
                              n \\
                              k \\
                            \end{array}
                          \right) \cdot
     \sqrt{1-\frac{\epsilon^2}{v_n^2}}^k \cdot \left(\frac{\epsilon}{v_n}\right)^{n-k} \cdot \sin\left(\frac{1}{2}(n-k)\cdot \pi\right) \nn
\ee
$\forall$ $\epsilon > 0$ and fixed $v_{(1)}\stackrel{!}{=}v_{\mathbf{1}}$ the admissible velocity $v_n(v_{\mathbf{1}},\epsilon)$ for left particle $\circMunit_{\:\mathbf{v}_n}$. For $\epsilon \ll v_{\mathbf{1}} < v_{(n)}$ we neglect terms of higher order $\mathcal{O} ( \frac{\epsilon}{v_{(n)}} )^2$ and keep the dominant for $k=n-1$
\[
   \frac{1}{v_{\mathbf{1}}} \;\; \stackrel{!}{=} \;\; \lim_{\epsilon\rightarrow0} \; n \cdot \sqrt{1-\frac{\epsilon^2}{v_n^2}}^{n-1} \cdot \frac{1}{v_n} \;\; = \;\;
   n \cdot \frac{1}{\lim_{\epsilon\rightarrow0} \; v_n} \;\; .
\]
In the limit - of refined steering kicks $\epsilon \rightarrow 0$ by reservoir elements $\circMunit_{\:\epsilon\cdot\mathbf{v}_{\mathbf{1}}}$ - we approximate (\ref{Abschnitt -- kin quant elast coll - associate n+1 impulse reversion processes}) the elastic head-on collision between one standard object $\circMunit_{\:\mathbf{v}_{n}}$ and a parallel beam of $n$ elements $\{\circMunit_{\:\mathbf{v}_{\mathbf{1}}}, \ldots , \circMunit_{\:\mathbf{v}_{\mathbf{1}}} \}\equiv \circMn_{\:\mathbf{v}_{\mathbf{1}}}$ (see figure \ref{pic_composition_coarse_graind}b).
Before and after the collision they fly with same velocity $\mathbf{v}_{\mathbf{1}}$ as if they were bound in a composite $\circMn_{\:\mathbf{v}_{\mathbf{1}}}$
\be
   \circMunit_{\:\mathbf{v}_n} \,,\: \circMn_{\:\mathbf{v}_{\mathbf{1}}} \;\;\stackrel{w_H}{\Rightarrow}\;\; \circMunit_{-\mathbf{v}_n} \,,\: \circMn_{-\mathbf{v}_{\mathbf{1}}} \nn \;\;.
\ee
In the limit, when the bundle becomes a ray, the initial velocities $\mathbf{v}_{\mathbf{1}}$,$\,\mathbf{v}_n$ satisfy relation
\be
   \lim_{\epsilon\rightarrow0} \; \mathbf{v}_n \;\; = \; - n \cdot \mathbf{v}_{\mathbf{1}} \;\; . \nn
\ee
\qed
We learn something about elastic collisions which we did not presuppose before. Our model provides a physical derivation of fundamental (collision) equation $m_1\cdot \Delta v_1 = m_2 \cdot \Delta v_2$ (including scope and limitations). Now we know more about elastic collisions than before.

\subsection{Calorimeter absorption model}\label{Kap - KM Dynamics - Basic Dynamical Measures - Measurement Means - Kinematic Quantification Calorimeter Action}

By controlled linkage of elastic head-on collisions and by relativity principle (view from the moving observer) we model an absorption process for a generic object $\circMO_{\:\mathbf{v}}$ in a calorimeter reservoir $\left\{ \circMunit_{\:\mathbf{v}=0} \right\}$. Physicists steer a series of deceleration kicks:
\begin{itemize}
    \item   They \emph{place} composites of resting reservoir elements $\circMunit_{\:\mathbf{0}}\ast\ldots\ast\circMunit_{\:\mathbf{0}}$ into the way of the incident object $\circMO_{\:\mathbf{v}}$. They \emph{generate} elastic head-on collisions $w_H$ (\ref{Abschnitt -- kin quant elast coll - elast head-on collision}) which rebound the object $\circMO_{\,\mathbf{v}'}$ with reduced velocity and kick the composite $\circMunit_{\:\mathbf{v}_{\mathbf{1}}}\ast\ldots\ast\circMunit_{\:\mathbf{v}_{\mathbf{1}}}$ into standard motion. Successively object $\circMO_{\:\mathbf{v}'}$ oscillates inside the deceleration cascade and kicks new initially resting elements out of the calorimeter reservoir $\left\{ \circMunit_{\:\mathbf{v}=0} \right\}$.
    \item The number of recoil particles $\sharp\left\{\circMunit_{\:\mathbf{v}_{\mathbf{1}}}\right\}$ quantifies the energy and momentum.
\end{itemize}

\begin{figure}    
  \begin{center}           
  \includegraphics[height=18cm]{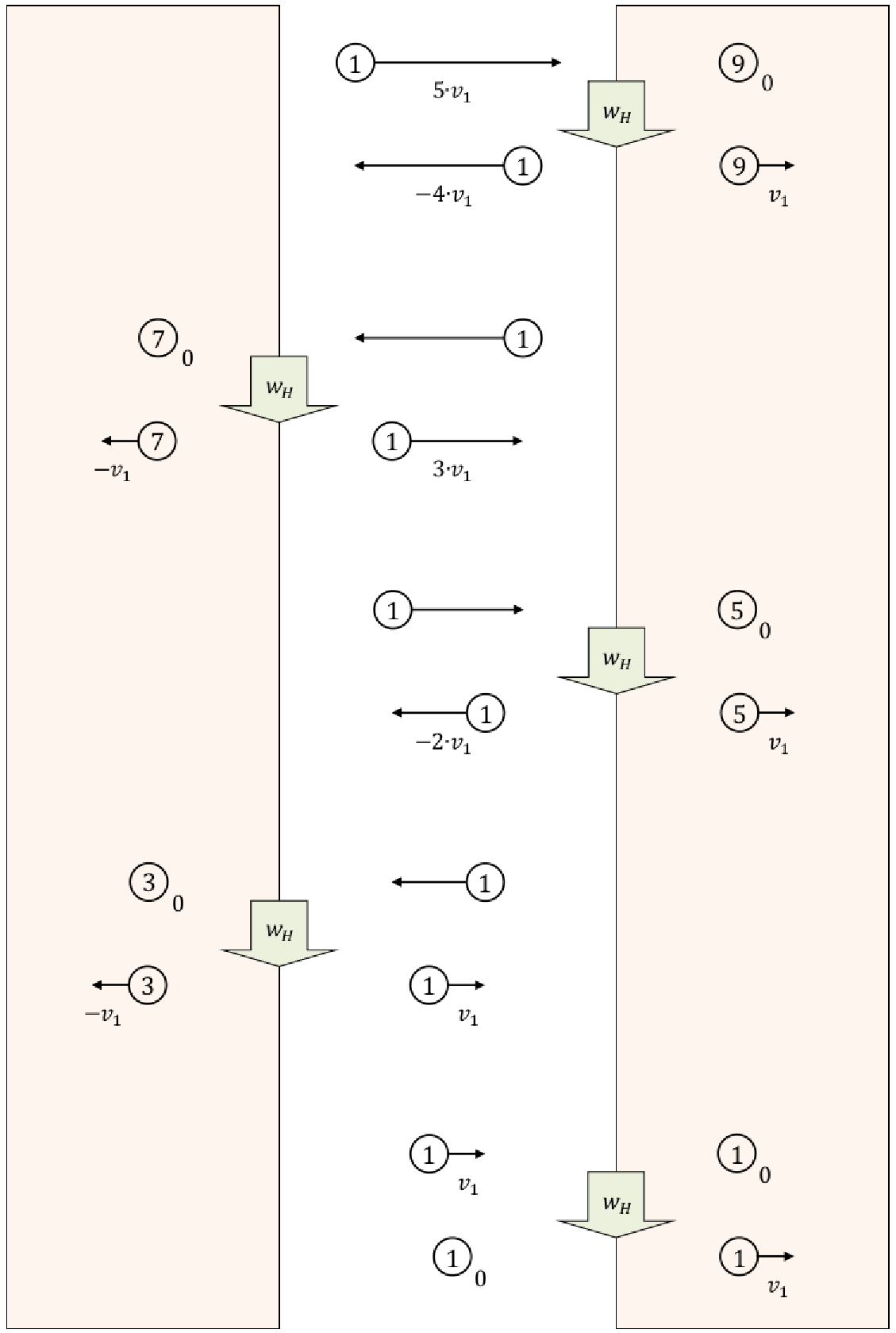}  
  \end{center}
  \vspace{-0cm}
  \caption{\label{pic_calorimeter_model} incident particle successively comes to rest by means of  elastic collisions with initially resting elements on the left resp. right side of the calorimeter reservoir $\left\{ \circMunit_{\:\mathbf{v}=\mathbf{0}} \right\}$
    }
  \end{figure}

Consider for example the elastic head-on collision (\ref{Abschnitt -- kin quant elast coll - elast head-on collision}) between one (fast) standard object and a composite of $9$ elements. For a drive-by observer the incident particle kicks a resting composite into motion $\mathbf{v}_{\mathbf{1}}$ and rebounds with reduced velocity to the left (see figure \ref{pic_calorimeter_model}). From those decelerations kicks we build our calorimeter model $W_{\mathrm{cal}}$. On the left we place again a suitable number of $7$ reservoir elements into the way, such that they get kicked out with the same standard velocity $\mathbf{v}_{\mathbf{1}}$. The incident particle successively rebounds with reduced velocity, until (after five right- and left-deceleration kicks) it stops inside the calorimeter. For the controlled deceleration of a particle $\circMunit_{\:\textcolor{cyan}{5}\cdot \mathbf{v}_{\mathbf{1}}}$ with velocity $5\cdot \mathbf{v}_{\mathbf{1}}$ we mobilize a total of $25$ initially resting reservoir elements. We kick $10$ particle pairs $\left\{\circMunit_{\:\mathbf{v}_{\mathbf{1}}}, \circMunit_{-\mathbf{v}_{\mathbf{1}}}\right\}$ with the same standard velocity $\pm\mathbf{v}_{\mathbf{1}}$ out of both sides of the calorimeter and $5$ single recoil particles $\circMunit_{\:\mathbf{v}_{\mathbf{1}}}$
\be
   \circMunit_{\:\textcolor{cyan}{5}\cdot \mathbf{v}_{\mathbf{1}}} \,,\: \textcolor{magenta}{25}\cdot \circMunit_{\:\mathbf{0}} \;\;\;
   \Rightarrow \;\;\;
   \circMunit_{\:\textcolor{cyan}{\mathbf{0}}} \,,\: \textcolor{magenta}{10} \cdot \left\{\circMunit_{\:\mathbf{v}_{\mathbf{1}}}, \circMunit_{-\mathbf{v}_{\mathbf{1}}}\right\} \,,\:  \textcolor{magenta}{5}\cdot \circMunit_{\:\mathbf{v}_{\mathbf{1}}} \nn \;\; .
\ee
We formulate the total balance of this process as ''reaction equation'' (along the language use among chemists).\footnote{Physicists formulate equations between \emph{measures} (\emph{Ma\ss{}}gleichungen); chemists on the contrary transitions between their \emph{carriers} (Ma\ss{}\emph{tr\"ager}) - we formulate both sides: carrier and its measure!} If we absorb the same standard particle $\circMunit_{\:\textcolor{cyan}{8}\cdot \mathbf{v}_{\mathbf{1}}}$ with higher velocity $8\cdot \mathbf{v}_{\mathbf{1}}$, we have to mobilize even $64$ initially resting reservoir elements. In the same series of deceleration kicks
\be
   \circMunit_{\:\textcolor{cyan}{8}\cdot \mathbf{v}_{\mathbf{1}}} \,,\: \textcolor{magenta}{64}\cdot \circMunit_{\:\mathbf{0}} \;\;\;
   \Rightarrow \;\;\;
   \circMunit_{\:\textcolor{cyan}{\mathbf{0}}} \,,\: \textcolor{magenta}{28} \cdot \left\{\circMunit_{\:\mathbf{v}_{\mathbf{1}}}, \circMunit_{-\mathbf{v}_{\mathbf{1}}}\right\} \,,\:  \textcolor{magenta}{8}\cdot \circMunit_{\:\mathbf{v}_{\mathbf{1}}}  \nn
\ee
we generate $28$ standard particle pairs and $8$ impulse carriers.

This illustrates the \emph{essence} of a basic measurement. On the billiard table in the beginning the ''capability to work'' and ''impact'' of moving bodies were only defined vaguely by pre-theoretic ordering relations ''more absorption effect against same pile of object balls than'' and ''overrunning in head-on collision''. This practical comparison $\circMunit_{\:\textcolor{cyan}{8}\cdot \mathbf{v}_{\mathbf{1}}} >_{E,\mathbf{p}} \circMunit_{\:\textcolor{cyan}{5}\cdot \mathbf{v}_{\mathbf{1}}}$ is now precisely determined by a \emph{number of equivalent reference elements} ($\textcolor{magenta}{28}$ resp. $\textcolor{magenta}{10}$ particle pairs $\left\{\circMunit_{\:\mathbf{v}_{\mathbf{1}}}, \circMunit_{-\mathbf{v}_{\mathbf{1}}}\right\}$ of equal capability to work and $\textcolor{magenta}{8}$ resp. $\textcolor{magenta}{5}$ recoil particles $\circMunit_{\:\mathbf{v}_{\mathbf{1}}}$ of equal impact). We will assess the physical meaning of the extracted calorimeter elements as units of energy and momentum by pre-theoretic ordering relations in Proposition \ref{Prop - kin quant Absorptions Wirkung - pre-theoretical characterization E unit and p unit}.
\begin{pr}
The calorimeter-deceleration-cascade $W_{\mathrm{cal}}$ is a physical model for absorbing unit object $\circMunit_{\:\textcolor{cyan}{n}\cdot \mathbf{v}_{\mathbf{1}}}$ with velocity $n\cdot \mathbf{v}_{\mathbf{1}}$ in an external calorimeter where it comes to rest $\circMunit_{\:\mathbf{0}}$
\be
   \circMunit_{\:\textcolor{cyan}{n}\cdot \mathbf{v}_{\mathbf{1}}}  \;\;
   \stackrel{W_{\mathrm{cal}}}{\Rightarrow} \;\;
   \circMunit_{\:\textcolor{cyan}{\mathbf{0}}} \,,\:
   \mathrm{RB} \nn \;\; .
\ee
In return we extract the reservoir balance for absorption
\be\label{Abschnitt -- kin quant Absorptions Wirkung - Reservoirbilanz - absorption}
   \mathrm{RB}\left[\circMunit_{\:n\cdot \mathbf{v}_{\mathbf{1}}} \Rightarrow \circMunit_{\:\mathbf{0}} \right] \;\; := \;\;
   \left( \frac{1}{2}\cdot n^2 - \frac{1}{2}\cdot n  \right)\cdot \left\{\circMunit_{-\mathbf{v}_{\mathbf{1}}}, \circMunit_{\:\mathbf{v}_{\mathbf{1}}}\right\} \;\; , \;\; n\cdot \circMunit_{\:\mathbf{v}_{\mathbf{1}}}
\ee
a certain number of standard particle pairs $\left\{\circMunit_{-\mathbf{v}_{\mathbf{1}}}, \circMunit_{\:\mathbf{v}_{\mathbf{1}}}\right\}$ and impulse carriers $\circMunit_{\:\mathbf{v}_{\mathbf{1}}}$ from a reservoir with resting standard elements $\left\{ \circMunit_{\:\mathbf{0}} \right\}$ (which we suppress in the notation).
\end{pr}
\textbf{Proof:}
Let the initial velocity be $(2 i+1)\cdot \mathbf{v}_{\mathbf{1}}$. Alice steers a cascade of elastic collisions with suitable packs of resting reservoir elements. We need to adjust the deceleration kicks.

Let Bob pick a suitable head-on collision $w_H$, which satisfies the relation $\mathbf{v} \stackrel{(\ref{Abschnitt -- kin quant elast coll - kinemtical relations elast collision two generic objects})}{=} - n\cdot \mathbf{w}$ (with initial velocities $\mathbf{v}:=(2i+\frac{1}{2})\cdot \mathbf{v}_{\mathbf{1}}$, $\mathbf{w}:=-\frac{1}{2}\cdot \mathbf{v}_{\mathbf{1}}$ and $n:=4i+1$). We pick the numerical values for later convenience. Thus Bob prepares a composite $\underbrace{\circMunit_{\:\mathbf{0}}\ast \ldots \ast \circMunit_{\:\mathbf{0}}}_{(4i+1)\times}$ of $(4i+1)$ elements such that in an elastic head-on collision (\ref{Abschnitt -- kin quant elast coll - elast head-on collision})
\be\label{Abschnitt -- kin quant Absorptions Wirkung - elastic longitudinal collision - antisym}
   \circMunit_{\:(2i+\frac{1}{2})\cdot \mathbf{v}_{\mathbf{1}}} \,,\: (4i+1)\cdot\circMunit_{-\frac{1}{2}\cdot\mathbf{v}_{\mathbf{1}}} \;\;\stackrel{w_H}{\Rightarrow}\;\; \circMunit_{-(2i+\frac{1}{2})\cdot \mathbf{v}_{\mathbf{1}}} \,,\: (4i+1)\cdot\circMunit_{\:\frac{1}{2}\cdot\mathbf{v}_{\mathbf{1}}}
\ee
the incident particle $\circMunit_{\:(2i+\frac{1}{2})\cdot \mathbf{v}_{\mathbf{1}}}$ with velocity $(2i+\frac{1}{2})\cdot \mathbf{v}_{\mathbf{1}}$ rebounds antiparallel from the composite $(4i+1)\cdot\circMunit_{-\frac{1}{2}\cdot\mathbf{v}_{\mathbf{1}}}$ with velocity $-\frac{1}{2}\cdot\mathbf{v}_{\mathbf{1}}$.

Let $\mathcal{B}$ob move relative to $\mathcal{A}$lice with constant velocity $\mathbf{v}_{\mathcal{B}}=\frac{1}{2}\cdot\mathbf{v}_{\mathbf{1}^{(\mathcal{A})}}$ to the right. She will see $\mathcal{B}$ob's collision with the same number of colliding elements, but different measured values of velocity (Galilei covariant transformation $v_{\circMunit}^{(\mathcal{A})}= v_{\circMunit}^{(\mathcal{B})} + v_{\mathcal{B}}^{(\mathcal{A})}$ for every element $\circMunit$). For $\mathcal{A}$lice the incident particle $\circMunit_{\:(2i+1)\cdot \mathbf{v}_{\mathbf{1}}}$ kicks into the right side of the calorimeter with velocity $(2i+1)\cdot \mathbf{v}_{\mathbf{1}}$ and rebounds with reduced velocity $-2i\cdot \mathbf{v}_{\mathbf{1}}$ to the left
\be\label{Abschnitt -- kin quant Absorptions Wirkung - elastic longitudinal collision - rechts}
   w_{H,r} : \;\; \circMunit_{\:(2i+1)\cdot \mathbf{v}_{\mathbf{1}}} \,,\: (4i+1)\cdot\circMunit_{\:\mathbf{0}} \;\;\stackrel{(\ref{Abschnitt -- kin quant Absorptions Wirkung - elastic longitudinal collision - antisym})}{\Rightarrow}\;\; \circMunit_{-2i\cdot \mathbf{v}_{\mathbf{1}}} \,,\: (4i+1)\cdot\circMunit_{\:\mathbf{v}_{\mathbf{1}}}
\ee
while the resting composite $(4i+1)\cdot\circMunit_{\:\mathbf{0}}$ gets kicked into standard velocity $\mathbf{v}_{\mathbf{1}}$. On the left side Alice places a new composite of $(4j+1)$ elements and generates the next analog deceleration kick
\be
   w_{H,l} : \;\; \circMunit_{-(2j+1)\cdot \mathbf{v}_{\mathbf{1}}} \,,\: (4j+1)\cdot\circMunit_{\:\mathbf{0}} \;\;\stackrel{(\ref{Abschnitt -- kin quant Absorptions Wirkung - elastic longitudinal collision - rechts})}{\Rightarrow}\;\; \circMunit_{\:2j\cdot \mathbf{v}_{\mathbf{1}}} \,,\: (4j+1)\cdot\circMunit_{-\mathbf{v}_{\mathbf{1}}} \;\; . \nn
\ee
with $j:=i-\frac{1}{2}$. After each round of right and left collisions $W := w_{H,r} \ast w_{H,l}$
\be\label{Abschnitt -- kin quant Absorptions Wirkung - elastic longitudinal collision - Umkehrabfolge}
   \circMunit_{\:(2i+1)\cdot \mathbf{v}_{\mathbf{1}}} \,,\: (4i+1)\cdot\circMunit_{\:\mathbf{0}} \,,\: (4i-1)\cdot\circMunit_{\:\mathbf{0}}
   \;\;\stackrel{W}{\Rightarrow}\;\;
   \circMunit_{\:(2i-1)\cdot \mathbf{v}_{\mathbf{1}}} \:,\;  (4i+1)\cdot\circMunit_{\:\mathbf{v}_{\mathbf{1}}} \,,\: (4i-1)\cdot\circMunit_{-\mathbf{v}_{\mathbf{1}}}
\ee
we add the extracted reservoir elements $(4i-1)\cdot\circMunit_{-\mathbf{v}_{\mathbf{1}}}\,,\: (4i+1)\cdot\circMunit_{\:\mathbf{v}_{\mathbf{1}}}$ from both sides of the calorimeter and the successive deceleration $\Delta v \stackrel{(\ref{Abschnitt -- kin quant Absorptions Wirkung - elastic longitudinal collision - Umkehrabfolge})}{:=} -2\cdot v_{\mathbf{1}}$. Each antiparallel particle pair $\circMunit_{-\mathbf{v}_{\mathbf{1}}} ,  \circMunit_{\:\mathbf{v}_{\mathbf{1}}} \stackrel{w_{\mathbf{1}}^{-1}}{\Rightarrow} \mathcal{S}_{\mathbf{1}}\big|_{\mathbf{0}} , \circMunit_{\:\mathbf{0}} , \circMunit_{\:\mathbf{0}}$ can charge a spring $\mathcal{S}_{\mathbf{1}}\big|_{\mathbf{0}}$ by standard process $w_{\mathbf{1}}^{-1}$  (\ref{Abschnitt -- basic dynamical measures - Einheitswirkung}). The resting elements stay in the calorimeter reservoir $\left\{ \circMunit_{\:\mathbf{0}} \right\}$. On each deceleration step $W_{(i)}$ $i=1,\dots,N$ Alice extracts
\be\label{Abschnitt -- kin quant Absorptions Wirkung - Reservoirbilanz - Deceleration}
   \mathrm{RB}\left[\circMunit_{\:(2i+1)\cdot \mathbf{v}_{\mathbf{1}}} \Rightarrow \circMunit_{\:(2i-1)\cdot \mathbf{v}_{\mathbf{1}}} \right] \; \stackrel{(\ref{Abschnitt -- kin quant Absorptions Wirkung - elastic longitudinal collision - Umkehrabfolge}) }{:=} \;
   (4i-1)\cdot  \mathcal{S}_{\mathbf{1}}\big|_{\mathbf{0}} \;\; , \;\; 2 \cdot \circMunit_{\:\mathbf{v}_{\mathbf{1}}}
\ee
$(4i-1)\cdot  \mathcal{S}_{\mathbf{1}}\big|_{\mathbf{0}}$ standard springs and $2\cdot \circMunit_{\:\mathbf{v}_{\mathbf{1}}}$ single elements (see figure \ref{pic_calorimeter_model}).\footnote{Two consecutive deceleration rounds and a final clean-up kick bring incident particle $\circMunit_{\:5\cdot \mathbf{v}_{\mathbf{1}}}$ with velocity $5\cdot \mathbf{v}_{\mathbf{1}}$ to rest. Alice counts $7+3$ particle pairs $\left\{\circMunit_{\:\mathbf{v}}, \circMunit_{-\mathbf{v}}\right\}$ and $2+2+1$ impulse carriers $\circMunit_{\:\mathbf{v}_{\mathbf{1}}}$.}

For the initial velocity $(2N+1)\cdot \mathbf{v}_{\mathbf{1}}$ Alice steers $N$ consecutive deceleration rounds $W_{\mathrm{cal}} \; := \; W_{(1)} \ast\dots\ast W_{(N)}$ inside the calorimeter until the incident object $\circMunit_{\:(2N+1)\cdot \mathbf{v}_{\mathbf{1}}}$ stops. Alice extracts the total reservoir balance for absorption  (\ref{Abschnitt -- kin quant Absorptions Wirkung - Reservoirbilanz - absorption})
\bea
   \mathrm{RB}\left[\circMunit_{\:(2N+1)\cdot \mathbf{v}_{\mathbf{1}}} \Rightarrow \circMunit_{\: \mathbf{0}} \right] & = &
   \sum_{i=1}^{N} \mathrm{RB}\left[\circMunit_{\:(2i+1)\cdot \mathbf{v}_{\mathbf{1}}} \Rightarrow \circMunit_{\:(2i-1)\cdot \mathbf{v}_{\mathbf{1}}} \right]
   \; + \; \mathrm{RB}\left[\circMunit_{\:\mathbf{v}_{\mathbf{1}}} \Rightarrow \circMunit_{\: \mathbf{0}} \right] \nn \\
   & \stackrel{(\ref{Abschnitt -- kin quant Absorptions Wirkung - Reservoirbilanz - Deceleration})}{=} & \sum_{i=1}^{N} \left( (4i-1) \cdot \mathcal{S}_{\mathbf{1}}\big|_{\mathbf{0}} \: , \; 2 \cdot \circMunit_{\:\mathbf{v}_{\mathbf{1}}} \right)
   \; + \; \left( 0\cdot \mathcal{S}_{\mathbf{1}}\big|_{\mathbf{0}} \: , \; 1 \cdot \circMunit_{\:\mathbf{v}_{\mathbf{1}}} \right) \nn \\
   & = & \!\!\! \underbrace{(2\cdot N^2+N)}_{\frac{1}{2}\cdot (2N + 1)^2 - \frac{1}{2}\cdot (2N +1)} \!\!\!\!
   \cdot \; \mathcal{S}_{\mathbf{1}}\big|_{\mathbf{0}} \;\: , \;\; (2N+1) \cdot \circMunit_{\:\mathbf{v}_{\mathbf{1}}}  \;\; .  \nn
\eea
\qed

Our calorimeter-collision-cascade $W_{\mathrm{cal}}$ absorbs the motion of a generic object $\circMO_{\:\mathbf{v}}$ by elastic collisions in return for the extraction of standard impulse carriers $\circMunit_{\:\mathbf{v}_{\mathbf{1}}}$.
\begin{lem}\label{Lem - kin quant Absorptions Wirkung - Reservoirbilanz - additivitaet}
The reservoir balance for absorbing an entire system is \underline{additive} in the number of elements $\circMi_{\:\mathbf{v}_{i}}$ $i=1,\dots,n$
\be
   \mathrm{RB} \left[ \circMunit_{\:\mathbf{v}_{1}} ,\dots, \circMn_{\:\mathbf{v}_{n}}
   \Rightarrow \circMunit_{\:\mathbf{0}} ,\dots, \circMn_{\:\mathbf{0}} \right]
   \; = \; \mathrm{RB} \left[ \circMunit_{\:\mathbf{v}_{1}} \Rightarrow \circMunit_{\:\mathbf{0}}  \right] \; + \ldots + \; \mathrm{RB} \left[ \circMn_{\:\mathbf{v}_{n}} \Rightarrow  \circMn_{\:\mathbf{0}} \right] \nn \;\; .
\ee
For the kinetic effect $\Delta\mathbf{v}_i$ of a generic interaction of motion we axtract
\be\label{Abschnitt -- kin quant Absorptions Wirkung - Reservoirbilanz - additivitaet}
   \mathrm{RB} \left[ \circMunit_{\:\mathbf{v}_{1}} ,\dots, \circMn_{\:\mathbf{v}_{n}}
   \Rightarrow \circMunit_{\:\mathbf{v}'_{1}} ,\dots, \circMn_{\:\mathbf{v}'_{n}} \right]
   \; = \; \sum_{i=1}^n \left( \mathrm{RB} \left[ \circMi_{\:\mathbf{v}_{i}} \Rightarrow \circMi_{\:\mathbf{0}} \right] \; - \mathrm{RB} \left[ \circMi_{\:\mathbf{v}'_i} \Rightarrow \circMi_{\:\mathbf{0}} \right] \right)
\ee
\end{lem}
\textbf{Proof:}
One can absorb every element $\circMi_{\:\mathbf{v}_{i}} \Rightarrow \circMi_{\:\mathbf{0}}$ in a separate absorption porcess $W_{\mathrm{cal}}^{(i)}$.
From $W_{\mathrm{cal}}^{(1)}, \ldots,W_{\mathrm{cal}}^{(n)}$ one extracts standard springs $\mathcal{S}_{\mathbf{1}}\big|_{\mathbf{v}=0}$ and impulse carriers $\circMunit_{\:\mathbf{v}_{\mathbf{1}}}$, which are all congruent with one another; their total number simply adds up. One can steer every step in the calorimeter-deceleration-cascade in the reverse way as an acceleration. By expending the absorption extract $- \mathrm{RB} \left[ \circMa_{\:\mathbf{v}'_a} \Rightarrow \circMa_{\:\mathbf{0}} \right]$ against the resting object $\circMO_{\:\mathbf{v}=0}$ we reproduce its initial state of motion
\be
   \circMa_{\:\mathbf{0}} \,,\: -\mathrm{RB}  \;\; \stackrel{W_{\mathrm{cal}}^{-1}}{\Rightarrow} \;\; \circMa_{\:\mathbf{v}'_a} \nn \;\; .
\ee
\qed

We show the universality of the calorimeter model. We construct the \emph{physical connection} between two calorimeters with a refined extraction velocity (Lemma \ref{Lem - kin quant Absorptions Wirkung - Reservoirbilanz - refinement}) and between standard calorimeters, which move relative to one another (Lemma \ref{Lem - kin quant Absorptions Wirkung - Reservoirbilanz - boost E and p units}).

Let $\mathcal{A}$lice and $\mathcal{B}$ob share a calorimeter reservoir $\left\{ \circMunit_{\:\mathbf{v}=0} \right\}$ with identically constituted standard elements $\circMunit^{(\mathcal{A})} \sim_{m} \circMunit^{(\mathcal{B})}$. $\mathcal{A}$lice calorimeter model ${W_{\mathrm{cal}}}^{(\mathcal{A})}$ is build from intrinsic standard actions (\ref{Abschnitt -- basic dynamical measures - Einheitswirkung})
$\mathcal{S}_{\mathbf{1}^{(\mathcal{A})} }\big|_{\mathbf{0}} , \circMunit_{\:\mathbf{0}} , \circMunit_{\:\mathbf{0}} \stackrel{w_{\mathbf{1}^{(\mathcal{A})}}}{\Rightarrow}  \circMunit_{\:\mathbf{v}_{\mathbf{1}^{(\mathcal{A})}}} , \circMunit_{-\mathbf{v}_{\mathbf{1}^{(\mathcal{A})}}}$; while $\mathcal{B}$ob builds his calorimeter ${W_{\mathrm{cal}}}^{(\mathcal{B})}$ from intrinsic actions
$\mathcal{S}_{\mathbf{1}^{(\mathcal{B})} }\big|_{\mathbf{0}} , \circMunit_{\:\mathbf{0}} \,,\: \circMunit_{\:\mathbf{0}} \stackrel{w_{\mathbf{1}^{(\mathcal{B})}}}{\Rightarrow} \circMunit_{\:\mathbf{v}_{\mathbf{1}^{(\mathcal{B})}}} , \circMunit_{-\mathbf{v}_{\mathbf{1}^{(\mathcal{B})}}}$. Let the $\mathcal{A}$lice extract impulse carriers $\circMunit_{\:\mathbf{v}_{\mathbf{1}^{(\mathcal{A})}}}$ with a standard velocity $\mathbf{v}_{\mathbf{1}^{(\mathcal{A})}} = k\cdot \mathbf{v}_{\mathbf{1}^{(\mathcal{B})}}\,$, $k\in\mathbb{N}$ which is a multiple of $\mathcal{B}$ob's extraction velocity $\mathbf{v}_{\mathbf{1}^{(\mathcal{B})}}$.
\begin{lem}\label{Lem - kin quant Absorptions Wirkung - Reservoirbilanz - refinement}
$\mathcal{B}$ob can \underline{refine} $\mathcal{A}$lice calorimeter measurements ${W_{\mathrm{cal}}}^{(\mathcal{A})}$ with standard elements $\circMunit$ of smaller extraction velocity $\mathbf{v}_{\mathbf{1}^{(\mathcal{B})}} =\frac{1}{k}\cdot \mathbf{v}_{\mathbf{1}^{(\mathcal{A})}}$. His calorimeter ${W_{\mathrm{cal}}}^{(\mathcal{B})}$ absorbs $\mathcal{A}$lice reference devices for standard momentum and energy transfer
\be\label{Abschnitt -- kin quant Absorptions Wirkung - Reservoirbilanz - refinement p}
   \mathrm{RB}^{(\mathcal{B})}\left[ \circMunit_{\:\mathbf{v}_{\mathbf{1}^{(\mathcal{A})}}}  \; \Rightarrow \; \circMunit_{\:\mathbf{0}}
   \right]
      \;\; \stackrel{(\ref{Abschnitt -- kin quant Absorptions Wirkung - Reservoirbilanz - absorption})}{=}
   \;\;  \left( \frac{1}{2}\cdot k^2 - \frac{1}{2}\cdot k  \right)\cdot \mathcal{S}_{\mathbf{1}^{(\mathcal{B})} }\big|_{\mathbf{0}}
   \;\; , \;\; k\cdot \circMunit_{\:\mathbf{v}_{\mathbf{1}^{(\mathcal{B})}}}
\ee
\be\label{Abschnitt -- kin quant Absorptions Wirkung - Reservoirbilanz - refinement E}
   \mathrm{RB}^{(\mathcal{B})}\left[ \mathcal{S}_{\mathbf{1}^{(\mathcal{A})} }\big|_{\mathbf{0}} \Rightarrow \emptyset \right]
     \;\; = \;\; k^2 \cdot \mathcal{S}_{\mathbf{1}^{(\mathcal{B})} }\big|_{\mathbf{0}}
\ee
in return for extracting refined energy and momentum carriers from $\mathcal{B}$ob.
\end{lem}
\textbf{Proof:}
\bea
   \mathrm{RB}^{(\mathcal{B})}\left[ \mathcal{S}_{\mathbf{1}^{(\mathcal{A})} }\big|_{\mathbf{0}} \Rightarrow \emptyset \right]
    & \stackrel{(\ref{Abschnitt -- basic dynamical measures - Einheitswirkung})}{=} &
    \mathrm{RB}^{(\mathcal{B})}\left[
    \circMunit_{\:-\mathbf{v}_{\mathbf{1}^{(\mathcal{A})}}} \,,\:
    \circMunit_{\:\mathbf{v}_{\mathbf{1}^{(\mathcal{A})}}} \; \Rightarrow \;
    \circMunit_{\:\mathbf{0}} \,,\: \circMunit_{\:\mathbf{0}} \right] \nn \\
    & \stackrel{(\ref{Abschnitt -- kin quant Absorptions Wirkung - Reservoirbilanz - additivitaet})}{=} &
    \mathrm{RB}^{(\mathcal{B})}\left[
    \circMunit_{\:-k\cdot\mathbf{v}_{\mathbf{1}^{(\mathcal{B})}}} \Rightarrow \circMunit_{\:\mathbf{0}}
    \right] +
    \mathrm{RB}^{(\mathcal{B})}\left[ \circMunit_{\:k\cdot\mathbf{v}_{\mathbf{1}^{(\mathcal{B})}}}
    \Rightarrow   \circMunit_{\:\mathbf{0}} \right] \nn \\
   & \stackrel{(\ref{Abschnitt -- kin quant Absorptions Wirkung - Reservoirbilanz - absorption})}{=} &
   \left( k^2 - k \right)\cdot \mathcal{S}_{\mathbf{1}^{(\mathcal{B})} }\big|_{\mathbf{0}} \:,\; k\cdot \circMunit_{-\mathbf{v}_{\mathbf{1}^{(\mathcal{A})}}}
   \:,\; k\cdot \circMunit_{\:\mathbf{v}_{\mathbf{1}^{(\mathcal{B})}}}  \;\;\; \stackrel{(\ref{Abschnitt -- basic dynamical measures - Einheitswirkung})}{=} \;\;\; k^2 \cdot \mathcal{S}_{\mathbf{1}^{(\mathcal{B})} }\big|_{\mathbf{0}}   \nn
\eea
\qed
\begin{co}
The refinement of $\mathcal{A}$lice calorimeter measurement with a high resolution calorimeter from $\mathcal{B}$ob is \underline{transitive}, i.e. equivalent to $\mathcal{B}$ob's direct measurement
\[
   {W_{\mathrm{cal}}}^{(\mathcal{B})} \left[  {W_{\mathrm{cal}}}^{(\mathcal{A})}  \left[\;\cdot\;\right]  \right] \;\; \equiv \;\; {W_{\mathrm{cal}}}^{(\mathcal{B})} \left[\;\cdot\;\right]  \;\; .
\]
\end{co}
\begin{figure}    
  \begin{center}           
  \includegraphics[height=7.0cm]{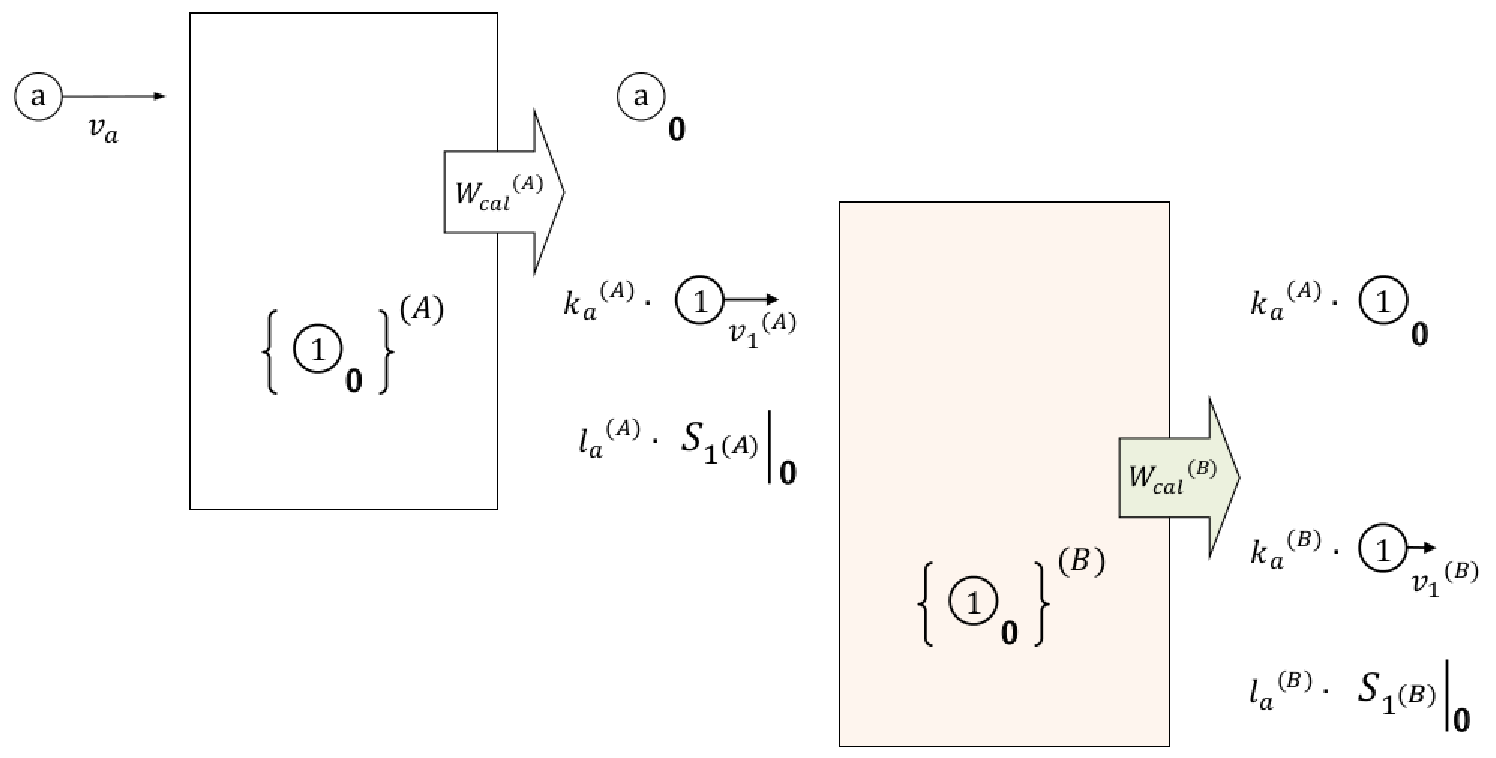}  
  \end{center}
  \vspace{-0cm}
  \caption{\label{pic_calorimeter_refinement} the ${W_{\mathrm{cal}}}^{(\mathcal{A})}$ output $\mathcal{S}_{\mathbf{1}^{(\mathcal{A})} }$, $\circMunit_{\:\mathbf{v}_{\mathbf{1}^{(\mathcal{A})}}}$ gets absorbed in calorimeter ${W_{\mathrm{cal}}}^{(\mathcal{B})}$
    }
  \end{figure}
\textbf{Proof:}
Let $\mathcal{A}$lice absorb a test particle $\circMa_{\:\mathbf{v}_a}$ with velocity $\mathbf{v}_a = n\cdot \mathbf{v}_{\mathbf{1}^{(\mathcal{A})}} = n\cdot (k\cdot \mathbf{v}_{\mathbf{1}^{(\mathcal{B})}})$
\be
   \mathrm{RB}^{(\mathcal{A})}\left[\circMa_{\:n\cdot \mathbf{v}_{\mathbf{1}^{(\mathcal{A})}}}  \Rightarrow \circMa_{\:\mathbf{0}} \right]
   \;\; \stackrel{(\ref{Abschnitt -- kin quant Absorptions Wirkung - Reservoirbilanz - absorption})}{=}
   \;\;  \left( \frac{1}{2}\cdot n^2 - \frac{1}{2}\cdot n  \right)\cdot \mathcal{S}_{\mathbf{1}^{(\mathcal{A})} }\big|_{\mathbf{0}}
   \;\; , \;\; n\cdot \circMunit_{\:\mathbf{v}_{\mathbf{1}^{(\mathcal{A})}}}  \nn \;\; .
\ee
$\mathcal{B}$ob absorbs her calorimeter extract in his high resolution calorimeter (see figure \ref{pic_calorimeter_refinement})
\bea
   & & \!\!\!\!\!\!\!\!\!\!\!\!\!\!\!\!\!\!\!\!\!\!\!\!\!\!\!\!\!\!\!   \!\!\!\!\!\!\!\!
   \mathrm{RB}^{(\mathcal{B})}\left[ \left( \frac{1}{2}\cdot n^2 - \frac{1}{2}\cdot n  \right)\cdot
   \mathcal{S}_{\mathbf{1}^{(\mathcal{A})} }\big|_{\mathbf{0}}
   \;\; , \;\; n\cdot \circMunit_{\:\mathbf{v}_{\mathbf{1}^{(\mathcal{A})}}} \right] \nn
   \\
   \;\;\;\;\;& \stackrel{(\ref{Abschnitt -- kin quant Absorptions Wirkung - Reservoirbilanz - additivitaet})}{=} & \!\!\!
   \left( \frac{1}{2}\cdot n^2 - \frac{1}{2}\cdot n  \right)\cdot \mathrm{RB}^{(\mathcal{B})}\left[
   \mathcal{S}_{\mathbf{1}^{(\mathcal{A})} }\big|_{\mathbf{0}} \Rightarrow \emptyset \right] \; + \; n\cdot \mathrm{RB}^{(\mathcal{B})}\left[
   \circMunit_{\:\mathbf{v}_{\mathbf{1}^{(\mathcal{A})}}} \Rightarrow \circMunit_{\:\mathbf{0}} \right] \nn \\
   \;\;\;\;\;& \stackrel{(\ref{Abschnitt -- kin quant Absorptions Wirkung - Reservoirbilanz - refinement E})(\ref{Abschnitt -- kin quant Absorptions Wirkung - Reservoirbilanz - refinement p})}{=} & \!\!\! \left( \frac{1}{2}\cdot (n\cdot k)^2 - \frac{1}{2}\cdot n\cdot k  \right)\cdot
   \mathcal{S}_{\mathbf{1}^{(\mathcal{B})} }\big|_{\mathbf{0}}
   \: , \; n\cdot k\cdot \circMunit_{\:\mathbf{v}_{\mathbf{1}^{(\mathcal{B})}}}
   \, \stackrel{(\ref{Abschnitt -- kin quant Absorptions Wirkung - Reservoirbilanz - absorption})}{=} \;\: \mathrm{RB}^{(\mathcal{B})}\left[\circMa_{\:n\cdot k\cdot \mathbf{v}_{\mathbf{1}^{(\mathcal{B})}}}  \Rightarrow \circMa_{\:\mathbf{0}} \right]
\nn
\eea
with the same output as from the direct absorption measurement of the test particle $\circMa_{\:\mathbf{v}_a}$.
\qed
\begin{lem}\label{Lem - kin quant Absorptions Wirkung - Reservoirbilanz - boost E and p units}
Let $\mathcal{A}$lice move relative to $\mathcal{B}$ob with constant velocity $\mathbf{v}_{\mathcal{A}}= n\cdot \mathbf{v}_{\mathbf{1}^{(\mathcal{B})}}$ for $n \in \mathbb{N}$. $\mathcal{B}$ob can reproduce the effect of $\mathcal{A}$lice \underline{boosted} units of energy and momentum
\be\label{Abschnitt -- kin quant Absorptions Wirkung - Reservoirbilanz - boost E unit}
   \mathrm{RB}^{(\mathcal{B})}\left[ \mathcal{S}_{\mathbf{1}^{(\mathcal{A})} }\big|_{n\cdot \mathbf{v}_{\mathbf{1}^{(\mathcal{B})}}} \Rightarrow \emptyset \right]
     \;\; = \;\; \mathcal{S}_{\mathbf{1}^{(\mathcal{B})} }\big|_{\mathbf{0}}
\ee
\be\label{Abschnitt -- kin quant Absorptions Wirkung - Reservoirbilanz - boost p unit}
   \mathrm{RB}^{(\mathcal{B})}\left[ \circMunit_{\:\mathbf{v}_{\mathbf{1}^{(\mathcal{A})}}}  \; \Rightarrow \; \circMunit_{\:\mathbf{0}^{(\mathcal{A})}}
   \right]
      \;\; = \;\;  n\cdot \mathcal{S}_{\mathbf{1}^{(\mathcal{B})} }\big|_{\mathbf{0}}
   \;\; , \;\; 1\cdot \circMunit_{\:\mathbf{v}_{\mathbf{1}^{(\mathcal{B})}}}
\ee
by expending \underline{resting} energy sources $\mathcal{S}_{\mathbf{1}^{(\mathcal{B})} }\big|_{\mathbf{0}}$ and impulse carriers $\circMunit_{\:\mathbf{v}_{\mathbf{1}^{(\mathcal{B})}}}$ from his calorimeter.
\end{lem}
\textbf{Proof:}
In $\mathcal{B}$ob's view (see remark \ref{Rem - covariant kinematical transformation}) $\mathcal{A}$lice' standard impulse carrier $\circMunit_{\:\mathbf{v}_{\mathbf{1}^{(\mathcal{A})}}}$ has the velocity
\be
   \mathbf{v}_{\mathbf{1}^{(\mathcal{A})}} \; = \; \pm 1\cdot \mathbf{v}_{\mathbf{1}^{(\mathcal{A})}} = (n\pm 1) \mathbf{v}_{\mathbf{1}^{(\mathcal{B})}} \nn
\ee
and her standard spring $\mathcal{S}_{\mathbf{1}^{(\mathcal{A})} }\big|_{\mathbf{0}^{(\mathcal{A})}}$ and reservoir elements $\circMunit_{\:\mathbf{0}^{(\mathcal{A})}}$ are in the state of motion
\be
   \mathbf{0}^{(\mathcal{A})} \; = \; 0\cdot \mathbf{v}_{\mathbf{1}^{(\mathcal{A})}} = n\cdot \mathbf{v}_{\mathbf{1}^{(\mathcal{B})}} \;\; . \nn
\ee
$\mathcal{B}$ob can absorb the effect of a boosted standard spring via two spectator particles, which are not effected in the end,
\bea
   \mathrm{RB} \!\left[ \mathcal{S}_{\mathbf{1}^{(\mathcal{A})}}\big|_{n\cdot \mathbf{v}_{\mathbf{1}}} \!\!\Rightarrow  \emptyset \right]
   & \!\!\stackrel{(\ref{Abschnitt -- kin quant Absorptions Wirkung - Reservoirbilanz - additivitaet})}{=}\!\! &   \mathrm{RB} \left[ \mathcal{S}_{\mathbf{1}^{(\mathcal{A})}}\big|_{n\cdot \mathbf{v}_{\mathbf{1}}} \,,\: \circMunit_{\:n\cdot \mathbf{v}_{\mathbf{1}}} \,,\: \circMunit_{\:n\cdot \mathbf{v}_{\mathbf{1}}}  \right] \; - \; \mathrm{RB} \left[ \circMunit_{\:n\cdot \mathbf{v}_{\mathbf{1}}} \,,\: \circMunit_{\:n\cdot \mathbf{v}_{\mathbf{1}}} \right] \nn \\
   & \!\!\stackrel{(\ref{Abschnitt -- basic dynamical measures - Einheitswirkung})}{=}\!\! & \mathrm{RB} \left[ \circMunit_{\:(n+1)\cdot \mathbf{v}_{\mathbf{1}}} \,,\: \circMunit_{\:(n-1)\cdot \mathbf{v}_{\mathbf{1}}}  \right] \; - \; \mathrm{RB} \left[ \circMunit_{\:n\cdot \mathbf{v}_{\mathbf{1}}} \,,\: \circMunit_{\:n\cdot \mathbf{v}_{\mathbf{1}}} \right] \nn \\
   & \!\!\stackrel{(\ref{Abschnitt -- kin quant Absorptions Wirkung - Reservoirbilanz - absorption})}{=}\!\! & \frac{1}{2} \left[ (n+1)^2 - (n+1) + (n-1)^2 - (n-1) \right] \cdot \mathcal{S}_{\mathbf{1}^{(\mathcal{B})}}\big|_{\mathbf{0}} \;,\; 2\cdot n \cdot \circMunit_{\:\mathbf{v}_{\mathbf{1}^{(\mathcal{B})}}} \,, \nn \\
   &  & \;\;\;\;\;\;\;\;\;\;\;\;\;\;\;\;\;\;\;\;\;\;\;\;\;\;\;\;\;\;\;\;\;\;\;\;\;\;\;\;\;\;\;\;\;\;\;\;\; - \;  ( n^2 - n ) \cdot \mathcal{S}_{\mathbf{1}^{(\mathcal{B})}}\big|_{\mathbf{0}} \;,\:  -2\cdot n \cdot \circMunit_{\:\mathbf{v}_{\mathbf{1}^{(\mathcal{B})}}} \nn \\
   & \!\!=\!\! & 1 \cdot \mathcal{S}_{\mathbf{1}^{(\mathcal{B})}}\big|_{\mathbf{0}} \;\: , \;\; 0 \cdot \circMunit_{\:\mathbf{v}_{\mathbf{1}^{(\mathcal{B})}}} \nn
\eea
in his own (resting) calorimeter reservoir. Thus by reversing this absorption process $\mathcal{B}$ob can convert one of his resting energy source $\mathcal{S}_{\mathbf{1}^{(\mathcal{B})} }\big|_{0\cdot \mathbf{v}_{\mathbf{1}^{(\mathcal{B})}}} \Rightarrow \mathcal{S}_{\mathbf{1}^{(\mathcal{A})} }\big|_{n\cdot \mathbf{v}_{\mathbf{1}^{(\mathcal{B})}}}$ into one energy source in the state of motion $n\cdot \mathbf{v}_{\mathbf{1}^{(\mathcal{B})}}$.

Similarly $\mathcal{B}$ob absorbs for $\mathcal{A}$lice (boosted) reference of impulse transfer $\circMunit_{\:1\cdot\mathbf{v}_{\:\mathbf{1}^{(\mathcal{A})}}} \Rightarrow \circMunit_{\:0\cdot\mathbf{v}_{\:\mathbf{1}^{(\mathcal{A})}}}$ the corresponding calorimeter reservoir balance for deceleration
\bea
    \mathrm{RB} \left[ \circMunit_{\:(n+1) \cdot \mathbf{v}_{\mathbf{1}}} \Rightarrow  \circMunit_{\:n\cdot \mathbf{v}_{\mathbf{1}}} \right] & \stackrel{(\ref{Abschnitt -- kin quant Absorptions Wirkung - Reservoirbilanz - additivitaet})}{=} & \mathrm{RB} \left[ \circMunit_{(n+1) \cdot \mathbf{v}_{\mathbf{1}}} \right] \;\; - \;\; \mathrm{RB} \left[ \circMunit_{n\cdot \mathbf{v}_{\mathbf{1}}} \right]
    \;\;\;\;\;\;\;\;\;\;\;\;\;\;\;\;\;\;\;\;\;\;\;\;\;\;\;\;\;\;\;\;\;\; \nn
\eea
\bea
   & \stackrel{(\ref{Abschnitt -- kin quant Absorptions Wirkung - Reservoirbilanz - absorption})}{=} & \frac{1}{2} \left[ (n+1)^2 - (n+1) \right] \cdot \mathcal{S}_{\mathbf{1}^{(\mathcal{B})} }\big|_{\mathbf{0}} \; , \; ( n+1 ) \cdot \circMunit_{\:\mathbf{v}_{\mathbf{1}^{(\mathcal{B})}}} \; , \;
   - \frac{1}{2} ( n^2 - n ) \cdot \mathcal{S}_{\mathbf{1}^{(\mathcal{B})} }\big|_{\mathbf{0}} \; , \; -n \cdot \circMunit_{\:\mathbf{v}_{\mathbf{1}^{(\mathcal{B})}}} \nn \\
   & = & n \cdot \mathcal{S}_{\mathbf{1}^{(\mathcal{B})} }\big|_{\mathbf{0}} \;\; , \;\; 1 \cdot \circMunit_{\:\mathbf{v}_{\mathbf{1}^{(\mathcal{B})}}} \nn
\eea
$\mathcal{B}$ob can also generate the effect of $\mathcal{A}$lice boosted impulse carrier by expending one of his impulse carriers $\circMunit_{\:1\cdot \mathbf{v}_{\mathbf{1}}}$ in a domino series of successively boosted energy sources $n \cdot \mathcal{S}_{\mathbf{1}^{(\mathcal{B})} }\big|_{\mathbf{0}}
\stackrel{(\ref{Abschnitt -- kin quant Absorptions Wirkung - Reservoirbilanz - boost E unit})}{=} \mathcal{S}_{\mathbf{1}^{(\mathcal{B})} }\big|_{1\cdot \mathbf{v}_{\mathbf{1}}}$, $\mathcal{S}_{\mathbf{1}^{(\mathcal{B})} }\big|_{2\cdot \mathbf{v}_{\mathbf{1}}}$ $\ldots$ $\mathcal{S}_{\mathbf{1}^{(\mathcal{B})} }\big|_{n\cdot \mathbf{v}_{\mathbf{1}}}$. $\mathcal{B}$ob can build equivalent physical models in many ways.
\qed

Now let us consider these models from an abstract physical perspective. We define the basic observables from elemental ordering relations \{\ref{Kap - KM Dynamics - Physical Measurement - Pre-theoretical Ordering Relation}\}. They fix the \emph{physical meaning} of our reference objects as units for energy and momentum and of our calorimeter extract.
\begin{pr}\label{Prop - kin quant Absorptions Wirkung - pre-theoretical characterization E unit and p unit}
Our standard spring $\mathcal{S}_{\mathbf{1}}\big|_{\mathbf{v}=\mathbf{0}}$ represents the unit energy and has no impulse.
\be\label{Abschnitt -- kin quant Absorptions Wirkung - pre-theoretical characterization E unit and p unit}
\begin{array}{rclrcl}
   E \left[ \mathcal{S}_{\mathbf{1}}\big|_{\mathbf{0}} \right] & &
   & \;\;\;\;\;\;\;\;\;\;\;\;\;
   E \left[ \circMunit_{\:\mathbf{v}_\mathbf{1}} \right] & = & \frac{1}{2}\cdot E \left[ \mathcal{S}_{\mathbf{1}}\big|_{\mathbf{0}} \right]
   \\
   \mathbf{p} \left[ \mathcal{S}_{\mathbf{1}}\big|_{\mathbf{0}} \right] & = & 0
   & \mathbf{p} \left[ \circMunit_{\:\mathbf{v}_\mathbf{1}} \right] &  &
\end{array}
\ee
The standard impulse carrier $\circMunit_{\:\mathbf{v}_\mathbf{1}}$ represents the unit momentum and also has energy.
\end{pr}
\textbf{Proof:}
The two dimensions energy and impulse are inseparably intertwined in unit action $\mathcal{S}_{\mathbf{1}}\big|_{\mathbf{0}} , \circMunit_{\:\mathbf{0}} , \circMunit_{\:\mathbf{0}} \stackrel{w_{\mathbf{1}}}{\Rightarrow} \circMunit_{-\mathbf{v}_{\mathbf{1}}} , \circMunit_{\:\mathbf{v}_{\mathbf{1}}}$ between our standard energy source and impulse carriers.

The resting energy source $\mathcal{S}_{\mathbf{1}}\big|_{\mathbf{v}=\mathbf{0}}$ can not overrun any moving object in a head-on collision test; if at all it will be overrun. It has no impact. Its abstract momentum vanishes $\mathbf{p} \left[ \mathcal{S}_{\mathbf{1}}\big|_{\mathbf{0}} \right] = 0$ (see Definition \ref{Def - vortheor Ordnungsrelastion - impulse}).

If two comoving impulse carriers $\circMunit_{\:\mathbf{v}_{\mathbf{1}}}, \circMunit_{\:\mathbf{v}_{\mathbf{1}}}$ generate the same \emph{absorption effect} on a test system like one standard spring $\mathcal{S}_{\mathbf{1}}\big|_{\mathbf{v}=\mathbf{0}}$, then by the equipollence of cause and effect (Definition \ref{Def - vortheor Ordnungsrelastion - energie}) their energy is the same $2\cdot E \left[ \circMunit_{\:\mathbf{v}_{\mathbf{1}}} \right] = E \left[ \mathcal{S}_{\mathbf{1}}\big|_{\mathbf{0}} \right]$. We test both energy sources with a high-resolution calorimeter $W^{(\epsilon)}_{\mathrm{cal}}$. It is built in the same reservoir of standard elements $\left\{\circMunit_{\:\mathbf{v}=0}\right\}$ from arbitrarily fine standard processes (\ref{Abschnitt -- kin quant elast coll - prepare congruent transversal impulse})
\be
   \;\;\;\;\;\;\;
   \mathcal{S}_{\epsilon}\big|_{\mathbf{0}} \,,\: \circMunit_{\:\mathbf{0}} \,,\:
   \circMunit_{\:\mathbf{0}} \;\; \stackrel{w_{\epsilon}}{\Rightarrow} \;\; \circMunit_{-\epsilon\cdot\mathbf{v}_{\mathbf{1}}} \,,\:
   \circMunit_{\:\epsilon\cdot\mathbf{v}_{\mathbf{1}}}  \;\;\;\;\;\;\; \epsilon > 0 \;\; . \nn
\ee
From the refined deceleration cascade we extract standard energy and momentum carriers $\mathcal{S}_{\epsilon}\big|_{\mathbf{0}}$ and $\circMunit_{\:\epsilon\cdot\mathbf{v}_{\mathbf{1}}}$
\be\label{Abschnitt -- kin quant Absorptions Wirkung - Reservoirbilanz - Einheit refinement p}
   \mathrm{RB}^{(\epsilon)} \left[  2\cdot \circMunit_{\:\mathbf{v}_{\mathbf{1}}}  \Rightarrow 2\cdot\circMunit_{\:\mathbf{0}}   \right] \;\; \stackrel{(\ref{Abschnitt -- kin quant Absorptions Wirkung - Reservoirbilanz - additivitaet})(\ref{Abschnitt -- kin quant Absorptions Wirkung - Reservoirbilanz - refinement p})}{=} \;\;
   \left( \frac{1}{\epsilon^2} - \frac{1}{\epsilon}  \right)\cdot \mathcal{S}_{\epsilon}\big|_{\mathbf{0}}
   \;\; , \;\; \frac{2}{\epsilon}\cdot \circMunit_{\:\epsilon\cdot\mathbf{v}_{\mathbf{1}}}
\ee
\be\label{Abschnitt -- kin quant Absorptions Wirkung - Reservoirbilanz - Einheit refinement E}
   \mathrm{RB}^{(\epsilon)} \left[ \mathcal{S}_{\mathbf{1}}\big|_{\mathbf{0}} \Rightarrow \emptyset \right] \;\; \stackrel{(\ref{Abschnitt -- kin quant Absorptions Wirkung - Reservoirbilanz - refinement E})}{=} \;\;  \frac{1}{\epsilon^2}\cdot \mathcal{S}_{\epsilon}\big|_{\mathbf{0}}
\ee
with smaller extraction velocity $\mathbf{v}_{\mathbf{1}^{(\epsilon)}}:=\epsilon\cdot\mathbf{v}_{\mathbf{1}}$. We probe the effect of both energy sources in the refined calorimeter extract. The finer reference devices $\mathcal{S}_{\epsilon}\big|_{\mathbf{0}}$ and $\circMunit_{\:\epsilon\cdot\mathbf{v}_{\mathbf{1}}}$ serve as a \emph{lowest common physical denominator} of $\mathcal{S}_{\mathbf{1}}\big|_{\mathbf{0}}$ and $\circMunit_{\:\mathbf{v}_{\mathbf{1}}}$. We compare the energy of one spring $E \left[ \mathcal{S}_{\mathbf{1}}\big|_{\mathbf{0}} \right]$ with the energy of one standard impulse carrier $E \left[ \circMunit_{\:\mathbf{v}_{\mathbf{1}}} \right]$ in high resolution calorimeter $W^{(\epsilon)}_{\mathrm{cal}}$ by the number of common (congruent) ''parts'' $\mathcal{S}_{\epsilon}\big|_{\mathbf{0}}$ and $\circMunit_{\:\epsilon\cdot\mathbf{v}_{\mathbf{1}}}$.

We add to both impulse carriers $2\cdot \circMunit_{\:\mathbf{v}_{\mathbf{1}}}$ (a heavy bouncing block) of $\frac{2}{\epsilon}$ high resolution impulse carriers $\circMunit_{-\epsilon\cdot\mathbf{v}_{\mathbf{1}}}$. In the refinement limit $\epsilon\rightarrow0$ their energy contribution disappears
\bea\label{Abschnitt -- kin quant Absorptions Wirkung - Reservoirbilanz - Einheit Erweiterung limit E}
   \lim_{\epsilon\rightarrow0} \;\;\; E \left[ \frac{2}{\epsilon}\cdot \circMunit_{-\epsilon\cdot\mathbf{v}_{\mathbf{1}}} \right] & < &
   \lim_{\epsilon\rightarrow0} \;\;\; E \left[ \frac{2}{\epsilon}\cdot \circMunit_{-\epsilon\cdot\mathbf{v}_{\mathbf{1}}} \,,\: \frac{2}{\epsilon}\cdot \circMunit_{\:\epsilon\cdot\mathbf{v}_{\mathbf{1}}} \right]   \nn \\
   & \stackrel{(\ref{Abschnitt -- basic dynamical measures - Einheitswirkung})}{=} &
   \lim_{\epsilon\rightarrow0} \;\;\; E \left[ \frac{2}{\epsilon}\cdot \mathcal{S}_{\epsilon}\big|_{\mathbf{0}} \right] \cdot \underbrace{\frac{\epsilon}{\epsilon}}_{=1} \;\; = \;\;
   \lim_{\epsilon\rightarrow0} \;\;\; E \left[ \underbrace{\frac{1}{\epsilon^2}\cdot \mathcal{S}_{\epsilon}\big|_{\mathbf{0}}}_{\stackrel{(\ref{Abschnitt -- kin quant Absorptions Wirkung - Reservoirbilanz - refinement E})}{\sim_E} \;\;  \mathcal{S}_{\mathbf{1}}\big|_{\mathbf{0}} } \; \cdot \; 2 \epsilon \right]   \nn \\
   & = & \lim_{\epsilon\rightarrow0} \;\;\; 2\epsilon\cdot E_{\mathbf{1}} \;\; = \;\; 0 \cdot E_{\mathbf{1}}   \;\; .
\eea
The larger system $2\cdot \circMunit_{\:\mathbf{v}_{\mathbf{1}}} \;\cup\; \frac{2}{\epsilon}\cdot \circMunit_{-\epsilon\cdot\mathbf{v}_{\mathbf{1}}}$ with two impulse carriers and the extra elements (from the bouncing block) extract from the high resolution calorimeter almost the same output
\bea\label{Abschnitt -- kin quant Absorptions Wirkung - Reservoirbilanz - Einheit refinement approximation}
   \mathrm{RB}^{(\epsilon)} \left[  2\cdot \circMunit_{\:\mathbf{v}_{\mathbf{1}}} \,,\:
   \frac{2}{\epsilon}\cdot \circMunit_{-\epsilon\cdot\mathbf{v}_{\mathbf{1}}}
   \Rightarrow
   2\cdot\circMunit_{\:\mathbf{0}} \,,\:
   \frac{2}{\epsilon}\cdot \circMunit_{\:\mathbf{0}} \right] & \!\!\stackrel{(\ref{Abschnitt -- kin quant Absorptions Wirkung - Reservoirbilanz - Einheit refinement p})}{=}\!\! &
   \left( \frac{1}{\epsilon^2} - \frac{1}{\epsilon}  \right)\cdot \mathcal{S}_{\epsilon}\big|_{\mathbf{0}}
   \;,\; \underbrace{\frac{2}{\epsilon}\cdot \circMunit_{\:\epsilon\cdot\mathbf{v}_{\mathbf{1}}} \;,\; \frac{2}{\epsilon}\cdot \circMunit_{-\epsilon\cdot\mathbf{v}_{\mathbf{1}}} }_{\stackrel{(\ref{Abschnitt -- basic dynamical measures - Einheitswirkung})\;\;\;}{\sim_E} \;\; \frac{2}{\epsilon}\cdot \mathcal{S}_{\epsilon}\big|_{\mathbf{0}} }   \nn \\
   & \!\!=\!\! & \left( \frac{1}{\epsilon^2} + \frac{1}{\epsilon}  \right)\cdot \mathcal{S}_{\epsilon}\big|_{\mathbf{0}}
   \nn \\
   & \!\!\stackrel{(\ref{Abschnitt -- kin quant Absorptions Wirkung - Reservoirbilanz - Einheit refinement E})}{=}\!\! & \mathrm{RB}^{(\epsilon)} \left[ \mathcal{S}_{\mathbf{1}}\big|_{\mathbf{0}} \Rightarrow \emptyset \right] \;\; + \;\; \frac{1}{\epsilon} \cdot \mathcal{S}_{\epsilon}\big|_{\mathbf{0}}
\eea
as for the absorption of one standard spring $\mathcal{S}_{\mathbf{1}}\big|_{\mathbf{0}}$.\footnote{Eventually we convert two comoving elements $\left\{\circMunit_{\:\mathbf{v}_{\mathbf{1}}}, \circMunit_{\:\mathbf{v}_{\mathbf{1}}}\right\} \Rightarrow \left\{\circMunit_{-\mathbf{v}_{\mathbf{1}}}, \circMunit_{\:\mathbf{v}_{\mathbf{1}}}\right\}$ into an antiparallel particle pair by letting, vividly spoken, one element repulse elastically $\circMunit_{\:\mathbf{v}_{\mathbf{1}}} \,,\: \circMM_{\:\mathbf{v}_M} \stackrel{(\ref{Abschnitt -- kin quant elast coll - elast head-on collision})}{\Rightarrow} \circMunit_{-\mathbf{v}_{\mathbf{1}}} \,,\: \circMM_{-\mathbf{v}_M}$ from a much heavier ''reservoir block''. The refined calorimeter measurements correspond to the limit $m[\circMunit] \ll m[\circMM]$ where the ''bouncing block'' $\mathbf{v}_M\rightarrow 0$ practically rests with a negligible contribution to the kinetic energy.} In the refinement limit $\epsilon\rightarrow0$ the absorption energy of one energy unit
\bea
   E \left[ \mathcal{S}_{\mathbf{1}}\big|_{\mathbf{0}} \right]
   & \stackrel{(\mathrm{Equip.})}{=} &
   E \left[  \mathrm{RB}^{(\epsilon)} \left[ \mathcal{S}_{\mathbf{1}}\big|_{\mathbf{0}} \Rightarrow \emptyset \right]   \right]   \nn \\
   & \stackrel{(\ref{Abschnitt -- kin quant Absorptions Wirkung - Reservoirbilanz - Einheit refinement approximation})}{=} &
   E \left[ \mathrm{RB}^{(\epsilon)} \left[  2\cdot \circMunit_{\:\mathbf{v}_{\mathbf{1}}} \,,\:
   \frac{2}{\epsilon}\cdot \circMunit_{-\epsilon\cdot\mathbf{v}_{\mathbf{1}}} \right] \right] \;\; - \;\;
   E \left[ \frac{1}{\epsilon} \cdot \mathcal{S}_{\epsilon}\big|_{\mathbf{0}} \right]
   \nn \\
   & = &
   \lim_{\epsilon\rightarrow0} \; E \left[ \mathrm{RB}^{(\epsilon)} \left[  2\cdot \circMunit_{\:\mathbf{v}_{\mathbf{1}}} \,,\:
   \frac{2}{\epsilon}\cdot \circMunit_{-\epsilon\cdot\mathbf{v}_{\mathbf{1}}} \right] \right]
   \;\; - \!\!\! \underbrace{\lim_{\epsilon\rightarrow0} \; E \left[ \frac{1}{\epsilon} \cdot \mathcal{S}_{\epsilon}\big|_{\mathbf{0}} \right]}_{\stackrel{(\ref{Abschnitt -- kin quant Absorptions Wirkung - Reservoirbilanz - refinement E})}{=} \; \lim_{\epsilon\rightarrow0} \; E \left[ \epsilon\cdot \mathcal{S}_{\mathbf{1}}\big|_{\mathbf{0}} \right] \; = \; 0\cdot E_{\mathbf{1}} }
   \nn \\
   & \stackrel{(\ref{Abschnitt -- kin quant Absorptions Wirkung - Reservoirbilanz - additivitaet})}{=} &
   \underbrace{\lim_{\epsilon\rightarrow0} \; E \left[ \mathrm{RB}^{(\epsilon)} \left[  2\cdot \circMunit_{\:\mathbf{v}_{\mathbf{1}}} \right] \right]}_{ \stackrel{(\mathrm{Equip.})}{=} \; 2\cdot E \left[ \circMunit_{\:\mathbf{v}_{\mathbf{1}}} \right] }
   \;\; + \;\; \underbrace{\lim_{\epsilon\rightarrow0} \; E \left[ \frac{2}{\epsilon} \cdot \circMunit_{-\epsilon\cdot\mathbf{v}_{\mathbf{1}}} \right]}_{\stackrel{(\ref{Abschnitt -- kin quant Absorptions Wirkung - Reservoirbilanz - Einheit Erweiterung limit E})}{=} \; 0\cdot E_{\mathbf{1}}} \nn
\eea
is two times the absorption energy of one impulse unit $\circMunit_{\:\mathbf{v}_{\mathbf{1}}}$. In particular one can conclude from (\ref{Abschnitt -- basic dynamical measures - Einheitswirkung}) that $E \left[ \circMunit_{-\mathbf{v}_{\mathbf{1}}} \right] = E \left[ \circMunit_{\:\mathbf{v}_{\mathbf{1}}} \right] $.
\qed
\begin{rem}\label{Rem - kin quant Absorptions Wirkung - inseparable units}
We presuppose an indivisible unit action $w_{\mathbf{1}}$ (of arbitrary internal structure) between the standard spring $\mathcal{S}_{\mathbf{1}}\big|_{\mathbf{0}}$ and impulse carrier $\circMunit_{\:\mathbf{v}_{\mathbf{1}}}$. From similar ''partial'' interactions $w_{\epsilon}$ we build a refined calorimeter model $W^{(\epsilon)}_{\mathrm{cal}}$. We compare the effect of the original reference devices $\mathcal{S}_{\mathbf{1}}\big|_{\mathbf{0}}$ and $\circMunit_{\:\mathbf{v}_{\mathbf{1}}}$ by the refined calorimeter extract. Thus $w_{\epsilon}$ is not ''part'' of $w_{\mathbf{1}}$.\footnote{The reference devices $\mathcal{S}_{\mathbf{1}}\big|_{\mathbf{0}}$ and $\circMunit_{\:\mathbf{v}_{\mathbf{1}}}$ are indivisible. We can substitute the effect of unit action $w_\mathbf{1}$ by the effect of an aggregate of finer standard actions $w_{\epsilon}$ in a reversion process or in the deceleration cascade (see figure \ref{pic_impulse_inversion_means} resp. \ref{pic_calorimeter_model}); but the original or refined units always enter the construction as a whole. We have the effect and impact of standard process $w_{\mathbf{1}}$ resp. $w_{\epsilon}$ in full or nothing at all.} It is building block of our calorimeter model; part in the construction of an engineer.
\end{rem}
\begin{lem}\label{Lem - kin quant Absorptions Wirkung - Reservoirbilanz - absorption proportional materiemenge}
Let a composite of $m$ bound unit elements $\underbrace{\circMunit \ast \dots \ast \circMunit}_{m\times}\;\sim_{m}\circMO$ have the same inertia as a generic object $\circMO_{\:\mathbf{v}}$. Then the reservoir extract for absorbing the object
\be\label{Abschnitt -- kin quant Absorptions Wirkung - Reservoirbilanz - absorption proportional materiemenge}
   \mathrm{RB} \left[ \circMO_{\:\mathbf{v}} \Rightarrow
   \circMO_{\:\mathbf{0}}  \right] \;\; = \;\; m \cdot \mathrm{RB} \left[ \circMunit_{\:\mathbf{v}} \Rightarrow \circMunit_{\:\mathbf{0}}  \right]
\ee
is $m$ times larger than for absorbing one unit element $\circMunit_{\:\mathbf{v}}$ with the same velocity $\mathbf{v}$.
\end{lem}
\textbf{Proof:}
Same inertia (Definition \ref{Def - vortheor Ordnungsrelastion - inertial mass}) implies, that in an elastic head-on collision test with same initial velocity $v'_{\mathcal{O}}:=\frac{1}{2}\cdot v_{\mathcal{O}}$ the composite $\circMunit\ast\dots\ast\circMunit$ and the generic object $\circMO$ must repulse in the same anti-symmetrical way $\circMO_{\:\mathbf{v}'_{\mathcal{O}}} \,,\; \circMunit\ast\dots\ast\circMunit_{\:-\mathbf{v}'_{\mathcal{O}}} \stackrel{w_H}{\Rightarrow} \circMO_{-\mathbf{v}'_{\mathcal{O}}} \,,\; \circMunit\ast\dots\ast\circMunit_{\;\mathbf{v}'_{\mathcal{O}}}$. For an observer moving with relative velocity $-\mathbf{v}'_{\mathcal{O}}$
\be
   \circMO_{\:\mathbf{v}_{\mathcal{O}}} \,,\; \circMunit\ast\dots\ast\circMunit_{\;\mathbf{0}} \;\; \stackrel{w_H}{\Rightarrow} \;\; \circMO_{\:\mathbf{0}} \,,\; \circMunit\ast\dots\ast\circMunit_{\;\mathbf{v}_{\mathcal{O}}} \nn
\ee
the generic object $\circMO_{\:\mathbf{v}_{a}}$ stops and kicks the initially resting composite into same motion.

We neutralize this elastic head-on collision by absorbing the (spectator) composite in our calorimeter
$\mathrm{RB} \left[ \circMunit\ast\dots\ast\circMunit_{\:\mathbf{v}_{a}} \Rightarrow \circMunit\ast\dots\ast\circMunit_{\:\mathbf{0}} \right] =: k_m\cdot \mathcal{S}_{\mathbf{1}}\big|_{\mathbf{0}} \,,\, l_m\cdot\,\circMunit_{\:\mathbf{v}_{\mathbf{1}}}$ and by catapulting the (temporarily) resting object in a reversed absorption $-\mathrm{RB} \left[ \circMO_{\:\mathbf{v}_{\mathcal{O}}} \Rightarrow \circMO_{\:\mathbf{0}} \right] =: k_{\mathcal{O}}\cdot \mathcal{S}_{\mathbf{1}}\big|_{\mathbf{0}} \,,\, l_{\mathcal{O}}\cdot\,\circMunit_{\:\mathbf{v}_{\mathbf{1}}}$ back into the original motion. The net effect is a circular process. The net reservoir extract
\be\label{Abschnitt -- kin quant Absorptions Wirkung - Reservoirbilanz - absorption proportional materiemenge - hilfssatz}
   \mathbf{p} \left[ (k_{\mathcal{O}}-k_m)\cdot\mathcal{S}_{\mathbf{1}}\big|_{\mathbf{0}} \:,\; (l_{\mathcal{O}}-l_m)\cdot\circMunit_{\:\mathbf{v}_{\mathbf{1}}} \right] \;\; \stackrel{(\ref{Abschnitt -- kin quant Absorptions Wirkung - pre-theoretical characterization E unit and p unit})}{=} \;\;  \mathbf{p} \left[ (l_{\mathcal{O}}-l_m)\cdot\circMunit_{\:\mathbf{v}_{\mathbf{1}}} \right] \;\; \stackrel{!}{=} \;\; 0
\ee
cannot have momentum (for conservation see Lemma \ref{Lem - kin quant Absorptions Wirkung - Reservoirbilanz - p conserved}). It also cannot have energy
\be
   E \;[ \:(k_{\mathcal{O}}-k_m)\cdot\mathcal{S}_{\mathbf{1}}\big|_{\mathbf{0}} \:,\; \underbrace{(l_{\mathcal{O}}-l_m)\cdot\circMunit_{\:\mathbf{v}_{\mathbf{1}}}}_{\stackrel{(\ref{Abschnitt -- kin quant Absorptions Wirkung - Reservoirbilanz - absorption proportional materiemenge - hilfssatz})}{=}\; 0} \:] \;\; \stackrel{!}{=} \;\; 0  \nn \;\; .
\ee
Thus the generic object $\circMO_{\:\mathbf{v}}$ generates the same reservoir extract ($l_{\mathcal{O}} \stackrel{!}{=} l_m$ impulse carriers
$\circMunit_{\:\mathbf{v}_{\mathbf{1}}}$ and $k_{\mathcal{O}} \stackrel{!}{=} k_m$ energy units $\mathcal{S}_{\mathbf{1}}\big|_{\mathbf{0}}$) as for absorbing the composite
\be
   \mathrm{RB} \left[ \circMO_{\:\mathbf{v}} \Rightarrow \circMO_{\:\mathbf{0}} \right] \;\;=\;\;
   \mathrm{RB} \left[ \circMunit\ast\dots\ast\circMunit_{\;\mathbf{v}} \Rightarrow \; \circMunit\ast\dots\ast\circMunit_{\;\mathbf{0}} \right] \;\;\stackrel{(\ref{Abschnitt -- kin quant Absorptions Wirkung - Reservoirbilanz - additivitaet})}{=}\;\;
   m \cdot \mathrm{RB} \left[ \circMunit_{\:\mathbf{v}} \Rightarrow \circMunit_{\:\mathbf{0}} \right] \nn \;\; .
\ee
From the pack we absorb every element one by one. All extracted energy-momentum units are congruent; their total number adds up to $m$ times the output of one unit element $\circMunit_{\:\mathbf{v}}$.
\qed

\subsection{Physical quantity}\label{Kap - KM Dynamics - Basic Dynamical Measures - Metrization - physical quantity}

With the calorimeter model we determine the magnitude of the energy of object $E\left[\circMO_{\:\mathbf{v}}\right]$. The incident object $\circMO_{\:\mathbf{v}}$ has the same capability to execute work \{\ref{Kap - KM Dynamics - Physical Measurement - Pre-theoretical Ordering Relation}\}
\be\label{Abschnitt -- kin quant Absorptions Wirkung - kin energy and momentum metrisiert - intro}
   \circMO_{\:\mathbf{v}} \;\; \sim_E  \;\; \mathcal{S}_{\mathbf{1}}\big|_{\mathbf{0}} \ast \ldots \ast \mathcal{S}_{\mathbf{1}}\big|_{\mathbf{0}}
\ee
as the calorimeter extract.\footnote{All individual energy sources $\mathcal{S}_{\mathbf{1}}\!\!\mid_{\mathbf{0}}$ \emph{act jointly against} the same cue ball (concatenation operation ''$\ast_E$''). Similarly we bound impulse resp. mass carriers $\circMunit$ together tightly, so that (in collision against third parties) their motion changes - passively - according to inertia of combined body (concatenation symbolized by ''$\ast_{\mathbf{p}}$'').} Therein each particle pair $\mathcal{S}_{\mathbf{1}}\big|_{\mathbf{0}}$ (standard spring) is completely equivalent with the next. According to the \emph{congruence principle}
\be \nn
   E\left[ \circMO_{\:\mathbf{v}} \right] \;\; \stackrel{(\ref{Abschnitt -- kin quant Absorptions Wirkung - kin energy and momentum metrisiert - intro})}{=} \;\;  E\left[ \,\mathcal{S}_{\mathbf{1}}\big|_{\mathbf{0}} \ast \ldots \ast \mathcal{S}_{\mathbf{1}}\big|_{\mathbf{0}} \, \right]
   \; \stackrel{(\mathrm{Congr.})}{=} \; E \cdot E\left[ \mathcal{S}_{\mathbf{1}}\big|_{\mathbf{0}} \right]
\ee
we find ''how many times larger'' its kinetic energy is, than the potential energy of one standard spring. The numerical factor $E := \sharp \left\{ \mathcal{S}_{\mathbf{1}}\big|_{\mathbf{0}} \right\}$ is a \emph{physical quantity}; our reference energy $E\left[ \mathcal{S}_{\mathbf{1}}\big|_{\mathbf{0}} \right]$ is a \emph{dimension}. We \emph{quantify the basic observable energy}. In the same way we conduct independent basic measurements of momentum and inertial mass.
\begin{theo}\label{Theorem - kin quant absorption action for free particle}
The object $\circMO_{\:\mathbf{v}}$ with inertial mass $m_{\circMO}=m\cdot m_{\circMunit}$ and velocity $\mathbf{v}_{\mathcal{O}}=n\cdot \mathbf{v}_{\mathbf{1}}$ has a \underline{kinetic} energy and momentum
\bea\label{Abschnitt -- kin quant Absorptions Wirkung - kin energy and momentum metrisiert}
   E \left[ \circMO_{\:\mathbf{v}} \right] & = & \left( \frac{1}{2} \cdot m \cdot n^2   \right) \;\cdot\; E \left[ \mathcal{S}_{\mathbf{1}}\big|_{\mathbf{0}} \right]
   \\
   \mathbf{p} \left[ \circMO_{\:\mathbf{v}} \right] & = & \left( m \cdot n \right)\; \cdot\; \mathbf{p} \left[ \circMunit_{\:\mathbf{v}_{\mathbf{1}}} \right]   \;\; . \nn
\eea
\end{theo}
\textbf{Proof:}
The calorimeter extract has the same capability to execute work as the incident object $\circMO_{\:\mathbf{v}}$, because our calorimeter model is reversible.\footnote{We steer the absorption by a series of elastic head-on collisions in system $\circMunit_{\:n\cdot \mathbf{v}_{\mathbf{1}}} \cup \left\{ \circMunit_{\:\mathbf{0}} \right\}$ of incident particle and calorimeter reservoir. Every step of the deceleration cascade is reversible, because it is build up from solely congruent standard actions $w_{\mathbf{1}}$ and because physicists can steer these processes both ways.} Its kinetic energy is transformed
\bea
   E \left[ \circMO_{\:\mathbf{v}_{\mathcal{O}}} \right] & \stackrel{(\mathrm{Equip.})}{=} & E \left[ \mathrm{RB} \left[ \circMO_{\:\mathbf{v}_{\mathcal{O}}} \Rightarrow \circMO_{\:\mathbf{0}}  \right] \right] \nn \\
   & \stackrel{(\ref{Abschnitt -- kin quant Absorptions Wirkung - Reservoirbilanz - absorption proportional materiemenge})}{=} & E \left[  m \cdot \mathrm{RB} \left[ \circMunit_{\:\mathbf{v}_{\mathcal{O}}} \Rightarrow \circMunit_{\:\mathbf{0}} \right] \right] \nn \\
   & \stackrel{(\ref{Abschnitt -- kin quant Absorptions Wirkung - Reservoirbilanz - absorption})(\mathrm{Congr.})}{=} &
   m \cdot \left\{ \left(\frac{1}{2} \cdot n^2 - \frac{1}{2}\cdot n \right)\cdot E \left[ \mathcal{S}_{\mathbf{1} }\big|_{\mathbf{0}} \right] \;\; + \;\; n\cdot E \left[ \circMunit_{\:\mathbf{v}_{\mathbf{1}}} \right] \right\} \nn \\
   & \stackrel{(\ref{Abschnitt -- kin quant Absorptions Wirkung - pre-theoretical characterization E unit and p unit})}{=} &  \left( \frac{1}{2} \cdot m \cdot n^2   \right) \;\cdot\; E \left[ \mathcal{S}_{\mathbf{1} }\big|_{\mathbf{0}} \right]   \nn
\eea
into the potential energy of $( \frac{1}{2} \cdot m \cdot n^2 )$ congruent energy units $\mathcal{S}_{\mathbf{1}}\big|_{\mathbf{0}}$. The calorimeter extract also has the same impact as the incident particle $\circMO_{\:\mathbf{v}_{\mathcal{O}}}$, since otherwise one could construct a perpetuum mobile (see Lemma \ref{Lem - kin quant Absorptions Wirkung - Reservoirbilanz - p conserved}). Its impulse
\bea
   \mathbf{p} \left[ \circMO_{\:\mathbf{v}_{\mathcal{O}}} \right] & = & \mathbf{p} \left[ \mathrm{RB} \left[ \circMO_{\:\mathbf{v}_{\mathcal{O}}} \Rightarrow \circMO_{\:\mathbf{0}}  \right] \right] \nn \\
   & \stackrel{(\ref{Abschnitt -- kin quant Absorptions Wirkung - Reservoirbilanz - absorption})(\ref{Abschnitt -- kin quant Absorptions Wirkung - pre-theoretical characterization E unit and p unit})}{=} & \mathbf{p} \left[ m \cdot n \cdot \circMunit_{\:\mathbf{v}_{\mathbf{1}}} \right] \;\; \stackrel{(\mathrm{Congr.})}{=} \;\; \left( m \cdot n \right) \;\cdot\; \mathbf{p} \left[ \circMunit_{\:\mathbf{v}_{\mathbf{1}}} \right]  \nn
\eea
is reproduced by $( m\cdot n )$ congruent impulse units $\circMunit_{\:\mathbf{v}_{\mathbf{1}}}$ from the calorimeter reservoir.
\qed

We construct the underlying calorimeter-deceleration-cascade from a reservoir with standard elements $\left\{\circMunit_{\:\mathbf{v}=0}\right\}$ and one single elementary standard process $w_{\mathbf{1}}$. The building block (of irrelevant inner structure) is well-defined by symmetry and relativity principle (without taking equations of motion etc. as a basis). In this reversible model we stop a moving object
\be
    \circMO_{\:\mathbf{v}} \;\; \stackrel{W_{\mathrm{cal}}}{\Rightarrow} \;\; \circMO_{\:\mathbf{0}} \;,\;\, \mathrm{RB} \nn
\ee
in return for generating standard reference devices from the reservoir $\left\{\circMunit_{\:\mathbf{v}=0}\right\}$. In the reservoir balance $\mathrm{RB}$ (\ref{Abschnitt -- kin quant Absorptions Wirkung - Reservoirbilanz - absorption}) we can count the individual
\begin{itemize}
\item   $\sharp \left\{ \mathcal{S}_{\mathbf{1} }\big|_{\mathbf{0}} \right\} \;\;$ standard springs resp. particle pairs (representing unit energy)
\item   $\sharp \left\{ \circMunit_{\:\mathbf{v}_{\mathbf{1}}} \right\} \;\;$ impulse carriers (representing unit impulse and half the unit energy)
\item   $\sharp \left\{ \circMunit \right\} \;\;$ amount of matter in an equally massive composite $\circMunit \ast \dots \ast \circMunit\;\sim_{m}\circMO$
\item   $\mathbf{v}_{\mathcal{O}} = n \cdot \mathbf{v}_{\mathbf{1}}\;$ multiplicity of unit velocity.
\end{itemize}
We measure ''how many times'' larger the inertial mass $m\left[ \circMO \,\right]=:m\cdot m\left[ \circMunit \,\right]$, kinetic energy $E \left[ \circMO_{\:\mathbf{v}} \right] =: E \cdot E \left[ \mathcal{S}_{\mathbf{1}}\big|_{\mathbf{0}} \right]$ and momentum $\mathbf{p} \left[ \circMO_{\:\mathbf{v}_{\mathcal{O}}} \right] =: p \cdot \mathbf{p} \left[ \circMunit_{\:\mathbf{v}_{\mathbf{1}}} \right]$ of the incident object $\circMO_{\:\mathbf{v}}$ is than in one standard spring $\mathcal{S}_{\mathbf{1} }\big|_{\mathbf{0}}$ and momentum carrier $\circMunit_{\:\mathbf{v}_{\mathbf{1}}}$ of reference process $w_{\mathbf{1}}$. From the layout of the building blocks in our measurement instrument we derive the relation between these physical quantities (\ref{Abschnitt -- kin quant Absorptions Wirkung - kin energy and momentum metrisiert}).

When we build the model in Galilei kinematics we derive primary dynamical equations
\[
   \frac{E_{\mathcal{O}}}{E_{\mathbf{1}}} = \frac{1}{2} \cdot \frac{m_{\mathcal{O}}}{m_{\mathbf{1}}} \cdot \left(\frac{\mathbf{v}_{\mathcal{O}}}{ \mathbf{v}_{\mathbf{1}}}\right)^2 \;\;\;\;\;\;\;\;
   \frac{\mathbf{p}_{\mathcal{O}}}{\mathbf{p}_{\mathbf{1}}} = \frac{m_{\mathcal{O}}}{m_{\mathbf{1}}} \cdot \frac{\mathbf{v}_{\mathcal{O}}}{ \mathbf{v}_{\mathbf{1}}} \;\;,
\]
in which all numerical values for kinetic energy $E =: \frac{E \left[ \circMO_{\:\mathbf{v}} \right]}{E \left[ \mathcal{S}_{\mathbf{1}}\big|_{\mathbf{0}} \right] }\:$, impulse $\mathbf{p} =: \frac{\mathbf{p} \left[ \circMO_{\:\mathbf{v}} \right]}{\mathbf{p} \left[ \circMunit_{\:\mathbf{v}_{\mathbf{1}}} \right]}\:$, inertial mass $m =: \frac{m\left[ \circMO \right]}{m \left[ \circMunit \,\right]}\:$ and velocity $n =: \frac{v_{\mathcal{O}}}{v_{\mathbf{1}}}$ occur in the form \emph{measure/unit measure}.\footnote{We write the numerical values $E$ formally as a ratio $E \left[ \circMO_{\:\mathbf{v}} \right] / E \left[ \mathcal{S}_{\mathbf{1}}\big|_{\mathbf{0}} \right]$ between energy of the moving cue ball (individual measure) and energy of the compressed standard spring (unit measure). This ''ratio'' symbolizes the result of a tangible operation; counting congruent units in the calorimeter model.} Wallot calls them ''tailored quantity equations'' (German: zugeschnittene Gr\"o{\ss}engleichungen) \cite{Wallot - Groessengleichungen Einheiten und Dimensionen}. When we steer the \emph{same} measurement process in Poincare kinematics, then we will derive all equations of relativistic dynamics \{\ref{Kap - Relativistic energy-momentum}\}.


\section{Momentum}\label{Kap - KM Dynamics - Basic Dynamical Measures - Momentum}

According to principle of inertia motion is preserved unless some body is effected by an external cause \cite{Euler Anleitung}. Sommerfeld \cite{Sommerfeld-Mechanik} defines the impulse of a moving body: ''Impulse means (with regards to direction and magnitude) that kick, which is capable of generating a given state of motion from the initial state of rest.''
Our calorimeter model provides a direct measurement: We generate this kick by a number of congruent standard kicks.

For collisions of irrelevant inner structure we define an elementary \emph{ordering criterion} for momentum (see Definition \ref{Def - vortheor Ordnungsrelastion - impulse}): Object $\circMa_{\:\mathbf{v}_{a}}$ has same impulse as object $\circMb_{\:\mathbf{v}_{b}}$
\be
   \circMa_{\:\mathbf{v}_a} \;\; \sim_{\mathbf{p}} \;\; \circMb_{\:\mathbf{v}_b} \nn
\ee
if in an inelastic head-on collision test the two bodies stick together and stop
\be
   \circMa_{\:\mathbf{v}_a} \,,\: \circMb_{\:\mathbf{v}_b} \;\; \Rightarrow \;\; \circMa\ast\circMb_{\;\mathbf{v}=0}   \;\; . \nn
\ee
The joint collision product $\circMa\ast\circMb_{\;\mathbf{v}=0}$ moves neither into the former direction of object $\circMa_{\:\mathbf{v}_{a}}$ to the right nor into the other direction of object $\circMb_{\:\mathbf{v}_{b}}$ to the left. If it continues moving into the direction of $\circMa_{\:\mathbf{v}_{a}}$ then the latter has more impact than object $\circMb_{\:\mathbf{v}_{b}}$
\be
   \circMa_{\:\mathbf{v}_{a}} \;\; >_{\mathbf{p}} \;\; \circMb_{\:\mathbf{v}_{b}} \nn  \;\; .
\ee

We \emph{concatenate} the momentum of two moving objects $\circMa_{\:\mathbf{v}_{a}}$ and $\circMb_{\:\mathbf{v}_{b}}$ by coupling two consecutive absorptions: We expend their impulses in the same calorimeter reservoir $\left\{\circMunit_{\:\mathbf{0}}\right\}$ and eventually against the same external element $\circMunit_{\:\mathbf{v}}$
\be
   \circMa_{\:\mathbf{v}_{a}} \,,\: \circMb_{\:\mathbf{v}_{b}} \,,\: \circMunit_{\:\mathbf{v}} \;\; \Rightarrow \;\; \circMa_{\:\mathbf{0}} \,,\: \circMb_{\:\mathbf{0}} \,,\: \circMunit_{\:\mathbf{v}'} \nn
\ee
such that the two original objects $\circMa_{\:\mathbf{0}}$ and $\circMb_{\:\mathbf{0}}$ stop.\footnote{Sommerfeld's defining kick is associated with generating motion (from rest). We examine kicks which annihilate motion (towards rest). In two-body collisions Sommerfeld regards the ''recipient''; we the ''giver''.} Our absorption model $W_{\mathrm{cal}}$ guarantees commutativity of the absorption order. We represent the impulse \emph{unit} $\circMunit_{\:\mathbf{v}_{\mathbf{1}}}$ by the impact of $\mathcal{A}$lice standard objects $\circMunit_{\:\mathbf{v}_{\mathbf{1}}}$ in the deceleration from unit motion $\mathbf{v}_{\mathbf{1}}$ to the state of rest.

Our calorimeter model $W_{\mathrm{cal}}$ provides a \emph{physical model} for the impact of particle $\circMa_{\:\mathbf{v}_{a}}$. In return for its absorption we extract $\mathrm{RB} \left[ \circMa_{\:\mathbf{v}_{a}} \Rightarrow \circMa_{\:\mathbf{0}}  \right] = k\cdot \mathcal{S}_{\mathbf{1}}\big|_{\mathbf{0}} , l\cdot\circMunit_{\:\mathbf{v}_{\mathbf{1}}}$ from the initially resting reservoir $\left\{\circMunit_{\:\mathbf{v}=0}\right\}$ $k$ particle pairs (energy units $\mathcal{S}_{\mathbf{1}}\big|_{\mathbf{0}}$ with no impulse (\ref{Abschnitt -- kin quant Absorptions Wirkung - pre-theoretical characterization E unit and p unit})) and a swarm $\circMunit_{\:\mathbf{v}_{\mathbf{1}}} ,\ldots, \circMunit_{\:\mathbf{v}_{\mathbf{1}}} =: l\cdot\circMunit_{\:\mathbf{v}_{\mathbf{1}}} $ of $l$ comoving impulse carriers $\circMunit_{\:\mathbf{v}_{\mathbf{1}}}$ with unit velocity $\mathbf{v}_{\mathbf{1}}$ into the direction of motion of the absorbed particle $\circMa_{\:\mathbf{v}_{a}}$. That is our physical model for momentum. The swarm of reservoir elements has the same impact
\be
   \circMunit_{\:\mathbf{v}_{\mathbf{1}}} ,\ldots, \circMunit_{\:\mathbf{v}_{\mathbf{1}}} \;\; \sim_{\mathbf{p}} \;\; \circMa_{\:\mathbf{v}_{a}} \nn
\ee
as incident object $\circMa_{\:\mathbf{v}_{a}}$ (see Lemma \ref{Lem - kin quant Absorptions Wirkung - Reservoirbilanz - p conserved}). $\mathcal{A}$lice measures the impulse of particle $\circMa_{\:\mathbf{v}_{a}}$
\be
   \mathbf{p}\left[ \circMa_{\:\mathbf{v}_{a}} \right] \;=:\; p\cdot \mathbf{p}\left[ \circMunit_{\:\mathbf{v}_{\mathbf{1}}} \right]   \nn
\ee
by the number $p:=\sharp \left\{ \circMunit_{\:\mathbf{v}_{\mathbf{1}}} \right\}$ of congruent impulse carriers and the reference impulse $\mathbf{p}\left[ \circMunit_{\:\mathbf{v}_{\mathbf{1}}} \right]$. This method of quantification is universally \emph{reproducible} and gives the same magnitudes (of momentum) independently from the individual physicist $\mathcal{A}$lice or $\mathcal{B}$ob.

\begin{lem}\label{Lem - kin quant Absorptions Wirkung - Reservoirbilanz - p conserved}
In our calorimeter model $W_{\mathrm{cal}}$ the extracted impulse carriers $\circMunit\ast\ldots\ast\circMunit_{\;\mathbf{v}_{\mathbf{1}}}$
\be
   \circMunit\ast\ldots\ast\circMunit_{\;\mathbf{v}_{\mathbf{1}}} \;\; \sim_{\mathbf{p}} \;\; \circMO_{\:\mathbf{v}_{\mathcal{O}}}   \nn
\ee
have the same impulse as the incident object $\circMO_{\:\mathbf{v}_{\mathcal{O}}}$. The transferred momentum is \underline{conserved}.
\end{lem}
\textbf{Proof:} Without restricting generality we consider the absorption (\ref{Abschnitt -- kin quant Absorptions Wirkung - Reservoirbilanz - absorption}) of one standard particle $\circMO\equiv\circMunit$ with velocity $\mathbf{v}_{\mathcal{O}}:=-n\cdot \mathbf{v}_{\mathbf{1}}$, $\;n\in \mathbb{N}$ which extracts standard impulse carriers $\circMunit\ast\ldots\ast\circMunit_{\;\mathbf{v}_{\mathbf{1}}} =: \circMn_{\:\mathbf{v}_{\mathbf{1}}}$ with velocity $\mathbf{v}_{\mathbf{1}}$. In a generic inelastic head-on collision test the incident object $\circMO_{\:\mathbf{v}_{\mathcal{O}}}$ and its impulse model $\circMn_{\:\mathbf{v}_{\mathbf{1}}}$
\be\label{Abschnitt -- Impuls - inelastic collision direct}
   \circMO_{\:\mathbf{v}_{\mathcal{O}}} \,,\: \circMn_{\:\mathbf{v}_{\mathbf{1}}} \;\; \stackrel{w_{(d)}}{\Rightarrow} \;\; \circMO\ast\circMn_{\:\mathbf{v}'} \;\; .
\ee
form a bound aggregate $\circMO\ast\circMn_{\:\mathbf{v}'}$ with velocity $\mathbf{v}'$ and bounding energy $E^{\ast}$. They have the same momentum if they collide, stick together and come to rest (see \emph{physical} Definition \ref{Def - vortheor Ordnungsrelastion - impulse}).\footnote{The head-on collision test of two objects is a tangible operation. It leads to unique measurement values.} Hence, we have to show, that the bound aggregate $\circMO\ast\circMn_{\:\mathbf{v}'}$ must stop $\mathbf{v}'\stackrel{!}{=}\mathbf{0}$.

Let us hypothetically assume the contrary. Then we could stop the aggregate $\circMO\ast\circMn_{\:\mathbf{v}'}$ in our calorimeter
\be\label{Abschnitt -- Impuls - inelastic collision - collision product absorbed}
   \circMO\ast\circMn_{\;\mathbf{v}'} \;\; \stackrel{W_{\mathrm{cal}}}{\Rightarrow} \;\; \circMO\ast\circMn_{\;\mathbf{0}} \,,\: k\cdot \mathcal{S}_{\mathbf{1}}\big|_{\mathbf{0}} \,,\: l\cdot \circMunit_{\:\mathbf{v}_{\mathbf{1}}}
\ee
and extract additional $k$ energy units $\mathcal{S}_{\mathbf{1}}\big|_{\mathbf{0}}$ and $l$ impulse carriers $\circMunit_{\:\mathbf{v}_{\mathbf{1}}}$ into the direction of $\mathbf{v}'$. One can steer the calorimeter-deceleration-cascade in the reverse way. Further let the direct inelastic collision $w_{(d)}$ (\ref{Abschnitt -- Impuls - inelastic collision direct}) be reversible. We assume that we can turn the resting aggregate $\circMO\ast\circMn_{\;\mathbf{0}}$ around arbitrary angle $\theta$
\be\label{Abschnitt -- Impuls - inelastic collision - collision product gedreht}
   \circMO\ast\circMn_{\;\mathbf{0}} \;\; \stackrel{\mathrm{R}_{\theta}}{\Rightarrow} \;\; \mathrm{R}_{\theta} \left[ \circMO\ast\circMn_{\;\mathbf{0}} \right]
\ee
so that after expending the bounding energy $E^{\ast}$ the release process inside aggregate $\circMO\ast\circMn$
\be\label{Abschnitt -- Impuls - inelastic collision direct - gedreht und inverse}
   \mathrm{R}_{\theta}\!\left[w_{(d)}^{-1}\right]: \;\; \circMO\ast\circMn_{\:\mathrm{R}_{\theta}\mathbf{v}'} \;\; \Rightarrow \;\; \circMO_{\:\mathrm{R}_{\theta}\mathbf{v}_{\mathcal{O}}} \,,\: \circMn_{\:\mathrm{R}_{\theta}\mathbf{v}_{\mathbf{1}}}
\ee
kicks the unbound objects into corresponding direction $\theta$ (like the spring in figure \ref{pic_Wirkungseinheit_Feder}).

We can also \emph{mediate} this inelastic collision $w_{(m)}$ by our reversible calorimeter model $W_{\mathrm{cal}}$. From separate absorption of incident object $\circMO_{\:\mathbf{v}_{\mathcal{O}}}$ and its impulse model $\circMunit\ast\ldots\ast\circMunit_{\;\mathbf{v}_{\mathbf{1}}}$ (we assume that its tight linkage can be opened and closed without practical consequences)
\be\label{Abschnitt -- Impuls - inelastic collision mediated}
   \circMO_{\:\mathbf{v}_{\mathcal{O}}=-n\cdot \mathbf{v}_{\mathbf{1}}} \,,\: \circMn_{\:\mathbf{v}_{\mathbf{1}}} \;\; \stackrel{w_{(m)}}{\Rightarrow} \;\; \circMO_{\:\mathbf{0}} \,,\: \circMn_{\:\mathbf{0}} \,,\: \left(\frac{1}{2}\cdot n^2 + \frac{1}{2}\cdot n \right)\cdot \mathcal{S}_{\mathbf{1}}\big|_{\mathbf{0}}
\ee
we extract $\frac{1}{2}\cdot n^2 + \frac{1}{2}\cdot n $ standard energy sources $\mathcal{S}_{\mathbf{1}}\big|_{\mathbf{0}}$
\be
\begin{array}{l}
   \mathrm{RB} \left[ \circMO_{\:\mathbf{v}_{\mathcal{O}}} \,,\: \circMunit\ast\ldots\ast\circMunit_{\;\mathbf{v}_{\mathbf{1}}} \;\Rightarrow\; \circMO_{\:\mathbf{0}}  \,,\:  \circMunit\ast\ldots\ast\circMunit_{\:\mathbf{0}}  \right] \nn \\
   \;\; \stackrel{(\ref{Abschnitt -- kin quant Absorptions Wirkung - Reservoirbilanz - additivitaet})}{=} \;\; \underbrace{\mathrm{RB} \left[ \circMO_{\:\mathbf{v}_{\mathcal{O}}}  \Rightarrow \circMO_{\:\mathbf{0}} \right]}_{\stackrel{(\ref{Abschnitt -- kin quant Absorptions Wirkung - Reservoirbilanz - absorption})}{=}\; \left(\frac{1}{2}\cdot n^2 - \frac{1}{2}\cdot n \right)\cdot\, \mathcal{S}_{\mathbf{1}}\big|_{\mathbf{0}} \,,\; n\cdot\, \circMunit_{-\mathbf{v}_{\mathbf{1}}} } \, + \;\;\;
\underbrace{\mathrm{RB} \left[ \circMunit\ast\ldots\ast\circMunit_{\;\mathbf{v}_{\mathbf{1}}} \Rightarrow \circMunit\ast\ldots\ast\circMunit_{\:\mathbf{0}}  \right]}_{=\;\mathrm{RB} \left[ \circMunit_{\:\mathbf{v}_{\mathbf{1}}} \,,\: \ldots \,,\: \circMunit_{\:\mathbf{v}_{\mathbf{1}}} \;\Rightarrow\; \circMunit_{\:\mathbf{0}} \,,\: \ldots \,,\: \circMunit_{\:\mathbf{0}} \right]\;\equiv\; n\cdot \,\circMunit_{\:\mathbf{v}_{\mathbf{1}}} }  \nn \\
   \;\; \stackrel{(\ref{Abschnitt -- basic dynamical measures - Einheitswirkung})}{=} \;\; \left(\frac{1}{2}\cdot n^2 + \frac{1}{2}\cdot n \right)\cdot \mathcal{S}_{\mathbf{1}}\big|_{\mathbf{0}}   \;\; .
\end{array}
\ee

We essentially use the \emph{isotropy} of an intrinsic interaction, like standard process $w_{\mathbf{1}}$, and concatenate the following series of \emph{separate} interactions
\be
   {w_{(m)}}^{-1} \;\ast\; w_{(d)} \;\ast\; W_{\mathrm{cal}} \;\ast\; \mathrm{R}_{\theta} \;\ast\; {W^{(\theta)}_{\mathrm{cal}}}^{-1} \;\ast\; {w_{(d)}^{(\theta)}}^{-1} \;\ast\; w^{(\theta)}_{(m)}
\nn
\ee
in the coinciding intermediate objects $\circMO_{\:\mathbf{v}_{\mathcal{O}}} \,,\: \circMn_{\:\mathbf{v}_{\mathbf{1}}} \,,\: \circMO\ast\circMn_{\;\mathbf{v}'}$ etc. which in between the consecutive interactions move freely
\be
\begin{array}{l}
   \circMO_{\:\mathbf{0}} \,,\: \circMn_{\:\mathbf{0}}  \,,\: \left(\frac{1}{2}\cdot n^2 + \frac{1}{2}\cdot n \right)\cdot \mathcal{S}_{\mathbf{1}}\big|_{\mathbf{0}} \;\; \stackrel{(\ref{Abschnitt -- Impuls - inelastic collision mediated})}{\Longrightarrow} \;\; \circMO_{\:\mathbf{v}_{\mathcal{O}}} \,,\: \circMn_{\:\mathbf{v}_{\mathbf{1}}}  \;\; \stackrel{(\ref{Abschnitt -- Impuls - inelastic collision direct})}{\Longrightarrow} \;\; \circMO\ast\circMn_{\;\mathbf{v}'}  \\
   \;\;\;\;\;
   \;\; \stackrel{(\ref{Abschnitt -- Impuls - inelastic collision - collision product absorbed})}{\Longrightarrow} \;\; \circMO\ast\circMn_{\;\mathbf{0}} \,,\: k\cdot \mathcal{S}_{\mathbf{1}}\big|_{\mathbf{0}} \,,\: l\cdot \circMunit_{\:\mathbf{v}_{\mathbf{1}}} \\
   \;\;\;\;\;\;\;\;\;\;\;\;
   \;\; \stackrel{(\ref{Abschnitt -- Impuls - inelastic collision - collision product gedreht})}{\Longrightarrow} \;\; \mathrm{R}_{\theta} \left[ \circMO\ast\circMn_{\;\mathbf{0}} \right] \,,\: k\cdot \mathcal{S}_{\mathbf{1}}\big|_{\mathbf{0}} \,,\: l\cdot \circMunit_{\:\mathbf{v}_{\mathbf{1}}} \,,\: \left(\,+\:l\cdot \circMunit_{\:\mathrm{R}_{\theta}\mathbf{v}_{\mathbf{1}}} \,-\: l\cdot \circMunit_{\:\mathrm{R}_{\theta}\mathbf{v}_{\mathbf{1}}} \right) \\
   \;\;\;\;\;\;\;\;\;\;\;\;\;\;\;\;\;\;\;
   \;\; \stackrel{(\ref{Abschnitt -- Impuls - inelastic collision - collision product absorbed})}{\Longrightarrow} \;\; \circMO\ast\circMn_{\:\mathrm{R}_{\theta}\mathbf{v}'} \,,\: l\cdot \circMunit_{\:\mathbf{v}_{\mathbf{1}}} \,-\: l\cdot \circMunit_{\:\mathrm{R}_{\theta}\mathbf{v}_{\mathbf{1}}} \\
   \;\;\;\;\;\;\;\;\;\;\;\;\;\;\;\;\;\;\;\;\;\;\;\;\;\;
   \;\; \stackrel{(\ref{Abschnitt -- Impuls - inelastic collision direct - gedreht und inverse})}{\Longrightarrow} \;\; \circMO_{\:\mathrm{R}_{\theta}\mathbf{v}_{\mathcal{O}}} \,,\: \circMn_{\:\mathrm{R}_{\theta}\mathbf{v}_{\mathbf{1}}} \,,\: l\cdot \circMunit_{\:\mathbf{v}_{\mathbf{1}}} \,-\: l\cdot \circMunit_{\:\mathrm{R}_{\theta}\mathbf{v}_{\mathbf{1}}} \\
   \;\;\;\;\;\;\;\;\;\;\;\;\;\;\;\;\;\;\;\;\;\;\;\;\;\;\;\;\;\;\;\;\;
   \;\; \stackrel{(\ref{Abschnitt -- Impuls - inelastic collision mediated})}{\Longrightarrow} \;\;
   \left(\frac{1}{2}\cdot n^2 + \frac{1}{2}\cdot n \right)\cdot \mathcal{S}_{\mathbf{1}}\big|_{\mathbf{0}} \,,\: \circMO_{\:\mathbf{0}} \,,\: \circMn_{\:\mathbf{0}}  \,,\;\;\; l\cdot \circMunit_{\:\mathbf{v}_{\mathbf{1}}} \,-\: l\cdot \circMunit_{\:\mathrm{R}_{\theta}\mathbf{v}_{\mathbf{1}}} \;\; .
\end{array} \nn
\ee
Both objects $\circMO_{\:\mathbf{0}} \,,\: \circMn_{\:\mathbf{0}}$ act as catalyzers. Throughout the process we - temporarily - expend energy units $\mathcal{S}_{\mathbf{1}}\big|_{\mathbf{0}}$ and bounding energy $E^{\ast}$ but in the end they are all recycled back into the reservoir. They have no net effect on the spectators $\circMO_{\:\mathbf{0}} \,,\: \circMn_{\:\mathbf{0}}$. In the end of this \emph{circular process} we expend $-l$ impulse carriers $\circMunit_{\:\mathrm{R}_{\theta}\mathbf{v}_{\mathbf{1}}}$ into arbitrarily rotated direction $\theta$ in return for generating $l$ impulse carriers $\circMunit_{\:\mathbf{v}_{\mathbf{1}}}$ without effecting anything else.

This hypothetical process would violate Euler's principle of sufficient reason; that every change in the state motion requires an external cause (physical reason) \cite{Euler Anleitung} and impossibility of a perpetuum mobile. A moving observer (with velocity $\mathbf{v}_{\mathbf{1}}$) could set initially resting reservoir particles $\{\circMunit_{\:\mathbf{0}}\}$ into motion with unit velocity $2\cdot\mathbf{v}_{\mathbf{1}}$ but also into opposite direction with velocity $-2\cdot\mathbf{v}_{\mathbf{1}}$ and thus generate particle pairs $\{\circMunit_{-2\cdot\mathbf{v}_{\mathbf{1}}} \,,\: \circMunit_{\:2\cdot\mathbf{v}_{\mathbf{1}}} \} \sim_{E} \mathcal{S}_{2}\big|_{\mathbf{0}}$ resp. energy sources without any reaction from nothing.
Therefore our hypothetical assumption $\mathbf{v}'\neq 0$ is invalid. In the direct inelastic head-on collision test $w_{(d)}$ the object $\circMO_{\:\mathbf{v}_{\mathcal{O}}}$ and its impulse model $\circMn_{\:\mathbf{v}_{\mathbf{1}}}$ stick together and come to rest. They have the same impact $\circMn_{\;\mathbf{v}_{\mathbf{1}}} \; \sim_{\mathbf{p}} \; \circMO_{\:\mathbf{v}_{\mathcal{O}}}$.
\qed

We measure the momentum from multi-partite systems by steering a separate absorption $W_{\mathrm{cal}}$ for each element (see Lemma \ref{Lem - kin quant Absorptions Wirkung - Reservoirbilanz - additivitaet}).
%
%
%
We extract impulse carriers $\circMunit_{\:\mathbf{v}_{\mathbf{1}}}$ and $\circMunit_{-\mathbf{v}_{\mathbf{1}}}$ - on the left and right side of the deceleration cascade - in the direction of its motion $\mathbf{v}_i$ (see figure \ref{pic_calorimeter_model}). For a generic many-particle system one extracts impulse carriers $\left\{\circMunit_{\:\mathbf{v}_{\!i}}\right\}_{i=1\ldots N}$ in various directions $\mathbf{v}_{i}\neq\mathbf{v}_{j}$.
\begin{theo}
Direction and magnitude of total momentum is calculable by \underline{vectorial addition}
\be
    \mathbf{p}\left[ \circMunit_{\:\mathbf{v}_1} , \dots , \circMunit_{\:\mathbf{v}_N} \right]  \;\; = \;\; \mathbf{p}\left[ \circMunit_{\:\mathbf{v}_1} \right] + \ldots + \; \mathbf{p}\left[ \circMunit_{\:\mathbf{v}_N} \right]   \;\; .
\ee
\end{theo}
\textbf{Proof:}
(In Galilei Kinematics) we construct a physical model $W$ for absorbing multiple standard elements $\circMunit_{\:\textcolor{cyan}{\mathbf{v}_i}}$ with velocities $\mathbf{v}_i$ into various directions $\mathbf{v}_i \nparallel \mathbf{v}_j$
\be
   \left\{\circMunit_{\:\textcolor{cyan}{\mathbf{v}_1}} , \dots , \circMunit_{\:\textcolor{cyan}{\mathbf{v}_N}}\right\} \,,\: \circMunit_{\:\mathbf{0}}
   \;\; \stackrel{W}{\Rightarrow} \;\; \left\{\circMunit_{\:\textcolor{cyan}{\mathbf{0}}} , \dots , \circMunit_{\:\textcolor{cyan}{\mathbf{0}}}\right\}  \,,\:
\circMunit_{\:\mathbf{v}_{(N)}} \;\; .
\nn
\ee
All elements $i=1,\ldots,N$ of the system $\left\{\circMunit_{\:\mathbf{0}} , \dots , \circMunit_{\:\mathbf{0}}\right\}$ stop; while one initially resting absorber particle $\circMunit_{\:\mathbf{v}_{(N)}}$ gets kicked, as we will show, into velocity $\mathbf{v}_{(N)} = \mathbf{v}_1 + \ldots + \mathbf{v}_N$.

\begin{figure}    
  \begin{center}           
  \includegraphics[height=7.3cm]{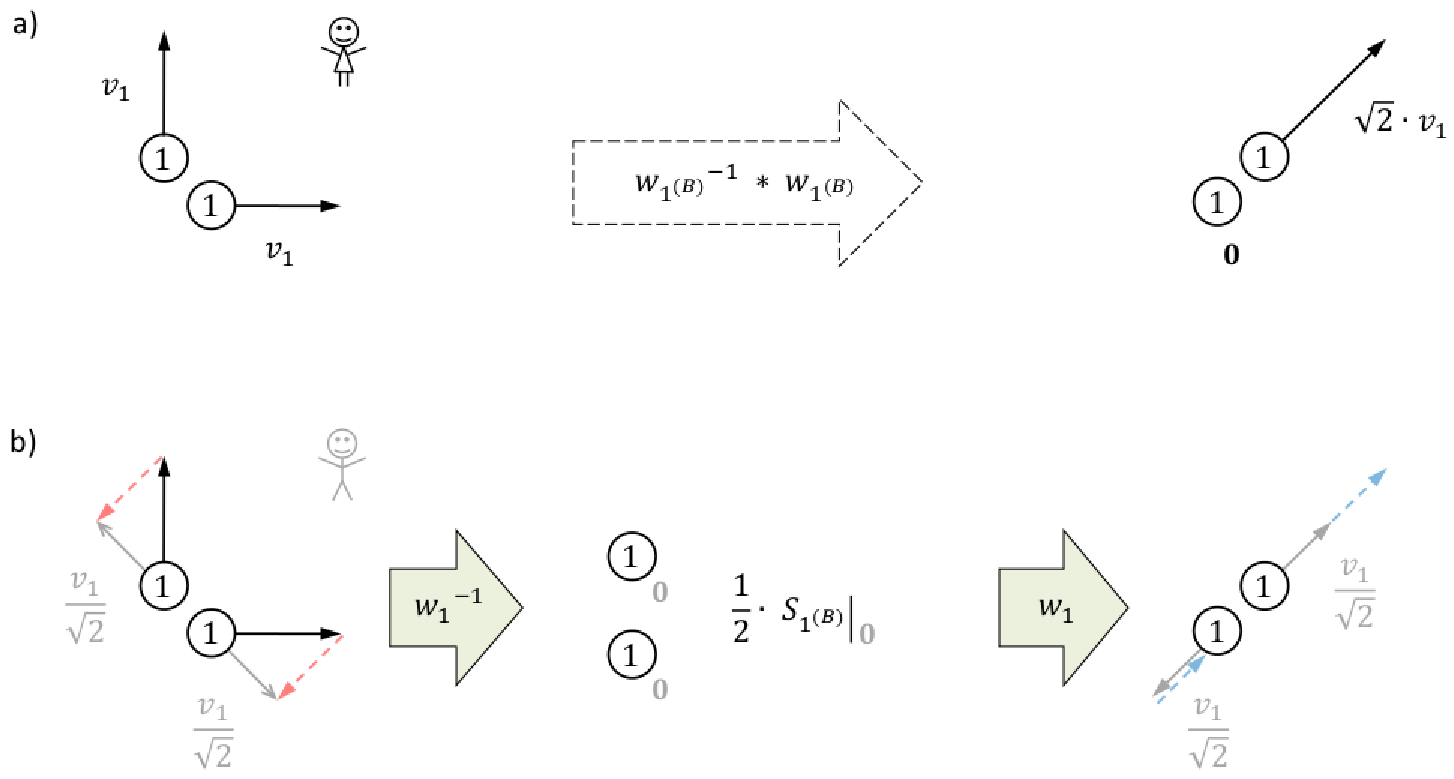}  
  \end{center}
  \vspace{-0cm}
  \caption{\label{pic_impuls_vektoriell_prinzip} a) vectorial impulse addition by underlying b) isotropic unit actions
    }
  \end{figure}

We illustrate $\mathcal{A}$lice complete momentum transfer from one moving particle $\circMunit_{\:\mathbf{v}_1}$ onto another particle $\circMunit_{\:\mathbf{v}_2}$ in figure \ref{pic_impuls_vektoriell_prinzip}.
Let the two standard objects $\circMunit_{\:\mathbf{v}_i}$ $i=1,2$ move with unit velocity into perpendicular directions
\be
   \mathbf{v}_{1}=\left(
                                               \begin{array}{c}
                                                 1 \\
                                                 0 \\
                                               \end{array}
                                             \right)\cdot \mathbf{v}_{\mathbf{1}^{(\mathcal{A})}}
\;\;\;\;\; \mathrm{resp.} \;\;\;\;\;
\mathbf{v}_{2}=\left(
                                               \begin{array}{c}
                                                 0 \\
                                                 1 \\
                                               \end{array}
                                             \right)\cdot \mathbf{v}_{\mathbf{1}^{(\mathcal{A})}} \;\; . \nn
\ee
Let $\mathcal{A}$lice move relative to $\mathcal{B}$ob with constant velocity $\mathbf{v}_{\mathcal{A}}=-\frac{1}{2}\cdot\left(
                                               \begin{array}{c}
                                                 1 \\
                                                 1 \\
                                               \end{array}
                                             \right)\cdot \mathbf{v}_{\mathbf{1}^{(\mathcal{B})}}$. For $\mathcal{B}$ob both particles $i=1,2$ move into opposite direction with measured values (see Remark \ref{Rem - covariant kinematical transformation})
\be
   \mathbf{v}_{i}^{(\mathcal{B})}= \mathbf{v}_{i}^{(\mathcal{A})} + \mathbf{v}_{\mathcal{A}}^{(\mathcal{B})}
   = \pm \left(
                                               \begin{array}{c}
                                                 1/2 \\
                                                 -1/2 \\
                                               \end{array}
                                             \right)
\;\;\;\;\;\;\;\; \mathrm{with} \;\;\; \mathbf{v}_{\mathcal{A}}^{(\mathcal{B})} = -\frac{1}{2}\cdot\left(
                                               \begin{array}{c}
                                                 1 \\
                                                 1 \\
                                               \end{array}
                                             \right)  \nn
\ee
and velocity $v_{i}^{(\mathcal{B})} = \frac{1}{\sqrt{2}}$. Let $\mathcal{B}$ob absorb them in his calorimeter $W_{\mathrm{cal}}$
\be
   \circMunit_{-\frac{1}{\sqrt{2}}\cdot\mathbf{v}_{\mathbf{1}^{(\mathcal{B})}}} \,,\: \circMunit_{\:\frac{1}{\sqrt{2}}\cdot\mathbf{v}_{\mathbf{1}^{(\mathcal{B})}}}
   \;\; \stackrel{w_1^{-1}}{\Rightarrow} \;\; \circMunit_{\:\mathbf{0}} \,,\: \circMunit_{\:\mathbf{0}} \,,\: \frac{1}{2}\cdot \mathcal{S}_{\mathbf{1}^{(\mathcal{B})} }\big|_{\mathbf{0}}   \nn  \;\; .
\ee
He extracts the equivalent of $\frac{1}{2}$ energy source $\mathcal{S}_{\mathbf{1}^{(\mathcal{B})} }\big|_{\mathbf{0}}$ and expends it in a consecutive standard action $w_1$ (suitably rotated by $90^{\circ}$) against the two temporary resting particles
\be
  \circMunit_{\:\mathbf{0}} \,,\: \circMunit_{\:\mathbf{0}} \,,\: \frac{1}{2}\cdot \mathcal{S}_{\mathbf{1}^{(\mathcal{B})} }\big|_{\mathbf{0}} \;\; \stackrel{w_1}{\Rightarrow} \;\; \circMunit_{-\frac{1}{\sqrt{2}}\cdot\mathrm{R}_{90^{\circ}}\mathbf{v}_{\mathbf{1}^{(\mathcal{B})}}} \,,\: \circMunit_{\:\frac{1}{\sqrt{2}}\cdot\mathrm{R}_{90^{\circ}}\mathbf{v}_{\mathbf{1}^{(\mathcal{B})}}}   \nn \;\; .
\ee
The particle pair $i=1,2$ repulses with velocity $\mathbf{v}'_{i}=\pm\frac{1}{2}\cdot\left(
                                               \begin{array}{c}
                                                 1 \\
                                                 1 \\
                                               \end{array}
                                             \right)\cdot \mathbf{v}_{\mathbf{1}^{(\mathcal{B})}}$
towards $\mathcal{A}$lice. She measures velocities ${\mathbf{v}'}_{i}^{(\mathcal{A})}= {\mathbf{v}'}_{i}^{(\mathcal{B})} + \mathbf{v}_{\mathcal{B}}^{(\mathcal{A})}$ with $\mathbf{v}_{\mathcal{B}}^{(\mathcal{A})} = \frac{1}{2}\cdot\left(
                                               \begin{array}{c}
                                                 1 \\
                                                 1 \\
                                               \end{array}
                                             \right) $
\be
   \mathbf{v}'_{1}=\left(
                                               \begin{array}{c}
                                                 1 \\
                                                 1 \\
                                               \end{array}
                                             \right)\cdot \mathbf{v}_{\mathbf{1}^{(\mathcal{A})}}
\;\;\;\;\; \mathrm{resp.} \;\;\;\;\;
\mathbf{v}'_{2}=0\cdot \mathbf{v}_{\mathbf{1}^{(\mathcal{A})}}  \nn \;\; .
\ee
In her view of $\mathcal{B}$ob's (elastic) association of reversible unit actions ${w_1^{-1}}^{(\mathcal{B})}\ast w_1^{(\mathcal{B})}$ particle $\circMunit_{\:\mathbf{v}'_{2}}$ appears to come to rest ${\mathbf{v}'}_{2}^{(\mathcal{A})}=0$. It has transferred all momentum (elastically) onto particle $\circMunit_{\:\mathbf{v}'_{1}}$ which moves with final velocity ${\mathbf{v}'}_{1}^{(\mathcal{A})}=\mathbf{v}_{1}^{(\mathcal{A})}+\mathbf{v}_{2}^{(\mathcal{A})}$.

For a system of $N$ inequivalent impulse carriers $\left\{\circMunit_{\:\mathbf{v}_1} , \circMunit_{\:\mathbf{v}_2} , \dots , \circMunit_{\:\mathbf{v}_N}\right\}$ $\mathcal{A}$lice successively transfers the impulse of all remaining elements $ \circMunit_{\:\mathbf{v}_i}$ with $i=2,\dots,N$ onto the absorber particle $ \circMunit_{\:\mathbf{v}_1}$. In the beginning of the induction it has velocity $\mathbf{v}_1=:\mathbf{v}_{(1)}$.

\begin{figure}    
  \begin{center}           
  \includegraphics[height=9.2cm]{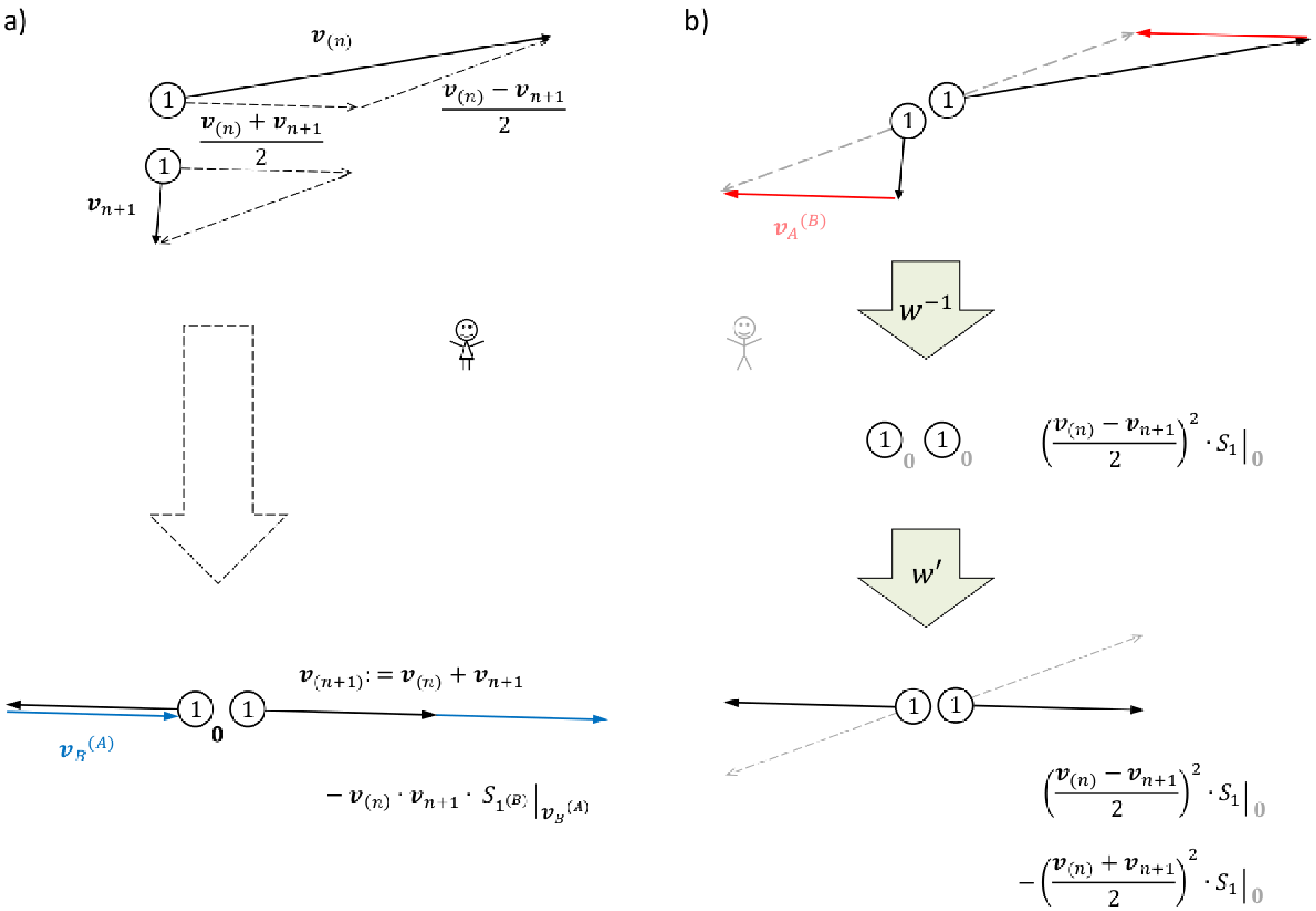}  
  \end{center}
  \vspace{-0cm}
  \caption{\label{pic_impuls_vektoriell_induktion} a) $\mathcal{A}$lice impulse transfer from
  $\circMunit _{\:\mathbf{v}_{n+1}}$ to $\circMunit _{\:\mathbf{v}_{(n)}}$ by b) standard actions of $\mathcal{B}$ob
    }
  \end{figure}

At each step $n\rightarrow n+1$ $\mathcal{A}$lice provides the absorber particle $\circMunit_{\:\mathbf{v}_{(n)}}$ with current velocity $\mathbf{v}_{(n)}$ and the next element $\circMunit_{\:\mathbf{v}_{n+1}}$ from the system (see figure \ref{pic_impuls_vektoriell_induktion}a).
Let $\mathcal{A}$lice move relative to $\mathcal{B}$ob with velocity $\mathbf{v}_{\mathcal{A}}=\mathbf{v}_{\mathcal{A}}^{(\mathcal{B})}\cdot \mathbf{v}_{\mathbf{1}^{(\mathcal{B})}}$ with measured value $\mathbf{v}_{\mathcal{A}}^{(\mathcal{B})}=-\frac{1}{2}\cdot \left( \mathbf{v}_{(n)}^{(\mathcal{A})} +  \mathbf{v}_{n+1}^{(\mathcal{A})} \right)$. Then, in $\mathcal{B}$ob's view both particles $\circMunit_{\:\mathbf{v}_{i}}$ move antiparallel
\be
   \mathbf{v}_{i}^{(\mathcal{B})}= \mathbf{v}_{i}^{(\mathcal{A})} + \mathbf{v}_{\mathcal{A}}^{(\mathcal{B})} = \left\{ \begin{array}{l}
           \;\;\: \frac{1}{2}\cdot \left( \mathbf{v}_{(n)}^{(\mathcal{A})} -  \mathbf{v}_{n+1}^{(\mathcal{A})} \right)\; , \;\; i=(n) \\
           -\frac{1}{2}\cdot \left( \mathbf{v}_{(n)}^{(\mathcal{A})} -  \mathbf{v}_{n+1}^{(\mathcal{A})} \right) \; , \;\; i=n+1 \;\; .
         \end{array}
 \right. \nn
\ee
$\mathcal{B}$ob absorbs both particles by a calorimeter-mediated inelastic collision analogous to (\ref{Abschnitt -- kin quant Absorptions Wirkung - Reservoirbilanz - refinement E})
\be
   \circMunit_{\:\mathbf{v}_{(n)}} \,,\: \circMunit_{\:\mathbf{v}_{n+1}}
   \;\; \stackrel{w^{-1}}{\Rightarrow} \;\; \circMunit_{\:\mathbf{0}} \,,\: \circMunit_{\:\mathbf{0}} \,,\: \left( \frac{\mathbf{v}_{(n)}^{(\mathcal{A})} -  \mathbf{v}_{n+1}^{(\mathcal{A})}}{2}  \right)^2 \cdot \mathcal{S}_{\mathbf{1}^{(\mathcal{B})} }\big|_{\mathbf{0}}
 \nn
\ee
in return for $\left( \frac{\mathbf{v}_{(n)}^{(\mathcal{A})} -  \mathbf{v}_{n+1}^{(\mathcal{A})}}{2}  \right)^2$ of $\mathcal{B}$ob's energy units $\mathcal{S}_{\mathbf{1}^{(\mathcal{B})} }\big|_{\mathbf{0}}$. By expending additional $\left( \frac{\mathbf{v}_{(n)}^{(\mathcal{A})} + \mathbf{v}_{n+1}^{(\mathcal{A})}}{2}  \right)^2$ energy sources $\mathcal{S}_{\mathbf{1}^{(\mathcal{B})} }\big|_{\mathbf{0}}$ from his reservoir
\be
   \circMunit_{\:\mathbf{0}} \,,\: \circMunit_{\:\mathbf{0}} \,,\: \left( \frac{\mathbf{v}_{(n)} +  \mathbf{v}_{n+1}}{2}  \right)^2 \cdot \mathcal{S}_{\mathbf{1}^{(\mathcal{B})} }\big|_{\mathbf{0}}
   \;\; \stackrel{w'}{\Rightarrow} \;\; \circMunit_{-\frac{\mathbf{v}_{(n)} +  \mathbf{v}_{n+1}}{2}\mathbf{v}_{\mathbf{1}^{(\mathcal{B})}}} \,,\: \circMunit_{\:\frac{\mathbf{v}_{(n)} +  \mathbf{v}_{n+1}}{2}\mathbf{v}_{\mathbf{1}^{(\mathcal{B})}}}
 \nn
\ee
$\mathcal{B}$ob kicks both temporarily resting particles with antiparallel velocity $\mathbf{v}'_{i}=\pm\frac{\mathbf{v}^{(\mathcal{A})}_{(n)} +  \mathbf{v}^{(\mathcal{A})}_{n+1}}{2}\cdot \mathbf{v}_{\mathbf{1}^{(\mathcal{B})}}$ into the direction of $\mathcal{A}$lice (see figure \ref{pic_impuls_vektoriell_induktion}b). For her they have velocity values
\be
   {\mathbf{v}'}_{i}^{(\mathcal{A})}= {\mathbf{v}'}_{i}^{(\mathcal{B})} + \mathbf{v}_{\mathcal{B}}^{(\mathcal{A})} = \left\{ \begin{array}{l}
            \mathbf{v}_{(n)}^{(\mathcal{A})} +  \mathbf{v}_{n+1}^{(\mathcal{A})} \;\; , \;\; i=(n) \\
            0 \;\; , \;\; i=n+1
         \end{array}
 \right. \nn
\ee
with $\mathbf{v}_{\mathcal{B}}^{(\mathcal{A})}=-\mathbf{v}_{\mathcal{A}}^{(\mathcal{B})}=\frac{1}{2}\cdot \left( \mathbf{v}_{(n)}^{(\mathcal{A})} +  \mathbf{v}_{n+1}^{(\mathcal{A})} \right)$ (see figure \ref{pic_impuls_vektoriell_induktion}a).

In $\mathcal{B}$ob's series of calorimeter-mediated inelastic collisions ${w^{-1}}^{(\mathcal{B})}\ast{w'}^{(\mathcal{B})}=:W^{(n)}$ particle $\circMunit_{\:\mathbf{v}_{n+1}}$ transfers all its momentum onto absorber $\circMunit_{\:\mathbf{v}_{(n)}}$
\be
   \circMunit_{\:\mathbf{v}_{(n)}} \,,\: \circMunit_{\:\mathbf{v}_{n+1}}
   \;\; \stackrel{W^{(n)}}{\Rightarrow} \;\; \circMunit_{\:\mathbf{v}'_{(n)}} \,,\: \circMunit_{\:\mathbf{0}} \,,\: - \mathbf{v}_{(n)}^{(\mathcal{A})} \cdot \mathbf{v}_{n+1}^{(\mathcal{A})} \; \cdot \mathcal{S}_{\mathbf{1}^{(\mathcal{B})} }\big|_{\mathbf{v}_{\mathcal{B}}}
 \nn
\ee
at the expense of $\left( \frac{\mathbf{v}_{(n)}^{(\mathcal{A})} -  \mathbf{v}_{n+1}^{(\mathcal{A})}}{2}  \right)^2-\left( \frac{\mathbf{v}_{(n)}^{(\mathcal{A})} +  \mathbf{v}_{n+1}^{(\mathcal{A})}}{2}  \right)^2 = -\mathbf{v}_{(n)}^{(\mathcal{A})} \cdot \mathbf{v}_{n+1}^{(\mathcal{A})}$ congruent energy units $\mathcal{S}_{\mathbf{1}^{(\mathcal{B})} }\big|_{\mathbf{v}_{\mathcal{B}}} \sim_E \mathcal{S}_{\mathbf{1}^{(\mathcal{A})} }\big|_{\mathbf{0}}$ (see Lemma \ref{Lem - kin quant Absorptions Wirkung - Reservoirbilanz - boost E and p units}). At every induction step $n\rightarrow n+1$ element $\circMunit_{\:\mathbf{v}_{n+1}}$ comes to rest while we couple the absorber $\circMunit_{\:\mathbf{v}'_{(n)}}$ with velocity $\mathbf{v}'_{(n)}=\mathbf{v}_{(n)}+\mathbf{v}_{n+1}=:\mathbf{v}_{(n+1)}$ into the next round of the process.

\begin{figure}    
  \begin{center}           
  \includegraphics[height=4cm]{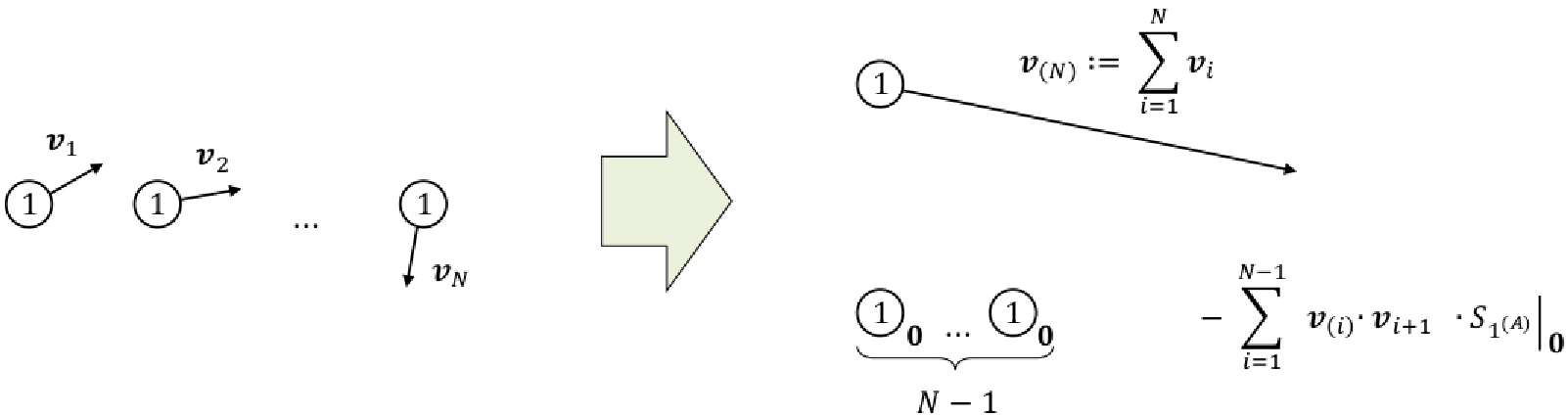}  
  \end{center}
  \vspace{-0cm}
  \caption{\label{pic_impuls_vektoriell_Endeffekt} end effect - initial momentum of
  $\left\{\circMunit _{\:\mathbf{v}_{2}} ,\ldots, \circMunit _{\:\mathbf{v}_{N}}\right\}$ transferred to $\circMunit _{\:\mathbf{v}_{1}}$
    }
  \end{figure}

After completing all steps of the impulse transfer process $W^{(1)}\ast\ldots\ast W^{(N-1)}:$
\be\label{Abschnitt -- Impuls - vollst Transfermodel}
     \circMunit_{\:\mathbf{v}_{1}} \,,\: \left\{ \circMunit_{\:\mathbf{v}_2} , \dots , \circMunit_{\:\mathbf{v}_N}\right\}
   \;\; \Rightarrow \;\; \circMunit_{\:\mathbf{v}_{(N)}} \,,\: \left\{ \circMunit_{\:\mathbf{0}} , \dots , \circMunit_{\:\mathbf{0}}\right\} \,,\: - \sum_{i=1}^{N-1} \mathbf{v}_{(i)} \cdot \mathbf{v}_{i+1} \: \cdot \mathcal{S}_{\mathbf{1}}\big|_{\mathbf{0}}
\ee
all elements $i=2,\ldots,N$ of the system stop while the absorber particle $\circMunit_{\:\mathbf{v}_{(N)}}$ moves on with final velocity $\mathbf{v}_{(N)}=\sum_{i=1}^{N} \mathbf{v}_{i}$ (see figure \ref{pic_impuls_vektoriell_Endeffekt}).

We transfer energy-momentum of system $\left\{\circMunit_{\:\mathbf{v}_1} , \dots , \circMunit_{\:\mathbf{v}_N}\right\}$ onto one absorber $\circMunit_{\:\mathbf{v}_{(N)}}$ and
\bea\label{Abschnitt -- Impuls - vollst Transfermodel - v Rechnerei}
   - \sum_{i=1}^{N-1} \mathbf{v}_{(i)} \cdot \mathbf{v}_{i+1} & \stackrel{(\ref{Abschnitt -- Impuls - vollst Transfermodel})}{=} & - \sum_{i=1}^{N-1} \left( \sum_{k=1}^{i} \mathbf{v}_{k}  \right) \cdot \mathbf{v}_{i+1}  \;\; = \;\; - \sum_{k<i=1}^{N}   \mathbf{v}_{k} \cdot  \mathbf{v}_{i} \nn \\
   & = & - \frac{1}{2} \cdot \sum_{i,k=1}^{N} \mathbf{v}_{i} \cdot  \mathbf{v}_{k} \; + \; \frac{1}{2} \sum_{i=1}^{N} \mathbf{v}_{i}^2  \nn \\
   & = & -\frac{1}{2} \cdot ( \mathbf{v}_{1}+ \ldots + \mathbf{v}_{N} ) \cdot ( \mathbf{v}_{1}+ \ldots + \mathbf{v}_{N} ) \;\; + \;\; \frac{1}{2} \sum_{i=1}^{N} \mathbf{v}_{i}^2
\eea
equivalent energy sources $\mathcal{S}_{\mathbf{1}}\big|_{\mathbf{0}}$. At each step of the calorimeter mediated process (\ref{Abschnitt -- Impuls - vollst Transfermodel}) momentum is conserved by Lemma \ref{Lem - kin quant Absorptions Wirkung - Reservoirbilanz - p conserved}. We measure the total momentum of the system $\mathbf{p}\left[ \circMunit_{\:\mathbf{v}_{1}} , \dots , \circMunit_{\:\mathbf{v}_N} \right] = \mathbf{p}\left[ \circMunit_{\:\mathbf{v}_{(N)}} \right]$ by one absorber particle and the latter in a calorimeter by the number of \emph{equivalent} impulse carriers $\circMunit_{\:\mathbf{v}_{\mathbf{1}}}$ which now all point into the \emph{same direction} of $\mathbf{v}_{(N)}:=\mathbf{v}_{1}+ \ldots + \mathbf{v}_N$. By linearity of the relation between velocity and momentum (\ref{Abschnitt -- kin quant Absorptions Wirkung - kin energy and momentum metrisiert}) for a single particle the total impulse $\mathbf{p}\left[ \circMunit_{\:\mathbf{v}_1}  \dots , \circMunit_{\:\mathbf{v}_N} \right] = \left( m_{\mathbf{1}} \cdot \mathbf{v}_1 + \ldots + m_{\mathbf{1}} \cdot \mathbf{v}_N \right) \cdot \mathbf{p}[\circMunit_{\:\mathbf{v}_1}]$ is the vector sum over all elements $\left\{\circMunit_{\:\mathbf{v}_{\!i}}\right\}_{i=1\ldots N}$.
\qed


\section{Inertial mass}\label{Kap - KM Dynamics - Basic Dynamical Measures - Inertial Mass}

We define the \emph{elemental order} as a special case of impulse comparison (see Definition \ref{Def - vortheor Ordnungsrelastion - inertial mass}): According to Galilei two bodies $\circMa \,, \circMb$ have same resistance against changes of their motion
\be
   \circMa \;\; \sim_{m} \;\;  \circMb   \nn
\ee
if in an inelastic head-on collision test $\circMa_{\:\mathbf{v}} , \circMb_{-\mathbf{v}} \stackrel{w}{\Rightarrow} \circMa\ast\circMb_{\;\mathbf{0}}$ with same initial velocity no one overruns the other \cite{Weyl - Philosophie der Mathematik und Naturwissenschaft}. If the aggregate $\circMa\ast\circMb_{\;\mathbf{v}'}$ moves along $\circMa_{\:\mathbf{v}}$ then the latter is more massive $\circMa >_{m} \circMb$ than object $\circMb$ and vice versa. After two reversible collision $w_H:=w\ast w^{-1}$ a particle pair $\circMa \sim_{m} \circMb$ with same inertia rebounds antiparallel with reversed velocities
\be\label{Abschnitt -- inertial mass - elastische Stosswirkung}
   \circMa_{\:\mathbf{v}} \,,\: \circMb_{-\mathbf{v}} \;\; \stackrel{w_H}{\Rightarrow} \;\; \circMa_{-\mathbf{v}} \,,\: \circMb_{\:\mathbf{v}}  \;\; .
\ee

We \emph{concatenate} the inertia of two objects $\circMa , \circMb$ by bounding ''$\ast$'' them together (by a practically massless sling in figure \ref{pic_BER_composite_collision}b). Then the aggregate $\circMa\ast\circMb$ acts like one single rigid body. We represent the \emph{unit} by the inertia of $\mathcal{A}$lice standard object $\circMunit$.

An equally massive composite of standard elements
\be\label{Abschnitt -- inertial mass - model same inertia}
   \circMunit \ast \dots \ast \circMunit  \;\; \sim_{m} \;\;  \circMO
\ee
is our \emph{physical model} for the inertia of generic object $\circMO$. After the collision test (\ref{Abschnitt -- inertial mass - elastische Stosswirkung}) $\mathcal{A}$lice \emph{counts} the amount of matter in her model.\footnote{The current mass standard is defined by one prototype ''$\mathrm{kg}$'' - the Ur-Kilogram in the French Bureau of Measures and Standards. In recent approaches it is replaced by one Si atom. For \emph{practical feasibility} one redefines this mass standard by means of - a certain number $N$ of - Si atoms: $\mathrm{kg} := N\cdot\mathrm{Si}$. Still each atom represents the inertial behavior of the mass unit in e.g. collisions. In a \emph{practical realization} of the Atom-counting approach one manufactures a single-crystal sphere of silicon atoms. Its radius (uncertainty on roughly a single atomic layer) and the lattice spacing between individual Si atoms (by X-ray spectroscopy) are precisely known. This \emph{reproducible} prototype is among the roundest \emph{man-made} objects in the world \cite{Brumfield replace kilo}.} By the congruence principle she finds
\be
   m\left[ \circMO \right] \;\;\stackrel{(\ref{Abschnitt -- inertial mass - model same inertia})}{=}\;\;
   m\left[ \circMunit \ast \dots \ast \circMunit \right]  \;\;\stackrel{(\mathrm{Congr.})}{=:}\;\;
   m\cdot m\left[ \circMunit \right]   \nn
\ee
''how many times'' more inertia the generic object $\circMO$ has than one reference body $\circMunit$. The number $m:= \sharp \left\{ \circMunit \right\} $ of standard elements $\circMunit$ is a \emph{physical quantity}; and the reference mass $m\left[ \circMunit \right]$ is a \emph{dimension}.\footnote{In reception of Hertz (third picture) for the foundation of mechanics \{\ref{Kap - KM Dynamics short Review - Hertz program}\} Sommerfeld \cite{Sommerfeld-Mechanik} remarks: ''(Hertz) succeeded in substituting all forces, by coupling into the system in question further (external) interactive systems. Thus Hertz could restrict to a force-free (treatment of) systems. He needed to comprehend ... all masses as multiples of, say, atomic unit masses.''
}
The operationalization is observer independent reproducible.

We measure the kinetic energy-momentum of generic object $\circMO_{\:\mathbf{v}}$ by the number of standard energy and impulse carriers from a calorimeter. An equally massive composite of unit elements $\circMunit \ast \dots \ast \circMunit \sim_{m}\circMO$ with same velocity $v_{\mathcal{O}}$ extracts the same (see Lemma \ref{Lem - kin quant Absorptions Wirkung - Reservoirbilanz - absorption proportional materiemenge}). The absorption extract for a composite $\circMunit \ast \dots \ast \circMunit$ is proportional to the number of elements (see Lemma \ref{Lem - kin quant Absorptions Wirkung - Reservoirbilanz - additivitaet}). Thus the kinetic energy-momentum (of object $\circMO_{\:\mathbf{v}}$) is proportional to the amount of matter (in the composite), i.e. to our physical quantity of inertial mass. The relation between the physical quantities of energy, momentum, mass becomes transparent.

\section{Energy}\label{Kap - KM Dynamics - Basic Dynamical Measures - Energy}

Leibniz specifies energy by its effect. We quantify the pre-theoretic comparison (\ref{Abschnitt -- vortheor Ordnungsrelastion - energetisches verhalten}).

\subsection{Kinetic energy}\label{Kap - KM Dynamics - Basic Dynamical Measures - kinetic Energy}

For the development of the concept of energy we follow the review of Schlaudt \cite{Schlaudt}. Mach characterizes the everyday pre-scientific notion ''driving force'': Soon after Galilei one did notice that behind the velocity of an object there is a certain capability to work. Something which allows to overcome force. How to measure this ''something'' was the subject of the ''vis viva'' dispute \cite{Mach - Mechanik in ihrer Entwicklung}. It was initially a vague, pre-theoretic notion. It has the peculiar feature - Schlaudt explains - that it cannot be quantified directly but solely by means of its effect. This is not a mathematical problem but a practical, whose solution entails the mathematical expression for force.
\begin{de}
Kinetic energy $E_{\mathrm{kin}}\left[\circMO_{\:\mathbf{v}}\right]$ is the capability to work associated with decelerating a moving object $\circMO_{\:\mathbf{v}}$.
\end{de}

According to Leibniz \emph{equipollence} principle ''il faut avoir recours \`{a} l'equipollence de la cause et de l'effect''. For quantification Leibniz further employs the principle of \emph{congruence}. To measure the cause $\mathcal{S}$ (Ursache) by its effect requires: (i) providing a precise standard action which successively consumes the source $\mathcal{S}$, (ii) the cumulative effect of formal repetitions reproduces the effect of $\mathcal{S}$ and (iii) guarantee that all copies of the standard action are congruent with one another \cite{Schlaudt}. In a practical test ''$\sim_{E}$'' they generate same effect \{\ref{Kap - KM Dynamics - Physical Measurement - Pre-theoretical Ordering Relation}\}.


Leibniz defines an elementary \emph{ordering criterion} (Definition \ref{Def - vortheor Ordnungsrelastion - energie}): Body $\circMa_{\:\mathbf{v}_{a}}$ has the same capability to work as body $\circMb_{\:\mathbf{v}_{b}}$
\be
   \circMa_{\:\mathbf{v}_{a}} \;\; \sim_{E} \;\;  \circMb_{\:\mathbf{v}_{b}}   \nn
\ee
if they generate the same effect against the same obstacle. We compare the absorption effect in our calorimeter $\left\{\circMunit_{\:\mathbf{v}=0}\right\}$ (until all motion stops). If particle $\circMa_{\:\mathbf{v}_{a}}$ can compress more standard springs than particle $\circMb_{\:\mathbf{v}_{b}}$ it has more kinetic energy $\circMa_{\:\mathbf{v}_{a}}  >_{E} \circMb_{\:\mathbf{v}_{b}}$ and vice versa.

We \emph{concatenate} the kinetic energy of particles $\circMa_{\:\mathbf{v}_{a}}$, $\circMb_{\:\mathbf{v}_{b}}$ by coupling them against the same external absorber system $\circMI_{\:\mathbf{v}}$ (collective index $\mathrm{I}:=1,\ldots,N$ for its elements)
\be
   \circMa_{\:\mathbf{v}_{a}} \,,\: \circMb_{\:\mathbf{v}_{b}} \,,\: \circMI_{\:\mathbf{v}_{\mathrm{I}}} \;\; \Rightarrow \;\; \circMa_{\:\mathbf{0}} \,,\: \circMb_{\:\mathbf{0}} \,,\: \circMI_{\:\mathbf{v}_{\mathrm{I}}'} \nn
\ee
such that the original particles $\circMa_{\:\mathbf{0}}$, $\circMb_{\:\mathbf{0}}$ stop. With Leibniz and D'Alembert we represent the \emph{unit} energy by the (potential of a) compressed standard springs $\mathcal{S}_{\mathbf{1} }\big|_{\mathbf{0}}$ by a fixed length in reference process $w_{\mathbf{1}}$ \{\ref{Kap - KM Dynamics - Basic Dynamical Measures - Quantification - Dynamical Unit}\}.

The absorption extract is also a \emph{physical model} for the kinetic energy of object $\circMO_{\:\mathbf{v}}$ (see Theorem \ref{Theorem - kin quant absorption action for free particle}). We measure it by the equally potent output of the calorimeter reservoir
\be\label{Abschnitt -- Energie - kinetic Energy}
   E_{\mathrm{kin}} \left[ \circMO_{\:\mathbf{v}} \right]
   \;\; \stackrel{(\mathrm{Equip.})}{=} \;\;
   E \left[ \mathrm{RB} \left[ \circMO_{\:\mathbf{v}} \Rightarrow \circMO_{\:\mathbf{v}=0}  \right] \right]
   \;\; \stackrel{(\mathrm{Congr.})}{=} \;\; \sharp \left\{ \mathcal{S}_{\mathbf{1}}\big|_{\mathbf{0}} \right\} \cdot E \left[ \mathcal{S}_{\mathbf{1}}\big|_{\mathbf{0}} \right]
\ee
and the latter according to the congruence principle by the number of extractable energy units $\mathcal{S}_{\mathbf{1}}\big|_{\mathbf{0}}$ and the reference energy $E\left[ \mathcal{S}_{\mathbf{1} }\big|_{\mathbf{0}} \right]$ (in an observer independent reproducible way).

\subsection{Potential energy}\label{Kap - KM Dynamics - Basic Dynamical Measures - potential Energy}

We measure the effect of an intrinsic interaction \{\ref{Kap - KM Dynamics - Physical Measurement - Interaction of Motion}\} in our calorimeter. At every moment we can capture individual elements $\circMi$ of a system $\circMunit\cup\ldots\cup\circMn$ and measure the kinetic energy-momentum $(E_{\mathrm{kin}}, \mathbf{p}) \left[ \circMi \,\right]$. This way we can analyze separate energy sources $\mathcal{S}_E$ or the implicit binding ''$\cup$'' in e.g. an electromagnetic or gravitational bound system (Systemdasein).
\begin{de}
The potential energy of a configuration transition in system $\circMunit\cup\ldots\cup\circMn$ generally changes the state of motion of all elements $\circMi\,$ with associated kinetic energy gain.
\end{de}

In e.g. an inelastic collision (\ref{Abschnitt -- Impuls - inelastic collision direct}) in two-partite system $\circMa\cup\circMb\,$ the incident particles
\be
   \circMa_{\:\mathbf{v}_{a}} \,,\: \circMb_{\:\mathbf{v}_{b}} \;\; \stackrel{w_{(d)}}{\Rightarrow} \;\; \circMa\ast\circMb_{\;\mathbf{v}=0}  \nn
\ee
form a bound aggregate $\circMa\ast\circMb_{\;\mathbf{0}}$ at rest (''$\ast$'' symbolizes the inner bound). In the reverse process we measure the separation energy; the potential energy $E^{\ast}:=V\left[ \circMa\ast\circMn_{\:\mathbf{0}} \Rightarrow \circMa_{\:\mathbf{0}} , \circMb_{\:\mathbf{0}}  \right]$ from the \emph{transition} from a bound to an unbound configuration, by the kinetic energy of the liberated elements $\circMa_{\:\mathbf{v}_{a}} , \circMb_{\:\mathbf{v}_{b}}\,$.

\begin{figure}    
  \begin{center}           
  \includegraphics[height=8.2cm]{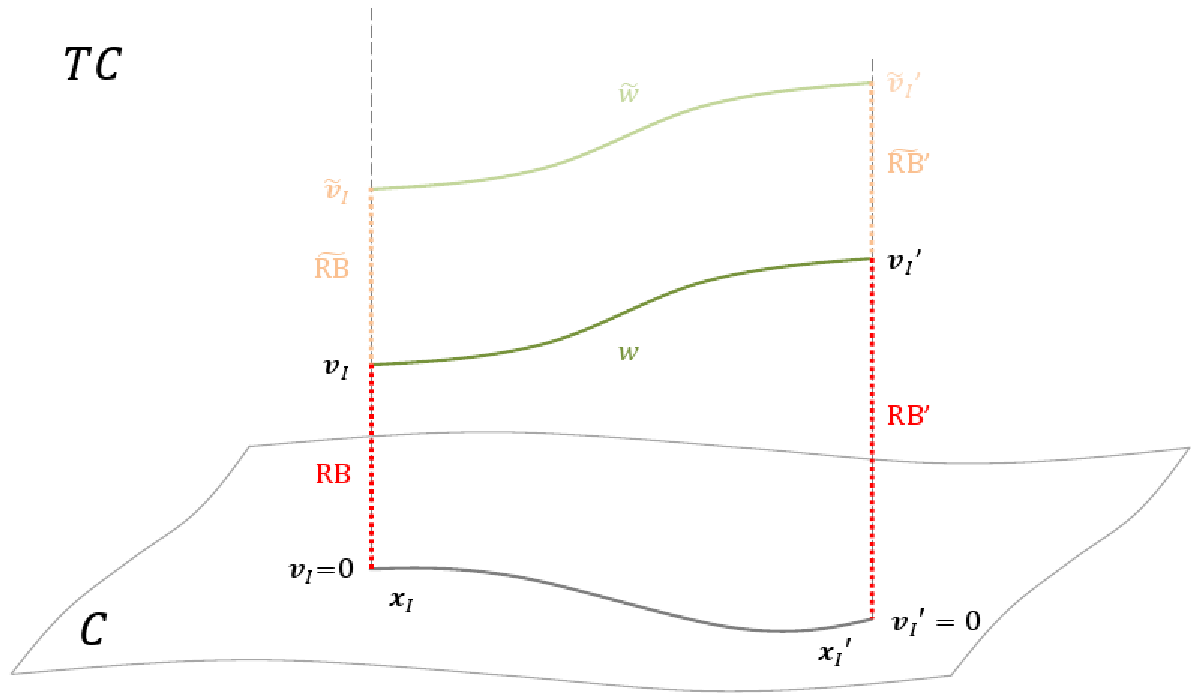}  
  \end{center}
  \vspace{-0cm}
  \caption{\label{pic_Wirkungs-Steuerung_selber-Pfad-Aquipollenz} potential energy (and momentum) extract from configuration transitions
    }
  \end{figure}

Consider a gradual configuration transition $\mathbf{x}_I\Rightarrow\mathbf{x}'_I$ in multi-partite systems $\circMunit\cup\ldots\cup\circMn$
\be
   \circMunit\cup\ldots\cup\circMn_{\;\mathbf{x}_I,\mathbf{v}_I}  \;\; \stackrel{w}{\Rightarrow} \;\; \circMunit\cup\ldots\cup\circMn_{\;\mathbf{x}'_I,\mathbf{v}'_I} \nn
\ee
with a kinetic effect $\mathbf{v}_I\Rightarrow\mathbf{v}'_I$ on the individual elements $I=1,\dots,n$. For standardization we mediate transitions between resting configurations $\mathbf{x}_I\Rightarrow\mathbf{x}'_I$ of the system
\be\label{Abschnitt -- Energie - Reservoirbilanz - reversible action in closed system}
   -\mathrm{RB} \big|_{\mathbf{x}_I} \ast w \ast \mathrm{RB}' \big|_{\mathbf{x}'_I}:
   \;\;\; \circMI_{\:\mathbf{x}_I,\mathbf{v}_I=\mathbf{0}}  \;\; \Rightarrow \;\; \circMI_{\:\mathbf{x}'_I,\mathbf{v}'_I=\mathbf{0}}
\ee
where before and after all elements rest. The calorimeter intervention $-\mathrm{RB} \big|_{\mathbf{x}_I}$ at the initial configuration $\mathbf{x}_I$ prepares the initial state of motion; then the intrinsic process $w$ evolves free (isolated from exterior steering interventions \{\ref{Kap - KM Dynamics - Potential of Mechanical System - Steering Action}\}) to the final configuration $\mathbf{x}'_I$, where we extract the kinetic effect by another calorimeter measurement $\mathrm{RB}' \big|_{\mathbf{x}'_I}$. We assume separate (and practically instantaneous \{\ref{Kap - KM Dynamics - Potential of Mechanical System - Steering Action}\}) interventions $\mathrm{RB}\left[ \circMi_{\:\mathbf{v}_i} \Rightarrow \circMi_{\:\mathbf{0}} \right]\big|_{\mathbf{x}_i}$ with outputs:
\begin{itemize}
  \item   from preparing the initial motion $\mathbf{0}\Rightarrow\mathbf{v}_I:=\left\{\mathbf{v}_1,\ldots,\mathbf{v}_n\right\}$ of all elements $\circMunit_{\:\mathbf{v}_1},\ldots,\circMn_{\:\mathbf{v}_n}$ in the starting configuration $\mathbf{x}_I:=\left\{\mathbf{x}_1,\ldots,\mathbf{x}_n\right\}$ of the system and
  \item   from absorbing the motion $\mathbf{v}'_I \Rightarrow \mathbf{0}$ of all elements in the final configuration $\mathbf{x}'_I$.
\end{itemize}
We illustrate the kinetic energy-momentum supply (extraction) to the system by rising (lowering) the momentary state of motion $\mathbf{v}_I$ along vertical fibres $T_{\mathbf{x}_I}\mathcal{C}$ in velocity space (see lower square in figure \ref{pic_Wirkungs-Steuerung_selber-Pfad-Aquipollenz}).

For the total extract from a configuration transition
\bea\label{Abschnitt -- Energie - Reservoirbilanz - configuration transition in closed system-simplified}
   \mathrm{RB}\left[ \circMI_{\:\mathbf{x}_I, \mathbf{0}}  \Rightarrow \circMI_{\;\mathbf{x}'_I, \mathbf{0}} \right]
   & \stackrel{(\ref{Abschnitt -- Energie - Reservoirbilanz - reversible action in closed system})}{=} & \mathrm{RB}\left[ \circMI_{\:\mathbf{0}} \Rightarrow  \circMI_{\:\mathbf{v}_I} \right] \big|_{\mathbf{x}_I} \;+\; \mathrm{RB}\left[ \circMI_{\:\mathbf{v}'_I} \Rightarrow  \circMI_{\:\mathbf{0}} \right] \big|_{\mathbf{x}'_I} \nn \\
   & \stackrel{(\ref{Abschnitt -- kin quant Absorptions Wirkung - Reservoirbilanz - additivitaet})}{=} & - \;\; \mathrm{RB}\left[ \circMI_{\:\mathbf{v}_I}  \Rightarrow \circMI_{\;\mathbf{v}'_I} \right]
\eea
we \emph{assume} (by superposition principle and compatibility requirements \{\ref{Kap - KM Dynamics - Potential of Mechanical System - Potential Field}\}) that intrinsic calorimeter measurements are independent from the bound state $\mathbf{x}_I$, $\mathbf{x}'_I$ of the individual elements and the same like for isolated elements. Also let the number of extractable measurement units $\mathcal{S}_{\mathbf{1}}\big|_{\mathbf{0}}$, $\circMunit_{\:\mathbf{v}_{\mathbf{1}}}$ from free action $\mathbf{x}_I,\mathbf{v}_I \stackrel{w}{\Rightarrow} \mathbf{x}'_I,\mathbf{v}'_I$ resp. $\mathbf{x}'_I,\tilde{\mathbf{v}}'_I \stackrel{\tilde{w}}{\Rightarrow} \mathbf{x}_I,\tilde{\mathbf{v}}_I$ along any two ways between fixed initial $\mathbf{x}_I$ and final configuration $\mathbf{x}'_I$ be the same. By suitable calorimeter interventions $\mathrm{RB}\big|_{\mathbf{x}_I}$ resp. $\mathrm{RB}'\big|_{\mathbf{x}'_I}$ we can steer a circular process (upper square in figure \ref{pic_Wirkungs-Steuerung_selber-Pfad-Aquipollenz})
\be
   \left(\tilde{\mathrm{RB}}-\mathrm{RB}\right) \ast w \ast \left(-\tilde{\mathrm{RB}}' + \mathrm{RB}' \right) \ast \tilde{w} \;\; .  \nn
\ee
In a closed \emph{conservative} system the total calorimeter extract $\tilde{\mathrm{RB}}-\mathrm{RB} \stackrel{!}{=} - \left(\tilde{\mathrm{RB}}' - \mathrm{RB}' \right)$ must vanish by the impossibility of a perpetuum mobile. Then, the calorimeter absorption extract from a configuration transition (\ref{Abschnitt -- Energie - Reservoirbilanz - configuration transition in closed system-simplified}) becomes independent from the initial condition $\mathbf{v}_I$ resp. $\tilde{\mathbf{v}}'_I$ under which the free process $w$ evolves.

By the equipollence principle we measure the potential energy from a configuration transition $\mathbf{x}_I\Rightarrow\mathbf{x}'_I$ by its kinetic effect $\mathbf{v}_I\Rightarrow\mathbf{v}'_I$ -- and the latter in our calorimeter model
\bea\label{Abschnitt -- Energie - potential Energy}
   V_{\mathrm{pot}} \left[ \circMI_{\:\mathbf{x}_I}  \Rightarrow \circMI_{\;\mathbf{x}'_I} \right]
   & \stackrel{(\mathrm{Equip.})}{:=} & E \left[ \mathrm{RB}\left[ \circMI_{\:\mathbf{x}_I, \mathbf{0}}  \Rightarrow \circMI_{\;\mathbf{x}'_I, \mathbf{0}} \right] \right] \nn \\
   & \stackrel{(\ref{Abschnitt -- Energie - Reservoirbilanz - configuration transition in closed system-simplified})}{=} & -\; E \left[ \mathrm{RB}\left[ \circMI_{\:\mathbf{v}_I}  \Rightarrow \circMI_{\;\mathbf{v}'_I} \right] \right] \;\; = \;\;
   - \; E_{\mathrm{kin}} \left[ \circMI_{\:\mathbf{v}_I}  \Rightarrow \circMI_{\;\mathbf{v}'_I} \right] \; . \;\;\;\;\;
\eea
The potential energy transforms into kinetic energy $E_{\mathrm{kin}}\left[ \circMI_{\:\mathbf{v}_I} \right] =  E_{\mathrm{kin}}\left[ \circMunit_{\:\mathbf{v}_1} \right] + \ldots + E_{\mathrm{kin}}\left[ \circMn_{\:\mathbf{v}_n} \right]$ of all elements. By this measurement principle the total energy of a closed system is \emph{conserved}
\be
   V_{\mathrm{pot}} \left[ \circMI_{\:\mathbf{x}_I} \right] \;+\; E_{\mathrm{kin}}\left[ \circMI_{\;\mathbf{v}_I} \right] \; \stackrel{(\ref{Abschnitt -- Energie - potential Energy})(\ref{Abschnitt -- Energie - kinetic Energy})}{=} \; V_{\mathrm{pot}} \left[ \circMI_{\;\mathbf{x}'_I} \right] \;+\; E_{\mathrm{kin}}\left[ \circMI_{\;\mathbf{v}'_I} \right] \nn
   \;\; .\footnote{We define potential energy of a \emph{configuration} $V_{\mathrm{pot}} \left[ \circMI_{\:\mathbf{x}_I} \right] := V_{\mathrm{pot}} \left[ \circMI_{\:\mathbf{x}_I}  \Rightarrow \circMI_{\;\mathrm{sep}} \right]$ by the extractable energy from its transition into a fixed reference configuration, e.g. the separated state $\circMI_{\;\mathrm{sep}}$ where all elements are liberated from the inner binding (\ref{Abschnitt -- Potential of Mechanical System - potential field - separation action}).}
\ee

\subsection{Quantification scheme}\label{Kap - KM Dynamics - Basic Dynamical Measures - Energy - quantification scheme}

By the equipollence principle the cause of an action has the same energy as its effect. We transform the liberated energy from a generic interaction $w$ into kinetic energy of the elements and then into standardized calorimeter portions. From well-defined (symmetry, relativity principle) building blocks $w_{\mathbf{1}}$ we construct
\be\label{Abschnitt -- Energie - models for quantification scheme}
   w_{\mathbf{1}}  \;\;\;\;\;\;\;\;\;\hookrightarrow\;\;\;\;\;\;\;\;\;  W_{\mathrm{cal}}^{(i)}  \;\;\;\;\;\;\;\;\;\hookrightarrow\;\;\;\;\;\;\;\;\;  -W_{\mathrm{cal}}^{(I)} \ast w \ast W_{\mathrm{cal}}^{(I)}
\ee
a physical model for absorbing the kinetic energy from individual particles $\circMi_{\:\mathbf{v}_i}$ and for extracting the potential energy from a system $\circMI_{\:\mathbf{x}_I}$.

All measurements refer to an elementary standard process $w_{\mathbf{1}}$ of irrelevant inner structure. The compression of a standard spring $\mathcal{S}_{\mathbf{1}}\big|_{\mathbf{0}}$ by a standard length provides a reproducible kinetic effect (kick unit bodies $\circMunit$ into unit velocity $\mathbf{v}_{\mathbf{1}}$). We define the unit energy $V_{\mathrm{pot}} \left[ w_{\mathbf{1}} \right]$ by the potential energy of one standard spring $\mathcal{S}_{\mathbf{1}}\big|_{\mathbf{0}}$ in reference process $w_{\mathbf{1}}$. We measure the kinetic energy of a particle $E_{\mathrm{kin}} \left[ \circMi_{\:\mathbf{v}_{i}} \Rightarrow \circMi_{\:\mathbf{0}}  \right]$ with the absorption model $W_{\mathrm{cal}}^{(i)}$. For stoping its motion we extract a number $\sharp\left\{ \mathcal{S}_{\mathbf{1}}\big|_{\mathbf{0}} \right\}$ of standard springs from a calorimeter reservoir $\{\circMunit_{\:\mathbf{0}}\}$. By the equipollence principle the absorption extract $\mathrm{RB} \left[ \circMi_{\:\mathbf{v}_{i}} \Rightarrow \circMi_{\:\mathbf{0}}  \right] \sim_{E} \circMi_{\:\mathbf{v}_{i}}$ has the same energy (\ref{Abschnitt -- Energie - kinetic Energy}). By repeated measurements for all elements $\circMI$ of the system we determine $-W_{\mathrm{cal}}^{(I)} \ast w \ast W_{\mathrm{cal}}^{(I)}$ the kinetic effect of a generic process $w$. By the equipollence principle the potential energy from a configuration transition $ V_{\mathrm{pot}}\left[ \circMI_{\:\mathbf{x}_I}  \Rightarrow \circMI_{\;\mathbf{x}'_I} \right] =  -  E_{\mathrm{kin}}\left[ \circMI_{\:\mathbf{v}_I}  \Rightarrow \circMI_{\;\mathbf{v}'_I} \right]$ transforms into kinetic energy of all elements (\ref{Abschnitt -- Energie - potential Energy}). Ultimately we \emph{quantify} the kinetic and potential energy by the number $\sharp\left\{ \mathcal{S}_{\mathbf{1}}\big|_{\mathbf{0}} \right\}$ of \emph{standard} energy sources (from one $W_{\mathrm{cal}}^{(i)}\,$ or multiple calorimeters $W_{\mathrm{cal}}^{(I)}\,$) and their reference energy $E\left[ \mathcal{S}_{\mathbf{1}}\big|_{\mathbf{0}} \right]$.
\begin{rem}
By means of our physical models (\ref{Abschnitt -- Energie - models for quantification scheme}) we make the transition
\begin{itemize}
  \item from \underline{units} of unquantified potential energy $V_{\mathrm{pot}} \left[ w_{\mathbf{1}} \right]$
  \item over the quantification of kinetic energy $E_{\mathrm{kin}} \left[ \circMi_{\:\mathbf{v}_{i}} \right]$
  \item to the quantification of potential energy $V_{\mathrm{pot}} \left[ \circMI_{\:\mathbf{x}_I} \right]$
\end{itemize}
\be
   \underbrace{V_{\mathrm{pot}} \left[ w_{\mathbf{1}} \right]}_{=:\;E_{\mathbf{1}}}
   \;\;\;\;\;\;\stackrel{(\ref{Abschnitt -- Energie - kinetic Energy})}{\hookrightarrow}\;\;\;\;\;\;  E_{\mathrm{kin}} \left[ \circMi_{\:\mathbf{v}_{i}} \Rightarrow \circMi_{\:\mathbf{0}}  \right]
   \;\;\;\;\;\;\stackrel{(\ref{Abschnitt -- Energie - potential Energy})}{\hookrightarrow}\;\;\;\;\;\;  V_{\mathrm{pot}} \left[ \circMI_{\:\mathbf{x}_I}  \Rightarrow \circMI_{\;\mathbf{x}'_I} \right]  \;\; .  \nn
\ee
\end{rem}
The quantification procedure is based on the equipollence of cause (kinetic or potential) and its \emph{standardized} effect in our calorimeter model (see Remark \ref{Rem - SRT Kin - inseparable unit}).
\\

Equipollence and conservation of energy are equivalent principles. The conservation of energy is very far from the status of an empirical law - as Schlaudt explains \cite{Schlaudt} - much more it is the basis for the quantification of ''vis viva'' $E_{\mathrm{kin}} \left[ \circMa_{\:\mathbf{v}_{a}} \right]$, the kinetic energy of individual particles and the \emph{condition} for measurements of potential energy $V_{\mathrm{pot}} \left[ \circMI_{\:\mathbf{x}_I} \right]$ in mechanical systems. The equipollence principle as the principle of equivalence of cause and effect resp. the conservation of energy resp. (in the practical form) the impossibility of a perpetuum mobile have their place not in physics but in a measurement theory.\footnote{In examination of Leibniz methodological principles Hecht \cite{Hecht - Ruben-Festschrift} remarks: ''Exclusion of a perpetuum mobile becomes for Leibniz the initial spark for introducing a method into natural science'' by which one can exactly determine (kinetic) energy. ''By an ideal machine (thought experiments with circular processes) one defines a measure... which facilitates quantitative comparison of motion. ... That such measure exists at all, can not be proven geometrically or physically; it requires a metaphysical explanation, which Leibniz expresses as (equipollence) principle, namely that the total cause must always equal the total effect.''} Lorenzen speaks of a \emph{measurement-theoretical a priori} \cite{Schlaudt}. Schlaudt explains: once one has accepted the conservation of energy in the sense of a measurement-theoretical a priori as the basis for determining the magnitude (quantification) of energy and energy seems to be lost - then one simply did not consider a \emph{closed system}.


\chapter{Physical analysis}\label{Kap - Analytical mechanics}

\begin{figure}    
  \begin{center}           
  \includegraphics[height=9.3cm]{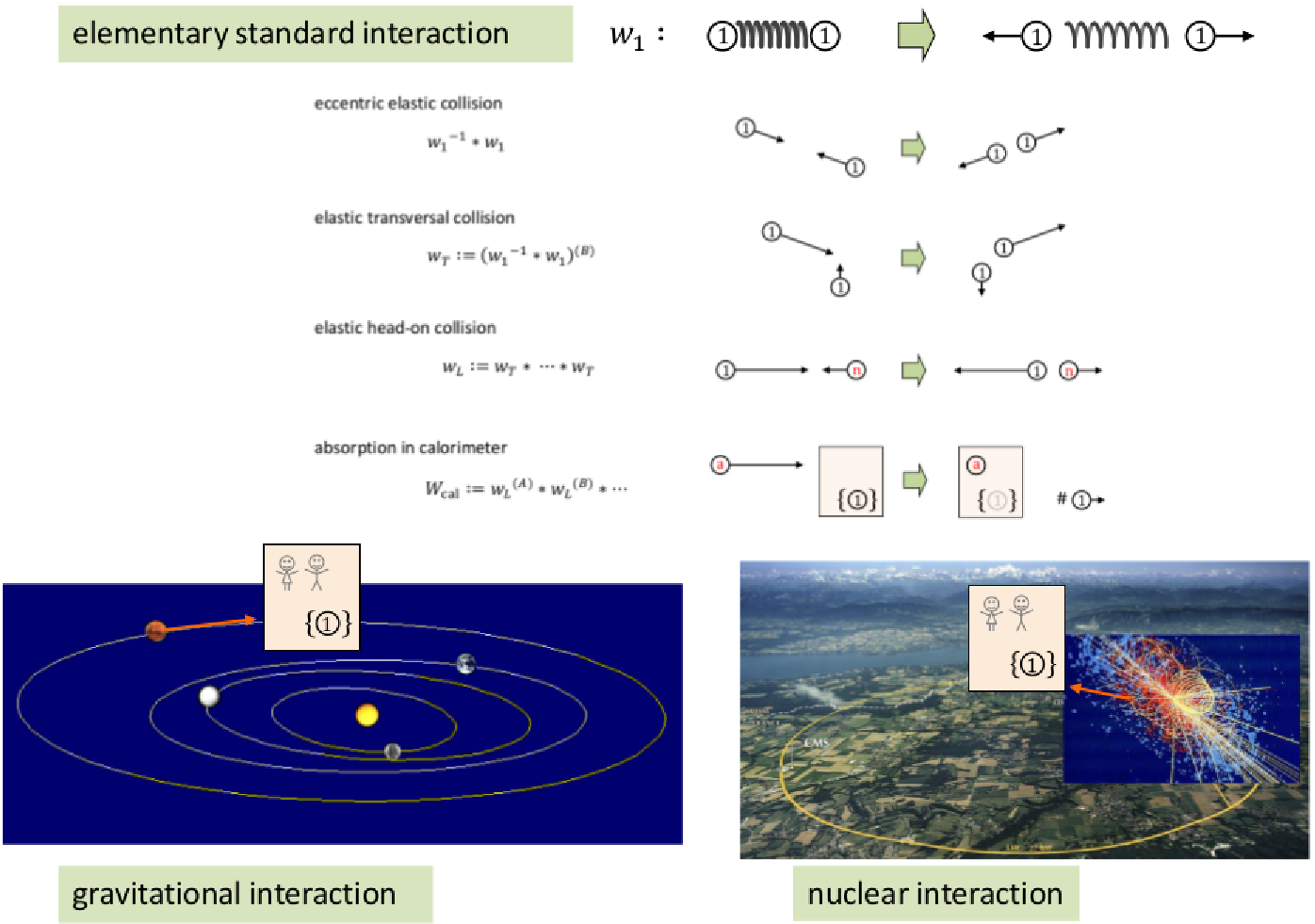}  
  \end{center}
  \vspace{-0cm}
  \caption{\label{pic_measurement_and_steering_instrument} calorimeter as measurement and steering instrument
    }
  \end{figure}

We will utilize the measurement instrument for analyzing generic e.g. gravitational or nuclear processes (see figure \ref{pic_measurement_and_steering_instrument}).
In Gedanken we can couple our calorimeter into the solar system and measure the energy and momentum of Mars along its orbit around the Sun. When in an elementary particle collision new particle generations develop, we can capture a jet with the calorimeter and measure energy and momentum of separate decay products.

In figure \ref{pic_measurement_and_steering_instrument} we link the elementary standard process $w_{\mathbf{1}}$, a gravitational $w_{\mathrm{grav}}$ and a nuclear process $w_{\mathrm{QM}}$. One can not tell which is basic and which more complex; even the internal structure of a spring is unknown. We presuppose solely the initial and final state. We pick the compression of a standard spring as elementary building block for our calorimeter because they are congruent and reproducible. Then we can measure the other processes $w_{\mathrm{grav}}$ and $w_{\mathrm{QM}}$ with reference process $w_{\mathbf{1}}$. That requires \emph{tangible} operations.

We build our calorimeter model from solely inelastic collisions $w_{\mathbf{1}}$ between standard particles $\circMunit$ (construction outlined in figure \ref{pic_Zusammensetzung_Kalorimeter}) which behave by symmetry and relativity principle in a well-defined way. We design the machinery to realize the practical norm for a basic measurement device: allow to count congruent reference units (activated standard springs, impulse carriers, standard bodies). We quantify the basic observables energy, momentum and inertial mass. From the layout of the standard actions $w_{\mathbf{1}}$ in the model $W_{\mathrm{cal}}$ we obtain a \emph{man-made} derivation for the fundamental equations between these quantities.

\section{Potential of mechanical system}\label{Kap - KM Dynamics - Potential of Mechanical System}

In a physical system $G_1\cup \ldots \cup G_N$ one can never get rid of the interaction between its elements. Though physicists can do the contrary; they can include additional actions into the system. One can temporarily couple the calorimeter into a system to stop individual elements and measure the energy and momentum $(E,\mathbf{p})\left[\circMa_{\:\mathbf{v}_a}\right]$. The stoping and boosting facilitates a steering action. We analyze the system in the presence of both actions. Every absorption process changes the initial conditions, prepare the state of motion of individual elements and \emph{steer} an undisturbed interaction: By controlled linkage of external steering interventions $\mathrm{RB}^{(i)}$ and consecutive segments of free intrinsic processes $w_i$ (without steering) physicists can drive the interacting system into any configuration (see figure \ref{pic_Wirkungs-Steuerung_diachrone-Verknuepfung}). We will steer the evolution of an intrinsic process $w$ through standardized configurations \{\ref{Kap - KM Dynamics - Potential of Mechanical System - Steering Action}\} and measure the potential energy and momentum gain \{\ref{Kap - KM Dynamics - Potential of Mechanical System - Potential Field}\}.

\subsection{Steering action}\label{Kap - KM Dynamics - Potential of Mechanical System - Steering Action}

For example consider the exploration of a gravitational interaction. We illustrate the practical interplay between brief steering actions $\mathrm{RB}^{(i)}$ and consecutive segments of undisturbed gravitational processes $w_i$ in recent GRAIL mission: NASA manufactured two satellites on Earth and launched them towards the Moon. In this physical system we can never get rid of the interaction e.g. by turning off gravity, making a virtual displacement $\delta \mathbf{x}_I$ of both satellites to the Moon and turning on the interaction in the final configuration. Every change in the relative configuration $\mathbf{x}_I \Rightarrow \mathbf{x}'_I$ of Earth, both satellites, Moon etc. happens under the mutual interaction of all elements.

For the physical examination of an interactive system we intervene by a series of steering actions. At the start of a satellite mission we - temporarily - \emph{couple} huge booster rockets against elements of the system. Two satellites are launched into escape velocity from Earth. Once most fuel is burnt those boosting rockets are \emph{decoupled} from our satellites and drop back in the atmosphere. The launching phase $\mathrm{RB}^{(1)}$ takes only a brief moment $\Delta t_{\mathrm{launch}} \ll \Delta T_{E\rightarrow M}$ compared to the duration of the following gravitational process $w_1$. We are still in practically the same configuration of the system (thin atmosphere). The first boost by rocket propulsion $\mathbf{x}^{(1)}_I , \mathbf{v}_I = 0  \stackrel{\mathrm{RB}^{(1)}}{\Rightarrow} \mathbf{x}^{(1)}_I , \mathbf{v}^{(1)}_I$ prepares suitable initial conditions (velocity $\mathbf{v}^{(1)}_I$ of the satellites). Instead of remaining bound to Earth both satellites propagate during a long stretch of gravitational interaction $\mathbf{x}^{(1)}_I , \mathbf{v}^{(1)}_I  \stackrel{w_1}{\Rightarrow} \mathbf{x}^{(2)}_I , {\mathbf{v}'_I}^{(1)}$ from Earth to Moon. We can practically \emph{separate} steering action $\mathrm{RB}^{(1)}$ from consecutive gravitational process $w_1$: During the short moment when rockets catapult our satellites into motion the effect of gravity is \emph{comparably negligible}. Once the steering device is decoupled the effect of gravitational interaction accumulates without further external interventions.

By a consecutive series of temporary boosts of steering rockets $\mathrm{RB}^{(i)}$ followed by the next segment of a purely gravitational process $w_i$ the physicist can navigate to any configuration of both satellites around the Moon. Without restricting generality we assume that every steering intervention takes a negligible time compared to the next segment of gravitational free fall. Each steering kick $\mathrm{RB}^{(i)}$ prepares the initial conditions (state of motion of satellites) such that the consecutive gravitational action $w_i$ evolves $\mathbf{x}^{(i)}_I \Rightarrow \mathbf{x}^{(i+1)}_I $ in a controlled way: NASA engineers set up both satellites in a standard formation around Moon. By analyzing the tidal effects on their orbits the GRAIL mission measures the gravitational potential and eventually maps out the distribution of gravitating matter on the Moon.

\begin{figure}    
  \begin{center}           
  \includegraphics[height=9.3cm]{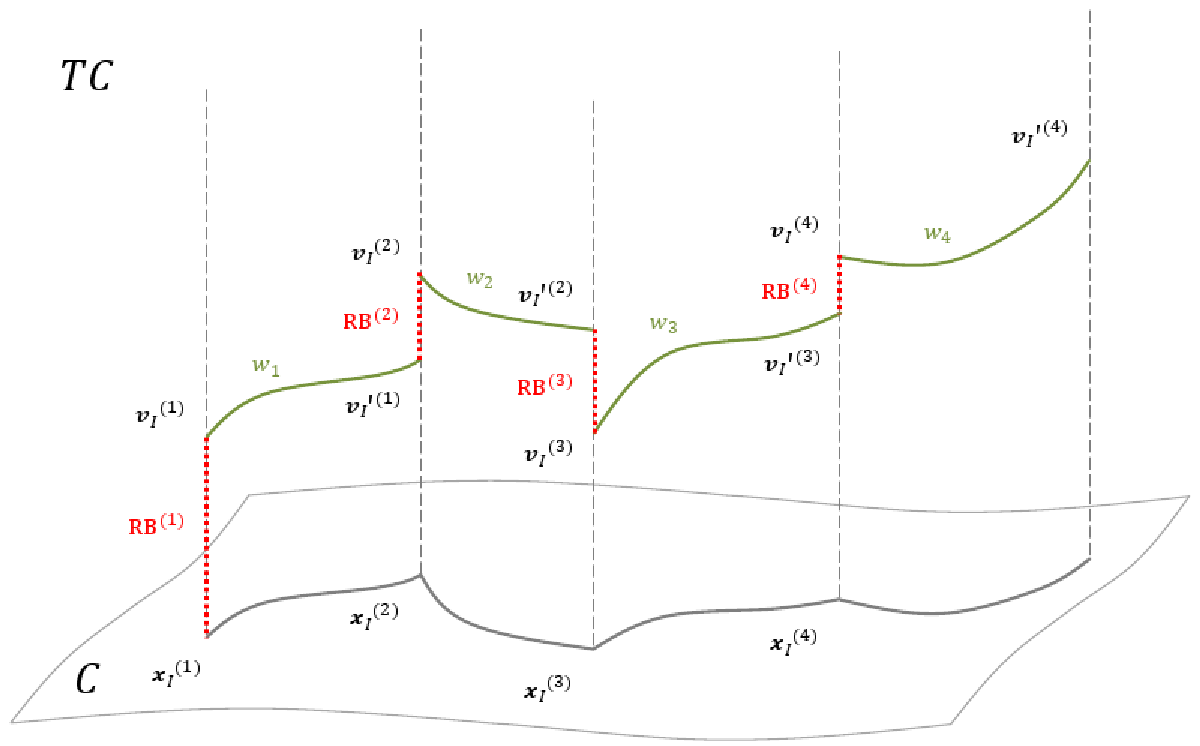}  
  \end{center}
  \vspace{-0cm}
  \caption{\label{pic_Wirkungs-Steuerung_diachrone-Verknuepfung} consecutive linkage of actions $\mathrm{RB}^{(i)} \ast w_i \ast \mathrm{RB}^{(i+1)} \ast \ldots$
    }
  \end{figure}

\begin{figure}    
  \begin{center}           
  \includegraphics[height=9cm]{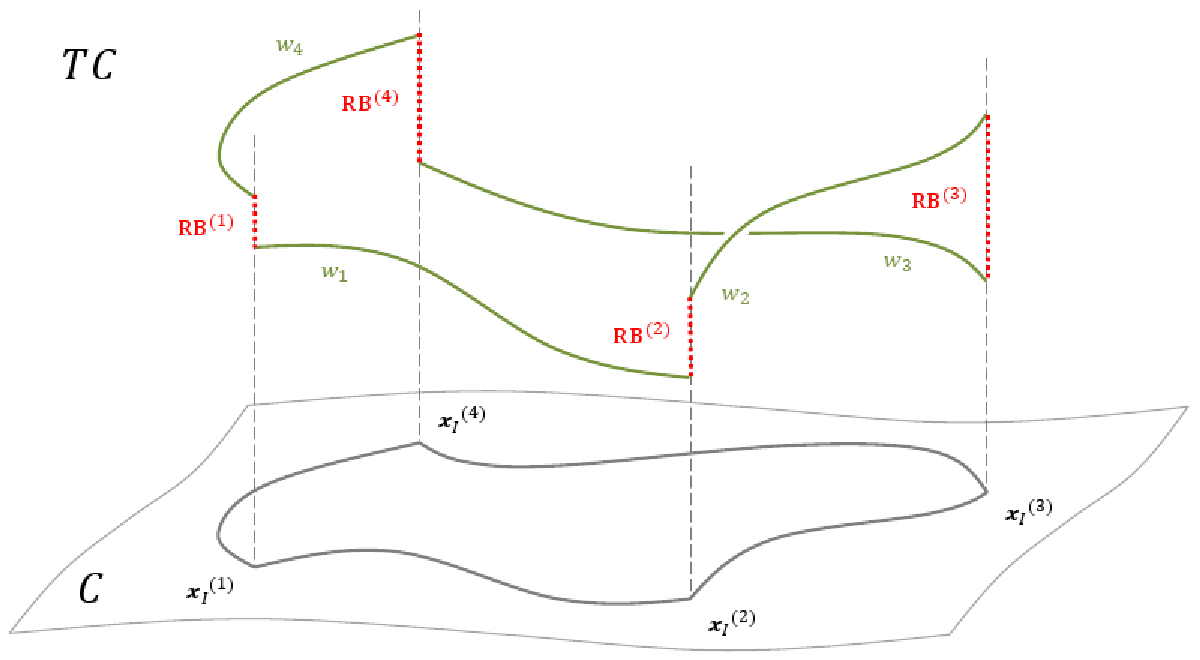}  
  \end{center}
  \vspace{-0cm}
  \caption{\label{pic_Wirkungs-Steuerung_Kreisprozess} steering a circular process
    }
  \end{figure}

Physicists \emph{control the process} of an interaction by repeated steering interventions $\mathrm{RB}^{(i)}$. With our (reversible) calorimeter model $W_{\mathrm{cal}}$ we can extract kinetic effects from gravitational interactions \{\ref{Kap - KM Dynamics - Basic Dynamical Measures - potential Energy}\}. But \emph{engine}ers can also couple additional energy-momentum carriers from an external reservoir against individual elements. Each steering action
\be\label{Abschnitt -- Potential of Mechanical System - Steering Action}
   \mathbf{x}^{(i)}_I , {\mathbf{v}'_I}^{(i-1)}  \;\;\stackrel{\mathrm{RB}^{(i)}}{\Rightarrow}\;\; \mathbf{x}^{(i)}_I , \mathbf{v}^{(i)}_I
\ee
only effects the state of motion at practically the same location $\mathbf{x}^{(i)}_I$. We regard it as a ''kick'' with known effect to the (collective) state of motion ${\mathbf{v}'_I}^{(i-1)}  \Rightarrow \mathbf{v}^{(i)}_I$ of respective elements but irrelevant timing structure. We illustrate each practically instantaneous steering kick $\mathrm{RB}^{(i)}$ as \emph{vertical lift} in velocity space $T_{\mathbf{x}^{(i)}_I}\mathcal{C}$ over the same configuration $\mathbf{x}^{(i)}_I$ of the system (see vertical fibres in figure \ref{pic_Wirkungs-Steuerung_diachrone-Verknuepfung}). Under new initial conditions $\mathbf{v}^{(i)}_I$ intrinsic process $w_i$ evolves
\be
   \mathbf{x}^{(i)}_I , \mathbf{v}^{(i)}_I  \;\;\stackrel{w_i}{\Rightarrow}\;\; \mathbf{x}^{(i+1)}_I , {\mathbf{v}'_I}^{(i)} \nn
\ee
to configuration $\mathbf{x}^{(i+1)}_I$ (see horizontal transition in figure \ref{pic_Wirkungs-Steuerung_diachrone-Verknuepfung}).
There steering kick $\mathrm{RB}^{(i+1)}$ \emph{prepares} new initial conditions for the next part of free action $w_{i+1}$ etc.
\begin{rem}
For the physical specification of an interacting system we draw on:
\begin{itemize}
  \item intrinsic processes $w_i$ in the isolated system
  \item external (practically instantaneous) steering interventions $\mathrm{RB}^{(i)}$ and
  \item consecutive coupling of both actions $\mathrm{RB}^{(1)}\ast w_1 \ast \mathrm{RB}^{(2)} \ast w_2 \ast \ldots$ in a controlled way.
\end{itemize}
\end{rem}
\begin{lem}\label{Lem - KM Dynamics - Potential of Mechanical System - Steering Action}
We can prepare standard paths for the piecewise analysis of a mechanical system
\begin{enumerate}
  \item   circular process $\gamma_1 \ast \gamma_2 \ast \ldots \ast \gamma_n$ which begins and ends $\gamma_1(0) = \gamma_n(t_n) = \mathbf{x}_I^{(1)}$ in the same configuration $\mathbf{x}_I^{(1)}$ of the system (see figure \ref{pic_Wirkungs-Steuerung_Kreisprozess})
  \item   reversion of a configuration path $\gamma\ast{\gamma}^{-1}$ (see figure \ref{pic_Wirkungs-Steuerung_selber-Pfad-Aquipollenz})
  \item   complete dynamical freezing in one configuration $\mathbf{x}_I$
  \item   partial freeze $\mathbf{x}_I \Rightarrow \mathbf{x}_I + \delta \mathbf{x}_n$ for everybody except unconstrained evolving element $\circMn\,$.
\end{enumerate}
\end{lem}
\textbf{Proof:}
Let a series of configurations of the system $\mathbf{x}_I^{(1)}, \mathbf{x}_I^{(2)},\ldots,\mathbf{x}_I^{(n)} \in \mathcal{C}$ be successively connected by $i=1,\ldots,n$ intrinsic processes $\mathbf{x}^{(i)}_I , \mathbf{v}^{(i)}_I  \stackrel{w_i}{\Rightarrow} \mathbf{x}^{(i+1)}_I , {\mathbf{v}'_I}^{(i)}$ so that the final segment $w_n$ leads back to starting configuration $\mathbf{x}^{(n+1)}_I\equiv \mathbf{x}^{(1)}_I$. Further let each steering ''kick'' ${\mathbf{v}'_I}^{(i-1)} \stackrel{\mathrm{RB}|_{\mathbf{x}^{(i)}_I}}{\Rightarrow} \mathbf{v}^{(i)}_I$ at intermediate configuration $\mathbf{x}^{(i)}_I$ absorb the final velocities ${\mathbf{v}'_I}^{(i-1)}$ from the previous segment $w_{i-1}$ and prepare the initial conditions $\mathbf{v}^{(i)}_I$ for the following undisturbed process $w_i$. By their consecutive coupling
\be
   \mathrm{RB}^{(1)}\ast w_1 \ast \mathrm{RB}^{(2)} \ast w_2 \ast \ldots \ast \mathrm{RB}^{(n)} \ast w_n   \nn
\ee
physicists steer a circular process $\gamma : (\mathbf{x}_I^{(1)},t_1) \Rightarrow (\mathbf{x}_I^{(2)},t_2) \Rightarrow \ldots \Rightarrow (\mathbf{x}_I^{(n)},t_n)$ through given configurations - generically in arbitrary duration $t_n$.\footnote{Every variation from an undisturbed action $w$ in fixed duration $t[w]$ is associated with extra ''steering effort'' (see Hamilton Principle \{\ref{Kap - KM Dynamics - Principle of Least Action}\}).}

A circular process $\mathrm{RB}^{(1)}\ast w \ast -\mathrm{RB}^{(2)} \ast \tilde{w}$ between two fixed configurations $\mathbf{x}_I^{(1)}, \mathbf{x}_I^{(2)} \in \mathcal{C}$ of the system arises from a pair of reversed internal processes
\bea
   \;\;\;\;\;\;\;\;
   & &  \mathbf{x}^{(1)}_I , \mathbf{v}^{(1)}_I   \;\; \stackrel{w}{\Rightarrow} \;\;   \mathbf{x}^{(2)}_I , \mathbf{v}_I^{(2)}   \;\;\;\;\;\;\;\;\;\; \mathrm{and} \nn \\
   & &  \mathbf{x}^{(2)}_I , \tilde{\mathbf{v}}_I^{(2)}    \;\;\stackrel{\tilde{w}}{\Rightarrow}\;\; \mathbf{x}^{(1)}_I , \tilde{\mathbf{v}}^{(1)}_I  \nn
\eea
(with initial velocities $\mathbf{v}^{(1)}_I$ resp. $\tilde{\mathbf{v}}^{(2)}_I$) and suitable steering kicks $\tilde{\mathbf{v}}^{(i)}_I  \stackrel{\mathrm{RB}|_{\mathbf{x}^{(i)}_I}}{\Rightarrow} \mathbf{v}_I^{(i)} \;\; i=1,2$. Then the system evolves between both configurations $\mathbf{x}_I^{(1)}, \mathbf{x}_I^{(2)}$ (see upper square in figure \ref{pic_Wirkungs-Steuerung_selber-Pfad-Aquipollenz}) in generally different durations $t[w]\neq t[\tilde{w}]$.

For arbitrarily close configurations $\mathbf{x}_I, \mathbf{x}_I^{(\epsilon)} \in \mathcal{C}$ we steer reversed internal processes
\bea\label{Abschnitt -- Potential of Mechanical System - Steering Action - element of fixation oscillation}
   \;\;\;\;\;\;\;\;
   \mathbf{x}_I , \mathbf{v}_I & \stackrel{w_{\epsilon}}{\Rightarrow} &  \mathbf{x}^{(\epsilon)}_I , \mathbf{v}_I^{(\epsilon)}   \;\;\;\;\;\;\;\;\;\; \mathrm{and}  \\
   \mathbf{x}^{(\epsilon)}_I , \tilde{\mathbf{v}}_I^{(\epsilon)} & \stackrel{\tilde{w}_{\epsilon}}{\Rightarrow} &  \mathbf{x}_I , \tilde{\mathbf{v}}_I  \nn
\eea
by interventions $\tilde{\mathbf{v}}_I \stackrel{\mathrm{RB}|_{\mathbf{x}_I}}{\Rightarrow} \mathbf{v}_I$ , $\:\tilde{\mathbf{v}}^{(\epsilon)}_I \stackrel{\mathrm{RB}|_{\mathbf{x}^{(\epsilon)}_I}}{\Rightarrow} \mathbf{v}_I^{(\epsilon)}$ in a circular process $\mathrm{RB} \ast w_{\epsilon} \ast -\mathrm{RB}^{(\epsilon)} \ast \tilde{w}_{\epsilon}$ between the initial $\mathbf{x}_I$ and nearby final configuration $\mathbf{x}^{(\epsilon)}_I$ . By repetition we approximately freeze all elements of the system around fixed configuration $\mathbf{x}_I$ (for the time of our interventions).

In a \emph{partial fixation} procedure $W_{I\setminus n} := \mathrm{RB}_{I\setminus n} \ast w_{\epsilon} \ast -\mathrm{RB}^{(\epsilon)}_{I\setminus n} \ast \tilde{w}_{\epsilon}$ we compensate the effect of reversible interaction $w_\epsilon$ (\ref{Abschnitt -- Potential of Mechanical System - Steering Action - element of fixation oscillation}) by suitable steering interventions $\mathrm{RB}_{I\setminus n} : \tilde{\mathbf{v}}_{I\setminus n}   \Rightarrow \mathbf{v}_{I\setminus n}$ on all elements in system $G_1\cup\ldots\cup G_N$ except for $G_n$. Repeated partial fixation oscillations $W_{I\setminus n}  \ast \ldots \ast W_{I\setminus n}$ freeze the configuration $\mathbf{x}_1=(\mathbf{x}_1,\ldots,\mathbf{x}_N)$ of all other elements except for the displacements $\delta \mathbf{x}_n$ from the internal process $w$ on the untouched element $G_n$.
\qed
From the operationalization of energy and momentum we obtain true physical quantities and derive basic equations of classical mechanics. Their scope and limitations stem from empirical approximations in our construction of material models for basic measurement operations. In gravity e.g. the spatiotemporal domain of gravitational interactions and the domain of calorimeter interventions $\mathrm{RB}^{(i)}$ justify a practical separation to arbitrary precision. The approximate \emph{separability} admits reproducible measurement procedures. By the extent to which each measurement action represents an instantaneous ''kick'' in practically the same place we can steer the gravitational process $w$ gradually along any path $\gamma\subset \mathcal{C}$ in the configuration space of the system and measure the potential energy and momentum gain.

\subsection{Potential field}\label{Kap - KM Dynamics - Potential of Mechanical System - Potential Field}

We generate each steering kick $\mathrm{RB}^{(i)}$ from elementary standard processes $w_{\mathbf{1}}$ when external calorimeter $\{\circMunit_{\:\mathbf{v}=0}\}$ is coupled against individual elements of the system $G_1\cup\ldots\cup G_N$. Otherwise the calorimeter does not effect the internal process $w$. Eventually we steer the evolution of an (electromagnetic or gravitational) process $w$ by (the orchestration of) external standard processes $w_{\mathbf{1}}$. We assume they satisfy \emph{compatibility conditions}:
\begin{itemize}
\item   superposition principle
\item   equivalence of intrinsic processes for boosted systems
\item   let the intrinsic evolution be determined by simultaneous initial conditions $\mathbf{x}_I$, $\mathbf{v}_I$.
\end{itemize}
By the superposition principle external steering actions effect individual elements in bound system $G_1\cup\ldots\cup G_N$ in the same way as they would effect isolated bodies $G_i$.\footnote{We assume that the effects of intrinsic action $w$ and external steering actions $w_{\mathbf{1}}$ are simply superposed. Effects of steering action $w_{\mathbf{1}}$ against elements of system $\circMa \cup \circMb\:$ and against free elements are practically \emph{indistinguishable}. Individual element $\circMa$ has same inertial behavior $m_{a}^{(\mathrm{bound})}\stackrel{!}{=}m_{a}^{(\mathrm{sep})}$ in a bound state and in separation. The effect of steered measurement interventions $W_{\mathrm{cal}}:= w_{\mathbf{1}} \ast \ldots \ast w_{\mathbf{1}}$ on individual elements of a bound system $\circMa \cup \circMb\:$ and the resulting physical quantities of kinetic energy-momentum satisfy the same quantity equations as determined for the absorbtion of isolated elements in Theorem \ref{Theorem - kin quant absorption action for free particle}.} While steering actions couple against passive elements of the system they do not directly effect the active source $\mathcal{S}$ of their intrinsic interaction $w$.\footnote{The cause of an interaction can be an external source or a ''binding field'' \{\ref{Kap - KM Dynamics - Basic Dynamical Measures - potential Energy}\}. The intrinsic process in a bound (e.g. electromagnetic or gravitational) system $\circMa \cup \circMb_{\:\mathbf{x}_I , \mathbf{v}_I}  \stackrel{w_1}{\Rightarrow}  \circMa \cup \circMb_{\:\mathbf{x}'_I , \mathbf{v}'_I}$ is caused by the presence of elements $\circMa$ and $\circMb$ in a system and the configuration change $\mathbf{x}_I \Rightarrow \mathbf{x}'_I$. The coupling of an external energy source $\mathcal{S}_E$ (e.g. compressed standard spring) against two isolated elements $\circMa\:$, $\circMb$ causes the interaction $\circMa_{\:\mathbf{0}}, \circMb_{\:\mathbf{0}}, \mathcal{S}_E \stackrel{w_2}{\Rightarrow} \circMa_{\:\mathbf{v}_a}, \circMb_{\:\mathbf{v}_b}, \mathcal{S}_0 \,$.} What is the absorption effect for an energy source? Is it also a source of extractable momentum?

\begin{figure}    
  \begin{center}           
  \includegraphics[height=8cm]{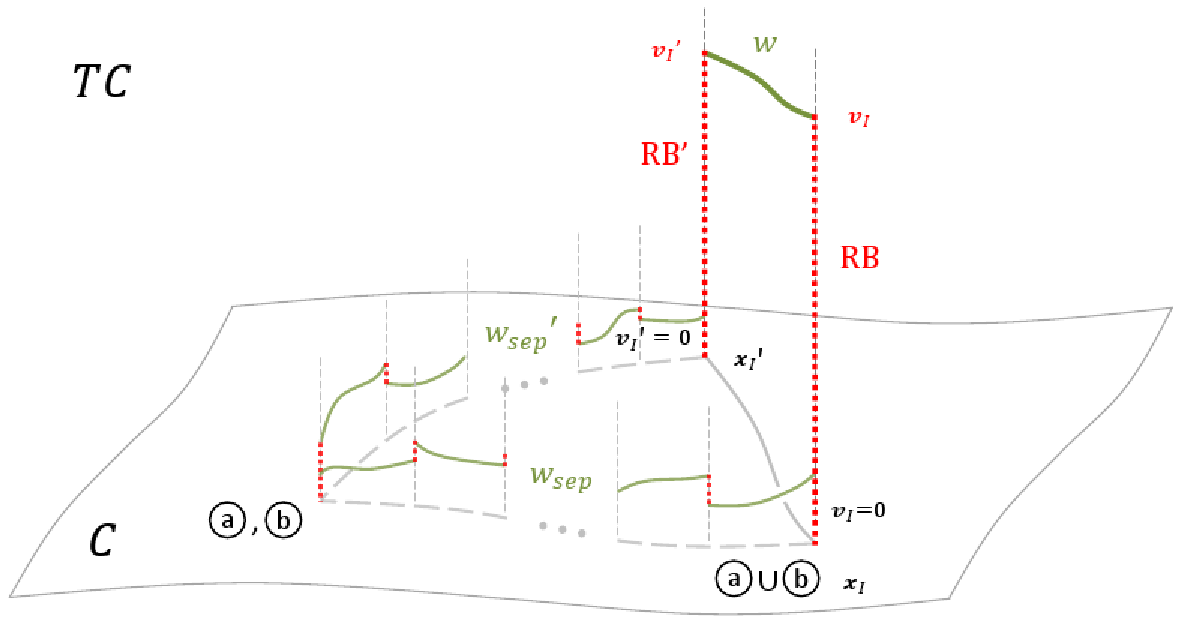}  
  \end{center}
  \vspace{-0cm}
  \caption{\label{pic_Wirkungs-Steuerung_Impulserhaltung} conservation of extractable momentum
    }
  \end{figure}

Our standardized extract from a resting configuration transition (\ref{Abschnitt -- Energie - Reservoirbilanz - reversible action in closed system}) absorbs potential energy-momentum from elements $G_i$ of the system $\{G_I\}$ onto neutral elements of an external calorimeter $\{\circMunit_{\:\mathbf{0}}\}$.\footnote{We can realize a neutral calorimeter reservoir for measuring (energy-momentum of) e.g. electromagnetic processes $w_{\mathrm{EM}}$. For gravitational interactions $w_{\mathrm{grav}}$ we cannot assume existence of one calorimeter for global extents of a gravitating system $G_1\cup\ldots\cup G_N$. We can only provide a calorimeter model $W_{\mathrm{cal}}\big|_{\:\mathcal{U}_i}$ in a local neighborhood of each element $G_i$. In the practice of intrinsic gravitational measurements we must examine the physical connection between calorimetric measurements at adjacent locations $W_{\mathrm{cal}}\big|_{\:\mathcal{U}_i}$ and $W_{\mathrm{cal}}\big|_{\:\mathcal{U}_j}$.} Each steering act $\mathrm{RB}\left[ {G_1\cup\ldots\cup G_N}_{\: \mathbf{v}} \Rightarrow {G_1\cup\ldots\cup G_N}_{\: \mathbf{0}} \right] \big|_{\mathbf{x}_I}$ in absorption model $-\mathrm{RB} \big|_{\mathbf{x}_I} \ast w \ast \mathrm{RB}' \big|_{\mathbf{x}'_I}$ extracts standard energy and momentum carriers from decelerating the aggregate in unchanged initial $\mathbf{x}_I$ resp. final configuration $\mathbf{x}'_I$.\footnote{We define momentum (\ref{Abschnitt -- vortheor Ordnungsrelastion - impulse verhalten}) by a head-on collision test provided the colliding objects remain preserved.} By the equipollence principle the total calorimeter extract (\ref{Abschnitt -- Energie - Reservoirbilanz - configuration transition in closed system-simplified}) has a unique energy. The potential energy from a system transforms by definition (\ref{Abschnitt -- Energie - potential Energy}) into kinetic energy of all elements. Finally we show that the absorption extract has no momentum. Thus in internal process $\mathbf{x}_I,\mathbf{v}_I \stackrel{w}{\Rightarrow} \mathbf{x}_I',\mathbf{v}_I'$ the \emph{total momentum of all passive elements} is conserved. There is no potential form of momentum associated with the energy source $\mathcal{S}$.
\begin{de}
The total balance for steering the evolution $w$ of a \underline{conservative} system $\{ G_I \}$ along a circular process back to the original configuration $\mathbf{x}_I^{(1)}$ and motion $\mathbf{v}_I$ vanishes
\be\label{Abschnitt -- Potential of Mechanical System - potential field - combined extract for steering along circular process}
   \mathrm{RB} \left[ \mathrm{RB}^{(1)} \ast w_1 \ast \ldots \ast \mathrm{RB}^{(n)} \ast w_n \right] \;\; = \;\;   \sum_{i=1}^{n} \mathrm{RB}^{(i)} \;\; = \;\;  0   \;\; .
\ee
\end{de}
\begin{lem}\label{Lem - Potential of Mechanical System - potential field - potential energy extract}
Then the potential energy between any two configurations does not depend from the steering process $\mathrm{RB}^{(1)} \ast w_1 \ast \ldots \ast \mathrm{RB}^{(n)} \ast w_n$ for a particular path $\gamma_1 \ast \ldots \ast \gamma_n: \mathbf{x}_I\Rightarrow\mathbf{x}'_I$ and simply adds up
\be\label{Abschnitt -- Potential of Mechanical System - potential field - wirkungs-weg-unabh}
   V_{\mathrm{pot}} \left[ \mathbf{x}_I\Rightarrow\mathbf{x}'_I \right] \;\; = \;\; \sum_{i=1}^n  V_{\mathrm{pot}} \left[ w_i \right] \;\; .
\ee
\end{lem}
\textbf{Proof:}
In a generic steering process $\mathrm{RB}^{(1)} \ast w_1 \ast \ldots \ast \mathrm{RB}^{(n)} \ast w_n$ the system evolves in segments of undisturbed intrinsic actions $w_i$ along configuration transitions $\gamma_i: \mathbf{x}^{(i)}_I\Rightarrow\mathbf{x}^{(i+1)}_I$ with corresponding steering actions ${\mathbf{v}_I'}^{(i-1)} \stackrel{\mathrm{RB}^{(i)}}{\Rightarrow} \mathbf{v}^{(i)}_I$ for matching their initial conditions. With starting kick $-\mathrm{RB} \left[ \mathbf{v}^{(0)}_I \right]$ we provide the initial condition for the generic steered process and with $\mathrm{RB} \left[ {\mathbf{v}'_I}^{(n)} \right]$ we absorb the kinetic effect in the final configuration $\mathbf{x}'_I$. From the standardized extraction into the external calorimeter we obtain
\be
\begin{array}{l}
   -\mathrm{RB} \left[ \mathbf{v}^{(0)}_I \right] \ast \left\{
   \mathrm{RB}^{(1)} \ast w_1 \ast \ldots \ast \mathrm{RB}^{(n)} \ast w_n
\right\} \ast \mathrm{RB} \left[ {\mathbf{v}'_I}^{(n)} \right]    \\
   \;\;\;\;\;\;\;\;\;\; =  \; -\mathrm{RB} \left[ \mathbf{v}^{(0)}_I \right] \ast
   \underbrace{\mathrm{RB} \left[ \mathbf{v}^{(0)}_I \right] -\mathrm{RB} \left[ \mathbf{v}^{(1)}_I \right]}_{\stackrel{(\ref{Abschnitt -- Potential of Mechanical System - Steering Action})}{=}\;\mathrm{RB}^{(1)}}   \ast \; w_1   \ast   \underbrace{\mathrm{RB} \left[ {\mathbf{v}'_I}^{(1)} \right]  -\mathrm{RB} \left[ {\mathbf{v}'_I}^{(1)} \right]}_{=\;\mathrm{Id}}       \\
\;\;\;\;\;\;\;\;\;\;\;\;\;\;\;\;\;\;\;\;\;\;\;\;\;\;\;\;\;\;\;\;
   \ast \underbrace{\mathrm{RB} \left[ {\mathbf{v}'_I}^{(1)} \right] -\mathrm{RB} \left[ \mathbf{v}^{(2)}_I \right]}_{\stackrel{(\ref{Abschnitt -- Potential of Mechanical System - Steering Action})}{=}\;\mathrm{RB}^{(2)}}   \ast \; w_2   \ast   \underbrace{\mathrm{RB} \left[ {\mathbf{v}'_I}^{(2)} \right] -\mathrm{RB} \left[ {\mathbf{v}'_I}^{(2)} \right]}_{=\;\mathrm{Id}} \; \ast \;\; \mathrm{etc.} \\
   \;\;\;\;\;\;\;\;\;\;= \; \sum_{i=1}^{n} -\mathrm{RB} \left[ \mathbf{v}_I^{(i)} \right] \ast w_i  \ast \mathrm{RB} \left[ {\mathbf{v}'_I}^{(i)} \right]
\end{array} \nn
\ee
with corresponding potential energy $V_{\mathrm{pot}} \left[ \mathbf{x}_I\Rightarrow\mathbf{x}'_I \right]  \stackrel{(\ref{Abschnitt -- Energie - potential Energy})}{=}  \sum_{i=1}^n  V_{\mathrm{pot}} \left[ w_i \right] $.
\qed
\begin{lem}\label{Lem - Potential of Mechanical System - potential field - no potential momentum extract}
In Galilei Kinematics no momentum is extractable from separating elements (from a bound system) or from an internal process (generic configuration transition)
\be
   \mathbf{p} \left[ \circMa \cup \circMb_{\: \mathbf{x}_I,\mathbf{0}} \Rightarrow \circMa_{\: \mathbf{0}} , \circMb_{\: \mathbf{0}} \right] \;\; = \;\; 0   \;\;\;\;\;\;\;\; \label{Abschnitt -- Potential of Mechanical System - potential field - no impulse extract from separation action}  \\
\ee
\be
   \mathbf{p} \left[ \circMa \cup \circMb_{\: \mathbf{x}_I,\mathbf{0}} \Rightarrow \circMa \cup \circMb_{\: \mathbf{x}'_I,\mathbf{0}}  \right] \;\; = \;\; 0 \;\; .  \label{Abschnitt -- Potential of Mechanical System - potential field - no impulse extract from steering configuration transition}
\ee
\end{lem}
\textbf{Proof:}
Consider a two-partite system $\circMa \cup \circMb\:$. Let a reversible separation process
\be\label{Abschnitt -- Potential of Mechanical System - potential field - separation action}
   \circMa \cup \circMb_{\:\mathbf{x}_I , \mathbf{0}} \;\; \stackrel{w_{\mathrm{sep}}}{\Rightarrow} \;\;
   \circMa_{\:\mathbf{0}} \,,\: \circMb_{\:\mathbf{0}} \,,\: \mathrm{RB}_{\mathrm{sep}}
\ee
liberate elements $\circMa$ and $\circMb\:$ from their bound state by a series of steering actions (see figure \ref{pic_Wirkungs-Steuerung_Impulserhaltung}). We store the calorimeter extract for separation $\mathrm{RB}_{\mathrm{sep}} := \mathrm{RB} \left[ \circMa \cup \circMb_{\:\mathbf{x}_I , \mathbf{0}} \Rightarrow \circMa_{\:\mathbf{0}} , \circMb_{\:\mathbf{0}}  \right]$ in an external reservoir $\{\circMunit_{\:\mathbf{v}=0}\}$. We can \emph{substitute} in a reversible way
\begin{itemize}
  \item   the original system $\circMa \cup \circMb$ with internal binding energy $E$ by
  \item   separated elements $\circMa\:, \circMb$ and sources $\mathcal{S}_{\mathbf{1}}\big|_{\mathbf{0}} , \circMunit_{\:\mathbf{v}_{\mathbf{1}}} \in \mathrm{RB}_{\mathrm{sep}}$ in the separation extract
\end{itemize}
with known boost behavior (\ref{Abschnitt -- kin quant Absorptions Wirkung - Reservoirbilanz - absorption}), (\ref{Abschnitt -- kin quant Absorptions Wirkung - Reservoirbilanz - boost E unit}), (\ref{Abschnitt -- kin quant Absorptions Wirkung - Reservoirbilanz - boost p unit}).\footnote{As alternative to separating the generic system ${G_1\cup \ldots \cup G_N}\big|_{\:\mathbf{x}_I}$ into isolated (elementary) particles $G_i$ also a transition $\mathbf{x}_I \Rightarrow \mathbf{x}_s$ into a standard configuration with symmetric reorientation and (effective) inertial behavior $m_{\mathbf{x}_s}$ (of bound aggregate ${G_1\cup \ldots \cup G_N}\big|_{\mathbf{x}_{s}}$) is sufficient, like the standard spring $\mathcal{S}_E\big|_{\mathbf{0}}$ in figure \ref{pic_Wirkungseinheit_Feder}.} We can boost the intrinsic separation process $w^{(\mathcal{B})}_{\mathrm{sep}} \Rightarrow w^{(\mathcal{A})}_{\mathrm{sep}}$ from $\mathcal{B}$ob's reference system to $\mathcal{A}$lice.

Let $\mathcal{A}$lice move relative to $\mathcal{B}$ob with constant velocity $v_{\mathcal{A}}$. According to our compatibility conditions for $\mathcal{B}$ob a boost of the bound system (including the elements and intrinsic sources of binding energy; aka binding ''field'' ) in arbitrary configuration $\mathbf{x}_I$
\be\label{Abschnitt -- Potential of Mechanical System - potential field - boosting bound and separately}
   \mathrm{RB} \left[ \circMa \cup \circMb_{\: \mathbf{0}} \Rightarrow \circMa \cup \circMb_{\: \mathbf{v}_{\mathcal{A}}} \right] \big|_{\mathbf{x}_I} \;\; / \;\, \mathrm{mod} \; \mathbf{x}_I \;\; = \;\; \mathrm{RB} \left[ \circMa_{\:\mathbf{0}} \,,\: \circMb_{\:\mathbf{0}} \Rightarrow  \circMa_{\:\mathbf{v}_{\mathcal{A}}} \,,\: \circMb_{\:\mathbf{v}_{\mathcal{A}}}  \right]
\ee
requires the same steering effort as the boost of all separated elements.\footnote{According to superposition principle $\mathcal{B}$ob's steering effort $\mathrm{RB} \left[ \circMa_{\:\mathbf{0}} \,,\: \circMb_{\:\mathbf{0}} \Rightarrow  \circMa_{\:\mathbf{v}_{\mathcal{A}}} \,,\: \circMb_{\:\mathbf{v}_{\mathcal{A}}}  \right]$ transfers  system ${\circMa_{\:\mathbf{0}} \cup \circMb_{\:\mathbf{0}}} \big|_{\mathbf{x}_I^{(\mathcal{B})}} \Rightarrow  {\circMa_{\:\mathbf{v}_{\mathcal{A}}} \cup \circMb_{\:\mathbf{v}_{\mathcal{A}}}} \big|_{\mathbf{x}_I^{(\mathcal{B})}} \stackrel{!}{\equiv} {\circMa_{\:\mathbf{0}} \cup \circMb_{\:\mathbf{0}}} \big|_{\mathbf{x}_I^{(\mathcal{A})}}$ - by equivalence of intrinsic actions $w^{(\mathcal{A})} \equiv w^{(\mathcal{B})}$ and same instantaneous initial configuration $\mathbf{x}_I^{(\mathcal{A})} \equiv \mathbf{x}_I^{(\mathcal{B})}$ - into an \emph{intrinsically equivalent} system for $\mathcal{A}$lice.}
$\mathcal{B}$ob can substitute the direct boost $\mathrm{RB}^{(B)}_{\mathrm{dir}}$ of the bound system
\be
\begin{diagram}
   \underbrace{{\circMa_{\:\mathbf{0}} \cup \circMb_{\:\mathbf{0}}} \big|_{\mathbf{x}_I^{(\mathcal{A})}}}_{\stackrel{!}{\equiv}\: {\circMa_{\:\mathbf{v}_{\mathcal{A}}} \cup \circMb_{\:\mathbf{v}_{\mathcal{A}}}} \big|_{\mathbf{x}_I^{(\mathcal{B})}}}    &   \; \lImplies^{{w^{(\mathcal{A})}_{\mathrm{sep}}}^{-1}} \; &  \circMa_{\:\mathbf{0}} \,,\: \circMb_{\:\mathbf{0}} \,,\: \mathrm{RB}^{(\mathcal{A})}_{\mathrm{sep}}  \\
   \uDotsto^{
   \begin{array}{c}
   \\
   \mathrm{RB}^{(B)}_{\mathrm{dir}}
   \end{array}
   \;\;\;\;
   }   &   &
   \uImplies_{\mathrm{RB}^{(B)}_{\mathrm{indir}}}   \\
    {\circMa_{\:\mathbf{0}} \cup \circMb_{\:\mathbf{0}}} \big|_{\mathbf{x}_I^{(\mathcal{B})}}   & \; \rImplies^{w^{(\mathcal{B})}_{\mathrm{sep}}} \;  &  \circMa_{\:\mathbf{0}} \,,\: \circMb_{\:\mathbf{0}} \,,\: \mathrm{RB}^{(\mathcal{B})}_{\mathrm{sep}}
\end{diagram}   \nn
\ee
%
by an indirect boost $\mathrm{RB}^{(B)}_{\mathrm{indir}}$ of the separate elements $\circMa\:$, $\circMb$ and steering ingredients $\mathrm{RB}_{\mathrm{sep}}^{(\mathcal{B})}$. Then $\mathcal{A}$lice can prepare ${w_{\mathrm{sep}}^{(\mathcal{A})}}^{-1}$ the intrinsically equivalent bound system ${\circMa_{\:\mathbf{0}} \cup \circMb_{\:\mathbf{0}}} \big|_{\mathbf{x}_I^{(\mathcal{A})}}$ from standard resources $\mathcal{S}_{\mathbf{1}^{(\mathcal{A})} }\big|_{\mathbf{0}} \, ,\, \circMunit_{\:\mathbf{v}_{\mathbf{1}^{(\mathcal{A})}}}$ in her own (boosted) calorimeter reservoir.

We can boost the entire system ${\circMa_{\:\mathbf{0}} \cup \circMb_{\:\mathbf{0}}} \big|_{\mathbf{x}_I^{(\mathcal{B})}}$ in both (connected and separated) ways
\be
   -\mathrm{RB}^{(B)}_{\mathrm{dir}} \;\ast\: \left( w^{(B)}_{\mathrm{sep}} \;\ast\: \mathrm{RB}^{(B)}_{\mathrm{indir}} \;\ast\: {w^{(A)}_{\mathrm{sep}}}^{-1}  \right)  \nn
\ee
along a circular process.\footnote{Without compatibility conditions between intrinsic process $w$ and external steering interventions $W_{\mathrm{cal}}$ the equivalence of steering bound system $\circMa \cup \circMb$ or isolated elements $\circMa \,,\: \circMb$ is lost. $\mathcal{B}$ob's steering effort $\mathrm{RB} \left[ \circMa_{\:\mathbf{0}} \,,\: \circMb_{\:\mathbf{0}} \Rightarrow  \circMa_{\:\mathbf{v}_{\mathcal{A}}} \,,\: \circMb_{\:\mathbf{v}_{\mathcal{A}}}  \right]$ on individual \emph{elements} ${\circMa_{\:\mathbf{0}} \cup \circMb_{\:\mathbf{0}}} \big|_{\mathbf{x}_I^{(\mathcal{B})}} \Rightarrow  {\circMa_{\:\mathbf{v}_{\mathcal{A}}} \cup \circMb_{\:\mathbf{v}_{\mathcal{A}}}} \big|_{\mathbf{x}_I^{(\mathcal{B})}} {\,\equiv\:\!\!\!\!\!\!\!/ \;\:} {\circMa_{\:\mathbf{0}} \cup \circMb_{\:\mathbf{0}}} \big|_{\mathbf{x}_I^{(\mathcal{A})}} $ not necessarily reproduces the intrinsically equivalent bound system for $\mathcal{A}$lice. In Poincare kinematics the boost of an intrinsic \emph{source} of (binding) energy requires additional steering effort \{\ref{Kap - SRT Dynamics - On the inertia of energy sources}\}. For electromagnetic interactions $w_{EM}$ the binding energy in system $\circMa \cup \circMb$ is determined by the retarded relative localization of its elements. In absence of (hypothetical absolute) initial conditions intrinsic action $w_{EM}$ is governed by the retarded Coulomb principle. The acceleration of an extended electromagnetically bound system $\circMa \cup \circMb_{\:\mathbf{x}_I^{(\mathcal{A})},\mathbf{0}^{(\mathcal{A})}}  \Rightarrow  \circMa \cup \circMb_{\:\mathbf{x}_I^{(\mathcal{B})},\mathbf{0}^{(\mathcal{B})}}$ requires additional subtle steering kicks (associated to ''radiation'') to reproduce an intrinsically equivalent configuration $\mathbf{x}_I^{(\mathcal{A})} \equiv \mathbf{x}_I^{(\mathcal{B})}$ for a moving observer.} Hence $\mathcal{B}$ob's steering expense for the direct boost of the system
\bea\label{Abschnitt -- Potential of Mechanical System - potential field - steering effort bound and separately}
   &  & \!\!\!\!\!\!\!\!\!\!\!\!\!\!\!\!\!\!\!
   \mathrm{RB} \left[  \circMa \cup \circMb_{\:\mathbf{0}} \Rightarrow \circMa \cup \circMb_{\:\mathbf{v}_{\mathcal{A}}}  \right] \; \stackrel{!}{=} \;
   \mathrm{RB} \left[ \circMa_{\:\mathbf{0}} \,,\: \circMb_{\:\mathbf{0}} \,,\: \mathrm{RB}^{(\mathcal{B})}_{\mathrm{sep}} \Rightarrow  \circMa_{\:\mathbf{v}_{\mathcal{A}}} \,,\: \circMb_{\:\mathbf{v}_{\mathcal{A}}} \,,\: \mathrm{RB}^{(\mathcal{A})}_{\mathrm{sep}} \big|_{\:\mathbf{v}_{\mathcal{A}}}  \right]   \nn \\
   &  & \;\;\;\;\;\;\;\;\;\;\;\;\;\;\;\;\;\;\;\;\;\;\; = \;
   \mathrm{RB} \left[ \circMa_{\:\mathbf{0}} \,,\: \circMb_{\:\mathbf{0}}  \Rightarrow  \circMa_{\:\mathbf{v}_{\mathcal{A}}} \,,\: \circMb_{\:\mathbf{v}_{\mathcal{A}}} \right] \; + \;
   \mathrm{RB} \left[ \mathrm{RB}^{(\mathcal{B})}_{\mathrm{sep}}  \Rightarrow  \mathrm{RB}^{(\mathcal{A})}_{\mathrm{sep}} \big|_{\:\mathbf{v}_{\mathcal{A}}} \right]
\eea
must match his expense for the indirect boost of all separated components, which implies that $\mathcal{B}$ob boosts the separation ingredients $\mathrm{RB}^{(\mathcal{B})}_{\mathrm{sep}} :=   k \cdot \mathcal{S}_{\mathbf{1}^{(\mathcal{B})} }\big|_{\mathbf{0}} \,,\: l \cdot \circMunit_{\:\mathbf{v}_{\mathbf{1}^{(\mathcal{B})}}} $ without expense
\be\label{Abschnitt -- Potential of Mechanical System - potential field - conservation boost separation ingredients}
   \mathrm{RB}^{(\mathcal{B})} \left[ \mathrm{RB}^{(\mathcal{B})}_{\mathrm{sep}}  \Rightarrow  \mathrm{RB}^{(\mathcal{A})}_{\mathrm{sep}} \big|_{\:\mathbf{v}_{\mathcal{A}}} \right] \; \stackrel{(\ref{Abschnitt -- Potential of Mechanical System - potential field - steering effort bound and separately})(\ref{Abschnitt -- Potential of Mechanical System - potential field - boosting bound and separately})}{=} \; 0    \;\; .
\ee
Hence separation action $(\ref{Abschnitt -- Potential of Mechanical System - potential field - separation action})$ can solely release energy souces (boosting $\mathcal{S}_{\mathbf{1}^{(\mathcal{B})} }\big|_{\mathbf{0}}$ is for free)
\be
   \mathrm{RB}^{(\mathcal{B})}_{\mathrm{sep}} \;\; := \;\;  k \cdot \mathcal{S}_{\mathbf{1}^{(\mathcal{B})} }\big|_{\mathbf{0}} \:,\; \underbrace{l}_{\stackrel{!}{=}\: 0} \cdot \circMunit_{\:\mathbf{v}_{\mathbf{1}^{(\mathcal{B})}}} \nn \;\; .
\ee
The boost $\circMunit_{\:\mathbf{v}_{\mathbf{1}^{(\mathcal{B})}}} \Rightarrow \circMunit_{\:\mathbf{v}_{\mathbf{1}^{(\mathcal{A})}}} $ of an impulse carrier would consume extra energy sources $\mathcal{S}_{\mathbf{1}^{(\mathcal{B})} }\big|_{\mathbf{0}}$ (see Lemma \ref{Lem - kin quant Absorptions Wirkung - Reservoirbilanz - boost E and p units}) in contradiction to conservation (\ref{Abschnitt -- Potential of Mechanical System - potential field - conservation boost separation ingredients}). Liberating binding energy $E$ of system $\circMa \cup \circMb$ generates no extractable momentum - which proves assertion (\ref{Abschnitt -- Potential of Mechanical System - potential field - no impulse extract from separation action}).

Next consider our standardized extract from an internal process $w$ in system $\circMa \cup \circMb$
\be
   \left( -\mathrm{RB} \big|_{\mathbf{x}_I} \right) \ast w \ast \mathrm{RB}' \big|_{\mathbf{x}'_I}:
   \;\;\; \circMa \cup \circMb_{\:\mathbf{x}_I,\mathbf{0}}  \;\; \Rightarrow \;\; \circMa \cup \circMb_{\:\mathbf{x}'_I,\mathbf{0}}   \nn
\ee
where steering kick $\mathrm{RB}$ prepares the initial state of motion and $\mathrm{RB}'$ absorbs the kinetic effect (see figure \ref{pic_Wirkungs-Steuerung_Impulserhaltung}).
%
Let separation actions $w_{\mathrm{sep}}$ and $w'_{\mathrm{sep}}$ prepare the initial $\mathbf{x}_I$ and final configuration $\mathbf{x}'_I$ of the system from separated elements $\circMa\:, \circMb$ and expense of separation extract $\mathrm{RB}_{\mathrm{sep}}$ resp. $\mathrm{RB}'_{\mathrm{sep}}$ (\ref{Abschnitt -- Potential of Mechanical System - potential field - separation action}). After steering the entire system along a circular process
\be
   W \;\; := \;\;  -\mathrm{RB}  \;\ast\; w \;\ast\; \mathrm{RB}' \;\ast\;  w'_{\mathrm{sep}}  \;\ast\; {w_{\mathrm{sep}}}^{-1} \nn
\ee
the total calorimeter extract $\mathrm{RB} \left[ W \right] \stackrel{!}{=}  0$ must vanish (\ref{Abschnitt -- Potential of Mechanical System - potential field - combined extract for steering along circular process}). Hence potential momentum is neither extractable from the complete circular process
\be
   \mathbf{p} \left[  \mathrm{RB} \left[ W \right]  \right] \; \stackrel{!}{=} \; 0
   \; = \; \mathbf{p} \left[ \left( -\mathrm{RB} \right) \ast w \ast \mathrm{RB}' \right] +
   \underbrace{\mathbf{p} \left[ w'_{\mathrm{sep}} \right]}_{\stackrel{(\ref{Abschnitt -- Potential of Mechanical System - potential field - no impulse extract from separation action})}{=}\: 0} + \underbrace{\mathbf{p} \left[ {w_{\mathrm{sep}}}^{-1} \right]}_{\stackrel{(\ref{Abschnitt -- Potential of Mechanical System - potential field - no impulse extract from separation action})}{=}\: 0}  \nn
\ee
nor from the kinetic effect of intrinsic action $w$ - which proves assertion (\ref{Abschnitt -- Potential of Mechanical System - potential field - no impulse extract from steering configuration transition}).
\qed
The extractable momentum from (stopping and separating) the entire system
\bea
   \mathbf{p} \left[ \circMa \cup \circMb_{\:\mathbf{x}_I, \mathbf{v}_I } \right]  & := &
   \mathbf{p} \left[ \circMa \cup \circMb_{\:\mathbf{x}_I, \mathbf{v}_I }  \Rightarrow  \circMa \cup \circMb_{\:\mathbf{x}_I, \mathbf{0} }  \right] \;\; + \;\;
   \mathbf{p} \left[ \circMa \cup \circMb_{\:\mathbf{x}_I, \mathbf{0} }  \Rightarrow  \circMa_{\:\mathbf{0} } , \circMb_{\:\mathbf{0} }  \right] \nn \\
   & \stackrel{(\ref{Abschnitt -- Potential of Mechanical System - potential field - boosting bound and separately})(\ref{Abschnitt -- Potential of Mechanical System - potential field - no impulse extract from separation action})}{=} & \mathbf{p} \left[ \circMa_{\:\mathbf{v}_a } \right] + \mathbf{p} \left[ \circMb_{\:\mathbf{v}_b } \right]   \label{Abschnitt -- Potential of Mechanical System - potential field - total momentum of system}
\eea
is the total momentum of all passive elements. In an intrinsic interaction $w$ it is conserved
\bea
   \!\!\!\!\!\!\! \mathbf{p} \left[ -\mathrm{RB}\big|_{\mathbf{x}_I} \ast w \ast \mathrm{RB}'\big|_{\mathbf{x}'_I} \right]  & \!=\! & \mathbf{p} \left[ -\mathrm{RB}\big|_{\mathbf{x}_I} \right] \;+\; \mathbf{p} \left[ \mathrm{RB}'\big|_{\mathbf{x}'_I} \right]
    \nn \\
   & \!\!\!\stackrel{(\ref{Abschnitt -- Potential of Mechanical System - potential field - boosting bound and separately})}{=}\!\!\! & - \mathbf{p} \left[ \circMa_{\:\mathbf{v}_a } \right] - \mathbf{p} \left[ \circMb_{\:\mathbf{v}_b } \right] + \mathbf{p} \left[ \circMa_{\:\mathbf{v}'_a } \right] + \mathbf{p} \left[ \circMb_{\:\mathbf{v}'_b } \right]  \stackrel{(\ref{Abschnitt -- Potential of Mechanical System - potential field - no impulse extract from steering configuration transition})}{=} \, 0 \; . \;\;\;\;\;\;
   \label{Abschnitt -- Potential of Mechanical System - potential field - conservation total momentum elements}
\eea
By the superposition principle we generalize to generic N-body systems $G_1\cup \ldots\cup G_N$.
\begin{co}\label{co - Potential of Mechanical System - conservation potential field and total momentum elements}
An internal interaction is determined by the \underline{potential field} $V_{\mathrm{pot}} \left[ \mathbf{x}_I\Rightarrow\mathbf{x}'_I \right]$ (\ref{Abschnitt -- Potential of Mechanical System - potential field - wirkungs-weg-unabh}) and \underline{momentum conservation} for all passive elements $\sum_{i\in I}\: \mathbf{p} \left[ \circMi_{\:\mathbf{v}_i } \right] \stackrel{!}{=} \mathrm{const.}$ (\ref{Abschnitt -- Potential of Mechanical System - potential field - conservation total momentum elements}).\footnote{In Poincare kinematics the kinetic energy-momentum $(E,\mathbf{p}) \left[ \mathcal{S}_{\mathbf{1}} \big|_{\mathbf{v}} \right]$ from a moving energy source depends on the velocity of the carrier \{\ref{Kap - Relativistic energy-momentum}\}.}
\end{co}


\section{Piecewise analysis}\label{Kap - KM Dynamics - Differentiated Analysis}

We analyze calorimeter measurements of \emph{infinitesimal segments} of an intrinsic interaction $w$ and with respect to the \emph{individual elements} of the (conservative) system $G_1\cup \ldots\cup G_N$. How does energy and momentum redistribute between the system and all passive elements $G_i$ throughout the spatiotemporal evolution of the process?

At first we analyze two-partite systems $\circMa \cup \circMb$ and then generalize by the superposition principle to n-body systems $G_1\cup \ldots\cup G_N$. Our standardized calorimeter extract from the configuration transition of an infinitesimal process $t,\mathbf{x}_I \stackrel{w}{\Rightarrow} t',\mathbf{x}'_I$
\be
   \mathrm{RB} \left[  \circMa \cup \circMb_{\:\mathbf{x}_I, \mathbf{0} }  \Rightarrow  \circMa \cup \circMb_{\:\mathbf{x}'_I, \mathbf{0} }  \right]
   \; \stackrel{(\ref{Abschnitt -- Energie - Reservoirbilanz - configuration transition in closed system-simplified})}{=} \: - \;\; \mathrm{RB} \left[ \circMa_{\:\mathbf{v}_a } \Rightarrow  \circMa_{\:\mathbf{v}_a + \Delta \mathbf{v}_a }  \right] \; - \; \mathrm{RB} \left[ \circMb_{\:\mathbf{v}_b } \Rightarrow  \circMb_{\:\mathbf{v}_b + \Delta \mathbf{v}_b }  \right]   \nn
\ee
is independent from steering path and initial conditions $\mathbf{v}_I=(\mathbf{v}_a , \mathbf{v}_b)$. Potential energy
\bea
   \underbrace{(E,\mathbf{p}) \left[ \mathrm{RB} \left[  \circMa \cup \circMb_{\:\mathbf{x}_I, \mathbf{0} }  \Rightarrow  \circMa \cup \circMb_{\:\mathbf{x}'_I, \mathbf{0} }  \right] \right]}_{\stackrel{(\mathrm{Cor.} \ref{co - Potential of Mechanical System - conservation potential field and total momentum elements} )}{=}\: \left( V_{\mathrm{pot}} \left[ \mathbf{x}_I\Rightarrow\mathbf{x}'_I \right] \;, \;\; \mathbf{0} \;\; \right)}
   & = & - \;\; \underbrace{(E,\mathbf{p})\left[\mathrm{RB} \left[ \circMa_{\:\mathbf{v}_a } \Rightarrow  \circMa_{\:\mathbf{v}_a + \Delta \mathbf{v}_a } \right] \right]}_{=\: \Delta \left( E_{\mathrm{kin}} \,,\:  \mathbf{p} \right)_{a}}    \nn \\
   &  &  \;\;\;\;\;\;\;\;\;\;\; - \; \underbrace{(E,\mathbf{p}) \left[ \mathrm{RB} \left[ \circMb_{\:\mathbf{v}_b } \Rightarrow  \circMb_{\:\mathbf{v}_b + \Delta \mathbf{v}_b } \right] \right]}_{=\: \Delta \left( E_{\mathrm{kin}} \,,\:  \mathbf{p} \right)_{b}}   \nn
\eea
transforms into kinetic energy of both elements $\circMi$ and their total momentum is conserved. In basic quantity equations (Theorem \ref{Theorem - kin quant absorption action for free particle}) we can suppress higher order terms $ \Delta\mathbf{v}_i  \ll \mathbf{v}_i$
\bea\label{Abschnitt -- differentiated analysis - E_kin p from particles in infinetesimal action}
   \Delta(E_{\mathrm{kin}},\mathbf{p})_i  &   \stackrel{(\ref{Abschnitt -- kin quant Absorptions Wirkung - kin energy and momentum metrisiert})}{=} & \left(   \left( \frac{m_i}{2} \cdot \left( \mathbf{v}_i + \Delta\mathbf{v}_i \right) ^2  - \frac{m_i}{2} \cdot { \mathbf{v}_i}^2  \right) \cdot E\left[ \mathcal{S}_{\mathbf{1}}\big|_{\mathbf{0}} \right]    \;,\: \left( m_i \cdot \Delta\mathbf{v}_i \right) \cdot \mathbf{p}\left[ \circMunit_{\:\mathbf{v}_{\mathbf{1}}} \right]     \right)   \nn \\
   & \simeq &  \left( \left( m_i \cdot \mathbf{v}_i \cdot \Delta\mathbf{v}_i \right) \cdot E\left[ \mathcal{S}_{\mathbf{1}}\big|_{\mathbf{0}} \right]  \;,\:    \left( m_i \cdot \Delta\mathbf{v}_i \right) \cdot \mathbf{p}\left[ \circMunit_{\:\mathbf{v}_{\mathbf{1}}} \right]      \right)  \;\; .
\eea

Provided physical quantities of length, duration \{\ref{Kap - Kinematics}\} and energy, momentum \{\ref{Kap - KM Dynamics - Basic Dynamical Measures}\} we introduce more differentiated termini suitable for analyzing continuous evolution. We define the ''force'' of an intrinsic action \{\ref{Kap - KM Dynamics - Differentiated Analysis - Force}\} and the ''displacement work'' in steered processes \{\ref{Kap - KM Dynamics - Differentiated Analysis - Displacement Work}\}. We determine properties and equations of the \emph{derived} physical quantities and ultimately derive the equations of motion \{\ref{Kap - KM Dynamics - Differentiated Analysis - Evolution}\} for interactions in a conservative system.


\subsection{Force}\label{Kap - KM Dynamics - Differentiated Analysis - Force}

\begin{de}\label{Def - differentiated analysis - force}
The \underline{force} $\mathbf{F}_a$ - of intrinsic action $w$ in system $\circMa_{\:\mathbf{v}_a} \cup \circMb_{\:\mathbf{v}_b} \big|_{\mathbf{x}_a,\mathbf{x}_b}$ with initial condition $\mathbf{v}_I$ at initial configuration $\mathbf{x}_I$ - against the element $\circMa$ is a derived physical quantity which specifies how its momentum evolves
\be\label{Abschnitt -- differentiated analysis - force - physical quantity and momentum_vector}
   \mathbf{F}_a^{(\mathcal{A})} \! \left[ w\big|_{\mathbf{x}_I,\mathbf{v}_I} \right] \cdot \Delta t_a^{(\mathcal{A})} \;\; := \;\; \Delta \mathbf{p}_a^{(\mathcal{A})}  \;\; .
\ee
\end{de}
\begin{co}
If element $\circMa$ moves - by its inertia and given initial conditions $\mathbf{v}_a$ - along infinitesimal way $\Delta \mathbf{s}_a \!= \mathbf{v}_a \cdot \Delta t$ the force $\mathbf{F}_a^{(\mathcal{A})}$ also determines how its kinetic energy evolves
\be\label{Abschnitt -- differentiated analysis - force - physical quantity and energy_scalar}
   \mathbf{F}_a^{(\mathcal{A})} \! \left[ w\big|_{\mathbf{x}_I,\mathbf{v}_I} \right]  \cdot \Delta \mathbf{s}_a^{(\mathcal{A})} \;\; = \;\; \Delta E_{\mathrm{kin}\: a}^{(\mathcal{A})}   \;\; .
\ee
\end{co}
\textbf{Proof:}
The force against element $\circMa_{\:\mathbf{v}_a}$ at velocity $\mathbf{v}_a$ - in an interaction $w$ with element $\circMb_{\:\mathbf{v}_b}$ at velocity $\mathbf{v}_b$ - satisfies
\be
   \mathbf{F}_a^{(\mathcal{A})}\big|_{\mathbf{v}_a,\mathbf{v}_b} \cdot \Delta \mathbf{s}_a^{(\mathcal{A})} \;\; \stackrel{(\ref{Abschnitt -- differentiated analysis - force - physical quantity and momentum_vector})(\ref{Abschnitt -- differentiated analysis - E_kin p from particles in infinetesimal action})}{=} \;\; \frac{m_a^{(\mathcal{A})} \cdot \Delta\mathbf{v}_a^{(\mathcal{A})}}{\Delta t^{(\mathcal{A})}} \;\cdot\; \mathbf{v}_a^{(\mathcal{A})} \cdot \Delta t^{(\mathcal{A})} \;\; \stackrel{(\ref{Abschnitt -- differentiated analysis - E_kin p from particles in infinetesimal action})}{=} \;\; \Delta E_{\mathrm{kin}\: a}^{(\mathcal{A})} \;\; . \nn
\ee
\qed
\begin{lem}\label{Lem - differentiated analysis - force - physical quantities counterforce and boost}
Let boosted inertial observers $\mathcal{A}$lice and $\mathcal{B}$ob measure the same interaction in system $\circMa \cup \circMb\,$. Their physical quantities for force against elements $\circMa$ and $\circMb$ satisfy
\bea
   \mathbf{F}_a^{(\mathcal{A})}  & = & - \mathbf{F}_b^{(\mathcal{A})}  \label{Abschnitt -- differentiated analysis - force - a b} \\
   \mathbf{F}_a^{(\mathcal{A})}  & = & \mathbf{F}_a^{(\mathcal{B})}  \;\; .  \label{Abschnitt -- differentiated analysis - force - Alice Bob}
\eea
\end{lem}
\textbf{Proof:}
By momentum conservation $\Delta \mathbf{p}_a + \Delta \mathbf{p}_b \stackrel{(\ref{Abschnitt -- Potential of Mechanical System - potential field - conservation total momentum elements})}{=} 0$ in two-partite system $\circMa \cup \circMb$
\[
   \mathbf{F}_a^{(\mathcal{A})} \;\; := \;\; \frac{\Delta \mathbf{p}_a^{(\mathcal{A})}}{\Delta t } \;\; = \;\; -  \frac{\Delta \mathbf{p}_b^{(\mathcal{A})}}{\Delta t } \;\;=:\;\; - \mathbf{F}_b^{(\mathcal{A})}
\]
the forces against $\circMa$ resp. $\circMb$ are oriented antiparallel with same strength.

Let $\mathcal{A}$lice move with constant velocity $\mathbf{v}_{\mathcal{A}} = v_{\mathcal{A}}^{(\mathcal{B})} \cdot \mathbf{v}_{\mathbf{1}^{(\mathcal{B})}}$ relative to $\mathcal{B}$ob. Their measured values for initial and final velocity of same bodies $\circMa\:,\circMb$ transform covariant $v_I^{(\mathcal{B})} = v_I^{(\mathcal{A})} + v_{\mathcal{A}}^{(\mathcal{B})}$ (see Remark \ref{Rem - covariant kinematical transformation}). In Galilei Kinematics they measure same values for acceleration
\[
   \Delta v_I^{(\mathcal{A})} \;\; = \;\; \Delta v_I^{(\mathcal{B})}
\]
in same duration $\Delta t^{(\mathcal{A})} = \Delta t^{(\mathcal{B})}$ and hence by same quantity equations (induced from intrinsically equivalent calorimeter measurements $\mathrm{RB}^{(\mathcal{A})}$ and $\mathrm{RB}^{(\mathcal{B})}$) the same physical quantity
\[
   \Delta p_I^{(\mathcal{A})} \; \stackrel{(\ref{Abschnitt -- differentiated analysis - E_kin p from particles in infinetesimal action})}{=} \; m_I \cdot \Delta v_I^{(\mathcal{A})} \;\; = \;\; m_I \cdot \Delta v_I^{(\mathcal{B})} \; \stackrel{(\ref{Abschnitt -- differentiated analysis - E_kin p from particles in infinetesimal action})}{=} \; \Delta p_I^{(\mathcal{B})}
\]
for infinitesimal momentum changes $\Delta \mathbf{p}_I = \Delta p_I^{(\mathcal{A})} \cdot \mathbf{p}\left[ \circMunit_{\:\mathbf{v}_{\mathbf{1}^{(\mathcal{A})}}} \right] = \Delta p_I^{(\mathcal{B})} \cdot \mathbf{p}\left[ \circMunit_{\:\mathbf{v}_{\mathbf{1}^{(\mathcal{B})}}} \right]$ and thus same force against element $\circMa$
\[
   F_a^{(\mathcal{A})}  \;\; := \;\; \frac{\Delta p_a^{(\mathcal{A})}}{\Delta t^{(\mathcal{A})} } \;\; = \;\; \frac{\Delta p_a^{(\mathcal{B})}}{\Delta t^{(\mathcal{B})} } \;\; =: \;\; F_a^{(\mathcal{B})} \;\; .
\]
\qed
\begin{theo}\label{Theorem - differentiated analysis - force - v-unabh}
The force of an interaction $w$ in conservative system $\circMa_{\:\mathbf{v}_a} \cup \circMb_{\:\mathbf{v}_b} \big|_{\mathbf{x}_a,\mathbf{x}_b}$ against element $\circMa$ is independent from the initial velocity $\mathbf{v}_I=(\mathbf{v}_a,\mathbf{v}_b)$ of both elements
\be
   \mathbf{F}_a^{(\mathcal{A})}  \;\; := \;\; \frac{\Delta \mathbf{p}_a^{(\mathcal{A})}}{\Delta t^{(\mathcal{A})} }   \;\, / \;\, \mathrm{mod} \; \mathbf{v}_a,\mathbf{v}_b   \;\; .
\ee
\end{theo}
\textbf{Proof:}
For an infinitesimal segment of their interaction $w$
\be\label{Abschnitt -- differentiated analysis - force - infinitesimal segment of intrinsic action}
   \circMa \cup \circMb_{\;t,\: \mathbf{x}_I,\: \mathbf{v}_I} \; \stackrel{w}{\Rightarrow} \;
   \circMa \cup \circMb_{\;t'=t+\Delta t\:,\; \mathbf{x}'_I\simeq\mathbf{x}_I+\mathbf{v}_I\cdot \Delta t\:,\; \mathbf{v}'_I=\mathbf{v}_I + \Delta \mathbf{v}_I }
\ee
we begin from well-defined potential energy $V_{\mathrm{pot}}[\mathbf{x}_a,\mathbf{x}_b]$ of the system (directly measurable by equipollence principle) and then analyze individual elements $\circMa_{\:\mathbf{v}_a}  \, / \mathrm{mod} \, \mathbf{v}_b$ and vice versa.

Let $\mathcal{A}$lice set up intrinsic process $w$ in system $\circMa \cup \circMb_{\:\mathbf{x}_I,\mathbf{v}_I/\mathbf{w}_I}$ at same initial configuration $\mathbf{x}_I = (\mathbf{x}_a,\mathbf{x}_b)$ once with initial conditions $\mathbf{v}_I = (\mathbf{v}_a,\mathbf{v}_b)$ and once with $\mathbf{w}_I = (\mathbf{w}_a,\mathbf{w}_b)$ (see figure \ref{pic_Wirkungs-Steuerung_selber-Pfad-Aquipollenz}). Without restricting generality $\mathcal{A}$lice \emph{prepares} both initial velocities so that total momentum of the system $m_a \cdot \mathbf{v}_a + m_b \cdot \mathbf{v}_b \stackrel{!}{=} 0$ resp. $m_a \cdot \mathbf{w}_a + m_b \cdot \mathbf{w}_b \stackrel{!}{=} 0$ vanishes (center of mass frame). In both processes $w$ resp. $\tilde{w}$ we let elements $\circMa\: , \circMb$ run through same configuration changes
\be\label{Abschnitt -- differentiated analysis - force - v-unabh - Kopplungsbedingung}
   \mathbf{v}_I\cdot \Delta t  \;\; \stackrel{!}{=} \;\; \mathbf{w}_I\cdot \Delta T
\ee
in durations $\Delta t \neq \Delta T$ corresponding to initial conditions $\mathbf{v}_I$ resp. $\mathbf{w}_I$.

Potential energy from the same configuration transition transforms into kinetic energy
\bea
   & & \!\!\!\!\!\!\!\!\!\!\!\!\!\!\!\!\!\!\!\!
   V_{\mathrm{pot}} \left[ \mathbf{x}_I\Rightarrow\mathbf{x}_I+\mathbf{v}_I\cdot \Delta t \right] \;\; = \;\; -\Delta E_{\mathrm{kin}\: a}^{(\mathcal{A})} - \Delta E_{\mathrm{kin}\: b}^{(\mathcal{A})}  \nn \\
   & &
   \stackrel{(\ref{Abschnitt -- differentiated analysis - force - physical quantity and energy_scalar})}{=} \;\;- \mathbf{F}_a^{(\mathcal{A})} \! \left[ w \right]  \cdot \Delta \mathbf{s}_a^{(\mathcal{A})} - \mathbf{F}_b^{(\mathcal{A})} \! \left[ w \right] \cdot \Delta \mathbf{s}_b^{(\mathcal{A})} \;\; \stackrel{(\ref{Abschnitt -- differentiated analysis - force - a b})}{=} \; - \mathbf{F}_a^{(\mathcal{A})} \! \left[ w \right]  \cdot (\mathbf{v}_a\cdot \Delta t - \mathbf{v}_b\cdot \Delta t)  \nn
\eea
and under modified initial conditions $\mathbf{w}_I$ into same kinetic energy
\bea
   V_{\mathrm{pot}} \left[ \mathbf{x}_I\Rightarrow\mathbf{x}_I+\mathbf{w}_I\cdot \Delta T \right] & = & \ldots \;\; = \;\; - \mathbf{F}_a^{(\mathcal{A})} \! \left[ \tilde{w} \right]  \cdot (\mathbf{w}_a\cdot \Delta T - \mathbf{w}_b\cdot \Delta T)  \nn \;\; .
\eea
By matching condition (\ref{Abschnitt -- differentiated analysis - force - v-unabh - Kopplungsbedingung}) in both processes the force $\mathbf{F}_a^{(\mathcal{A})} \left[ w \right] = \mathbf{F}_a^{(\mathcal{A})} \left[ \tilde{w} \right]$ against element $\circMa$ is the same.

Consider an \emph{active} boost of the entire system $\circMa_{\:\mathbf{v}_a} \cup \circMb_{\:\mathbf{v}_b} \big|_{\mathbf{x}_a,\mathbf{x}_b}$ from $\mathcal{A}$lice towards $\mathcal{B}$ob. Let $\mathcal{A}$lice move relative to $\mathcal{B}$ob with constant velocity
$\mathbf{v}_{\mathcal{A}} = \left( - v_a^{(\mathcal{A})} - \frac{\Delta v_a^{(\mathcal{A})}}{2}  \right) \cdot \mathbf{v}_{\mathbf{1}^{(\mathcal{B})}}$. In $\mathcal{B}$ob's frame intrinsic action $w_{\mathcal{B}}$ (\ref{Abschnitt -- differentiated analysis - force - infinitesimal segment of intrinsic action}) evolves in an \emph{intrinsically equivalent} way - with same intrinsic duration $\Delta t^{(\mathcal{B})}$, configuration changes $\Delta\mathbf{x}_I^{(\mathcal{B})}$ and acceleration $\Delta\mathbf{v}_I^{(\mathcal{B})}$. In Galilei kinematics $\mathcal{A}$lice measures for $\mathcal{B}$ob's boosted action $w_{\mathcal{B}}$ \footnote{This is a \emph{passive} transformation of physical quantities of same objects in $w_{\mathcal{B}}$ with regard to different observers $\mathcal{A}$lice and $\mathcal{B}$ob.}
\be\label{Abschnitt -- differentiated analysis - force - infinitesimal segment of Bob}
   \circMa_{\: - \frac{\Delta\mathbf{v}_a}{2}} \cup  \circMb_{\:
   \mathbf{v}_b - \mathbf{v}_a - \frac{\Delta\mathbf{v}_a}{2}} \;\; \stackrel{w_{\mathcal{B}}}{\Rightarrow} \;\;
   \circMa_{\: + \frac{\Delta\mathbf{v}_a}{2}} \cup  \circMb_{\:
   \left(\mathbf{v}_b - \mathbf{v}_a - \frac{\Delta\mathbf{v}_a}{2}\right) + \Delta\mathbf{v}_b}
\ee
the same forces against element $\circMa$ like $\mathcal{B}$ob and like in her intrinsic action $w_{\mathcal{A}}$ (\ref{Abschnitt -- differentiated analysis - force - infinitesimal segment of intrinsic action})
\be\label{Abschnitt -- differentiated analysis - force - v-unabh - Kraft Alice und Bob gegen a}
   \mathbf{F}_a^{(\mathcal{A})} \left[ w_{\mathcal{B}} \right] \;\; \stackrel{(\ref{Abschnitt -- differentiated analysis - force - Alice Bob})}{=} \;\; \mathbf{F}_a^{(\mathcal{B})} \left[ w_{\mathcal{B}} \right]
   \;\; \stackrel{(\mathrm{Equiv.})}{=} \;\; \mathbf{F}_a^{(\mathcal{A})} \left[ w_{\mathcal{A}} \right]  \;\; .
\ee
Similarly let $\mathcal{A}$lice move relative to $\mathcal{C}$harlie with velocity $\mathbf{v}_{\mathcal{A}} =\! \left( - \mathbf{w}_a^{(\mathcal{A})} - \frac{\Delta\mathbf{w}_a^{(\mathcal{A})}}{2}  \right) \cdot \mathbf{v}_{\mathbf{1}^{(\mathcal{C})}}$. Again $\mathcal{A}$lice measures for $\mathcal{C}$harlie's boosted action $\tilde{w}_{\mathcal{C}}$
\be\label{Abschnitt -- differentiated analysis - force - infinitesimal segment of Charlie}
   \circMa_{\: - \frac{\Delta\mathbf{w}_a}{2}} \cup  \circMb_{\:
   \mathbf{w}_b - \mathbf{w}_a - \frac{\Delta\mathbf{w}_a}{2}} \;\; \stackrel{\tilde{w}_{\mathcal{C}}}{\Rightarrow} \;\;
   \circMa_{\: + \frac{\Delta\mathbf{w}_a}{2}} \cup  \circMb_{\:
   \left(\mathbf{w}_b - \mathbf{w}_a - \frac{\Delta\mathbf{w}_a}{2}\right) + \Delta\mathbf{w}_b}
\ee
the same forces against element $\circMa$ as for $\mathcal{B}$ob's boosted action $ w_{\mathcal{B}}$
\be\label{Abschnitt -- differentiated analysis - force - v-unabh - Kraft Alice und Charlie gegen a}
   \mathbf{F}_a^{(\mathcal{A})} \left[ \tilde{w}_{\mathcal{C}} \right] \;\; \stackrel{(\ref{Abschnitt -- differentiated analysis - force - v-unabh - Kraft Alice und Bob gegen a})}{=} \;\; \mathbf{F}_a^{(\mathcal{A})} \left[ w_{\mathcal{B}} \right]  \;\; .
\ee

$\mathcal{A}$lice measures two intrinsic processes $ w_{\mathcal{B}}$ and $ \tilde{w}_{\mathcal{C}}$ in same system $\circMa \cup  \circMb_{\:\mathbf{x}_I}$. At same initial configuration $\mathbf{x}_I = \left( \mathbf{x}_a , \mathbf{x}_b  \right)$ she prepares suitable initial conditions
\bea
   \mathbf{v}_I \left[ w_{\mathcal{B}} \right]  & \stackrel{(\ref{Abschnitt -- differentiated analysis - force - infinitesimal segment of Bob})}{=} &
   ( \; - \underbrace{\frac{\Delta\mathbf{v}_a}{2}}_{\simeq\: \mathbf{0}}  \;\;,\;\; \underbrace{\mathbf{v}_b - \mathbf{v}_a - \frac{\Delta\mathbf{v}_a}{2}}_{\simeq\: \mathbf{v}_b - \mathbf{v}_a} \;  ) \nn \\
   \mathbf{v}_I \left[ \tilde{w}_{\mathcal{C}} \right]  & \stackrel{(\ref{Abschnitt -- differentiated analysis - force - infinitesimal segment of Charlie})}{=} &
   ( \; - \underbrace{\frac{\Delta\mathbf{w}_a}{2}}_{\simeq\: \mathbf{0}}  \;\;,\;\; \underbrace{\mathbf{w}_b - \mathbf{w}_a - \frac{\Delta\mathbf{w}_a}{2}}_{\simeq\: \mathbf{w}_b - \mathbf{w}_a} \;  ) \nn   \;\; .
\eea
In both cases initially element $\circMa_{\: \mathbf{0}}$ is (practically) at rest while element $\circMb$ is set up with different initial velocity $\mathbf{v}_b - \mathbf{v}_a$ resp. $\mathbf{w}_b - \mathbf{w}_a$.\footnote{For infinitesimal segments of intrinsic action $w$ we have $\Delta\mathbf{v}_{a/b} \ll \mathbf{v}_{a/b}$ and in so called center of mass frame $\mathbf{v}_{a}$ and $\mathbf{v}_{b}$ are oriented antiparallel (\ref{Abschnitt -- Potential of Mechanical System - potential field - total momentum of system}), (\ref{Abschnitt -- kin quant Absorptions Wirkung - kin energy and momentum metrisiert}).} Therefore the force against same (resting) element $\circMa_{\: \mathbf{0}}$ (\ref{Abschnitt -- differentiated analysis - force - v-unabh - Kraft Alice und Charlie gegen a}) in system $\circMa_{\:\mathbf{0}} \cup \circMb_{\:\mathbf{v}_b - \mathbf{v}_a} \big|_{\mathbf{x}_I}$ and in system $\circMa_{\:\mathbf{0}} \cup \circMb_{\:\mathbf{w}_b - \mathbf{w}_a} \big|_{\mathbf{x}_I}$ does not depend on initial velocity of element $\circMb\:$. And in reverse the force against element $\circMb_{\:\mathbf{v}_b}$ (in intrinsic action $w$) in system $\circMa_{\:\mathbf{0}} \cup \circMb_{\:\mathbf{v}_b} \big|_{\mathbf{x}_I}$ with same resting element $\circMa_{\: \mathbf{0}}$ in configuration $\mathbf{x}_I$ does not depend on its velocity $\mathbf{v}_b$ (see Lemma \ref{Lem - differentiated analysis - force - physical quantities counterforce and boost}). In common words: For \emph{same physical ''source''} $\circMa_{\:\mathbf{0}}$ (at rest) the force against different ''test-particles'' $\circMb_{\:\mathbf{v}_b}\big|_{\mathbf{x}_I}$ is velocity $\mathbf{v}_b$ independent.
\qed
In Galilei kinematics the force of intrinsic action $w$ in system $\circMa_{\:\mathbf{v}_a} \cup \circMb_{\:\mathbf{v}_b}$ is independent from the actual velocity of both elements $\mathbf{v}_a$ and $\mathbf{v}_b$. In Definition \ref{Def - differentiated analysis - force} we can skip to specify its initial condition $\mathbf{v}_I$. Hence the invariant numerical ratio ''force''
\be
   \mathbf{F}_a^{(\mathcal{A})} \;\; := \;\; \frac{\Delta \mathbf{p}_a^{(\mathcal{A})}}{\Delta t_a^{(\mathcal{A})}} \left[ w\big|_{\mathbf{x}_I,\mathbf{v}_I} \right]    \;\, / \;\, \mathrm{mod} \; \mathbf{v}_I  \nn
\ee
is a meaningful \emph{derived} physical quantity. Provided the physical conditions for deriving force are justified, this property is more significant than a formal abbreviation in the formalism.


\subsection{Displacement work}\label{Kap - KM Dynamics - Differentiated Analysis - Displacement Work}

\begin{de}
\underline{Displacement work} $\nabla^{(a)} V_{\mathrm{pot}} \cdot \,\delta \mathbf{s}_a $ is the steering energy for generating a partial configuration variation $\mathbf{s}_a, \mathbf{s}_b \Rightarrow \mathbf{s}_a + \delta \mathbf{s}_a, \mathbf{s}_b$ for element $\circMa$ and preserving frozen position for element $\circMb\,$.
\end{de}
\begin{pr}\label{Prop - differentiated analysis - force - grad Verschiebungsarbeit}
In Galilei kinematics the displacement work for steering (all elements of) a conservative system $\circMa \cup \circMb\: \big|_{\mathbf{s}_I}$ into a partial configuration variation $\mathbf{s}_I \Rightarrow \mathbf{s}_I + \delta \mathbf{s}_a$
\be\label{Abschnitt -- differentiated analysis - displacement work - displacement work}
   - \nabla^{(a)} V_{\mathrm{pot}} \cdot \delta \mathbf{s}_a \;\; = \;\; \mathbf{F}_a \cdot \delta \mathbf{s}_a
\ee
determines the force (of free evolving internal interaction $w$) against element $\circMa\,$.
\end{pr}
\textbf{Proof:}
We also generate the partial configuration transition by a generic steering process $\mathrm{RB}^{(1)} \ast w_1 \ast \ldots \ast \mathrm{RB}^{(n)} \ast w_n$ with segments of undisturbed intrinsic actions $w_i$ along $\mathbf{s}^{(i)}_I\Rightarrow\mathbf{s}^{(i)}_I + \delta \mathbf{s}_a^{(i)}$ and suitable steering interventions $\mathrm{RB}^{(i)}$ for preparing initial conditions (see Lemma \ref{Lem - KM Dynamics - Potential of Mechanical System - Steering Action}). Potential energy $V_{\mathrm{pot}} \!\left[ \mathbf{s}_I\!\Rightarrow\!\mathbf{s}'_I \right]$ is only configuration dependent (\ref{Abschnitt -- Potential of Mechanical System - potential field - wirkungs-weg-unabh}); along each infinitesimal segment we liberate $V_{\mathrm{pot}} \!\left[ w \right] = \nabla^{(a)} V_{\mathrm{pot}}\big|_{\mathbf{s}_I} \cdot \delta \mathbf{s}_a $. Let the free evolving internal interaction
\[
   \circMa_{\:\mathbf{s}_a}\! \cup \circMb_{\:\mathbf{s}_b}\big|_{\mathbf{v}_I}  \;\; \stackrel{w}{\Rightarrow} \;\;  \circMa_{\:\mathbf{s}_a + \Delta \mathbf{s}_a}\! \cup \circMb_{\:\mathbf{s}_b + \Delta \mathbf{s}_b}\big|_{\mathbf{v}_I'}
\]
generate the same total intrinsic configuration transition\footnote{By momentum conservation internal forces in two-partite system $\circMa \cup \circMb$ are head-on. Without restricting generality in the center of mass frame mutual forces and displacements $\Delta \mathbf{s}_a, \Delta \mathbf{s}_b$ are collinear with $\delta \mathbf{s}_a$.}
\be\label{Abschnitt -- differentiated analysis - displacement work - matching configuration transition}
   \Delta \mathbf{s}_a - \Delta \mathbf{s}_b \;\; = \;\; \delta \mathbf{s}_a  \;\; .
\ee
Then the same potential energy transforms into kinetic energy
\bea
   - \nabla^{(a)} V_{\mathrm{pot}} \cdot \delta \mathbf{s}_a \;\; \stackrel{(\ref{Abschnitt -- Energie - potential Energy})}{=} \;\;
   \underbrace{\Delta E_{\mathrm{kin} \: a}}_{\stackrel{(\ref{Abschnitt -- differentiated analysis - force - physical quantity and energy_scalar})}{=}\: \mathbf{F}_a \cdot \Delta \mathbf{s}_a } \; + \; \underbrace{\Delta E_{\mathrm{kin} \: b}}_{\stackrel{(\ref{Abschnitt -- differentiated analysis - force - physical quantity and energy_scalar})}{=}\: \mathbf{F}_b \cdot \Delta \mathbf{s}_b } \;\; \stackrel{(\ref{Abschnitt -- differentiated analysis - force - a b})}{=} \;\; \mathbf{F}_a \cdot \underbrace{( \Delta \mathbf{s}_a - \Delta \mathbf{s}_b )}_{\stackrel{(\ref{Abschnitt -- differentiated analysis - displacement work - matching configuration transition})}{=}\: \delta \mathbf{s}_a } \nn  \;\; .
\eea
In a two-partite system the force is parallel to the intrinsic separation $\mathbf{s}_a - \mathbf{s}_b\,$; hence also the partial gradient of potential energy $\nabla^{(a)} V_{\mathrm{pot}}\!\left[ \mathbf{s}_a , \mathbf{s}_b \right]$ is oriented in the same way
\be
   - \nabla^{(a)} V_{\mathrm{pot}} [\mathbf{s}_a, \mathbf{s}_b] \;\; = \;\; \mathbf{F}_a \;\; / \;\, \mathrm{mod} \; \mathbf{v}_a, \!\mathbf{v}_b    \nn
\ee
and by Theorem \ref{Theorem - differentiated analysis - force - v-unabh} the same also for generic relative motion of both elements in $\circMa \cup \circMb\,$.

For n-body system $G_1\cup\ldots\cup G_N$ we divide the potential energy $V_{G_1\cup\ldots\cup G_N} := \sum_{i<j} V_{G_i\cup G_j}$ for two-partite subsystems $V_{G_i\cup G_j}$ - which by superposition principle do not depend on other elements $G_k$ for $k\neq i,j$. Then the force of intrinsic action $w$ against element $G_i$ is given by
\be
   \mathbf{F}_i \;\; = \;\; - \sum_{j \neq i}  \; \nabla^{(i)} V_{G_i\cup G_j} \nn  \;\; .
\ee
\qed


\subsection{Equation of motion}\label{Kap - KM Dynamics - Differentiated Analysis - Evolution}

We specify an intrinsic interaction $w$ in a conservative system $\circMa_{\:\mathbf{v}_a} \!\cup \circMb_{\:\mathbf{v}_b} \big|_{\mathbf{x}_a,\mathbf{x}_b}$ of arbitrary internal structure (based on pre-theoretic ordering relations \{\ref{Kap - KM Dynamics - Physical Measurement - Pre-theoretical Ordering Relation}\}, quantification scheme \{\ref{Kap - KM Dynamics - Basic Dynamical Measures - Energy - quantification scheme}\} and action principles \{\ref{Kap - KM Dynamics - Potential of Mechanical System}\}) by basic physical quantities of energy and momentum. The force $\mathbf{F}_i$ specifies their distribution between all elements $i\in I$ of the system along infinitesimal segments of the process $w: t,\mathbf{s}_I\Rightarrow t+\Delta t, \mathbf{s}_I + \Delta \mathbf{s}_I$.
\be\label{Abschnitt -- differentiated analysis - evolution - differentzierte impuls Verlauf}
\begin{diagram}
   \mathbf{p}_{i}\big|_{t+\Delta t}    &   \!\!\!\!\!\!\!\!\!\!
      \stackrel{(\mathrm{Theo.} \ref{Theorem - differentiated analysis - force - v-unabh})}{=:} \;\;  \underbrace{\mathbf{F}_i \! \left[ w\big|_{\mathbf{v}_I} \right]  \: / \: \mathrm{mod} \; \mathbf{v}_I}_{\stackrel{(\mathrm{Prop.}\ref{Prop - differentiated analysis - force - grad Verschiebungsarbeit})}{=}\: - \nabla^{(i)} V_{\mathrm{pot}}\big|_{\mathbf{x}_I}}  \;\; \cdot \;\;  \Delta t \;\;\;\; +  & \;\;\; \mathbf{p}_{i}\big|_{t} \;\;\;  \\
\dTo^{\mathrm{RB}_i \;}   &   &
   \dTo_{\; \mathrm{RB}_i}   \\
    m_i \cdot \left( \mathbf{v}_i + \Delta\mathbf{v}_i \right)   &     &  m_i \cdot \mathbf{v}_i
\end{diagram}
\ee
Since the force of an intrinsic interaction is independent of the initial condition $\mathbf{v}_I$ and given by the gradient of the displacement work $- \nabla^{(i)} V_{\mathrm{pot}}$ at the momentary configuration $\mathbf{x}_I$ we can determine the momentum evolution. With the basic relation between impact velocity and momentum in a calorimeter (\ref{Abschnitt -- differentiated analysis - E_kin p from particles in infinetesimal action}) we inherit the equation of motion
\be\label{Abschnitt -- differentiated analysis - evolution - inherited equations of motion}
   \;\;\;\;\;\;\; m_i \cdot \frac{\mathrm{d}^2 \mathbf{s}_i }{{\mathrm{d} t}^2} \;\; = \;\; - \nabla^{(i)} V_{\mathrm{pot}} \;\;\;\;\;\;\;\;\; \forall \; i\in I
\ee
for every element. From the infinitesimal analysis of momentum evolution (\ref{Abschnitt -- differentiated analysis - evolution - differentzierte impuls Verlauf}) we \emph{induce} Newton's equations for the evolution of motion. The displacement work (\ref{Abschnitt -- differentiated analysis - displacement work - displacement work}) associated with the momentary inertial motion in the system $(\Delta t, \Delta \mathbf{s}_I)\big|_{t}$ successively causes infinitesimal variations of the evolving  state of motion $(\Delta t, \Delta \mathbf{s}_I)\big|_{t+\Delta t}$.

Next consider mechanical systems with \emph{built in} constraints. We analyze the spatiotemporal evolution of intrinsic action $w$ in closed system $E\cup G_1 \cup \ldots \cup G_N$. In addition we subdivide into external element $E$ (e.g. earth) and ''inner'' elements $G_1 \cup \ldots \cup G_N$ (e.g. bound parts of a physical pendulum). The latter have fixed rigid connections among one another. They enforce partial fixation of sought after solution to the equations of motion $\mathbf{s}_i - \mathbf{s}_j \stackrel{!}{=} \mathrm{const.}$ for all inner elements $i,j\in I := \{1,\ldots ,N\}$. The motion of rigid subsystem $G_1 \cup \ldots \cup G_N$ must be compatible with $n$ (holonomic) constraints. We parameterize admissible displacements (degrees of freedom of the subsystem) by $3 N-n$ generalized coordinates $q$
\[
   \Delta \mathbf{s}_I \;\; = \;\; \sum_{k=1}^{3 N-n} \; \frac{\partial \mathbf{s}_I}{\partial q_k } \cdot \, \Delta q_k \;\; .
\]

D'Alembert postulates: (\emph{inner}) constraint forces do not contribute to displacement work. For (admissible) inertial displacements - due to given initial velocity $\dot{q} \big|_t$ - the transformation of kinetic energy (\ref{Abschnitt -- differentiated analysis - force - physical quantity and energy_scalar})
\be\label{Abschnitt -- differentiated analysis - evolution - differentzierte Energie Verlauf}
   \Delta E_{\mathrm{kin} \; I} \;\; = \;\; - \nabla^{(q)} V_{\mathrm{pot}} \cdot \Delta q
\ee
is determined by (\emph{applied} forces from) the potential $ V_{\mathrm{pot}} = \sum_{i\in I} V_{E\cup G_i}$ of the inner parts of the subsystem $G_i$ with the external element $E$. With the basic kinetic energy-velocity relation (\ref{Abschnitt -- differentiated analysis - E_kin p from particles in infinetesimal action}) we obtain (Lagrange's form of) the equations of motion
\bea
   0  &  \stackrel{(\ref{Abschnitt -- differentiated analysis - evolution - differentzierte Energie Verlauf})}{=}  & \sum_{i=1}^{N} \; m_i \cdot \underbrace{\mathbf{v}_i}_{\approx\: \frac{\Delta \mathbf{s}_i}{\Delta t}} \cdot \underbrace{\Delta\mathbf{v}_i}_{=\: \mathbf{a}_i \cdot \Delta t}  \;\;+\:\; \underbrace{\nabla^{(q_K)} V_{\mathrm{pot}}}_{=\: \nabla^{(\mathbf{s}_I)} V_{\mathrm{pot}} \cdot \frac{\partial \mathbf{s}_I}{\partial q_K } }  \cdot \; \Delta q_K  \nn  \\
   & = & \sum_{i=1}^{N} \left( m_i \cdot \mathbf{a}_i \; + \; \nabla^{(i)} V_{\mathrm{pot}}  \right) \cdot \underbrace{\frac{\partial \mathbf{s}_i}{\partial q_K } \cdot \Delta q_K}_{=:\: \delta \mathbf{s}_i} \nn
\eea
for all admissible - so called - ''virtual displacements'' $\delta \mathbf{s}_I$. D'Alembert's ''principle of virtual work'' is an additional postulate to account for the collective effect of constraint forces. Without determining the details of unknown inner binding actions this method provides \emph{reduced equations of motion} for rigid bound subsystem - in generalized coordinates.\footnote{Ruben outlines the historic development \cite{Peter '67 - zum Streit um das wahre Mass der Kraft}: Newton's mechanics is essentially formulated for the free point mass. The formation of mechanics for systems was stimulated by the question of Mersenne 1646, to determine the center of oscillation for a physical pendulum. The difficulty was that (constraint) forces - due to rigid spatial connections - were unknown. They had to be determined from their ''effect''. Huygens 1673 recognized the law according to which different elements of the composite pendulum mutually influence their motion which is driven by respective force of
gravity: If several weights which are attached to a pendulum fall then their center of gravity swings back to same height independently whether the rigid spatial connections are separated or not. Huygens solution is based on the principle of impossibility of a perpetuum mobile. D'Alembert 1743 introduced the differential formulation of this principle and Lagrange - using the concept of work - brought it into the familiar form: Constraint forces do not provide work! \cite{Mach - Mechanik in ihrer Entwicklung}}


\section{Principle of least action functional}\label{Kap - KM Dynamics - Principle of Least Action}

\subsection{Historic development}\label{Kap - KM Dynamics - Principle of Least Action - Historic Development}

Szab\'o \cite{Szabo Geschichte der Mechanischen Prinzipien} outlines the historic development. Fermat 1629 has been first to take a general principle as a basis for motion: ''the only requirement is that nature always proceeds on the way of least resistance... but not, as people generally assume, that nature always chooses the shortest way''. Leibniz 1708 introduced a quantity of action and explained: ''The action is not what you think, here the consideration of time is inevitable; the action is like the product of $(m,\mathbf{s},\mathbf{v})$ or of $(t, E_{\mathrm{kin}})$. I have noticed that during changes of motion it always turns into a maximum or minimum.'' Euler 1743 gave this proposition a mathematically immaculate form. He recognized that: ''all actions in nature obey some law of maximum or minimum... Some property of maximum or minimum is localized in the trajectory of thrown objects... The nature of this property can not be seen easily from metaphysical principles. The trajectories in question
are ascertainable by direct methods (with less calculus)... so that one can determine what of them is maximal or minimal. (Euler particularly regards) the resulting effect of acting forces on the state of motion of bodies... I did not discover these interesting connections a priori but only a posteriori... after several attempts... I found the expression for a quantity which turns into a minimum during natural motion'' \cite{Suisky - Euler}.

While Leibniz (just as Maupertuis) suggests a teleological guiding principle: ''that the actual world is the best of all possible worlds'' - Euler ascribes to his quantity of action no further validity beyond the examined cases. According to Euler no general principle is found. Bavink \cite{Bavink Ergebnisse und Probleme der Naturwissenschaften} describes the teleological interpretation: ''It seems as if nature selects from \emph{many per se possible} motions the one which achieves a largest possible effect through least possible means... To the present day Hamilton's principle has to serve for lines of thought... which see processes evolve, as if so to say \emph{nature had to consider} at the beginning of time period $t_2-t_1$ how to keep the value of an \emph{integral} $\int_{t_1}^{t_2} E \mathrm{d}t$ as low as possible.'' Bavink diagnoses ''these propositions essentially contain nothing but the statement, that under certain circumstances something certain happens: the principle of causality. What really
happens is determined by differential equations and every differential equation can be regarded as a condition that a certain function turns into a minimum or maximum - in the latter one even has a wide freedom of choice. With the validity of equations of motion for mechanics one can theoretically state many other functions of this sort.''

What in Bavink's purely mathematical point of view appears - as a sober statement of facts, as complete equivalence of differential principle and integral principle (about \emph{physical quantities}) - we reconsider taking into account underlying \emph{physical conditions}. In reality what happens is determined by equations of motion \{\ref{Kap - KM Dynamics - Differentiated Analysis - Evolution}\}. Given initial conditions $\mathbf{s}_I,\mathbf{v}_I$ permit - not many but - exactly one internal process. To realize other possible processes requires external steering interventions. In order to run along different paths $\gamma$ resp. $\gamma +\delta\gamma$ through same end point configurations $\mathbf{s}_I\Rightarrow\mathbf{s}'_I$ in fixed duration $\Delta t$ - not nature but - steering physicist needs to consider how to couple temporary steering actions $\mathrm{RB}^{(i)}$. He has the choice between many possible (types of) steering options. The integral named ''coercion'' or ''action'' has physical meaning only in its role as variation functional. A variation essentially analyzes the difference between two - differently steered - processes of an interaction.

Planck \cite{Planck - Wege zur physikalischen Erkenntnis} emphasizes, to define a variational principle requires stipulation of
\begin{itemize}
  \item   \emph{conditions} for ''virtual motions'', i.e. the repertory of processes \emph{from which to choose} (resp. what steerable processes of the interaction we compare with one another)
  \item   quantity of ''action'', i.e. the \emph{characteristic with regard to which the selection is made}.
\end{itemize}
''The former is of exactly same importance as the quantity of action itself - Planck explains - because depending on the type of stipulated variation conditions the content of the principle takes a completely different meaning!\footnote{The quantity of ''action'' has no absolute meaning without above selection criteria.} It took a long time until that - long disregarded - circumstance was understood clearly and let after precursors Leibniz, Maupertuis, Euler to the first correct version of the principle of least action.''

Lagrange 1760 compares motions in a system of material points $\gamma$ resp. $\gamma +\delta\gamma^{(\mathrm{Lagr})}$ between fixed endpoint configurations $\mathbf{s}_I$ and $\mathbf{s}'_I$ under the condition that the total energy $E_{\mathrm{tot}}\stackrel{!}{=}\mathrm{const}$ does not change. On the contrary he permits arbitrary variations in duration $t\!\left[\gamma + \delta\gamma^{(\mathrm{Lagr})} \right]$. Then - Lagrange postulates - his quantity of action
\be\label{Abschnitt -- Principle of Least Action - Lagranges Ausdruck}
   S_{\mathrm{Lagr}}\left[ \gamma \right] \;\;
:= \;\; \int_{\gamma\big|_{E_{\mathrm{tot}}}} E_{\mathrm{kin} \, I} \:\mathrm{d}t \;\; = \;\;
   \int_{\gamma\big|_{E_{\mathrm{tot}}}} \frac{1}{2}\: m_I\: \mathbf{v}_I \cdot \mathbf{v}_I \:\mathrm{d}t \;\; = \;\;
   \frac{1}{2} \int_{\gamma\big|_{E_{\mathrm{tot}}}} \mathbf{p}_I \cdot \mathrm{d}\mathbf{s}_I
\ee
becomes minimal for the true trajectory (i.e. free running process $\gamma$ of intrinsic action $w$). An engineer can generate Lagrange variations $\delta\gamma^{(\mathrm{Lagr})}$ of the course of intrinsic action $w$ by suitable steering actions: e.g. instantaneous redistribution (absorbing here, expending there) of energy units $\mathcal{S}_{\mathbf{1}}\big|_{\mathbf{0}}$ between different elements of the system. He can temporarily couple standard kicks $\circMunit_{\:\mathbf{v}_{\mathbf{1}}}$ from his reservoir against the system with no net-energy transfer.

Similarly Hamilton 1834 compares trajectories $\gamma$ resp. $\gamma +\delta\gamma^{(\mathrm{Ham})}$ between fixed endpoint configurations $\mathbf{s}_I$ and $\mathbf{s}'_I$ but requires preservation of durations $\Delta t\stackrel{!}{=}\mathrm{const}$. Instead he allows temporary variations of total energy $E_{\mathrm{tot}}$. Then - Hamilton postulates - his quantity of action
\be\label{Abschnitt -- Principle of Least Action - Hamiltons Ausdruck}
   S_{\mathrm{Ham}}\left[ \gamma \right] \;\;
:= \;\; \int_{\gamma\big|_{\Delta t}} \left( E_{\mathrm{kin} \, I} - V_{\mathrm{pot}} \right) \mathrm{d}t   \ee
becomes minimal for the true (undisturbed, isolated) motion of the system.\footnote{Helmholtz called Hamilton's integrand (\ref{Abschnitt -- Principle of Least Action - Hamiltons Ausdruck}) ''kinetic potential'' - nowadays it is called ''Lagrangian''.} The generation of Hamilton type variations $\delta\gamma^{(\mathrm{Ham})}$ require other steering interventions. The engineer can temporarily expend additional standard energy sources $\mathcal{S}_{\mathbf{1}}\big|_{\mathbf{0}}$ (from his external calorimeter reservoir) but he needs to keep an eye on the steering duration.\footnote{In the case without forces the free mass point runs through fixed endpoint configuration $\mathbf{s}_I \Rightarrow \mathbf{s}'_I$ - according to Lagrange (\ref{Abschnitt -- Principle of Least Action - Lagranges Ausdruck}) with constant velocity $|\mathbf{v}_I|$ in minimum duration $t$ and according to Hamilton (\ref{Abschnitt -- Principle of Least Action - Hamiltons Ausdruck}) with minimum velocity $|\mathbf{v}_I|$ in fixed duration $\Delta t$ - along the shortest path, a \emph{straight line}.}


\subsection{Minimal steering effort}

We \emph{analyze} the variation of the free course of an interaction $w$ from a practical perspective. We examine familiar mathematical formulation based on measurement methodical principles. We regard ''virtual displacements'' $\gamma + \delta \gamma$ as real process from external steering interventions $\mathrm{RB}^{(1)}\ast w_1 \ast \mathrm{RB}^{(2)}\ast w_2 \ast \ldots$ (Lemma \ref{Lem - KM Dynamics - Potential of Mechanical System - Steering Action}) and compare the free evolving and the steered process with regard to the \emph{steering effort} required for generating the (temporary) deviation $\delta \gamma$.

\begin{figure}    
  \begin{center}           
  \includegraphics[height=19.8cm]{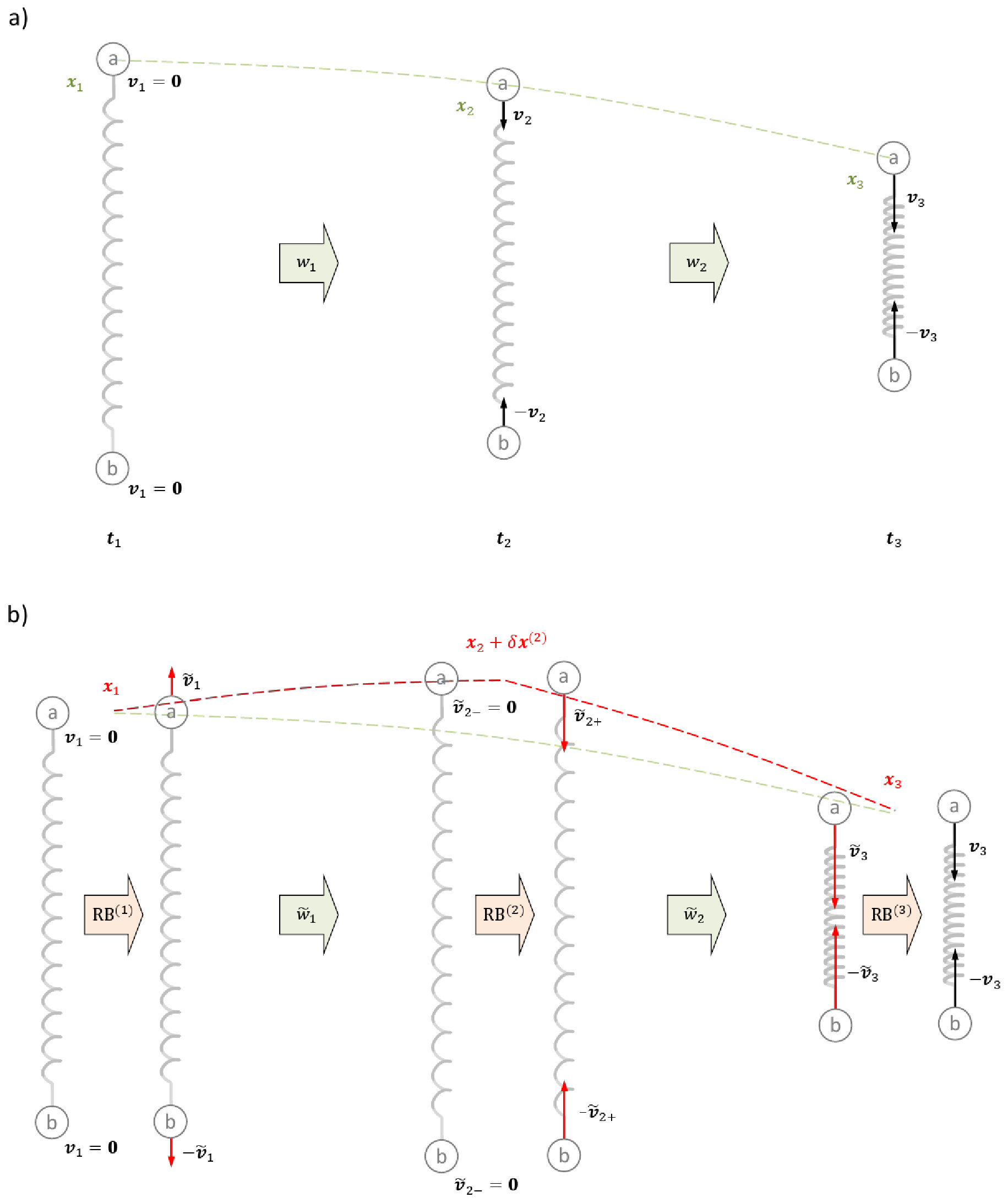}  
  \end{center}
  \vspace{-0cm}
  \caption{\label{pic_Wirkungs-Steuerung_Feder-Beispiel} a) contraction action b) course steered by instantaneous calorimeter interventions
    }
  \end{figure}

For illustration consider the contraction of a spring with massive bodies $\circMa, \circMb$ attached on both ends (see figure \ref{pic_Wirkungs-Steuerung_Feder-Beispiel}a).
%
The undisturbed process $w$ would evolve from an initially expanded configuration $(x_1,t_1) \Rightarrow (x_2,t_2) \Rightarrow (x_3,t_3) $ with increasing velocity towards a less expanded final configuration $(x_3,t_3)$. We can also steer the contraction process $w$ through a varied intermediate configuration $(x_1,t_1) \Rightarrow (x_2 + \delta x_2 ,t_2) \Rightarrow (x_3,t_3)$. Let a practically instantaneous intervention $\circMa_{\:\mathbf{0}}, \circMb_{\:\mathbf{0}} \stackrel{\mathrm{RB}^{(1)}}{\Rightarrow} \circMa_{\:\mathbf{v}}, \circMb_{-\mathbf{v}}$ at same initial configuration $(x_1,t_1)$ catapult both bodies into opposite motion. Under modified initial conditions the system evolves towards an even more expanded intermediate configuration $(x_2 + \delta x_2 ,t_2)$. Another kick of the right strength $\mathrm{RB}^{(2)}: \circMa_{\:\mathbf{0}}, \circMb_{\:\mathbf{0}} \Rightarrow \circMa_{\:\mathbf{v}'}, \circMb_{-\mathbf{v}'}$ provides enough momentum such that the next segment of the contraction $w_2$ evolves to the same final configuration $(x_3,t_3)$ - with more velocity though. With final steering intervention $\mathrm{RB}^{(3)}$ we extract the surplus kinetic energy and momentum from both elements $\circMa, \circMb$ such that the system continues evolving like the original contraction $w$ in the same undisturbed way.

By coupling three external steering resp. absorption kicks $\mathrm{RB}^{(1)} , \mathrm{RB}^{(2)} , \mathrm{RB}^{(3)}$ at the right moment and strength we generate a Hamilton type variation $\gamma +\delta\gamma^{(\mathrm{Ham})}$ of the course of intrinsic action $w$ (see figure \ref{pic_Wirkungs-Steuerung_Feder-Beispiel}b). The two-partite system $\circMa\cup\circMb$ runs through fixed endpoints $\mathbf{x}_1 := (\mathbf{x}_a,\mathbf{x}_b)\big|_{t_1}$ and $\mathbf{x}_3 := (\mathbf{x}_a,\mathbf{x}_b)\big|_{t_3}$ in same fixed duration $\Delta t := t_3 - t_1$. To drive the contraction process of an expanded spring through varied intermediate configuration $\mathbf{x}_2 +\delta\mathbf{x}_2 := (\mathbf{x}_a,\mathbf{x}_b)\big|_{t_2} + (\delta\mathbf{x}_a,\delta\mathbf{x}_b)$ we temporarily expend steering energy $E_{ \mathrm{RB}^{(1)}} \big|_{t_1}$ and $E_{\mathrm{RB}^{(2)}} \big|_{t_2}$ from the calorimeter reservoir, which we fully retrieve $E_{\mathrm{RB}^{(3)}} \big|_{t_3}$ at the end (\ref{Abschnitt -- Potential of Mechanical System - potential field - combined extract for steering along circular process}).

\begin{figure}    
  \begin{center}           
  \includegraphics[height=9.3cm]{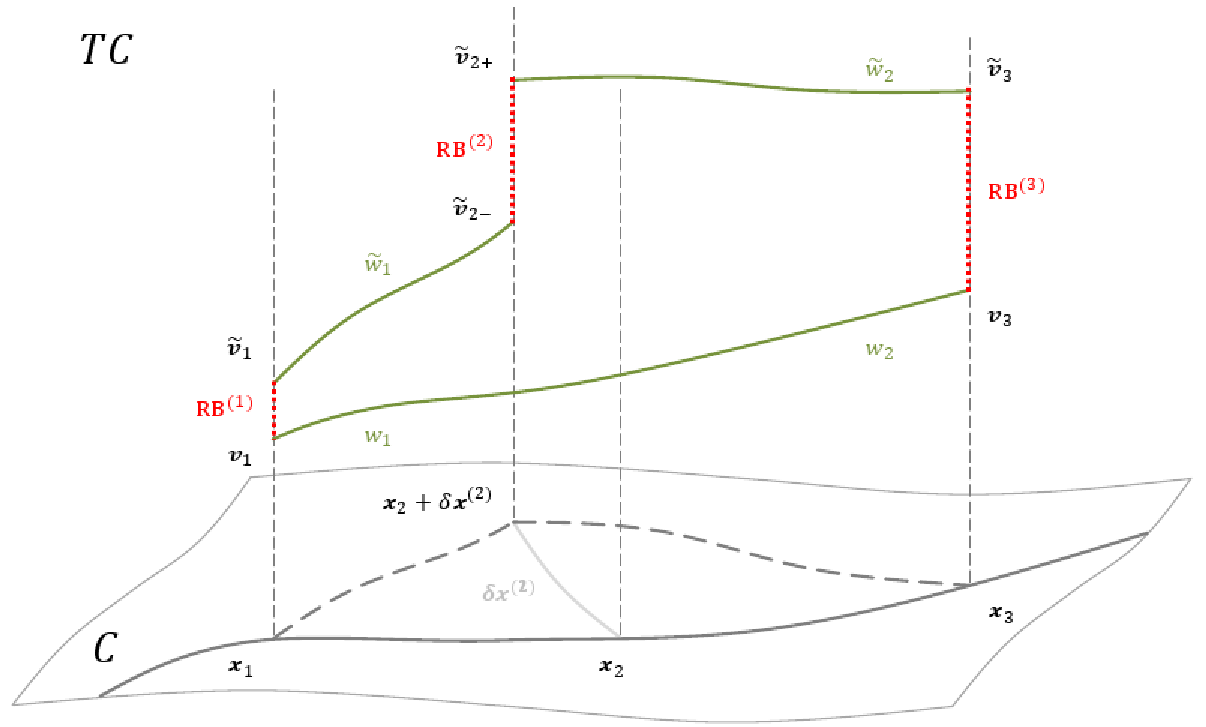}  
  \end{center}
  \vspace{-0cm}
  \caption{\label{pic_Wirkungs-Steuerung_Allgemein} steered variation of action $w$
    }
  \end{figure}

Our steering maneuver $\mathrm{RB}^{(1)}\ast \tilde{w}_1 \ast \mathrm{RB}^{(2)}\ast \tilde{w}_2 \ast \mathrm{RB}^{(3)}$ is a simple physical process for the Hamilton type variation of an action $w$. Both processes run through fixed endpoints $\mathbf{x}_1$ and $\mathbf{x}_3$ in fixed duration $\Delta t$. Despite varied intermediate configuration $\mathbf{x}_2 + \delta \mathbf{x}_2$ they continue evolving (after the temporary steering intervention) in the same undisturbed way. We can successively substitute segments of free intrinsic action $\tilde{w}_i$ with ''three-step'' steering actions and thus generate every form for a local Hamilton type variation $\delta\gamma^{(\mathrm{Ham})}$. Bernoulli summarized the basic idea of variation: ''The extremal property of sought after curve is also contained in all segments of that curve, in particular also in all of its (infinitesimal) elements'' \cite{Szabo Geschichte der Mechanischen Prinzipien}. Hence for our variational analysis (of the steering effort) it is sufficient to examine the physical process behind basic ''three-step''
variation maneuver of figure \ref{pic_Wirkungs-Steuerung_Allgemein}.
%
\begin{theo}
For every local Hamilton type variation $\delta\gamma^{(\mathrm{Ham})}$ (through fixed endpoints $\mathbf{x}_I$ and $\mathbf{x}'_I$ in fixed duration $\Delta t$) of the free evolution $\gamma$ of intrinsic action $w$ the variation of Hamilton's quantity of action (\ref{Abschnitt -- Principle of Least Action - Hamiltons Ausdruck}) is \underline{positive definite}
\be\label{Abschnitt -- Principle of Least Action - Hamiltons Variation pos definite}
   \;\;\;\;\;\;\;\;\;\;\;\;\;\;\;\;\;\; 0 \;\; < \;\; \delta S_{\mathrm{Ham}} \left[ \gamma \right] \;\; := \;\;    S_{\mathrm{Ham}}\left[ \gamma + \delta\gamma^{(\mathrm{Ham})} \right] \:-\: S_{\mathrm{Ham}}\left[ \gamma \right]  \;\;\;\;\;\;\;\;\; \forall \;\; \delta\gamma^{(\mathrm{Ham})}     .
\ee
\end{theo}
\textbf{Proof:} In the spirit of Euler \{\ref{Kap - KM Dynamics - Principle of Least Action - Historic Development}\} we examine the extremal characteristic - the quantity which turns into a minimum for undisturbed actions - from the direct method. In a system of point masses $\{G_I\}$ the trajectory of an interaction $w$ is determined by equations of motion
\be
   m_I \cdot \frac{\mathrm{d}^2 \mathbf{x}_I }{{\mathrm{d} t}^2} \;\; \stackrel{(\ref{Abschnitt -- differentiated analysis - evolution - inherited equations of motion})}{=} \;\; - \nabla^{(I)} V_{\mathrm{pot}} \;\; .   \nn
\ee
For variations of Bernoulli's infinitesimal nature the force against respective elements $i\in I$
\[
   -\nabla^{(i)} V_{\mathrm{pot}} \big|_{\mathcal{U}_{\mathbf{x}_I}} \;\; \simeq \;\; -\nabla^{(i)} V_{\mathrm{pot}} \big|_{\mathbf{x}_I}
\]
is constant in local neighborhood $\mathcal{U}$ of initial configuration $\mathbf{x}_I$. Without restricting generality we analyze the ''three-step'' steering maneuver for a Hamilton type variation with particular focus on temporarily expended steering interventions $\mathrm{RB}^{(i)}$ and steering duration $\Delta T_i$.\footnote{We analyze intrinsic action $w$ in closed system $\{G_I\}$. Potential energy and force do not depend on time.}

For given initial configuration $\left(T_1 , \mathbf{x}_I \right)$ and initial velocity $\left\{ \frac{\mathrm{d}\mathbf{x}_i}{\mathrm{d}t} \right\}_{i \in I} =: \frac{\mathrm{d}\mathbf{x}_I}{\mathrm{d}t} =: \mathbf{v}_I$ resp. $\tilde{\mathbf{v}}_I$ the \emph{free course} of intrinsic action $w$ has a local trajectory for elements $i\in I$ of the system
\be\label{Abschnitt -- Principle of Least Action - Hamiltons Variation pos definite - local trajectory}
   (t, \mathbf{x}_I)\,[\,t\,] \;\; = \;\; \left(T_1 , \mathbf{x}_I \right) \;+\; \left( 1 , \mathbf{v}_I \right)\cdot (t-T_1) \;-\; \left( 0 , \frac{\nabla^{I} V_{\mathrm{pot}}}{m_I} \right)\cdot \frac{1}{2} \cdot (t-T_1)^2 \;\; .
\ee
Let the steering start in moment $T_1=0$. We define the \emph{''three-step'' variation} of intrinsic action $w$ - through fixed endpoints $\mathbf{x}_I \big|_{T_1}$, $\mathbf{x}_I \big|_{T_3}$ in fixed duration $(T_3 - T_1)$ - in terms of given deviation $\mathbf{x}_I \big|_{T_2} + \delta \mathbf{x}_I^{(2)}$ from the free trajectory at intermediate moment $T_2$ (see figure \ref{pic_Wirkungs-Steuerung_Allgemein}).
\\

An engineer determines required steering kicks backwards from the equations of motion
\begin{enumerate}
\item   Given configuration of ''three-step'' variation $\delta\gamma^{(\mathrm{Ham})} (T_1, T_2, T_3, \delta \mathbf{x}_I^{(2)})$ implies matching conditions between segments of undisturbed intrinsic actions $w$, $\tilde{w}_1$ and $\tilde{w}_2$.
\item   Their trajectories $\gamma$, $\tilde{\gamma}_1$ and $\tilde{\gamma}_2$ specify the velocity transitions $\mathbf{v}_I^{(1)} \Rightarrow \tilde{\mathbf{v}}_I^{(1)}$ and ultimately the strength of kicks $\mathrm{RB}^{(1)}$ at steering moment $T_1$ and analogous for $T_2$, $T_3$.
\end{enumerate}
We find the steering intervention $\mathbf{v}_I^{(1)} \stackrel{\mathrm{RB}^{(1)}}{\Rightarrow} \tilde{\mathbf{v}}_I^{(1)}$ for the initial velocity of process $\tilde{w}_1$ from the matching condition $\tilde{\mathbf{x}}_I \big|_{T_2}\stackrel{!}{=} \mathbf{x}_I \big|_{T_2} +\: \delta \mathbf{x}_I^{(2)}$. The corresponding local free trajectories
\bea\label{Abschnitt -- Principle of Least Action - Hamiltons Variation pos definite - velocity left of T_2}
   \mathbf{x}_I + \tilde{\mathbf{v}}_I^{(1)} \cdot T_2 - \frac{\nabla^{I} V_{\mathrm{pot}}}{m_I} \cdot \frac{1}{2} \cdot {T_2}^2 & \stackrel{(\ref{Abschnitt -- Principle of Least Action - Hamiltons Variation pos definite - local trajectory})}{=} & \mathbf{x}_I + \mathbf{v}_I^{(1)} \cdot T_2 - \frac{\nabla^{I} V_{\mathrm{pot}}}{m_I} \cdot \frac{1}{2} \cdot {T_2}^2  \;+\;  \delta \mathbf{x}_I^{(2)}  \nn \\
   \tilde{\mathbf{v}}_I^{(1)} & \stackrel{!}{=} & \mathbf{v}_I^{(1)} + \frac{\delta \mathbf{x}_I^{(2)}}{T_2}
\eea
fix the necessary initial velocities and hence the steering energy at starting moment $T_1$
\be\label{Abschnitt -- Principle of Least Action - Hamiltons Variation pos definite - steering energy E_1}
   E_{ \mathrm{RB}^{(1)}} \;\; \stackrel{(\ref{Abschnitt -- kin quant Absorptions Wirkung - kin energy and momentum metrisiert})}{=} \;\; \frac{1}{2} \cdot m_I \cdot \left\{  \left(\mathbf{v}_I^{(1)}\right)^2 - \left(\tilde{\mathbf{v}}_I^{(1)}\right)^2 \right\}
\ee
where we \emph{sum} all contributions for individual elements $i\in I$ of the system.\footnote{We do not need to keep track of associated steering momentum $\mathbf{p}_{ \mathrm{RB}^{(1)}} := \mathbf{p}\left[ \mathrm{RB} [ \mathbf{v}_I^{(1)} \Rightarrow \tilde{\mathbf{v}}_I^{(1)} ] \right]$ here.}

Similarly we find the final steering kick $\tilde{\mathbf{v}}_I^{(3)} \stackrel{\mathrm{RB}^{(3)}}{\Rightarrow} \mathbf{v}_I^{(3)}$ for absorbing the surplus kinetic energy and momentum from $\tilde{w}_2$. We reexpress the local trajectories (\ref{Abschnitt -- Principle of Least Action - Hamiltons Variation pos definite - local trajectory}) for $w$ and $\tilde{w}_2$ in terms of the \emph{joint} configuration $\left(T_3 , \mathbf{x}_I \right)$ and final velocity $\mathbf{v}_I^{(3)}$ resp. $\tilde{\mathbf{v}}_I^{(3)}$
\be\label{Abschnitt -- Principle of Least Action - Hamiltons Variation pos definite - local trajectory rechts}
   (t, \mathbf{x}_I)\,[\,t\,] \;\; = \;\; \left(T_3 , \mathbf{x}_I \right) \;+\; \left( 1 , \mathbf{v}_I^{(3)} \right)\cdot (t-T_3) \;-\; \left( 0 , \frac{\nabla^{I} V_{\mathrm{pot}}}{m_I} \right)\cdot \frac{1}{2} \cdot (t-T_3)^2 \;\; .
\ee
Again displacement condition $\tilde{\mathbf{x}}_I \big|_{T_2}\stackrel{!}{=} \mathbf{x}_I \big|_{T_2} +\: \delta \mathbf{x}_I^{(2)}$ on corresponding undisturbed trajectories
\bea\label{Abschnitt -- Principle of Least Action - Hamiltons Variation pos definite - velocity right of T_2}
   \mathbf{x}_I^{(3)} + \tilde{\mathbf{v}}_I^{(3)} \cdot (T_2-T_3) - \frac{\nabla^{I} V_{\mathrm{pot}}}{m_I} \cdot \frac{1}{2} \cdot (T_2-T_3)^2 \!\!\!\!\!\!\!\!\!\!\!\!\!\!\!\!\!\!\!\!\!\!\!\!\!\!\!\!
   \!\!\!\!\!\!\!\!\!\!\!\!\!\!\!\!\!\!\!\!\!\!\!\!\!\!\!\!\!\!\!\!\!\!\!\!\!
   & & \nn \\
   & = & \mathbf{x}_I^{(3)} + \mathbf{v}_I^{(3)} \cdot (T_2-T_3) - \frac{\nabla^{I} V_{\mathrm{pot}}}{m_I} \cdot \frac{1}{2} \cdot (T_2-T_3)^2  \;+\;  \delta \mathbf{x}_I^{(2)}  \nn \\
   \tilde{\mathbf{v}}_I^{(3)} & \stackrel{!}{=} & \mathbf{v}_I^{(3)} + \frac{\delta \mathbf{x}_I^{(2)}}{T_2-T_3}
\eea
specifies the velocity transition and corresponding steering energy at final moment $T_3$
\be\label{Abschnitt -- Principle of Least Action - Hamiltons Variation pos definite - steering energy E_3}
   E_{ \mathrm{RB}^{(3)}} \;\; = \;\; \frac{1}{2} \cdot m_I \cdot \left\{ \left(\tilde{\mathbf{v}}_I^{(3)}\right)^2  - \left(\mathbf{v}_I^{(3)}\right)^2 \right\}  \;\; .
\ee

Finally we determine the steering kick $\tilde{\mathbf{v}}_I^{(2-)} \stackrel{\mathrm{RB}^{(2)}}{\Rightarrow} \tilde{\mathbf{v}}_I^{(2+)}$ in the middle turning moment $T_2$ (see figure \ref{pic_Wirkungs-Steuerung_Allgemein}). Derivation of left $\tilde{\gamma}_1$ resp. right $\tilde{\gamma}_2$ coming trajectories gives the velocities
\bea
   \tilde{\mathbf{v}}_I^{(2-)} & \stackrel{(\ref{Abschnitt -- Principle of Least Action - Hamiltons Variation pos definite - velocity left of T_2})}{=} & \mathbf{v}_I^{(2)} + \frac{\delta \mathbf{x}_I^{(2)}}{T_2}  \nn \\
   \tilde{\mathbf{v}}_I^{(2+)} & \stackrel{(\ref{Abschnitt -- Principle of Least Action - Hamiltons Variation pos definite - velocity right of T_2})}{=} & \mathbf{v}_I^{(2)} + \frac{\delta \mathbf{x}_I^{(2)}}{T_2-T_3}  \nn  \;\;.
\eea
and steering energy
\be\label{Abschnitt -- Principle of Least Action - Hamiltons Variation pos definite - steering energy E_2}
   E_{ \mathrm{RB}^{(2)}} \;\; = \;\; \frac{1}{2} \cdot m_I \cdot \left\{  \left(\tilde{\mathbf{v}}_I^{(2-)}\right)^2 - \left(\tilde{\mathbf{v}}_I^{(2+)}\right)^2 \right\} \;\; .
\ee

With known trajectories $\gamma$, $\tilde{\gamma}_1$ and $\tilde{\gamma}_2$ of undisturbed intrinsic processes $w$, $\tilde{w}_1$ and $\tilde{w}_2$ we examine Hamilton's quantity of action
\[
   S_{\mathrm{Ham}}\left[ \gamma \right] \;\; \stackrel{(\ref{Abschnitt -- Principle of Least Action - Hamiltons Ausdruck})}{=} \;\; \int_{\gamma\big|_{\Delta t}} \mathrm{d}t \; \left\{ \frac{m_I}{2}\cdot \left( \mathbf{v}_I\big|_t \right)^2
   \;-\: V_{\mathrm{pot}} \big|_{\mathbf{x}_I(t)} \right\}
\]
with summation convention for the kinetic energy of all elements $i\in I$. \emph{Locally} around initial configuration $\mathbf{x}_I^{(1)}:=\mathbf{x}_I\big|_{T_1}$ the potential field $V_{\mathrm{pot}} \big|_{\mathbf{x}_I} \simeq V_{\mathrm{pot}}\big|_{\mathbf{x}_I^{(1)}} + \nabla^{I} V_{\mathrm{pot}} \cdot [\mathbf{x}_I - \mathbf{x}_I^{(1)} ]$ is linear. For the \emph{undisturbed course} $\gamma$ of intrinsic action $w$ we obtain
\[
   S_{\mathrm{Ham}}\left[ \gamma \right] \;\; = \;\;
\int_{T_1}^{T_3} \mathrm{d}t \; \left\{ \frac{m_I}{2}\cdot \left( \mathbf{v}_I\big|_t \right)^2
   \;-\: \left( V_{\mathrm{pot}}^{(1)}  +  \nabla^{I} V_{\mathrm{pot}}\cdot [\mathbf{x}_I\big|_t - \mathbf{x}_I^{(1)} ]  \right)
   \right\}
\]
and similarly for the \emph{steered course} $\tilde{\gamma}_1\ast\tilde{\gamma}_2$ of ''three-step'' variation $\mathrm{RB}^{(1)}\ast \tilde{w}_1 \ast \mathrm{RB}^{(2)}\ast \tilde{w}_2 \ast \mathrm{RB}^{(3)}$
\bea
   S\left[ \tilde{\gamma}_1\ast\tilde{\gamma}_2 \right] \! & = & \!
\int_{T_1}^{T_2} \! \mathrm{d}t  \left\{ \frac{m_I}{2}\cdot \! \left( \mathbf{v}_I\big|_t + \frac{\delta \mathbf{x}_I^{(2)}}{T_2} \right)^2
   - \left( V_{\mathrm{pot}}^{(1)}  +  \nabla^{I} V_{\mathrm{pot}}\cdot [\mathbf{x}_I\big|_t + \frac{\delta \mathbf{x}_I^{(2)}}{T_2} \cdot t - \mathbf{x}_I^{(1)} ]  \right) \!  \right\}  \nn  \\
   & &  \!\!\!\!\!\!\!\!\!\!\!\!\!\!\!\!\!\!\!\!\!\!\!\!\!\!\!\!\!\!
   + \: \int_{T_2}^{T_3} \!\! \mathrm{d}t  \left\{ \frac{m_I}{2}\!\cdot \!\! \left( \mathbf{v}_I\big|_t + \frac{\delta \mathbf{x}_I^{(2)}}{T_2-T_3} \right)^2 \!\!
   - \left( \! V_{\mathrm{pot}}^{(1)}  +  \nabla^{I} V_{\mathrm{pot}}\cdot [\mathbf{x}_I\big|_t + \frac{\delta \mathbf{x}_I^{(2)}}{T_2-T_3} \cdot (t\!-\!T_3) - \mathbf{x}_I^{(1)} ] \! \right) \!  \right\}  \nn
\eea
with varied trajectories $\tilde{\mathbf{x}}_I$, $\tilde{\mathbf{v}}_I$ substituted in terms of
\be
   \tilde{\mathbf{x}}_I\big|_{[T_1,T_2]}  \stackrel{(\ref{Abschnitt -- Principle of Least Action - Hamiltons Variation pos definite - local trajectory})(\ref{Abschnitt -- Principle of Least Action - Hamiltons Variation pos definite - velocity left of T_2})}{=}  \mathbf{x}_I\big|_t \:+\: \frac{\delta \mathbf{x}_I^{(2)}}{T_2} \cdot t    \;\;\;\;\;\; \mathrm{resp.} \;\;\;\;\;\;
    \tilde{\mathbf{x}}_I\big|_{[T_2,T_3]}  \stackrel{(\ref{Abschnitt -- Principle of Least Action - Hamiltons Variation pos definite - local trajectory rechts})(\ref{Abschnitt -- Principle of Least Action - Hamiltons Variation pos definite - velocity right of T_2})}{=}  \mathbf{x}_I\big|_t \:+\: \frac{\delta \mathbf{x}_I^{(2)}}{T_2-T_3} \cdot (t-T_3)   \nn  \;\; .
\ee

The variation of Hamilton's quantity of action
\bea\label{Abschnitt -- Principle of Least Action - Hamiltons Variation pos definite - variation S_Ham}
   \delta S_{\mathrm{Ham}}\left[ \gamma \right]  & := &  S_{\mathrm{Ham}}\left[ \tilde{\gamma}_1\ast\tilde{\gamma}_2 \right] \;-\; S_{\mathrm{Ham}}\left[ \gamma \right]  \nn \\
   & = &  \! \int_{T_1}^{T_2}  \mathrm{d}t  \left\{ \frac{m_I}{2}\cdot \!\left( \! \left( \frac{\delta \mathbf{x}_I^{(2)}}{T_2}\right)^2  \!+  2\cdot \mathbf{v}_I\big|_t \cdot \frac{\delta \mathbf{x}_I^{(2)}}{T_2}  \right) - \; \nabla^{I} V_{\mathrm{pot}} \cdot \frac{\delta \mathbf{x}_I^{(2)}}{T_2} \cdot t   \right\} \nn \\
   &  & \!\!\!\!\!\!\!\!\!\!\!\!\!\!\!\!\!
+ \int_{T_2}^{T_3}  \mathrm{d}t  \left\{ \frac{m_I}{2}\cdot \!\left( \! \left( \frac{\delta \mathbf{x}_I^{(2)}}{T_2-T_3}\right)^2  \!+  2\cdot \mathbf{v}_I\big|_t \cdot \frac{\delta \mathbf{x}_I^{(2)}}{T_2-T_3}  \right) - \; \nabla^{I} V_{\mathrm{pot}} \cdot \frac{\delta \mathbf{x}_I^{(2)}}{T_2-T_3} \cdot (t-T_3)  \right\} \nn  \\
   & =: &  \underbrace{\frac{m_I}{2}\cdot \! \left( \frac{\delta \mathbf{x}_I^{(2)}}{T_2}\right)^2 \!\!\cdot (T_2-T_1)}_{>\: 0} \;\;+\;\; \underbrace{\frac{m_I}{2}\cdot \! \left( \frac{\delta \mathbf{x}_I^{(2)}}{T_2-T_3}\right)^2 \!\!\cdot (T_3-T_2)}_{>\: 0} \;\;+\;\; R
\eea
decomposes into two positive terms and a residue $R$ which - using $\mathbf{v}_I\big|_t \! \stackrel{(\ref{Abschnitt -- Principle of Least Action - Hamiltons Variation pos definite - local trajectory})}{=} \mathbf{v}_I^{(1)} - \frac{\nabla^{I} V_{\mathrm{pot}}}{m_I} \cdot t\;$ -
\bea
   R  & \stackrel{(\ref{Abschnitt -- Principle of Least Action - Hamiltons Variation pos definite - variation S_Ham})}{:=} &  \! \int_{T_1=0}^{T_2}  \mathrm{d}t  \left\{ m_I\cdot \mathbf{v}_I^{(1)} \cdot \frac{\delta \mathbf{x}_I^{(2)}}{T_2}  \; - \; m_I\cdot \frac{\delta \mathbf{x}_I^{(2)}}{T_2} \cdot \frac{\nabla^{I} V_{\mathrm{pot}}}{m_I} \cdot t  \; - \; \nabla^{I} V_{\mathrm{pot}} \cdot \frac{\delta \mathbf{x}_I^{(2)}}{T_2} \cdot t   \right\}  \nn \\
   &  &  \!\!\!\!\!\!\!
+  \int_{T_2}^{T_3}  \mathrm{d}t  \left\{ m_I\cdot \mathbf{v}_I^{(1)} \cdot \frac{\delta \mathbf{x}_I^{(2)}}{T_2-T_3}  \; - \; \frac{\delta \mathbf{x}_I^{(2)}}{T_2-T_3} \cdot \nabla^{I} V_{\mathrm{pot}} \cdot t  \; - \; \nabla^{I} V_{\mathrm{pot}} \cdot \frac{\delta \mathbf{x}_I^{(2)}}{T_2-T_3} \cdot (t-T_3)   \right\}  \nn
\eea
\bea
   R  & = &   \left\{ m_I\cdot \mathbf{v}_I^{(1)} \cdot \frac{\delta \mathbf{x}_I^{(2)}}{T_2} \cdot T_2 \; - \; \frac{\delta \mathbf{x}_I^{(2)}}{T_2} \cdot \nabla^{I} V_{\mathrm{pot}} \cdot T_2^2 \right\} \:+\: \left\{ m_I\cdot \mathbf{v}_I^{(1)} \cdot \frac{\delta \mathbf{x}_I^{(2)}}{T_2-T_3} \cdot (T_3-T_2) \right. \nn \\
   &  &  \;\;\;\;\;\;\;\;\;\;\;\;\;\;\;\;
  \left.- \; \frac{\delta \mathbf{x}_I^{(2)}}{T_2-T_3} \cdot \nabla^{I} V_{\mathrm{pot}} \cdot (T_3^2-T_2^2)  \; - \; \nabla^{I} V_{\mathrm{pot}} \cdot \frac{\delta \mathbf{x}_I^{(2)}}{T_2-T_3} \cdot (-T_3)\cdot(T_3-T_2) \right\} \nn \\
   & & \!\!\!\!\!\!\!\!\!\!\!\!\!\!
   = \left\{ m_I\cdot \mathbf{v}_I^{(1)} \!\cdot \delta \mathbf{x}_I^{(2)} - \delta \mathbf{x}_I^{(2)} \!\cdot \nabla^{I} V_{\mathrm{pot}} \cdot T_2 \right\} + \left\{ - m_I\cdot \mathbf{v}_I^{(1)} \!\cdot \delta \mathbf{x}_I^{(2)} + \delta \mathbf{x}_I^{(2)} \!\cdot \nabla^{I} V_{\mathrm{pot}} \cdot T_2 \right\} \;=\; 0 \nn
\eea
vanishes exactly. Therefore the variation of Hamilton's quantity of action
\[
   \;\;\;\;\;\;\;\;\;\;\;\;\;\;\;\;\;\;  \delta S_{\mathrm{Ham}}\left[ \gamma \right]  \; \stackrel{(\ref{Abschnitt -- Principle of Least Action - Hamiltons Variation pos definite - variation S_Ham})}{>} \;  0  \;\;\;\;\;\;\;\;\;\;\;\; \forall \;\;\; T_2, \delta \mathbf{x}_I^{(2)}
\]
is positive definite for all basic ''three-step'' variations $\mathrm{RB}^{(1)}\ast \tilde{w}_1 \ast \mathrm{RB}^{(2)}\ast \tilde{w}_2 \ast \mathrm{RB}^{(3)} \!\left[ T_2, \delta \mathbf{x}_I^{(2)} \right]$ - and hence for all Hamilton type variations $\delta \gamma^{(\mathrm{Ham})}$ of free course $\gamma$ of intrinsic action $w$.
\qed
\begin{pr}
For the variation (of free course $\gamma$) of intrinsic action $w$ Hamilton's quantity
\be
   0 \; < \; \delta S_{\mathrm{Ham}} \left[ \gamma \right] \; \stackrel{(\ref{Abschnitt -- Principle of Least Action - Hamiltons Variation pos definite})}{=} \, - \:E_{\mathrm{RB}^{(1)}} \cdot T_3 \; - \; E_{\mathrm{RB}^{(2)}} \cdot (T_3 - T_2) \; - \; \nabla^I V_{\mathrm{pot}} \cdot \delta \mathbf{x}_I^{(2)} \cdot T_3  \nn
\ee
is determined by the - external - \underline{steering effort} in terms of
\begin{enumerate}
  \item   temporarily expended steering energy $- E_{\mathrm{RB}^{(1)}}$ and $- E_{\mathrm{RB}^{(2)}}$ into system $\{G_I\}$
  \item   corresponding activation duration $T_3-T_1$ resp. $T_3-T_2$ of the system until extraction at final steering moment $T_3$ and
  \item   passively coupled and extracted additional potential steering energy from temporary displacement along $\delta \mathbf{x}_I^{(2)}$.
\end{enumerate}
\end{pr}
\textbf{Proof:}
To generate a \emph{temporary} displacement $\delta \mathbf{x}_I^{(2)}$ from the free evolution of intrinsic action $w$ the engineer steers energy from his external reservoir $\{\circMunit_{\:\mathbf{0}}\}$ into the system $\{G_I\}$
\bea
   -E_{\mathrm{RB}^{(1)}} & \!\!\!\stackrel{(\ref{Abschnitt -- Principle of Least Action - Hamiltons Variation pos definite - steering energy E_1})(\ref{Abschnitt -- Principle of Least Action - Hamiltons Variation pos definite - velocity left of T_2})}{=} &
\frac{m_I}{2}\cdot \! \left( \frac{\delta \mathbf{x}_I^{(2)}}{T_2}\right)^2 \;+\;  m_I\cdot \mathbf{v}_I^{(1)} \cdot \frac{\delta \mathbf{x}_I^{(2)}}{T_2}  \nn \\
   -E_{\mathrm{RB}^{(2)}} &  &  \!\!\!\!\!\!\!\!\!\!\!\!\!\!\!\!\!\!\!\!\!\!\!\!\!
\stackrel{(\ref{Abschnitt -- Principle of Least Action - Hamiltons Variation pos definite - steering energy E_2})}{=}
- \: \frac{m_I}{2}\cdot \! \left( \frac{\delta \mathbf{x}_I^{(2)}}{T_2}\right)^{\!\!2} -\, m_I\cdot \!\!\!\!\!\!\!\!\!\!\!\!\!
\underbrace{\mathbf{v}_I^{(2)}}_{\stackrel{(\ref{Abschnitt -- Principle of Least Action - Hamiltons Variation pos definite - local trajectory})}{=} \: \mathbf{v}_I^{(1)} - \frac{\nabla^{I} V_{\mathrm{pot}}}{m_I} \cdot T_2} \!\!\!\!\!\!\!\!\!\!\!
\cdot \frac{\delta \mathbf{x}_I^{(2)}}{T_2}
\; + \; \frac{m_I}{2}\cdot \! \left( \frac{\delta \mathbf{x}_I^{(2)}}{T_2-T_3}\right)^{\!\!2} +\, m_I\cdot \underbrace{\mathbf{v}_I^{(2)}}_{\mathrm{etc.}} \cdot \frac{\delta \mathbf{x}_I^{(2)}}{T_2-T_3}   \nn
\eea
which together with activation duration $T_3$ resp. $T_3-T_2$ proves straightforward
\bea
   & &  \!\!\!\!\!\!\!\!\!
   - \:E_{\mathrm{RB}^{(1)}} \cdot T_3 \; - \; E_{\mathrm{RB}^{(2)}} \cdot (T_3 - T_2)  \nn \\
   & & \;\;\;\;\;\;\;\;\;\;\;\; = \;\;
   \nabla^I V_{\mathrm{pot}} \cdot \delta \mathbf{x}_I^{(2)} \cdot T_3 \;\;+\;
\underbrace{\frac{m_I}{2}\cdot \! \left( \frac{\delta \mathbf{x}_I^{(2)}}{T_2}\right)^{\!2} \cdot T_2 \;+\; \frac{m_I}{2}\cdot \! \left( \frac{\delta \mathbf{x}_I^{(2)}}{T_2-T_3}\right)^{\!2} \cdot (T_3-T_2)}_{\stackrel{(\ref{Abschnitt -- Principle of Least Action - Hamiltons Variation pos definite - variation S_Ham})}{=}\: \delta S_{\mathrm{Ham}}[\gamma]} \;\;.  \nn
\eea
The system is activated with the additional energy $-E_{\mathrm{RB}^{(1)}}$ and $-E_{\mathrm{RB}^{(2)}}$ for duration $T_3-T_1$ resp. $T_3-T_2$ until all extra energy is absorbed
\[
   E_{\mathrm{RB}^{(3)}} \;\; \!\!\!\stackrel{(\ref{Abschnitt -- Principle of Least Action - Hamiltons Variation pos definite - steering energy E_3})(\ref{Abschnitt -- Principle of Least Action - Hamiltons Variation pos definite - velocity right of T_2})}{=} \;\;
\frac{m_I}{2}\cdot \! \left( \frac{\delta \mathbf{x}_I^{(2)}}{T_2-T_3}\right)^2 \;+\;  m_I\cdot
\!\!\!\!\!\!\!\!
\underbrace{\mathbf{v}_I^{(3)}}_{\stackrel{(\ref{Abschnitt -- Principle of Least Action - Hamiltons Variation pos definite - local trajectory})}{=} \: \mathbf{v}_I^{(1)} - \frac{\nabla^{I} V_{\mathrm{pot}}}{m_I} \cdot T_3}
\!\!\!\!\!
\cdot \frac{\delta \mathbf{x}_I^{(2)}}{T_2-T_3}
\]
again back into external calorimeter reservoir $\{\circMunit_{\:\mathbf{0}}\}$. The engineer retrieves all temporarily expended steering energy from system $\{G_I\}$ at final moment $T_3$ of his steering intervention
\[
   E_{\mathrm{RB}^{(1)}} + E_{\mathrm{RB}^{(2)}} + E_{\mathrm{RB}^{(3)}} \;\; = \;\; 0   \;\; .
\]
\qed
\begin{co}
In free running (no exterior steering actions $\mathrm{RB}^{(i)}$) course of intrinsic action $w$ steering effort is absent. Hamilton's (positive definite) quantity of action is \underline{minimal}.
\end{co}
The variational analysis compares two physical processes - steered action and free evolving action. Steered actions are associated with extra steering effort. We leave analogous physical examination of the steering effort which generates Lagrange type variations  $\delta\gamma^{(\mathrm{Lagr})}$ of the free course $\gamma$ of intrinsic action $w$ as a future exercise.


\chapter{Relativistic mechanics}\label{Kap - Relativistic energy-momentum}

We introduce a novel foundation of physics from the operationalization of its basic observables. We begun with classical and relativistic kinematics \{\ref{Kap - Kinematics}\}. Seizing on a programmatic proposal by Heinrich Hertz \cite{Hertz - Einleitung zur Mechanik} we did arrive via quantification of energy-momentum at the fundamental equations and conservation laws \{\ref{Kap - Mechanics}\}. We define energy, momentum and inertial mass from the classical comparison ''more capability to work than'' (against same test system) and ''more impact than'' (in a collision). To find ''how many times'' more we develop Helmholtz analysis \cite{Helmholtz - Zaehlen und Messen} of basic measurements. For a relativistic revision we additionally assume the light principle. From the same thought experiments and physical operations we derive the relativistic equations between these quantities. We begin from vivid pictures, undisputed measurement principles and the colloquial description without any mathematical presupposition.


\section{Pre-theoretic elements}\label{Kap - SRT Dynamics - Pre-theoretic elements}

By the construction of a gedanken-calorimeter \cite{Hartmann-KM_Dyn} from a single \emph{elementary reference process} (inelastic collision of irrelevant inner structure) and \emph{symmetry principles} we can measure the energy and momentum of all other more complex processes. Before we develop the relativistic construction \{\ref{Kap - SRT_Dyn - Basic Measurement}\} we show, that this pre-theoretic building block is crucial already in the traditional approaches to the relativistic formalism, where origin and status of the four-momentum variables remain unclear.

Einstein's 1905 foundation of relativistic kinematics (simultaneity, duration, length and the Lorentz transformation) grew out of the light principle and relativity principle \cite{Einstein '05 - Zur ED bewegter Koerper}.\footnote{Sexl Urbantke \cite{Sexl Urbantke Relativity} explain the light principle in everyday language: The speed of light is constant and independent of the motion of the source. They also give a \emph{pre-theoretic} definition of the principle of relativity: Consider two (e.g. scattering) experiments, set up in exactly the same manner in the inertial frame of Alice resp. Bob (e.g. a collision test or the scattering between two charged particles). The result is the same for both systems. All processes of nature with identical initial conditions lead to identical results.} His early derivation of the mechanical expressions (force, mass, energy) did require one extra postulate: the electromagnetic equations. Though, ''because the Lorentz transformations, the real basis of the special theory of relativity, in itself has nothing to do with the Maxwell theory'' in 1935 Einstein provides an independent derivation of relativistic dynamics which ''except for the Lorentz transformation, will depend only on the assumption of the conservation principles for impulse and energy'' \cite{Einstein '35 - mass energy equivalence}.

\begin{figure}    
  \begin{center}           
  \includegraphics[height=3.6cm]{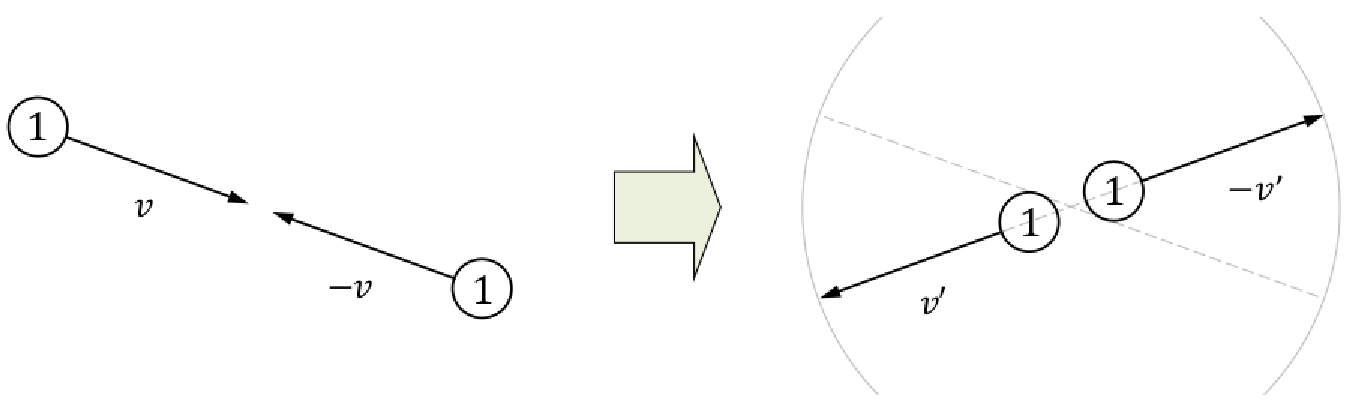}  
  \end{center}
  \vspace{-0cm}
  \caption{\label{pic_elastic_collision} particle pair before and after an elastic collision
    }
  \end{figure}

Einstein \emph{postulates} energy and momentum expressions for a material particle. He verifies the ansatz in an elastic collision between ''identically constituted material particles'' (equal mass). Before and after an eccentric collision the velocities must be equal and opposite (see figure \ref{pic_elastic_collision}).
%
''This holds independently of the particular law of dependence of impulse and energy on the velocity'' \cite{Einstein '35 - mass energy equivalence}. That is a pre-theoretic way of expressing the conservation of the associated ''capability to work'' (energy) and ''impact'' (momentum). For different observers the numerical values from the beginning do not change in the end. Are they also admissible beyond the simple case? The necessary condition is not sufficient.

In much the same way Mermin \cite{Mermin '89 Space and Time in Special Relativity} begins from collisions as ''an encounter between particles which begins and ends with several particles all very far apart from one another, well beyond their interactions, and hence moving uniformly in straight lines''. He takes ''as an experimental fact the conservation of ordinary non-relativistic energy and momentum'' expressions (classical momentum-velocity relation $\mathbf{p}=m\cdot\mathbf{v}$ etc.) for low speed particles and ''then tries to \emph{guess} what a relativistic definition of energy and momentum should be by'' requirements about the non-relativistic limit and the conservation for any inertial observer.\footnote{Relativistic conservation laws must reproduce the known non-relativistic expressions when all particles have speeds much less than $c$ and ''\emph{if} any inertial observer \emph{calculates} the (supposed) energy and momentum of a group of particles and finds they are conserved in a collision, so will any other inertial observer.''} He finds a consistent form for relativistic equations though ''has two shortcomings in mind'': He can ''not show that it is the only one'' and why \emph{must} energy and momentum be conserved?\footnote{Mermin only specifies a consistent form they have \emph{if} they are conserved (which remains his assumption).}

Beside uniqueness also the physical interpretation remained problematic to this day. The inadmissible notion ''relativistic inertial mass'' was an attempt to understand mass from the un-natural assumption of the Newtonian relation between force and acceleration $\mathbf{F}=m\cdot \mathbf{a}$.\footnote{Lorentz 1899 introduced the ''velocity dependence of mass'' by applying the non-relativistic formula $p=m\cdot v$ in the relativistic region, where this formula is not valid. First Einstein 1905 \cite{Einstein '05 - Zur ED bewegter Koerper} adopted the notion ''relativistic inertial mass'', later 1935 \cite{Einstein '35 - mass energy equivalence} avoided it as a basis for the physical theory and ultimately 1948 abandoned it as a meaningful physical notion ''for which no clear definition can be given'' \cite{Adler - Does mass really depend on velocity dad?}.} In general both are not even parallel $\mathbf{F} \nparallel \mathbf{a}$. Already Hertz \cite{Hertz - Einleitung zur Mechanik} stood for eliminating the fundamental role of ''force'' as it grew out of Newton's axiomatic system.\footnote{Hertz outlined \{\ref{Kap - KM Dynamics short Review - Hertz program}\} (and we completed \cite{Hartmann-KM_Dyn}) a foundation of the physical theory from the ''energy'' of directly observable phenomena (independent of Newton's equations of motion and broader).} Adler \cite{Adler - Does mass really depend on velocity dad?}, Okun \cite{Okun - concept of mass} reject the relativistic mass; they prefer an invariant ''rest mass'' which is consistent with a four-vector formulation.\footnote{Their criticism invokes a formal argument, that ''relativistic mass'' obscures the four-dimensional symmetry of the four-vector formulation \cite{Okun - concept of mass}, and a pedagogical claim, that the ''use of relativistic mass can mislead students into believing that the structure of moving objects is actually affected by their motion as was actually the case in the Lorentz theory'' as examined in recent work by Brown \cite{Brown - Physical Relativity}.} Despite repeated attacks from the elegant formalism \cite{Oas - abuse and use of relativistic mass} still many textbooks approach the basic concepts in these two contradicting ways \cite{Oas - use of realtivistic mass in various published works}. From a pure mathematical perspective one can transform the basic equations in arbitrary ways; the physical meaning though fades away.\footnote{Okun \cite{Okun - concept of mass} appeals: ''Every year millions of boys and girls throughout the world are taught special relativity in such a way that they miss the essence of the subject. Archaic and confusing notions are hammered into their heads. It is our duty - the duty of professional physicists - to stop this process''.} The \emph{formal} definition of ''relativistic mass'' from Newton's equations or of ''rest mass'' in the four-vector formalism forfeits the understanding of ''mass'' as a basic observable. Sole focus on the formalism neglects the empirical origin.

Can one reveal the meaning of mass, energy from just \emph{arithmetic} operations?\footnote{Wilczek admits: ''there is much less interpretation involved in the foundation of modern physics... The equations really do speak for themselves'' \cite{Wilczek - concept of mass}. One seeks guidance by formal principles (postulated symmetries, action formalism etc.). Without physical justification that terminology remains mysterious too, intangible to the experience and language of students and teachers (despite original interest in physics).} For kinematics Einstein scrutinizes the underlying vivid \emph{measurement} operations. The entire mathematical formalism was known to Lorentz and Poincar\'{e} before; though Lorentz admits ''the most radical and most important step towards the theory of relativity (was) the elimination of absolute time'' \cite{Sexl Urbantke Relativity}. From the point of view of philosophy of science Sexl, Urbantke regard: ''Poincare's work (as) a partially uninterpreted formalism in which the assignment between theoretical terms and empirical terms is partially absent.''  Einstein begins from the definition of an intrinsic length, duration and undisputed principles of a physical (light, relativity) and methodical character (laser ranging procedure, construct simultaneity) to quantify these basic observables and derive the Lorentz transformation. The tangible operations lead to a mathematical formulation, not the other way around. We take Einstein's view
\begin{quote}
   ''One is ordinarily accustomed to study geometry divorced from any relation between its concepts and experience.'' While the ''pure mathematician... is satisfied if he can deduce his theorems from axioms correctly, that is, without errors of logic. The question as to whether Euclidean geometry is \emph{true} or not does not concern him. But for our purpose it is necessary to \emph{associate} the fundamental concepts of geometry with natural objects; without such an association geometry is worthless for the physicist.'' \cite{Einstein-Grundlagen der ART}
\end{quote}
in place of all mathematical structure.

While everybody knows the basic observables ''duration'', ''length'' from everyday life, for dynamics one only adopts a formal view: ''that energy and momentum are nothing else than functions of mass and velocity that, under suitable conditions, happen to be conserved'' \cite{Sonego - Deriving relativistic momentum and energy}. In the prominent approaches there is no guarantee a priori, that the expressions are conserved in general \cite{Sonego - Deriving relativistic momentum and energy}. Whether Einstein \cite{Einstein '35 - mass energy equivalence}, Mermin \cite{Mermin '89 Space and Time in Special Relativity} guess the energy-momentum expressions or whether Born \cite{Born - Einstein's GR}, Feynman \cite{Feynman lectures I} derive them entirely from pre-theoretic requirements in the simple cases or whether one stipulates the four-momentum formally, for the verification and interpretation in the ''math first - physics second'' approach everyone refers to the same \emph{physical test}: An (in)elastic collision between identically constituted material particles represents the conservation of energy and momentum, whatever the correct formula may be. The velocities must be equal and opposite before and after the collision \cite{Einstein '35 - mass energy equivalence}; and a moving observer determines his velocities for the same process by the Lorentz transformation.\footnote{We assume that Poincare kinematics (measuring motion in \emph{absence of interactions}) is known from light and relativity principles \cite{Einstein '05 - Zur ED bewegter Koerper} and given before the next issue, how to measure present interactions, arises.}

We use these pre-theoretic elements for a complementary ''physics first - math second'' foundation of dynamics. We use the elastic collision not as consistency test \emph{after} guessing nebulous four-momentum expressions; that process becomes our building block for an independent measurement instrument \{\ref{Kap - SRT_Dyn - Basic Measurement}\}. We start from principles of measurement practice and formulate the mathematical description of the tangible operations second. We show how - in principle and without mathematical preassumption - one can \emph{define} reliable physical quantities of energy and momentum from direct measurement operations.

\section{Basic measurement}\label{Kap - SRT_Dyn - Basic Measurement}

First we define the basic observables. One knows the spatiotemporal order ''longer than'' by testing whether one object or process covers the other \cite{Hartmann-SRT-Kin}. Similarly one defines energy, momentum and inertial mass from practical comparisons \cite{Hartmann-KM_Dyn}. According to Leibniz one has ''more capability to work'' if against the same test system the effect of one source \emph{exceeds} the effect of another source (we specify the absorption effect in a calorimeter later). According to Galilei one has more ''impact'' if in an head-on collision test one body \emph{overruns} the other (they collide, stick together and move jointly with one or the other or stop). One can conduct the basic comparison ''$>$'' directly. Next one wants to express the value also numerically. We develop Helmholtz program for finding ''how many times'' more. For standardization of the conduct of reproducible experiments one can manufacture equally long meter sticks, uniform running clocks etc. and concatenate ''$\ast$'' them in suitable ways.

We provide a reservoir $\left\{\circMunit\,, \mathcal{S}\right\}$ with standard bodies ''$\circMunit$'' with same inertia and with equally charged springs ''$\mathcal{S}$'' (they simply must catapult the test objects in the same way). As a reference for ''capability to work'' and ''impact'' we pick the same standard process as for classical measurements (see figure \ref{pic_Wirkungseinheit_Feder}): Let a compressed standard spring catapult a resting particle pair (with standard velocity $\pm\mathbf{v}_{\mathbf{1}}$) into diametrically opposed directions
\be\label{Abschnitt -- Basic Measurement - Einheitswirkung}
   \mathcal{S}_{\mathbf{1}}\big|_{\mathbf{v}=\mathbf{0}} \,,\: \circMunit_{\:\mathbf{v}=\mathbf{0}} \,,\: \circMunit_{\:\mathbf{v}=\mathbf{0}}  \;\;\; \stackrel{w_{\mathbf{1}}}{\Rightarrow} \;\;\;  \circMunit_{\:\mathbf{v}_{\mathbf{1}}} \,,\: \circMunit_{-\mathbf{v}_{\mathbf{1}}}
\ee
or vice versa, let the particle pair compress a neutral spring (we suppress empty springs in the notation). The initial and final state of an inelastic collision (of irrelevant inner structure) are well-defined.

By consecutive compression and decompression of the spring we can generate an eccentric elastic collision between two standard bodies (see figure \ref{pic_SRT_elastic_transversal_collision}c). If Alice drives by with same horizontal velocity, she will see the process as a transversal kick (see figure \ref{pic_SRT_elastic_transversal_collision}a). The kinematics is well-defined by symmetry and relativity principle (view from a moving observer).\footnote{We start from the original physical grounds like Huygens, who did study the elastic collisions between identically constituted bodies to derive the collision laws for billiard balls, and like Einstein \cite{Einstein '35 - mass energy equivalence} and Feynman \cite{Feynman lectures I}, who for the same reason investigate the interaction between bodies which collide and stick together.} Though compared to the classical case the geometric form is slightly modified. Due to relativistic corrections given impact velocities generate smaller deflection angles (Lemma \ref{Lem - SRT Dyn - elast collisions - Winkel und Geschwindigkeit}). From these standard kicks we construct increasingly complex collision models. The coupling ''$\ast$'' involves the same steering task as in Galilei kinematics \cite{Hartmann-KM_Dyn}. We design a machinery to realize the practical norm for a basic measurement device: allow to \emph{count congruent units}. We assemble the relativistic building blocks to model the same elastic head-on collision between two generic (non equivalent!) bodies \{\ref{Kap - SRT_Dyn - Collision model}\} and ultimately an absorption process for a generic particle $\circMa_{\:\mathbf{v}_{\!a}}$ in a calorimeter reservoir \{\ref{Kap - SRT Dyn - Calorimeter model}\}.

Finally we change to an abstract physical perspective. We regard the elements therein solely as representatives of the associated ''capability to work'' and ''impact''. We count, how many standard obstacles the moving body $\circMa_{\:\mathbf{v}_{\!a}}$ can overcome before it stops and thus determine the \emph{magnitude} of its energy. We transfer its impact onto a certain number of standard impulse carriers and thus quantify the momentum \{\ref{Kap - SRT Dyn - Quantification of energy-momentum}\}. We build the calorimeter from one single elementary reference process (\ref{Abschnitt -- Basic Measurement - Einheitswirkung}) in which the three dimensions: unit energy $E\left[ \mathcal{S}_{\mathbf{1}}\big|_{\mathbf{v}=\mathbf{0}} \right]$ (of a compressed standard spring at rest), unit mass $m\left[\circMunit\right]$ (of a standard body) and standard velocity $\mathbf{v}_{\mathbf{1}}$ are inseparably intertwined. From matching the form and layout of the building blocks in the machinery we can count the numbers of activated standard springs, standard bodies, velocity units. We obtain a geometric proof for the kinetic energy-velocity relation and similarly for the generic momentum-velocity relation, the energy-mass relation etc. We develop the physical quantities and derive their fundamental equations.

\subsection{Elastic collisions}\label{Kap - SRT_Dyn - Collision model}

\begin{figure}    
  \begin{center}           
  \includegraphics[height=3.6cm]{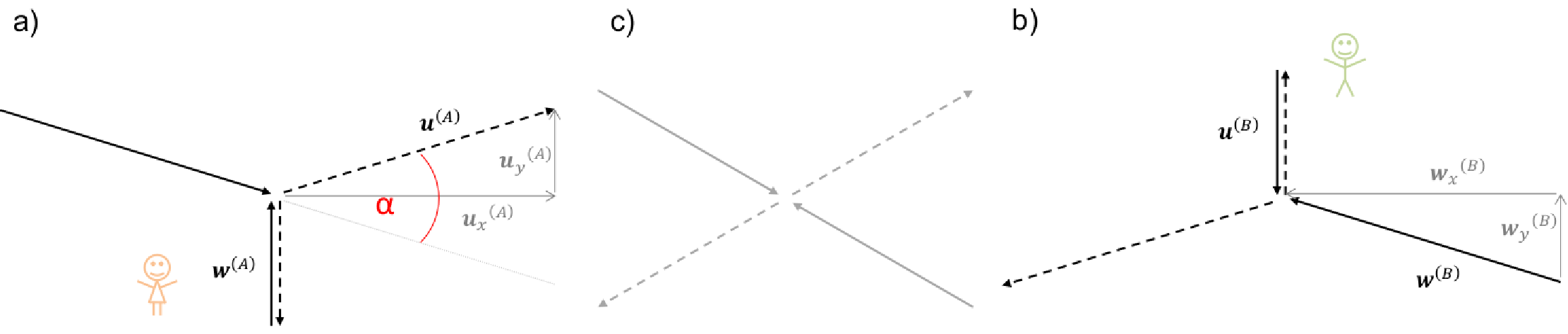}  
  \end{center}
  \vspace{-0cm}
  \caption{\label{pic_SRT_elastic_transversal_collision} c) symmetric elastic collision a) seen as elastic transversal collision when $\mathcal{A}$lice drives by with same horizontal velocity to the left and b) when $\mathcal{B}$ob drives to the right
    }
  \end{figure}

\begin{lem}\label{Lem - SRT Dyn - elast collisions - Winkel und Geschwindigkeit}
Let in an elastic transversal collision between equivalent objects (see figure \ref{pic_SRT_elastic_transversal_collision}a)
\be\label{Abschnitt -- SRT Dynamics - elstic collision model - elastic transversal collision}
     \circMunit_{\:\mathbf{v}} \,,\: \circMunit_{\:\epsilon\cdot\mathbf{v}_{\mathbf{1}}} \;\;\;\Rightarrow\;\;\; \circMunit_{\:\mathbf{v}'} \,,\: \circMunit_{-\epsilon\cdot\mathbf{v}_{\mathbf{1}}}
\ee
reservoir particle $\circMunit_{\:\epsilon\cdot \mathbf{v}_{\mathbf{1}}}$ kick in from below with fixed velocity $\epsilon\cdot \mathbf{v}_{\mathbf{1}}$ and rebound antiparallel. Then the incident object $\circMunit_{\:\mathbf{v}}$ moves on with same velocity $\mathbf{v}'=\mathrm{R}_{\alpha}\!\mathbf{v}$ deflected by an angle $\alpha$
\be\label{Formel - SRT_v-alpha-elastic transversal collision}
   \sin\left( \frac{\alpha}{2} \right) = \frac{\sqrt{1-\frac{v_{x}^2}{c^2}}}{v} \cdot \epsilon \;\; .
\ee
\end{lem}
\textbf{Proof:} The elastic collision of identically constituted bodies $\circMunit$ is well-defined by symmetry and Poincare covariance. One can freely adjust the scattering angle of an eccentric elastic collision (see figure \ref{pic_SRT_elastic_transversal_collision}c), e.g. with a suitable impact parameter between two billiard balls. Let Alice and Bob fly by horizontally into opposite direction and observe the same process. With Feynman's trick \cite{Feynman lectures I} one can determine the kinematical description.

Let $\mathcal{A}$lice drive by in a car with same horizontal velocity to the left. In her view the lower ball moves up and down vertically with the same velocity $\pm\epsilon\cdot v_{\mathbf{1}^{(\mathcal{A})}}$
\[
   \mathbf{w} = \left(
               \begin{array}{c}
                 w^{(\mathcal{A})}_x  \\
                 w^{(\mathcal{A})}_y \\
               \end{array}
             \right) \cdot v_{\mathbf{1}^{(\mathcal{A})}}
             = \left(
               \begin{array}{c}
                 0  \\
                 \pm \epsilon \\
               \end{array}
             \right) \cdot v_{\mathbf{1}^{(\mathcal{A})}} \;\;\;\;\;\;\;\;\;\;\;\;\;\;
   \mathbf{u} = \left(
               \begin{array}{c}
                 u^{(\mathcal{A})}_x  \\
                 \mp u^{(\mathcal{A})}_y \\
               \end{array}
             \right) \cdot v_{\mathbf{1}^{(\mathcal{A})}}
\]
while the upper ball flies on with horizontal component $u^{(\mathcal{A})}_x$ unvaried and a swap in the vertical component $\mp u^{(\mathcal{A})}_y$ (we denote $\mathcal{A}$lice intrinsic reference velocity $v_{\mathbf{1}^{(\mathcal{A})}}$ and similarly $\mathcal{B}$ob's units $v_{\mathbf{1}^{(\mathcal{B})}}$). Let $\mathcal{B}$ob move relative to $\mathcal{A}$lice with same horizontal velocity to the right
\be\label{Abschnitt -- SRT Dynamics - elstic collision model - Lorentz_trafo-Geschw_Bob}
   \mathbf{v}_{\mathcal{B}} = \left(
               \begin{array}{c}
                 u^{(\mathcal{A})}_x  \\
                 0 \\
               \end{array}
             \right) \cdot v_{\mathbf{1}^{(\mathcal{A})}}   \;\; .
\ee
Their measured values for duration, length transform by the Lorentz transformation
\be\label{Abschnitt -- SRT Dynamics - elstic collision model - Lorentz_trafo}
   t^{(\mathcal{B})} = \frac{t^{(\mathcal{A})} - \frac{v}{c^2} \cdot x^{(\mathcal{A})}}{\sqrt{1 - \frac{v^2}{c^2}}} \;\;\;\;\;\;\;\;
   x^{(\mathcal{B})} = \frac{x^{(\mathcal{A})} - v \cdot t^{(\mathcal{A})}}{\sqrt{1 - \frac{v^2}{c^2}}} \;\;\;\;\;\;\;\;
   y^{(\mathcal{B})} = y^{(\mathcal{A})}  \;\;\;\;\;\;\;\;
   z^{(\mathcal{B})} = z^{(\mathcal{A})}  \;\; .
\ee
From these basic quantities one induces the transformation for the derived quantity of velocity. One derives $\mathcal{B}$ob's velocity values for the colliding bodies
\bea
   u^{(\mathcal{B})}_x & := & \frac{\Delta x^{(\mathcal{B})}}{\Delta t^{(\mathcal{B})}} \;\;\stackrel{(\ref{Abschnitt -- SRT Dynamics - elstic collision model - Lorentz_trafo})}{=}\;\; \frac{\Delta x^{(\mathcal{A})} - v \cdot \Delta t^{(\mathcal{A})}}{\Delta t^{(\mathcal{A})} - \frac{v}{c^2} \cdot \Delta x^{(\mathcal{A})}} \;\;=\;\; \frac{u_x - v}{1-\frac{u_x\cdot v}{c^2}} \label{Abschnitt -- SRT Dynamics - elstic collision model - induc_velocity_trafo - horizontal} \\
   u^{(\mathcal{B})}_y & := & \frac{\Delta y^{(\mathcal{B})}}{\Delta t^{(\mathcal{B})}} \;\;\stackrel{(\ref{Abschnitt -- SRT Dynamics - elstic collision model - Lorentz_trafo})}{=}\;\; \frac{\Delta y^{(\mathcal{A})} \cdot \sqrt{1 - \frac{v^2}{c^2}} }{\Delta t^{(\mathcal{A})} - \frac{v}{c^2} \cdot \Delta x^{(\mathcal{A})}} \;\;=\;\; \frac{u_y \cdot \sqrt{1 - \frac{v^2}{c^2}} }{1-\frac{u_x\cdot v}{c^2}} \label{Abschnitt -- SRT Dynamics - elstic collision model - induc_velocity_trafo}   \;\; .
\eea
From $\mathcal{B}$ob's perspective (on the same collision) the lower ball flies to the left with fixed horizontal velocity $w^{(\mathcal{B})}_x$ and a reversion in the vertical component $\pm w^{(\mathcal{B})}_y$
\[
   \mathbf{w} = \left(
               \begin{array}{c}
                 w^{(\mathcal{B})}_x  \\
                 \pm w^{(\mathcal{B})}_y \\
               \end{array}
             \right) \cdot v_{\mathbf{1}^{(\mathcal{B})}}
   \;\;\;\;\;\;\;\;\;\;\;\;\;\;
   \mathbf{u} = \left(
               \begin{array}{c}
                 u^{(\mathcal{B})}_x  \\
                 u^{(\mathcal{B})}_y \\
               \end{array}
             \right) \cdot v_{\mathbf{1}^{(\mathcal{B})}}
             = \left(
               \begin{array}{c}
                 0  \\
                 \mp \epsilon \\
               \end{array}
             \right) \cdot v_{\mathbf{1}^{(\mathcal{B})}}
\]
while the upper ball moves up and down vertically with the same velocity $\epsilon\cdot v_{\mathbf{1}^{(\mathcal{B})}}$. For $\mathcal{A}$lice and $\mathcal{B}$ob the roles of the upper and lower ball are reversed (see figure \ref{pic_SRT_elastic_transversal_collision}b). $\mathcal{A}$lice and $\mathcal{B}$ob's perspectives are equivalent; they observe equal ''vertical'' impact velocities
\be\nn
   u^{(\mathcal{B})} \;\; \stackrel{!}{=} \;\; - w^{(\mathcal{A})} \;\; = \;\; \epsilon  \;\; .
\ee
Finally $\mathcal{A}$lice can express the vertical component of her ''horizontal'' particle
\be\nn
   u^{(\mathcal{A})}_y \;\; \stackrel{(\ref{Abschnitt -- SRT Dynamics - elstic collision model - Lorentz_trafo-Geschw_Bob})(\ref{Abschnitt -- SRT Dynamics - elstic collision model - induc_velocity_trafo})}{=} \;\; \epsilon \cdot \sqrt{1 - \frac{{u^{(\mathcal{A})}_x}^2}{c^2}}
\ee
in terms of her impact velocity $\epsilon$. Then the deflection angle $\alpha$ follows from figure \ref{pic_SRT_elastic_transversal_collision}a.
\qed
In the following construction outline we fix the transversal impact velocity $\epsilon\cdot v_{\mathbf{1}}$ of all reservoir elements. With given initial velocity $v$ we can determine the deflection angle $\alpha(v,\epsilon)$ and vice versa, provided the latter we find the necessary initial velocity $v(\alpha, \epsilon)$.

\begin{figure}    
  \begin{center}           
  \includegraphics[height=20cm]{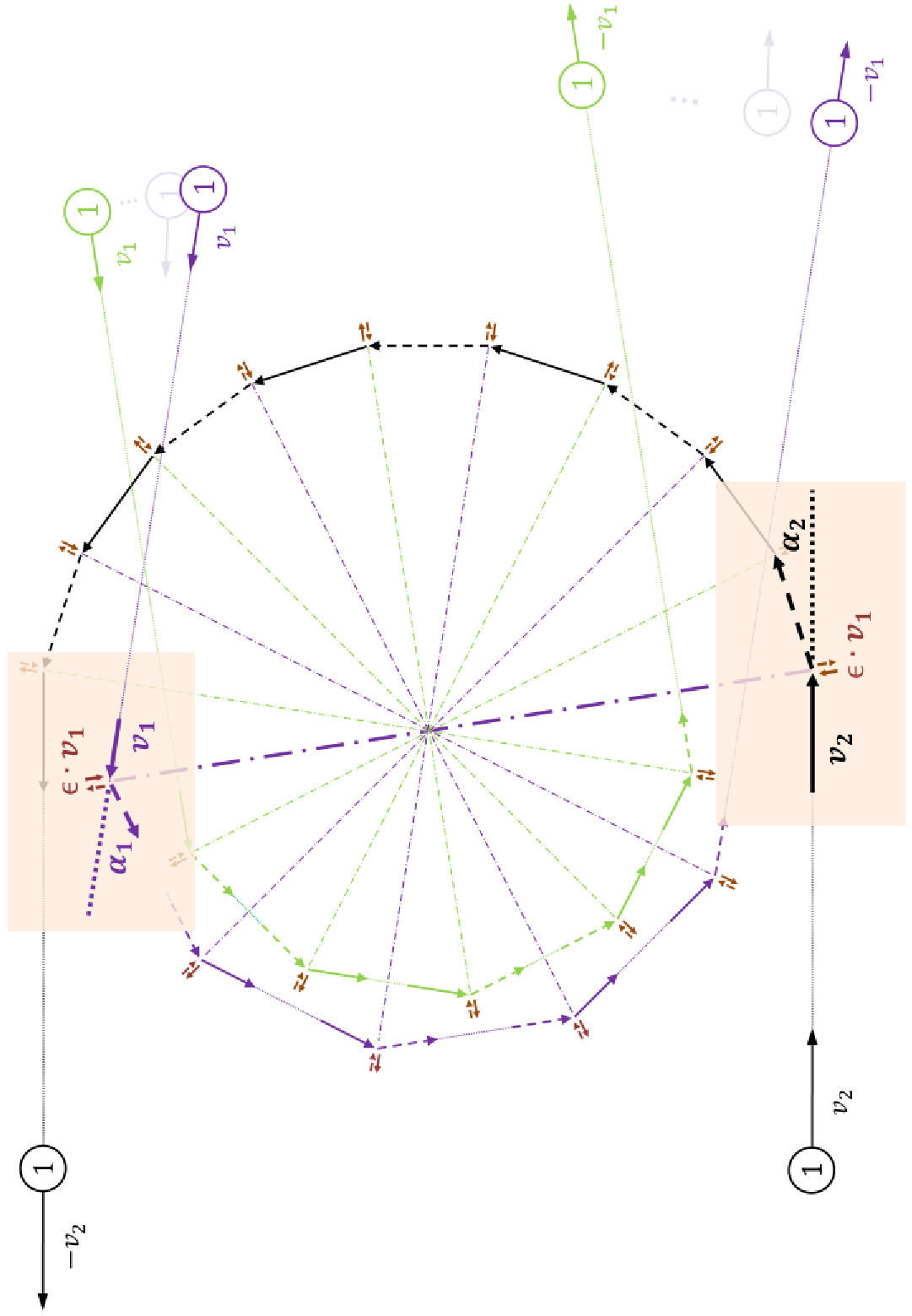}  
  \end{center}
  \vspace{-0cm}
  \caption{\label{pic_elast_collision_2-5} align standard collisions
    }
  \end{figure}

Let one transversal standard kick (highlighted black in figure \ref{pic_elast_collision_2-5}) deflect an incident particle $\circMunit_{\:v_{2}}$ with velocity $v_{2}$ around angle $\alpha_{2}=18^{\circ}$; after a series of $9$ more kicks of the same strength its direction is reversed; and similar for a slower particle $\circMunit_{\:v_{\mathbf{1}}}$ which requires half the standard kicks (highlighted purple in figure \ref{pic_elast_collision_2-5}).
In every transversal kick (\ref{Abschnitt -- SRT Dynamics - elstic collision model - elastic transversal collision}) we generate recoil particles $\circMunit_{\:\epsilon\cdot\mathbf{v}_{\mathbf{1}}}$ from an external reservoir with \emph{same} velocity $\epsilon\cdot v_{\mathbf{1}}$ (depicted brown in figure \ref{pic_elast_collision_2-5}). To capture and recycle all the temporarily mobilized steering elements from the center, we align all ''radial'' standard kicks in the depicted way (pairwise along the dashed lines at diametrically opposed locations). In the total balance the reservoir particles do not appear. In the net result only the motion of the one particle $\circMunit_{\:v_{2}}$ from bottom left and the (bundle of) two particles $\circMunit_{\:v_{\mathbf{1}}}$, $\circMunit_{\:v_{\mathbf{1}}}$ from the top right is reversed. We determine the relation between their admissible velocities $v_{2}$ and $v_{\mathbf{1}}$ from matching \emph{form} and \emph{number} of the radial kicks $w_{\mathrm{rad}} \left[ \circMunit_{\:\mathbf{v}_{2}} , \circMunit_{\:\epsilon\cdot v_{\mathbf{1}}} \right]$ and $w_{\mathrm{rad}} \left[ \circMunit_{\:\mathbf{v}_{\mathbf{1}}} , \circMunit_{\:\epsilon\cdot v_{\mathbf{1}}} \right]$ (so that the total machinery functions).\footnote{Every radial standard kick must deflect the incident particle $\circMunit_{\:\mathbf{v}_{2}}$ (highlighted black in figure \ref{pic_elast_collision_2-5}) around one half of the deflection angle $\alpha_{2} \stackrel{!}{=} \frac{1}{2} \cdot \alpha_{1}$ than for the slower particle $\circMunit_{\:\mathbf{v}_{\mathbf{1}}}$ (highlighted purple).}

By refinement of the building blocks one can construct similar models for the elastic collision between one particle $\circMunit_{\:\mathbf{v}_{n}}$ with high velocity $\mathbf{v}_{n}$ (from bottom left) and a spreading bundle of $n$ standard particles $\circMunit_{\:\mathbf{v}_{\mathbf{1}}},\ldots,\circMunit_{\:\mathbf{v}_{\mathbf{1}}}$ (from top right). By construction (see figure \ref{pic_elast_collision_2-5}) the incident bundle inherits an opening angle of order $\alpha_{1}$. In the refinement limit $\epsilon \rightarrow 0$ (of an increasing number of radial kicks with negligible impact velocity $\epsilon\cdot v_{\mathbf{1}}$) the spreading of the bundle $\mathrm{lim}_{\epsilon \rightarrow 0} \: \alpha_1(v_{\mathbf{1}},\epsilon) \stackrel{(\ref{Formel - SRT_v-alpha-elastic transversal collision})}{=} 0$ narrows: We get an elastic head-on collision (\ref{Abschnitt -- SRT kin quant elast coll - elast head-on collision}).
\begin{pr}\label{Prop - SRT Dynamics - elstic collision model - refinement limit elast n+1 Stoss}
Let one standard body $\circMunit_{\:\mathbf{v}_{n}}$ (from left) and a rigid composite $\circMn := \circMunit\ast\ldots\ast\circMunit$ of $n$ standard elements (from right) collide and repulse from one another with reversed velocities, symbolized
\be\label{Abschnitt -- SRT kin quant elast coll - elast head-on collision}
   \circMunit_{\:\mathbf{v}_n} \,,\: \circMn_{\:\mathbf{v}_{\mathbf{1}}} \;\;\Rightarrow\;\;\circMunit_{-\mathbf{v}_n} \,,\: \circMn_{-\mathbf{v}_{\mathbf{1}}}  \;\; .
\ee
Then in Poincare kinematics the initial velocities $\mathbf{v}_{n}$, $\mathbf{v}_{\mathbf{1}}$ satisfy the relation
\be\label{Abschnitt -- SRT Dynamics - elstic collision model - kinemtical relations elast collision n+1}
   \frac{\mathbf{v}_n}{\sqrt{1-\frac{{v_n}^2}{c^2}}} \;\; = \;\; - n \cdot \frac{\mathbf{v}_{\mathbf{1}}}{\sqrt{1-\frac{{v_{\mathbf{1}}}^2}{c^2}}} \;\; .
\ee
\end{pr}
\textbf{Proof:}
We approximate the elastic collision between two generic objects. Without restricting generality let them be composites of unit objects $\circMunit$. We do not presuppose how the velocities change in more complex collisions. The trick is to mediate the direct interaction by an indirect replacement process with an external \emph{reservoir}, which \emph{in the end is not effected}! Our model solely consists of elastic collisions between standard objects $\circMunit$ which must behave in a symmetrical way. We know the collision law for $1+1$ equivalent objects by symmetry and relativity principle (Lemma \ref{Lem - SRT Dyn - elast collisions - Winkel und Geschwindigkeit}). Based on it we construct the collision model for $n+1$ equivalent objects and ultimately for $n+m$ composites (Corollary \ref{Cor - SRT Dynamics - elstic collision model - kinematical relation m/n+1 Stoss}). We assemble intrinsically well-defined building blocks to a \emph{functioning} configuration (criterium above). From the number and form we derive the amount of matter-velocity relation (\ref{Abschnitt -- SRT Dynamics - elstic collision model - kinemtical relations elast collision n+1}).

\begin{figure}    
  \begin{center}           
  \includegraphics[height=14cm]{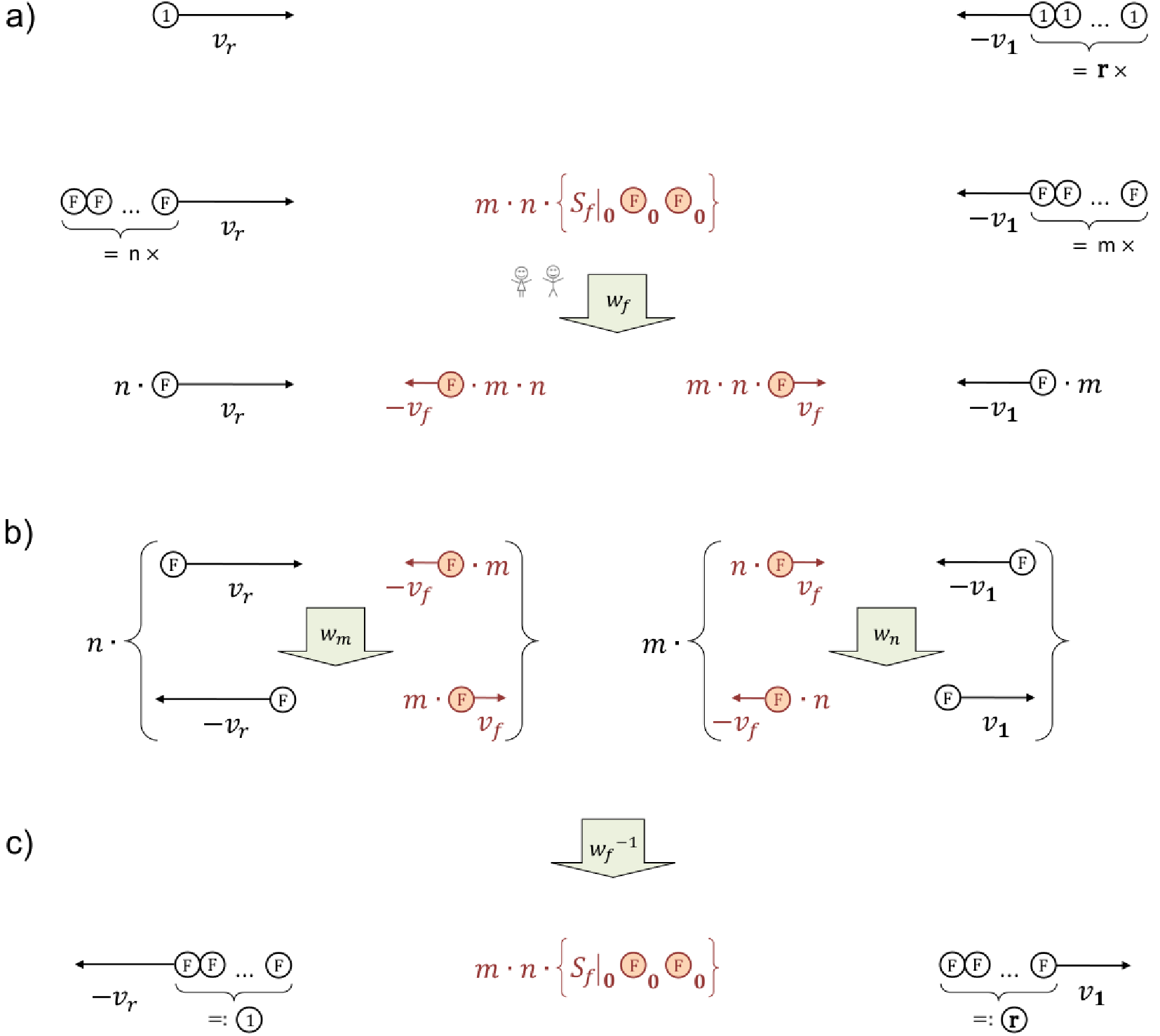}  
  \end{center}
  \vspace{-0cm}
  \caption{\label{pic_SRT_collision_model_refined} a) The \emph{direct} elastic head-on collision of two composites is b) \emph{mediated} by temporarily activated fragment pairs $\circMF_{-\mathbf{v}_f} \,,\: \circMF_{\:\mathbf{v}_f}$. In the end c) all steering resources are recycled back into the fragment reservoir $\left\{ \mathcal{S}_{f}\big|_{\mathbf{v}=\mathbf{0}} , \circMF_{\:\mathbf{v}=\mathbf{0}} \right\}$.
    }
  \end{figure}

The construction outlined before follows the same steps as in Galilei kinematics \cite{Hartmann-KM_Dyn} except that the relativistic building blocks obey modified impact velocity-deflection angle relations $v_{n}(\alpha_n, \epsilon)$ and $v_{\mathbf{1}}(\alpha_1, \epsilon)$ (\ref{Formel - SRT_v-alpha-elastic transversal collision}). As one can see from figure \ref{pic_elast_collision_2-5} we ultimately align the radial standard kicks (highlighted black resp. purple) against the fast incident object $\circMunit_{\:v_n}$ and against the $n$ incident elements $\circMunit_{\:v_{\mathbf{1}}}, \ldots, \circMunit_{\:v_{\mathbf{1}}}$ with matching deflection angles $\alpha_1\stackrel{!}{=}n\cdot\alpha_n$. In the refinement limit $\epsilon\rightarrow0$ with $\alpha_i\rightarrow0$, ${v_{i}}_y\rightarrow0$, ${v_i}_x\rightarrow v_i$, $\sin \alpha_i \rightarrow \alpha_i$ the matching condition is equivalent to $\sin(\frac{\alpha_1}{2}) \stackrel{!}{=} \sin(n\cdot \frac{\alpha_n}{2}) \rightarrow n\cdot\sin(\frac{\alpha_n}{2})$ and implies with Lemma \ref{Lem - SRT Dyn - elast collisions - Winkel und Geschwindigkeit}
\be
   \frac{\sqrt{1-\frac{v_{\mathbf{1}}^2}{c^2}}}{v_{\mathbf{1}}} \cdot \epsilon \;\; \stackrel{(\ref{Formel - SRT_v-alpha-elastic transversal collision})}{=} \;\; n\cdot \frac{\sqrt{1-\frac{v_{n}^2}{c^2}}}{v_{n}} \cdot \epsilon  \nn
\ee
the relation (\ref{Abschnitt -- SRT Dynamics - elstic collision model - kinemtical relations elast collision n+1}) between the admissible initial velocities $v_{\mathbf{1}}$, $v_{n}$. Our physical model mediates the elastic head-on collision between one unit object $\circMunit_{\:v_{n}}$ with velocity $v_{n}(v_{\mathbf{1}},n)$ and a parallel beam of $n$ elements $\circMunit_{\:v_{\mathbf{1}}}, \ldots ,\circMunit_{\:v_{\mathbf{1}}}$ with velocity $v_{\mathbf{1}}$. Before and after the collision its elements fly with same velocity $\mathbf{v}_{\mathbf{1}}$ as if they were bound in a rigid composite $\circMn_{\:\mathbf{v}_{\mathbf{1}}}$.
\qed
\begin{co}\label{Cor - SRT Dynamics - elstic collision model - kinematical relation m/n+1 Stoss}
In an elastic head-on collision $\circMunit_{\:\mathbf{v}_r} \,,\: \circMr_{-\mathbf{v}_{\mathbf{1}}} \Rightarrow \circMunit_{-\mathbf{v}_r} \,,\: \circMr_{\:\mathbf{v}_{\mathbf{1}}}$ between unit object $\circMunit_{\:\mathbf{v}_{r}}$ and a composite $\circMr$ of $r:=\frac{m}{n}$ unit elements $\circMunit_{\:\mathbf{v}_{\mathbf{1}}}$ with standard velocity $\mathbf{v}_{\mathbf{1}}$ the admissible velocity $\mathbf{v}_{r}$ satisfies
\be\label{Abschnitt -- SRT Dynamics - elstic collision model - kinemtical relations elast collision m/n+1}
    \gamma_{v_r}\cdot\mathbf{v}_r \;\; = \;\; \frac{m}{n} \cdot \gamma_{v_{\mathbf{1}}}\cdot\mathbf{v}_{\mathbf{1}} \;\; .\footnote{In our calorimeter model \{\ref{Kap - SRT Dyn - Calorimeter model}\} we will process \emph{fragments} of standard impulse carriers $\circMunit_{\:\mathbf{v}_\mathbf{1}}$. We abbreviate the common term $\frac{1}{\sqrt{1-\frac{v^2}{c^2}}} =: \gamma$ in all arithmetic expressions.}
\ee
\end{co}
\textbf{Proof:}
Let the fragment $\circMF$ play the role of a \emph{common physical denominator} of the standard body $\circMunit=:\underbrace{\circMF\ast\ldots\ast\circMF}_{n\times}$ and its rational multiple $\circMr=\underbrace{\circMunit\ast\ldots\ast\circMunit}_{\frac{m}{n}\times}:=\underbrace{\circMF\ast\ldots\ast\circMF}_{m\times}$. We express the head-on collision between the standard body $\circMunit_{\:\mathbf{v}_r}$ and the (rational) composite $\circMr_{\:\mathbf{v}_{\mathbf{1}}}$ in terms of their common fragments: the composite $n\cdot\circMF_{\:\mathbf{v}_{r}}$ of $n$ fragments with admissible velocity $\mathbf{v}_{r}$ from left runs into the other composite $m\cdot\circMF_{\:\mathbf{v}_{\mathbf{1}}}$ of $m$ fragments with standard velocity $\mathbf{v}_{\mathbf{1}}$ from right (see figure \ref{pic_SRT_collision_model_refined}a).

Let Alice have access to an external ''fragment reservoir'' $\left\{ \mathcal{S}_{f}\big|_{\mathbf{0}} , \circMF_{\:\mathbf{0}} \right\}$ and standard processes ''$w_f$'' analogous to (\ref{Abschnitt -- basic dynamical measures - Einheitswirkung}).\footnote{By our pre-theoretic construction principle identically constituted bodies behave symmetrical in elastic head-on collisions. The size of the standard object $\circMunit$ is variable or composable from smaller fragments $\circMF$.} By temporarily expending $m\cdot n$ energy sources $\mathcal{S}_{f}\big|_{\mathbf{0}}$
\[
   m\cdot n\cdot \left(  \;\mathcal{S}_{f}\big|_{\mathbf{0}} \,,\: \circMF_{\:\mathbf{0}} \,,\: \circMF_{\:\mathbf{0}}
   \;\;\stackrel{w_f}{\Rightarrow}\;\; \circMF_{-\mathbf{v}_f} \,,\: \circMF_{\:\mathbf{v}_f}   \; \right)
\]
Alice generates $m\cdot n$ anti-parallel fragment-pairs. We chose fragment velocity $v_f(v_{\mathbf{1}},n)$
\be\label{Abschnitt -- SRT Dynamics - elstic collision model - kinemtical relations elast collision m/n+1 - rechts}
    \gamma_{v_{\mathbf{1}}}\cdot\mathbf{v}_{\mathbf{1}} \;\; \stackrel{(\ref{Abschnitt -- SRT Dynamics - elstic collision model - kinemtical relations elast collision n+1})}{=} \;\; n \cdot \gamma_{v_f}\cdot\mathbf{v}_{f}
\ee
such that in an elastic head-on collision $n\cdot\circMF_{\:\mathbf{v}_f} \,,\: \circMF_{-\mathbf{v}_{\mathbf{1}}} \;\stackrel{w_n}{\Rightarrow}\; n\cdot\circMF_{-\mathbf{v}_f}  \,,\: \circMF_{\:\mathbf{v}_{\mathbf{1}}}$ the composite $n\cdot\circMF$ of $n$ equivalent elements $\circMF_{\:\mathbf{v}_{f}}$ with fragment velocity $\mathbf{v}_{f}$ rebounds antiparallel from one single fragment $\circMF_{\:\mathbf{v}_{\mathbf{1}}}$ with standard velocity $\mathbf{v}_{\mathbf{1}}$ (see figure \ref{pic_SRT_collision_model_refined}b). Further we fix the admissible velocity $v_r(v_{f},m)$
\be\label{Abschnitt -- SRT Dynamics - elstic collision model - kinemtical relations elast collision m/n+1 - links}
    \gamma_{v_r}\cdot\mathbf{v}_{r} \;\; \stackrel{(\ref{Abschnitt -- SRT Dynamics - elstic collision model - kinemtical relations elast collision n+1})}{=} \;\; m \cdot \gamma_{v_f}\cdot\mathbf{v}_{f}
\ee
such that in an elastic head-on collision $\circMF_{\:\mathbf{v}_r} \,,\: m\cdot\circMF_{-\mathbf{v}_{f}} \;\stackrel{w_m}{\Rightarrow}\; \circMF_{-\mathbf{v}_r}  \,,\: m\cdot\circMF_{\:\mathbf{v}_{f}}$ one single fragment $\circMF_{\:\mathbf{v}_r}$ with initial velocity $\mathbf{v}_r$ rebounds antiparallel from the composite $m\cdot\circMF$ of $m$ equivalent elements $\circMF_{\:\mathbf{v}_{\mathbf{1}}}$ with fragment velocity $\mathbf{v}_{f}$.

The fragment pairs $\left\{\circMF_{-\mathbf{v}_f} , \circMF_{\:\mathbf{v}_f}\right\}$ mediate an elastic head-on collision between an incident composite $n\cdot\circMF_{\:\mathbf{v}_{r}}$ from left and another composite $m\cdot\circMF_{-\mathbf{v}_{\mathbf{1}}}$ from right
\[
   \left( m\cdot n\cdot  w_f  \right) \ast \left( n\cdot w_m , m\cdot w_n \right) \ast \left( m\cdot n\cdot  w_f^{-1} \right):\;\: \underbrace{n\cdot\circMF_{\:\mathbf{v}_{r}}}_{\circMunit_{\mathbf{v}_r}} \,,\: \underbrace{m\cdot\circMF_{-\mathbf{v}_{\mathbf{1}}}}_{\circMr_{-\mathbf{v}_{\mathbf{1}}}} \;\Rightarrow\; \underbrace{n\cdot\circMF_{-\mathbf{v}_{r}}}_{\circMunit_{-\mathbf{v}_r}} \,,\: \underbrace{m\cdot\circMF_{\:\mathbf{v}_{\mathbf{1}}}}_{\circMr_{\mathbf{v}_{\mathbf{1}}}}
\]
Each fragment rebounds antiparallel with same velocity. Alice can recycle $(m\cdot n)\times w_{f}^{-1}$ all temporarily expended resources back into the fragment reservoir $\left\{ \mathcal{S}_{f}\big|_{\mathbf{0}} , \circMF_{\:\mathbf{0}} \right\}$ (see figure \ref{pic_SRT_collision_model_refined}c). The reservoir mediated collision has the same initial and final state as the direct elastic collision of both composites. The admissible initial velocity $v_r(v_{\mathbf{1}},\frac{m}{n})$ is given from
\be \nn
    \gamma_{v_r}\cdot\mathbf{v}_{r} \;\; \stackrel{(\ref{Abschnitt -- SRT Dynamics - elstic collision model - kinemtical relations elast collision m/n+1 - links})(\ref{Abschnitt -- SRT Dynamics - elstic collision model - kinemtical relations elast collision m/n+1 - rechts})}{=} \;\; \frac{m}{n} \cdot \gamma_{v_{\mathbf{1}}}\cdot\mathbf{v}_{\mathbf{1}} \;\; .
\ee
\qed

\subsection{Absorption model}\label{Kap - SRT Dyn - Calorimeter model}

By controlled linkage of elastic head-on collisions and by relativity principle (same process seen from a moving perspective) we have developed an absorption process for a generic particle in a calorimeter (see figure \ref{pic_calorimeter_model}). Let for example one fast standard particle collide with a composite of $n$ elements: $\circMunit_{\:n\mathbf{v}_{\mathbf{1}}} , \circMn_{-\mathbf{v}_{\mathbf{1}}} \stackrel{(\ref{Abschnitt -- SRT kin quant elast coll - elast head-on collision})}{\Rightarrow} \circMunit_{-n\mathbf{v}_{\mathbf{1}}} , \circMn_{\:\mathbf{v}_{\mathbf{1}}}$. For an observer who is initially comoving $-\mathbf{v}_{\mathbf{1}}$ with the composite: $\circMunit_{\:(n+1)\mathbf{v}_{\mathbf{1}}} , \circMn_{\:(-1+1)\cdot\mathbf{v}_{\mathbf{1}}} \Rightarrow \circMunit_{\:(-n+1)\mathbf{v}_{\mathbf{1}}} , \circMn_{\:(1+1)\mathbf{v}_{\mathbf{1}}}$ the particle kicks a resting composite into motion $2\mathbf{v}_{\mathbf{1}}$ and rebounds with reduced velocity to the left. From those deceleration kicks we build the calorimeter model. On the left we place again a suitable number of reservoir elements into the way, such that they get kicked out with the same standard velocity $-2\mathbf{v}_{\mathbf{1}}$. The incident particle successively rebounds with reduced velocity, until it stops inside the calorimeter (for details see \{\ref{Kap - KM Dynamics - Basic Dynamical Measures - Measurement Means - Kinematic Quantification Calorimeter Action}\}).

When we build the same models in Poincare kinematics, then for the individual right- and left-deceleration kicks additional Lorentz terms appear. We have derived the generic collision law (Proposition \ref{Prop - SRT Dynamics - elstic collision model - refinement limit elast n+1 Stoss}) by assembling well-defined transversal kicks between pairs of standard bodies. In the same way we assemble intrinsically well-defined head-on collisions (\ref{Abschnitt -- SRT kin quant elast coll - elast head-on collision}) to a \emph{functioning} calorimeter configuration, which (whatever comes in) generates only standard energy and momentum carriers. We construct the deceleration-cascade with suitable fragments of standard bodies, so that they all get kicked out again with the same standard velocity $v_s$ on both sides of the calorimeter. Then we can integrate all fragments of congruent energy and momentum units for the entire deceleration.
\begin{pr}\label{Prop - SRT Dynamics - calorimeter model - reservoir balance for absorption}
The calorimeter-deceleration-cascade is a physical model for absorbing unit object $\circMunit_{\:\mathbf{v}_{\mathcal{O}}}$ with velocity $\mathbf{v}_{\mathcal{O}}=v^{(S)}_{\mathcal{O}}\cdot\,\mathbf{v}_S$ in an external calorimeter where it comes to rest
\be
   \circMunit_{\:v\cdot \mathbf{v}_S}  \;\;\;
   \Rightarrow \;\;\;
   \circMunit_{\:\mathbf{0}} \,,\:  \mathrm{RB} \nn \;\; .
\ee
In return we extract the reservoir balance for absorption
\be\label{Abschnitt -- SRT Dynamics - calorimeter model - reservoir balance for absorption}
   \mathrm{RB}\left[\circMunit_{\:v\cdot \mathbf{v}_S} \Rightarrow \circMunit_{\:\mathbf{0}} \right] \;\; := \;\;
   \left(c^2\cdot (\gamma - 1) - \gamma \cdot v   \right) \cdot \left\{\circMunit_{-\mathbf{v}_S}, \circMunit_{\:\mathbf{v}_S}\right\} \;\; , \;\; \left( \gamma \cdot v \right) \cdot \circMunit_{\:\mathbf{v}_S}
\ee
a certain number of standard particle pairs $\left\{\circMunit_{-\mathbf{v}_S}, \circMunit_{\:\mathbf{v}_S}\right\}$ and impulse carriers $\circMunit_{\:\mathbf{v}_S}$ from a reservoir with resting standard elements $\left\{ \circMunit_{\:\mathbf{v}=\mathbf{0}} \right\}$ (which we suppress in the notation).\footnote{Let measurements refer to an intrinsic standard velocity $\mathbf{v}_S$ (of a clockhand). They specify \emph{how many times} larger the original velocity $\mathbf{v}_{\mathcal{O}}=\underbrace{v^{(S)}_{\mathcal{O}}}_{=:\:v}\!\cdot\,\mathbf{v}_S$, speed of light $c_{\mathcal{L}}=\underbrace{c^{(S)}_{\mathcal{L}}}_{=:\:c}\!\cdot\,\mathbf{v}_S$ etc. is than the reference $\mathbf{v}_S$. We abbreviate the numerical values $v$ resp. $c$; the full notation distinguishes values for different observers.}
\end{pr}
\textbf{Proof:}
Bob steers a series of elastic collisions with suitable fragments of resting reservoir elements. For every deceleration kick he adjusts the amount of matter in the fragment to the momentary impact velocity. Alice and Charlie help with the preparation.

\begin{figure}    
  \begin{center}           
  \includegraphics[height=18cm]{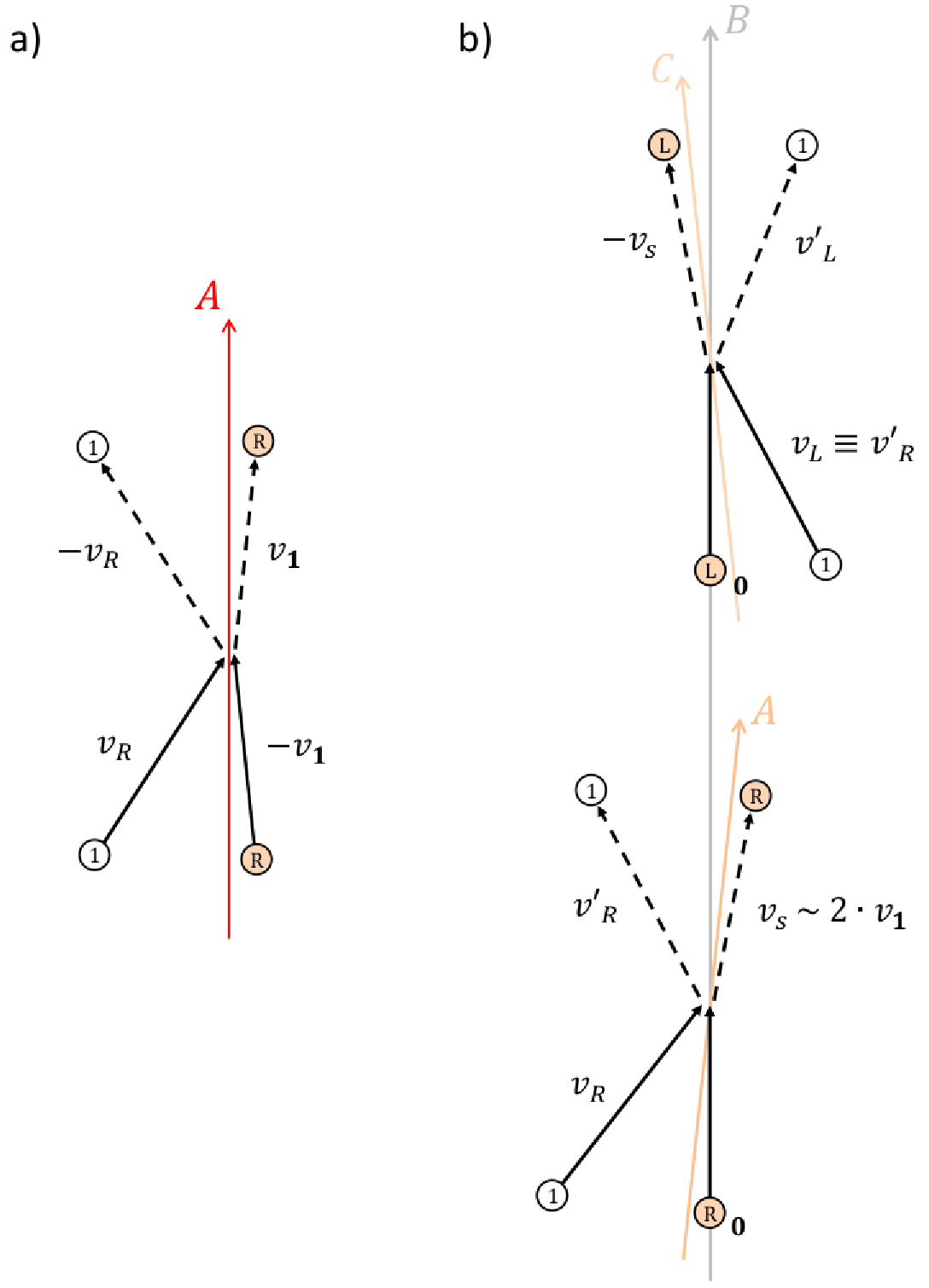}  
  \end{center}
  \vspace{-0cm}
  \caption{\label{pic_SRT_deceleration_kicks} a) elastic head-on collision as b) deceleration kicks in a spacetime diagram
    }
  \end{figure}

Let particle $\circMunit_{\:\mathbf{v}_R}$ with initial velocity $\mathbf{v}_R$ strike into the right side of the calorimeter (see bottom in figure \ref{pic_SRT_deceleration_kicks}b).
For given velocity $v_R$ and \emph{fixed} unit velocity $v_{\mathbf{1}}$ $\mathcal{A}$lice prepares a composite $n_R \cdot \circMunit =: \circMR$ of $n_R(v_R,v_{\mathbf{1}})$ standard bodies
\be\label{Abschnitt -- SRT Dynamics - calorimeter model - reservoir balance for absorption - n von Rechtskick linear}
    n_R \;\; \stackrel{(\ref{Abschnitt -- SRT Dynamics - elstic collision model - kinemtical relations elast collision n+1})}{:=} \;\; \gamma_{v_R} \cdot v_{R}^{(\mathcal{A})}  \cdot (\gamma_{v_{\mathbf{1}}} \cdot v_{\mathbf{1}})^{-1}
    \;\; \simeq \;\; \gamma_{v_R} \cdot v_{R}^{(\mathcal{A})}  \cdot \left(
   {\underbrace{v_{\mathbf{1}}}_{=\:1} \;+\; \underbrace{\mathcal{O}\:(\,\frac{v_{\mathbf{1}}}{c}\,)^2}_{\ll 1}}\right)
\ee
(for small unit velocity $v_{\mathbf{1}} \ll c$ we approximate with the dominant term) such that in head-on collision $\circMunit_{\:\mathbf{v}_R} , \circMR_{-\mathbf{v}_{\mathbf{1}}} \Rightarrow \circMunit_{-\mathbf{v}_R}  , \circMR_{\:\mathbf{v}_{\mathbf{1}}}$ the particle $\circMunit_{\:\mathbf{v}_R}$ and the composite $\circMR_{-\mathbf{v}_{\mathbf{1}}}$ \emph{rebound antiparallel} with velocities $v_R$ and $v_{\mathbf{1}}$ (see figure \ref{pic_SRT_deceleration_kicks}a). Let $\mathcal{B}$ob comove with the composite with velocity $\mathbf{v}_{\mathcal{B}}=\underbrace{v_{\mathcal{B}}^{(\mathcal{A})}}_{-1}\cdot\mathbf{v}_{\mathbf{1}^{(\mathcal{A})}}$ relative to $\mathcal{A}$lice to the left. $\mathcal{B}$ob views the objects $\circMi$ in her process
$\circMunit_{\:\mathbf{v}_R} , \circMR_{\:\mathbf{0}} \Rightarrow \circMunit_{\:\mathbf{v}'_R}  , \circMR_{\:\mathbf{v}_S}$ (see figure \ref{pic_SRT_deceleration_kicks}b) with transformed velocities $v_i^{(\mathcal{B})} \left(v_i^{(\mathcal{A})} , v_{\mathcal{B}}^{(\mathcal{A})} \right)$
\be\label{Abschnitt -- SRT Dynamics - calorimeter model - reservoir balance for absorption - v Trafo}
   v_i^{(\mathcal{B})} \; \stackrel{(\ref{Abschnitt -- SRT Dynamics - elstic collision model - induc_velocity_trafo - horizontal})}{=} \;
   \frac{v_i^{(\mathcal{A})} - v_{\mathcal{B}}^{(\mathcal{A})}}{1-\frac{v_i^{(\mathcal{A})} \cdot v_{\mathcal{B}}^{(\mathcal{A})}}{c^2}}
   \;\; \simeq \;\;
   v_i^{(\mathcal{B})}\big|_{v_{\mathcal{B}}^{(\mathcal{A})}\,=\,0} \:+\: \frac{\partial v_i^{(\mathcal{B})}}{\partial v_{\mathcal{B}}^{(\mathcal{A})}}\bigg|_{0} \cdot v_{\mathcal{B}}^{(\mathcal{A})}
   \;=\; v^{(\mathcal{A})}_i - \left( 1 - \frac{ {v_{i}^{(\mathcal{A})}}^2 \,}{c^2}  \right)\cdot \underbrace{v_{\mathcal{B}}^{(\mathcal{A})}}_{-1}
\ee
in terms of $\mathcal{A}$lice velocity values (we approximate for small relative velocity $v_{\mathcal{B}}^{(\mathcal{A})} = -1 \ll c$). For $\mathcal{B}$ob the incident particle $\circMunit_{\:\mathbf{v}_R}$ kicks into the right with velocity $v_R^{(\mathcal{B})}$ and rebounds with reduced velocity ${v_R'}^{(\mathcal{B})}$ to the left
\be\label{Abschnitt -- SRT Dynamics - calorimeter model - reservoir balance for absorption - v_Bob von Rechtskick}
   v_R^{(\mathcal{B})} \, \stackrel{(\ref{Abschnitt -- SRT Dynamics - calorimeter model - reservoir balance for absorption - v Trafo})}{\simeq} \, v_R^{(\mathcal{A})} + \left( 1 - \frac{ v_R^2 }{c^2} \right)
   \;\;\;\;\;\;\;\;\;\;\;\;
   {v_R'}^{(\mathcal{B})} \, \simeq \, -v_R^{(\mathcal{A})} + \left( 1 - \frac{ v_R^2 }{c^2} \right)
\ee
while the initially resting composite $\circMR_{\:\mathbf{0}}$ gets kicked into standard velocity $\mathbf{v}_S^{(\mathcal{B})} := \frac{ \mathbf{v}_{\mathbf{1}} + \mathbf{v}_{\mathbf{1}}}{1 + \frac{\mathbf{v}_{\mathbf{1}} \cdot \mathbf{v}_{\mathbf{1}}}{c^2}}$.

Similarly on the left side $\mathcal{C}$harly prepares the next deceleration kick with a composite $n_L\cdot\circMunit =: \circML$ of $n_L$ additional reservoir elements, so that the previous recoil particle $\circMunit_{\:\mathbf{v}'_R}$ with velocity $\mathbf{v}'_R =: \mathbf{v}_L$ and the new composite $\circML_{\:\mathbf{v}_{\mathbf{1}}}$ with unit velocity $\mathbf{v}_{\mathbf{1}}$ repulse antiparallel $\circML_{\:\mathbf{v}_{\mathbf{1}}} , \circMunit_{-\mathbf{v}_L} \Rightarrow \circML_{-\mathbf{v}_{\mathbf{1}}} , \circMunit_{\:\mathbf{v}_L}$. Let $\mathcal{B}$ob move relative to Charlie with velocity $\mathbf{v}_{\mathcal{B}}=\underbrace{v_{\mathcal{B}}^{(\mathcal{C})}}_{+1}\cdot\mathbf{v}_{\mathbf{1}^{(\mathcal{C})}}$ to the right (see figure \ref{pic_SRT_deceleration_kicks}b). In $\mathcal{C}$harlie's frame the incident particle $\circMunit_{\:\mathbf{v}_L}$ has initial velocity
\be\label{Abschnitt -- SRT Dynamics - calorimeter model - reservoir balance for absorption - v_L C substituted}
   v_L^{(\mathcal{C})} \;\: \stackrel{!}{=} \;\; {v'_R}^{(\mathcal{C})}
   \; \stackrel{(\ref{Abschnitt -- SRT Dynamics - calorimeter model - reservoir balance for absorption - v Trafo})}{\simeq} \;
   {v'_R}^{(\mathcal{B})} - \left( 1 - \frac{ {{v'_R}^{(\mathcal{B})}}^2 }{c^2} \right) \cdot \underbrace{v^{(\mathcal{B})}_{\mathcal{C}}}_{-1}
   \; \stackrel{(\ref{Abschnitt -- SRT Dynamics - calorimeter model - reservoir balance for absorption - v_Bob von Rechtskick})}{\simeq} \;
   - \:v_R^{(\mathcal{A})} + 2\cdot \left( 1 - \frac{ v_R^2 }{c^2} \right)  + \underbrace{\mathcal{O}\:(\,\frac{1}{c}\,)^2}_{\ll 1}
\ee
and rebounds with reversed velocity ${v'_L}^{(\mathcal{C})}\stackrel{!}{=}-v_L^{(\mathcal{C})}$; hence Charlie requires $n_L(v_L,v_{\mathbf{1}})$
\bea\label{Abschnitt -- SRT Dynamics - calorimeter model - reservoir balance for absorption - n von Linkskick}
    n_L \;\: \stackrel{(\ref{Abschnitt -- SRT Dynamics - elstic collision model - kinemtical relations elast collision n+1})}{:=} \;\; \left(\gamma \cdot v \right) \big|_{v_{L}^{(\mathcal{C})}}  \,\cdot\, \underbrace{\left(\gamma \cdot v\right)^{-1} \big|_{v_{\mathbf{1}}}}_{\rightarrow 1}
    & \!\simeq\! & \left(\gamma \cdot v \right) \big|_{v_{R}^{(\mathcal{A})}} \;+\; \frac{\partial}{\partial v} \left(\gamma \cdot v \right) \!\bigg|_{v_{R}^{(\mathcal{A})}} \!\!\cdot \left(  |v_{L}^{(\mathcal{C})}| - v_{R}^{(\mathcal{A})}  \right) \nn \\
    & \stackrel{(\ref{Abschnitt -- SRT Dynamics - calorimeter model - reservoir balance for absorption - v_L C substituted})}{\simeq} &
    \gamma_{v_R} \cdot v_{R}^{(\mathcal{A})} \;\; - \;\; 2\cdot \gamma_{v_R}
\eea
standard elements for the new composite to \emph{function} (they must repulse antiparallel with same velocity). For later convenience we note the useful relations
\be\label{Abschnitt -- SRT Dynamics - calorimeter model - reservoir balance for absorption - Ableitung-standard}
   \frac{\partial}{\partial v} \; \left(\gamma \cdot v \right) \; = \; \gamma^{3}
   \;\;\;\;\;\;\;\;\;\;\;\;\;\;
   \frac{\partial}{\partial v} \; \gamma \; = \; \frac{1}{c^2} \cdot \gamma^{3}\cdot v \;\; .
\ee
Eventually $\mathcal{B}$ob observes $\mathcal{C}$harlie's left kick
$\circML_{\:\mathbf{0}} , \circMunit_{\:\mathbf{v}_L} \Rightarrow \circML_{\:\mathbf{v}_S} , \circMunit_{\:\mathbf{v}'_L} $ with final velocity
\be\label{Abschnitt -- SRT Dynamics - calorimeter model - reservoir balance for absorption - v_Bob von Linkskick}
   {v'}_L^{(\mathcal{B})} \,\stackrel{(\ref{Abschnitt -- SRT Dynamics - calorimeter model - reservoir balance for absorption - v Trafo})}{\simeq}\, {v'_L}^{(\mathcal{C})} - \left( 1 - \frac{ {{v'_L}^{(\mathcal{C})}}^2 }{c^2} \right) \cdot \underbrace{v^{(\mathcal{C})}_{\mathcal{B}}}_{+1}
   \;\; \stackrel{(\ref{Abschnitt -- SRT Dynamics - calorimeter model - reservoir balance for absorption - v_L C substituted})}{=} \;\;
   v_R^{(\mathcal{A})} - 3\cdot \left( 1 - \frac{ v_R^2 }{c^2} \right) \;+\; \underbrace{\mathcal{O}\:(\,\frac{1}{c}\,)^2}_{\ll 1}
\ee
which for Charlie appear antiparallel ${v'_L}^{(\mathcal{C})}\stackrel{!}{=}-v_L^{(\mathcal{C})}$.

After each round of right and left collisions
\be
   \circMunit_{\: \mathbf{v}_{R}}  \:,\;  n_R \cdot \circMunit_{\: \mathbf{0}} \:,\; n_L \cdot \circMunit_{\:\mathbf{0}} \;\;\Rightarrow\;\;  \circMunit_{\:\mathbf{v}'_L} \:,\;  n_R \cdot \circMunit_{\: \mathbf{v}_{S}} \:,\; n_L \cdot \circMunit_{-\mathbf{v}_{S}}  \nn
\ee
$\mathcal{B}$ob adds the extracted reservoir elements $n_L \cdot \circMunit_{-\mathbf{v}_{S}} \,,\: n_R \cdot \circMunit_{\: \mathbf{v}_{S}}$ from both sides of the calorimeter and the successive deceleration
\be\label{Abschnitt -- SRT Dynamics - calorimeter model - reservoir balance for absorption - fuer integration Delta v deceleration step linear}
   \Delta v^{(\mathcal{B})} \;\; := \;\; v_R^{(\mathcal{B})} - {v'}_L^{(\mathcal{B})}
   \;\; \stackrel{(\ref{Abschnitt -- SRT Dynamics - calorimeter model - reservoir balance for absorption - v_Bob von Rechtskick})(\ref{Abschnitt -- SRT Dynamics - calorimeter model - reservoir balance for absorption - v_Bob von Linkskick})}{\simeq} \;\;
   4\cdot \left( 1 - \frac{ v_R^2 }{c^2} \right) \;\; .
\ee
Each antiparallel particle pair $\circMunit_{-\mathbf{v}_S} ,  \circMunit_{\:\mathbf{v}_S} \Rightarrow \mathcal{S}\big|_{\mathbf{v}=\mathbf{0}} , \circMunit_{\:\mathbf{v}=\mathbf{0}} , \circMunit_{\:\mathbf{v}=\mathbf{0}}$ can charge a neutral spring $\mathcal{S}$ by standard process (\ref{Abschnitt -- Basic Measurement - Einheitswirkung}). The resting elements return back into the calorimeter reservoir $\left\{ \circMunit_{\:\mathbf{v}=\mathbf{0}} \right\}$. From every deceleration step Bob extracts the reservoir balance
\be
   \mathrm{RB}\left[\circMunit_{\:\mathbf{v}_R} \Rightarrow \circMunit_{\:\mathbf{v}_R-\Delta \mathbf{v}} \right]
   \;\; := \;\;   n_L \cdot \mathcal{S}\!\!\mid_{\mathbf{v}=\mathbf{0}} \:,\;  (n_R - n_L)\cdot \circMunit_{\:\mathbf{v}_S} \nn
\ee
$n_L$ (fragments of) resting standard springs $\mathcal{S}\!\!\mid_{\mathbf{0}}$ and $n_R - n_L$ left over recoil particles $\circMunit_{\:\mathbf{v}_S}$. In the limit of refined extraction velocity $\mathbf{v}_S \rightarrow 0$ Bob integrates an increasing number of deceleration rounds for the incident particle $\circMunit_{\: \mathbf{v}}$ with momentary velocity $\mathbf{v}=\mathbf{v}_R$ since the deceleration size $\Delta \mathbf{v} \rightarrow 0$ diminishes. Bob extracts the number of (fragments of) particle pairs and single elements per deceleration step
\bea
   \frac{\mathrm{d} N_{\mathrm{pair}}}{\mathrm{d} v} & := & \lim_{v_S\rightarrow \,0} \:
   \frac{n_L}{\Delta v} \;\; \stackrel{(\ref{Abschnitt -- SRT Dynamics - calorimeter model - reservoir balance for absorption - n von Linkskick})(\ref{Abschnitt -- SRT Dynamics - calorimeter model - reservoir balance for absorption - fuer integration Delta v deceleration step linear})}{=} \;\; \frac{1}{4}\cdot\gamma^3\cdot v \;-\; \frac{1}{2}\cdot \gamma^3  \label{Abschnitt -- SRT Dynamics - calorimeter model - reservoir balance for absorption - integration dN_pair / dv}   \\
   \frac{\mathrm{d} N_{\mathrm{sing}}}{\mathrm{d} v} & := & \lim_{v_S\rightarrow \,0} \:
   \frac{n_R-n_L}{\Delta v} \;\; \stackrel{(\ref{Abschnitt -- SRT Dynamics - calorimeter model - reservoir balance for absorption - n von Rechtskick linear})(\ref{Abschnitt -- SRT Dynamics - calorimeter model - reservoir balance for absorption - fuer integration Delta v deceleration step linear})}{=} \;\;  \frac{1}{2}\cdot \gamma^3  \label{Abschnitt -- SRT Dynamics - calorimeter model - reservoir balance for absorption - integration dN_sing / dv}  \;\; .
\eea

For the entire deceleration Bob accumulates
\bea\label{Abschnitt -- SRT Dynamics - calorimeter model - reservoir balance for absorption - integral N_sing}
   N_{\mathrm{sing}} & := & \!\int_0^v \mathrm{d}v \: \frac{\mathrm{d} N_{\mathrm{sing}}}{\mathrm{d}v}
   \; \stackrel{(\ref{Abschnitt -- SRT Dynamics - calorimeter model - reservoir balance for absorption - integration dN_sing / dv})}{=} \;
   \int_0^v \mathrm{d}v \;\;\frac{1}{2}\cdot\!\!\!\!\underbrace{\gamma^3}_{\stackrel{(\ref{Abschnitt -- SRT Dynamics - calorimeter model - reservoir balance for absorption - Ableitung-standard})}{=}\: \frac{\mathrm{d}}{\mathrm{d}v} (\gamma \cdot v) }
    = \;\; \frac{1}{2}\cdot \gamma \cdot v   \\
   N_{\mathrm{pair}} & := & \!\int_0^v \mathrm{d}v \: \frac{\mathrm{d} N_{\mathrm{pair}}}{\mathrm{d}v}
   \; \stackrel{(\ref{Abschnitt -- SRT Dynamics - calorimeter model - reservoir balance for absorption - integration dN_pair / dv})}{=} \;
   \int_0^v \mathrm{d}v \left( \:\frac{1}{4} \right.
   \cdot\!\!\!\!\underbrace{\gamma^3\cdot v}_{\stackrel{(\ref{Abschnitt -- SRT Dynamics - calorimeter model - reservoir balance for absorption - Ableitung-standard})}{=}\: \frac{\mathrm{d}}{\mathrm{d}v}\:  c^2\cdot\gamma } - \left. \frac{1}{2} \cdot \gamma^3 \right)
   \;=\;\; \frac{1}{4}\cdot c^2 \cdot (\gamma -1) \:-\: \frac{1}{2}\cdot \gamma\cdot v  \nn
\eea
fragments of standard impulse carriers $\circMunit_{\:\mathbf{v}_S}$ and particle pairs (standard springs $\mathcal{S}\!\!\mid_{\mathbf{v}=\mathbf{0}}$). By the construction (see figure \ref{pic_SRT_deceleration_kicks}) the numerical velocity values $\mathbf{v}_{\mathcal{O}}=v^{(\mathcal{B})}_{\mathcal{O}}\cdot\mathbf{v}_{\mathbf{1}^{(\mathcal{B})}}$, $c_{\mathcal{L}}=c^{(\mathcal{B})}_{\mathcal{L}}\cdot\mathbf{v}_{\mathbf{1}^{(\mathcal{B})}}$ refer to unit velocity $\mathbf{v}_{\mathbf{1}^{(\mathcal{B})}}$; however we extract a number of reservoir particles $\circMunit_{\pm\mathbf{v}_S}$ with standard velocity $\mathbf{v}_{S} \stackrel{(\ref{Abschnitt -- SRT Dynamics - calorimeter model - reservoir balance for absorption - v Trafo})}{\simeq} 2\cdot \mathbf{v}_{\mathbf{1}^{(\mathcal{B})}}$. In these new reference units
\be\label{Abschnitt -- SRT Dynamics - calorimeter model - reservoir balance for absorption - standard units}
   \mathbf{v}_{\mathcal{O}} \: = \: \underbrace{v_{\mathcal{O}}^{(\mathcal{B})} \cdot \frac{1}{2} }_{=:\:v_{\mathcal{O}}^{(S)}} \:\cdot\: \underbrace{2 \cdot \mathbf{v}_{\mathbf{1}^{(\mathcal{B})}}}_{\mathbf{v}_S}
   \;\;\;\;\;\;\;\;\;\;\;\;\;\;
   c_{\mathcal{L}} \:=\:  \underbrace{c_{\mathcal{L}}^{(\mathcal{B})} \cdot \frac{1}{2} }_{=:\:c_{\mathcal{L}}^{(S)}} \:\cdot\: \underbrace{2 \cdot \mathbf{v}_{\mathbf{1}^{(\mathcal{B})}}}_{\mathbf{v}_S}
\ee
the relation between the impact velocity and the number of activated calorimeter elements transforms accordingly
\bea\label{Abschnitt -- SRT Dynamics - calorimeter model - reservoir balance for absorption - integral N_sing RECALIBRATED}
   N_{\mathrm{sing}} & \stackrel{(\ref{Abschnitt -- SRT Dynamics - calorimeter model - reservoir balance for absorption - integral N_sing})}{=} &  \frac{1}{2}\cdot\!\frac{v^{(\mathcal{B})}}{\sqrt{1-\frac{{v^{(\mathcal{B})}}^2}{{c^{(\mathcal{B})}}^2}  }}
   \;\; \stackrel{(\ref{Abschnitt -- SRT Dynamics - calorimeter model - reservoir balance for absorption - standard units})}{=} \;\;
   \frac{v^{(S)}}{\sqrt{1-\frac{{v^{(S)}}^2}{{c^{(S)}}^2}  }}   \\
   N_{\mathrm{pair}} & \stackrel{(\ref{Abschnitt -- SRT Dynamics - calorimeter model - reservoir balance for absorption - integral N_sing})}{=} &
   \frac{1}{4}\cdot {c^{(\mathcal{B})}}^2 \cdot (\gamma - 1) \: - \: \frac{1}{2}\cdot\gamma\cdot v^{(\mathcal{B})}
   \;\; \stackrel{(\ref{Abschnitt -- SRT Dynamics - calorimeter model - reservoir balance for absorption - standard units})}{=} \;\;
   {c^{(\mathcal{S})}}^2 \cdot (\gamma - 1) \: - \: \gamma\cdot v^{(\mathcal{S})}  \nn  \;\; .
\eea
\qed

\subsection{Quantification of energy-momentum}\label{Kap - SRT Dyn - Quantification of energy-momentum}

In the calorimeter model we can count the activated standard springs and impulse carriers. Practical comparisons of Leibniz and Galilei \{\ref{Kap - SRT_Dyn - Basic Measurement}\} fix the \emph{physical meaning} of our reference objects as units for energy and momentum and of the calorimeter extract. It has the ''same impact as'' the incident particle, since otherwise one could instrumentalize the comparison behavior for a perpetuum mobile (see Lemma \ref{Lem - kin quant Absorptions Wirkung - Reservoirbilanz - p conserved}). It also has the ''same capability to work'' because our calorimeter model is reversible, we can steer every deceleration step both ways.\footnote{In calorimeter model $W_{\mathrm{cal}}$ all energy carriers $\mathcal{S}\big|_{\mathbf{0}}$ \emph{act jointly}. They expend their capability to work against the same incident object (concatenation operation ''$\ast_E$''). By steering the reverse calorimeter-collision-cascade we reproduce the initial motion and transform back the original kinetic energy.}
\begin{theo}\label{Theorem - SRT Dynamics - Energy-momentum of generic particle}
An object $\circMO_{\:\mathbf{v}}$ with inertial mass $m[\circMO] = m \cdot m[\circMunit]$ and velocity $\mathbf{v}_{\mathcal{O}}=v \cdot \mathbf{v}_S$ has kinetic energy and momentum
\be\label{Abschnitt -- SRT Dynamics - calorimeter model - Quantification energy-momentum - generic particle}
\begin{array}{rcl}
   E_{\mathrm{kin}} \left[ \circMO_{\:\mathbf{v}} \right] & = & \left( m\cdot c^2 \cdot (\gamma -1) \right) \cdot E \left[ \mathcal{S}\big|_{\mathbf{v}=\mathbf{0}} \right]
   \\
   \mathbf{p} \left[ \circMO_{\:\mathbf{v}} \right] & = & \left( m\cdot \gamma \cdot v \right) \cdot \mathbf{p} \left[ \circMunit_{\:\mathbf{v}_S} \right]     \;\; .
\end{array}
\ee
\end{theo}
\textbf{Proof:}
We measure the inertia of the generic body $\circMO_{\:\mathbf{v}_{\mathcal{O}}}$ with an equally massive composite of standard elements (in an head-on collision test with same initial velocity no one overruns the other \{\ref{Kap - SRT_Dyn - Basic Measurement}\}) and the latter according to the congruence principle
\be
   m\left[ \circMO \right] \;\;\stackrel{(\mathrm{Galilei})}{=}\;\;
   m\left[ \circMunit \ast \dots \ast \circMunit \right]  \;\;\stackrel{(\mathrm{Congr.})}{=:}\;\;
   m\cdot m\left[ \circMunit \right]   \nn
\ee
by the number $m:= \sharp \left\{ \circMunit \right\} $ of standard elements and their unit mass $m\left[ \circMunit \right]$. When we unlock the inner binding, the composite $\circMunit\ast\dots\ast\circMunit_{\:\mathbf{v}_{\mathcal{O}}} \Rightarrow \circMunit_{\:\mathbf{v}_{\mathcal{O}}},\ldots,\circMunit_{\:\mathbf{v}_{\mathcal{O}}}$ turns into a swarm of $m$ unit objects with velocity $\mathbf{v}_{\mathcal{O}}$. The generic object $\circMO_{\:\mathbf{v}}$ generates the same reservoir extract as for absorbing the $m$ unit elements $\circMunit_{\:\mathbf{v}_{\mathcal{O}}}$ one by one; the contrary would allow a perpetuum mobile (see Lemma \ref{Lem - kin quant Absorptions Wirkung - Reservoirbilanz - absorption proportional materiemenge}).

Let us absorb one unit element $\circMunit_{\:\mathbf{v}_{\mathcal{O}}}$ with same initial velocity $\mathbf{v}_{\mathcal{O}}$. We transfer its impact onto $N_{\mathrm{pair}}$ particle pairs and $N_{\mathrm{sing}}$ individual elements
\bea
   \mathbf{p} \left[ \circMunit_{\:\mathbf{v}_{\mathcal{O}}} \right] \;\; \stackrel{(\mathrm{Galilei})}{=} \;\; \mathbf{p} \left[ \mathrm{RB} \left[ \circMunit_{\:\mathbf{v}_{\mathcal{O}}} \Rightarrow \circMunit_{\:\mathbf{0}}  \right] \right] & \stackrel{(\ref{Abschnitt -- SRT Dynamics - calorimeter model - reservoir balance for absorption})}{=} & \mathbf{p} \left[ \: N_{\mathrm{pair}}\,\cdot \right. \left\{\circMunit_{-\mathbf{v}_S}, \circMunit_{\:\mathbf{v}_S}\right\} \,,\: \left. N_{\mathrm{sing}}\cdot  \circMunit_{\:\mathbf{v}_S}  \: \right] \nn \\
   & \stackrel{(\mathrm{Congr.})}{=} & N_{\mathrm{sing}} \cdot \mathbf{p} \left[ \circMunit_{\:\mathbf{v}_S} \right] \;\; \stackrel{(\ref{Abschnitt -- SRT Dynamics - calorimeter model - reservoir balance for absorption - integral N_sing RECALIBRATED})}{=} \;\;
   \gamma \cdot v \cdot \mathbf{p} \left[ \circMunit_{\:\mathbf{v}_S} \right] \nn  \;\; .
\eea
The particle pair $\mathbf{p} \left[ \circMunit_{-\mathbf{v}_S}, \circMunit_{\:\mathbf{v}_S} \right] \stackrel{(\ref{Abschnitt -- Basic Measurement - Einheitswirkung})}{=}  \mathbf{p} \left[ \mathcal{S}\big|_{\mathbf{v}=\mathbf{0}} \right] = 0$ can charge a resting spring and hence has no impulse. Each impulse carrier $\circMunit_{\:\mathbf{v}_S}$ is equivalent with the next and represents unit momentum $\mathbf{p} \left[ \circMunit_{\:\mathbf{v}_S} \right]$. We measure the momentum of the incident particle by the extracted impulse carriers and the latter, according to the congruence principle, by the number $\sharp \left\{ \circMunit_{\:\mathbf{v}_S} \right\} = \gamma \cdot v $ of standard elements and their reference momentum $\mathbf{p} \left[ \circMunit_{\:\mathbf{v}_S} \right]$. For the generic object $\circMO_{\:\mathbf{v}}$ we measure $m$ times more.

The kinetic energy (capability to work associated with the motion) of the incident particle $E\left[\circMunit_{\:\mathbf{v}_{\mathcal{O}}}\right]$ is completely transformed
\bea\label{Abschnitt -- SRT Dynamics - calorimeter model - Quantif en-mom - standard particle - kinetic energy v Beziehung unbereinigt}
   E \left[ \circMunit_{\:\mathbf{v}_{\mathcal{O}}} \right] & \!\!\stackrel{(\mathrm{Leibniz})(\ref{Abschnitt -- SRT Dynamics - calorimeter model - reservoir balance for absorption})}{=}\!\! & E \left[ \: N_{\mathrm{pair}}\,\cdot  \left\{\circMunit_{-\mathbf{v}_S}, \circMunit_{\:\mathbf{v}_S}\right\} \,,\: \frac{N_{\mathrm{sing}}}{2} \cdot  \left\{\circMunit_{\:\mathbf{v}_S}, \circMunit_{\:\mathbf{v}_S}\right\}  \: \right]  \nn \\
   & \stackrel{(\mathrm{Congr.})}{=} & \left( N_{\mathrm{pair}} + \frac{N_{\mathrm{sing}}}{2}\right) \:\cdot\: E \left[ \circMunit_{-\mathbf{v}_S}, \circMunit_{\:\mathbf{v}_S}  \right] \nn \\
   & \!\!\stackrel{(\ref{Abschnitt -- SRT Dynamics - calorimeter model - reservoir balance for absorption - integral N_sing RECALIBRATED})}{=}\!\! & \left(c^2\cdot (\gamma - 1) \;\; - \;\; \frac{1}{2} \cdot \gamma \cdot v   \right)  \:\cdot\: E \left[ \mathcal{S}\big|_{\mathbf{v}=\mathbf{0}} \right]   \;\;\;\;\;\;\;
\eea
into potential energy of the absorber material. In the second step each pair of standard impulse carriers $2\cdot \circMunit_{\:\mathbf{v}_S} \sim_{E} \left\{\circMunit_{-\mathbf{v}_S}, \circMunit_{\:\mathbf{v}_S}\right\} $ has the same capability to work as an antiparallel pair (for the physical construction see Proposition \ref{Prop - kin quant Absorptions Wirkung - pre-theoretical characterization E unit and p unit}). The latter $E \left[ \circMunit_{-\mathbf{v}_S}, \circMunit_{\:\mathbf{v}_S} \right] \stackrel{(\ref{Abschnitt -- Basic Measurement - Einheitswirkung})}{=}  E \left[ \mathcal{S}\big|_{\mathbf{v}=\mathbf{0}} \right]$ can charge a standard spring and represents the unit energy. Hence we measure ''how many times larger'' the kinetic energy is, than the potential energy of one resting standard spring $\mathcal{S}\big|_{\mathbf{v}=\mathbf{0}}$.

We derive the number (\ref{Abschnitt -- SRT Dynamics - calorimeter model - reservoir balance for absorption}) of extracted energy and momentum carriers $\circMunit_{\:\mathbf{v}_S}$ in the limit of an arbitrarily small $\epsilon\rightarrow 0$ extraction velocity $\mathbf{v}_{S'} := \epsilon \cdot \mathbf{v}_{S}$; then for every impact velocity $\mathbf{v}_{\mathcal{O}} = v_{\mathcal{O}}^{(S)} \cdot \frac{1}{\epsilon} \:\cdot\: \underbrace{\epsilon \cdot \mathbf{v}_{S}}_{\mathbf{v}_{S'}}$ the numerical value $v_{\mathcal{O}}^{(S')} := v_{\mathcal{O}}^{(S)} \cdot \frac{1}{\epsilon}\rightarrow\infty$ appears arbitrarily large. Hence in comparison to the dominant term in (\ref{Abschnitt -- SRT Dynamics - calorimeter model - Quantif en-mom - standard particle - kinetic energy v Beziehung unbereinigt}) (following Mermin \cite{Mermin '89 Space and Time in Special Relativity})
\be
   {c^{(\mathcal{S'})}}^2 \cdot (\gamma - 1) \; = \;
   c^2 \cdot \frac{1-\sqrt{ 1 - \frac{{v}^2}{c^2} }}{\sqrt{ 1 - \frac{{v}^2}{c^2} }} \cdot \frac{1+\sqrt{ 1 - \frac{{v}^2}{c^2} }}{1+\sqrt{ 1 - \frac{{v}^2}{c^2} }} \:=\:
   \frac{v_{\mathcal{O}}^{(\mathcal{S'})}}{\sqrt{1 - \frac{{v}^2}{c^2}}} \cdot \underbrace{\frac{v_{\mathcal{O}}^{(\mathcal{S'})}}{1+\sqrt{ 1 - \frac{{v}^2}{c^2} }}}_{\rightarrow\infty} \; \gg \; \gamma \cdot v_{\mathcal{O}}^{(\mathcal{S'})} \nn
\ee
the second term is negligibly small; thus we derive the expression (\ref{Abschnitt -- SRT Dynamics - calorimeter model - Quantification energy-momentum - generic particle}).\footnote{A calorimeter measurement with a refined extraction velocity $\mathbf{v}_{S'}$ reveals the meaning of the negligible part of the kinetic energy (\ref{Abschnitt -- SRT Dynamics - calorimeter model - Quantif en-mom - standard particle - kinetic energy v Beziehung unbereinigt}): With a ''heavier absorber'' one extracts single impulse carriers $\gamma \cdot \mathbf{v}_{S'} \cdot \circMunit_{\:\mathbf{v}_{S'}}$ though with less total kinetic energy. The kinetic energy $E \left[ \circMO_{\mathbf{v}} \right]$ of the incident object transforms mostly into potential energy ${c^{(\mathcal{S'})}}^2 \cdot (\gamma - 1) \cdot E \left[ \circMunit_{-\mathbf{v}_{S'}}, \circMunit_{\:\mathbf{v}_{S'}}  \right]$
of the particle pairs resp. refined standard springs.}
\qed
For the absorption of a moving object $\circMO_{\:\mathbf{v}}$ we extract standard springs and impulse carriers with the same energy and momentum. By counting them we quantify ''how many times'' larger the inertial mass $m\left[ \circMO \,\right]=:m_{\mathcal{O}}^{(S)}\cdot m\left[ \circMunit \,\right]$, kinetic energy $E \left[ \circMO_{\:\mathbf{v}} \right] =: E_{\mathcal{O}}^{(S)} \cdot E \left[ \mathcal{S}\big|_{\mathbf{v}=\mathbf{0}} \right]$ and momentum $\mathbf{p} \left[ \circMO_{\:\mathbf{v}} \right] =: p_{\mathcal{O}}^{(S)} \cdot \mathbf{p} \left[ \circMunit_{\:\mathbf{v}_{S}} \right]$ of the incident object $\circMO_{\:\mathbf{v}}$ is than in the standard energy source $\mathcal{S}\big|_{\mathbf{v}=\mathbf{0}}$ and impulse carrier $\circMunit_{\:\mathbf{v}_{S}}$ of our reference process $w_{S}$ (\ref{Abschnitt -- Basic Measurement - Einheitswirkung}). We measure the basic observables \emph{independently} from one another. From the number of the respective standard elements in the calorimeter model we derive the relation between the impact velocity and the kinetic energy-momentum. We derive equations (\ref{Abschnitt -- SRT Dynamics - calorimeter model - Quantification energy-momentum - generic particle})
\be \nn
   \frac{\mathbf{p}_{\mathcal{O}}}{\mathbf{p}_S} = \frac{m_{\mathcal{O}}}{m_S} \cdot  \frac{(\mathbf{v}_{\mathcal{O}}/ \mathbf{v}_S)}{\sqrt{1-\frac{(v_{\mathcal{O}} / v_S)^2}{(c / v_S)^2}}}
   \;\;\;\;\;\;\;\;\;\;
   \frac{E_{\mathrm{kin}\: \mathcal{O}}}{E_S} =\frac{m_{\mathcal{O}}}{m_S} \cdot  \left(\frac{c}{v_S}\right)^2 \!\! \cdot  (\frac{1}{\sqrt{1-\frac{(v_{\mathcal{O}} / v_S)^2}{(c / v_S)^2}}}-1)  \;\;,
\ee
in which all numerical values for kinetic energy $E_{\mathcal{O}}^{(S)} =: \frac{E \left[ \circMO_{\:\mathbf{v}} \right]}{E \left[ \mathcal{S}\big|_{\mathbf{0}} \right] }\:$, impulse $p_{\mathcal{O}}^{(S)} =: \frac{\mathbf{p} \left[ \circMO_{\:\mathbf{v}} \right]}{\mathbf{p} \left[ \circMunit_{\:\mathbf{v}_S} \right]}\:$, inertial mass $m_{\mathcal{O}}^{(S)} =: \frac{m\left[ \circMO \right]}{m \left[ \circMunit \,\right]}\:$ and velocity $v_{\mathcal{O}}^{(S)} =: \frac{v_{\mathcal{O}}}{v_S}$ occur in the form \emph{measure/unit measure}. Each formal ratio symbolizes the result of a physical operation; counting standard units in the calorimeter model.\footnote{We measure each basic observable as \emph{unity of quantity and quality}. We find a relation between coinciding numbers of reference devices $E_{\mathcal{O}}^{(S)}$, $p_{\mathcal{O}}^{(S)}$; but not between their dimensions $E \left[ \mathcal{S}\big|_{\mathbf{v}=\mathbf{0}} \right]$, $\mathbf{p} \left[ \circMunit_{\:\mathbf{v}_S} \right]$ etc. Those we define directly by the tangible carriers (''capability to work'' of an intrinsically resting standard spring $\mathcal{S}\big|_{\mathbf{v}=\mathbf{0}}$, ''impact'' of a decelerating standard body $\circMunit_{\:\mathbf{v}_{S}}\Rightarrow\circMunit_{\:\mathbf{v}=0}$ etc.) in the practical comparison \{\ref{Kap - SRT_Dyn - Basic Measurement}\}.} Without the extra assumption of the light principle in Galilei kinematics the \emph{same} construction leads to the basic equations of classical mechanics \cite{Hartmann-KM_Dyn}.

Commonly measurements in dynamics rely on postulated equations.
%
%
One calculates the quantity of a dynamical observable as a derived value $F=m\cdot a$, $p=m\cdot v$ etc. Our program presents a novel method for introducing basic observables. The operationalization is complementary to a formal foundation \{\ref{Kap - KM Dynamics - Discussion - Principles}\}. We assess faithful physical quantity of energy by counting congruent reference units. Because according to Wallot \cite{Wallot - Groessengleichungen Einheiten und Dimensionen} ''the physical measure is \emph{not} defined by the product of quantity (of the measure) times unit measure. From the physical measure we arrive via the unit measure at the measured value, not reversely.'' Only if a standard energy source is defined, one can count how many  standard obstacles the incident object can overcome, before it stops in the calorimeter; only then therefore comes the \emph{counting}! The basic quantities have a precise physical meaning; under arbitrary reparametrization of the fundamental equations it will be concealed \{\ref{Kap - SRT Dynamics - Pre-theoretic elements}\}.

\section{Quantity equations}\label{Kap - SRT Dynamics - quantitative equations}

\subsection{Transformation law}

\begin{co}\label{Cor - SRT Dynamics - calorimeter model - phys connection dun units - E,p_B (E,p_A)}
Let $\mathcal{A}$lice move relative to $\mathcal{B}$ob with constant velocity $\mathbf{v}_{\mathcal{A}}= v_{\mathcal{A}}^{(\mathcal{B})} \cdot \mathbf{v}_{S^{(\mathcal{B})}}$. $\mathcal{A}$lice energy-momentum values transform into $\mathcal{B}$ob's values (for the same process) by
\be\label{Abschnitt -- SRT Dynamics - calorimeter model - phys connection dun units - E,p_B (E,p_A)}
   \left(\!\!
     \begin{array}{c}
       E \\
       \mathbf{p} \\
     \end{array}
   \!\!\right)^{\!\!(\mathcal{B})} \;\; = \;\;\; \gamma \cdot
   \left(
     \begin{array}{ccc}
       1 & \!\!& \mathbf{v}_{\mathcal{A}}^{(\mathcal{B})} \\
       & \!\!& \\
       \mathbf{v} / c^2 & \!\!& 1 \\
     \end{array}
   \right) \:
   \left(\!\!
     \begin{array}{c}
       E \\
       \mathbf{p} \\
     \end{array}
   \!\!\right)^{\!\!(\mathcal{A})} \;\; .
\ee
\end{co}
\textbf{Proof:}
$\mathcal{A}$lice measures an effect on e.g. object $\circMO_{\:\mathbf{v}_{\mathcal{O}}} \Rightarrow \circMO_{\:\mathbf{v}'_{\mathcal{O}}}$ by counting standard impulse carriers $\circMunit_{\:\mathbf{v}_{S^{(\mathcal{A})}}}$ from her calorimeter reservoir $\left\{ \circMunit_{\: \mathbf{v}_{\mathcal{A}}} \right\}$. $\mathcal{B}$ob can measure this effect from either the original $\circMO$ or from her units one by one; he counts the impulse units $\circMunit_{\:\mathbf{v}_{S^{(\mathcal{B})}}}$ in his comoving calorimeter $\left\{ \circMunit_{\: \mathbf{v}_{\mathcal{B}}} \right\}$. We want to find the transformation between these numbers. We construct a \emph{physical connection} between the effect of $\mathcal{A}$lice (boosted) energy and impulse standards $\mathcal{S}\big|_{\mathbf{v}_{\mathcal{A}}}$, $\circMunit_{\:\mathbf{v}_{S^{(\mathcal{A})}}}$ and the (resting) units of $\mathcal{B}$ob $\mathcal{S}\big|_{\mathbf{v}_{\mathcal{B}}}$, $\circMunit_{\:\mathbf{v}_{S^{(\mathcal{B})}}}$.

$\mathcal{B}$ob views $\mathcal{A}$lice reference devices $i$ with a transformed velocity $v_i^{(\mathcal{B})} \left(v_{\mathcal{B}}^{(\mathcal{A})}, v_i^{(\mathcal{A})} \right)$
\be\label{Abschnitt -- SRT Dynamics - calorimeter model - phys connection dun units - velocity derivative}
   v_i^{(\mathcal{B})} \; \stackrel{(\ref{Abschnitt -- SRT Dynamics - elstic collision model - induc_velocity_trafo - horizontal})}{=} \;
   \frac{-v_{\mathcal{B}}^{(\mathcal{A})} + v_i^{(\mathcal{A})}}{1-\frac{v_{\mathcal{B}}^{(\mathcal{A})} \cdot v_i^{(\mathcal{A})}}{c^2}}
   \;\simeq\; v_i^{(\mathcal{B})}\big|_{v_i^{(\mathcal{A})}\,=\,0} \;+\; \frac{\partial v_i^{(\mathcal{B})}}{\partial v_i^{(\mathcal{A})}}\bigg|_{0} \cdot v_i^{(\mathcal{A})}
   \;=\; v_{\mathcal{A}}^{(\mathcal{B})} + \left( 1 - \frac{ {v_{\mathcal{A}}^{(\mathcal{B})}}^2 \,}{c^2}  \right)\cdot v_i^{(\mathcal{A})}
\ee
in terms of $\mathcal{A}$lice velocity values (now we approximate for small standard velocity $v_{S^{(\mathcal{A})}} \ll c$). To $\mathcal{B}$ob her standard impulse carriers $\circMunit_{\pm\mathbf{v}_{S^{(\mathcal{A})}}}$ appear with velocity
\be
   \mathbf{v}_i \; = \; \pm 1\cdot \mathbf{v}_{S^{(\mathcal{A})}} \; = \;
   \left( v_{\mathcal{A}}^{(\mathcal{B})} + \left( 1 - \frac{ {v_{\mathcal{A}}^{(\mathcal{B})}}^2 \,}{c^2} \right) \cdot (\pm 1) \right) \cdot \mathbf{v}_{S^{(\mathcal{B})}} \nn
\ee
and her intrinsically resting reservoir elements $\circMunit_{\:\mathbf{v}_{\mathcal{A}}}$ and standard springs $\mathcal{S}\big|_{\mathbf{v}_{\mathcal{A}}}$ with velocity
\be
   \mathbf{v}_{\mathcal{A}} \; = \; 0\cdot \mathbf{v}_{S^{(\mathcal{A})}} \;=\; v_{\mathcal{A}}^{(\mathcal{B})}\cdot \mathbf{v}_{S^{(\mathcal{B})}} \;\; . \nn
\ee

$\mathcal{B}$ob measures $\mathcal{A}$lice impulse transfer $\circMunit_{\:\mathbf{v}_{S^{(\mathcal{A})}}} \Rightarrow \circMunit_{\:\mathbf{0^{(\mathcal{A})}}}$ as a deceleration with energy
\bea\label{Abschnitt -- SRT Dynamics - calorimeter model - phys connection dun units - E_B (p_A)}
   & & \!\!\!\!\!\!\!\!\!\!\!\!\!\!\!\!\!\!\!\!\!\!
   E^{(\mathcal{B})} \left[ \circMunit_{\:v_i^{(\mathcal{B})}} \Rightarrow \circMunit_{\:v_{\mathcal{A}}^{(\mathcal{B})}} \right]
   \;\; = \;\; E_{\mathrm{kin}}^{(\mathcal{B})} \left[ \circMunit_{\:v_i^{(\mathcal{B})}} \right] \;\; - \;\; E_{\mathrm{kin}}^{(\mathcal{B})} \left[ \circMunit_{\:v_{\mathcal{A}}^{(\mathcal{B})}} \right]   \nn \\
   & & \!\!\!\!\!\!\!\!\!\!\!\!\!\!\!\!\!\!
   =\:  \frac{\mathrm{d}}{\mathrm{d} v_i^{(\mathcal{A})} } \, E_{\mathrm{kin}}^{(\mathcal{B})}   \cdot \underbrace{{v_i^{(\mathcal{A})}}}_{=\:1} \:+\:
   \frac{1}{2} \cdot \frac{\mathrm{d}^2}{ {\mathrm{d} v_i^{(\mathcal{A})}}^2 } \, E_{\mathrm{kin}}^{(\mathcal{B})}  \cdot \underbrace{{v_i^{(\mathcal{A})}}^2}_{=\:1} \:+\: \mathcal{O} \left( \frac{1}{c^2} \right)
   \stackrel{(\ref{Abschnitt -- SRT Dynamics - calorimeter model - phys connection dun units - d E_B / d v_A})(\ref{Abschnitt -- SRT Dynamics - calorimeter model - phys connection dun units - d^2 E_B / d v_A^2})}{\simeq}  \gamma \cdot v_{\mathcal{A}}^{(\mathcal{B})}  + \frac{1}{2} \cdot \gamma
\eea
where in the Taylor series expansion with derivatives (and $m^{(\mathcal{B})}[\circMunit]=1$)
\be\label{Abschnitt -- SRT Dynamics - calorimeter model - phys connection dun units - d E_B / d v_A}
   \frac{\mathrm{d}}{\mathrm{d} v_i^{(\mathcal{A})} } \, E_{\mathrm{kin}}^{(\mathcal{B})} \left[ \circMunit_{\: v_i^{(\mathcal{B})}} \right] =
   \frac{\mathrm{d} E_{\mathrm{kin}}^{(\mathcal{B})} }{\mathrm{d} v_i^{(\mathcal{B})} } \cdot
   \frac{\mathrm{d} v_i^{(\mathcal{B})} }{\mathrm{d} v_i^{(\mathcal{A})} }
   \stackrel{(\ref{Abschnitt -- SRT Dynamics - calorimeter model - Quantification energy-momentum - generic particle})(\ref{Abschnitt -- SRT Dynamics - calorimeter model - phys connection dun units - velocity derivative})}{=}
   \frac{\mathrm{d}}{\mathrm{d} v } \left( m\cdot c^2 \!\cdot (\gamma-1)  \right)
   \!\cdot \!\left( 1 - \frac{ v^2 \,}{c^2}  \right)  \stackrel{(\ref{Abschnitt -- SRT Dynamics - calorimeter model - reservoir balance for absorption - Ableitung-standard})}{=} \: \gamma \cdot v_{\mathcal{A}}^{(\mathcal{B})}
\ee
\be\label{Abschnitt -- SRT Dynamics - calorimeter model - phys connection dun units - d^2 E_B / d v_A^2}
   \frac{\mathrm{d}^2 }{ {\mathrm{d} v_i^{(\mathcal{A})}}^2 } \, E_{\mathrm{kin}}^{(\mathcal{B})} \; \stackrel{(\ref{Abschnitt -- SRT Dynamics - calorimeter model - phys connection dun units - d E_B / d v_A})}{=} \;
   \frac{\mathrm{d}}{\mathrm{d} v_i^{(\mathcal{B})} } \left(  \gamma \cdot v  \right) \cdot \frac{\mathrm{d} v_i^{(\mathcal{B})} }{\mathrm{d} v_i^{(\mathcal{A})} } \; \stackrel{(\ref{Abschnitt -- SRT Dynamics - calorimeter model - reservoir balance for absorption - Ableitung-standard})(\ref{Abschnitt -- SRT Dynamics - calorimeter model - phys connection dun units - velocity derivative})}{=} \; \gamma^3 \cdot \left( 1 - \frac{ v^2 \,}{c^2}  \right) \;=\; \gamma
\ee
and by induction
\be
   \frac{\mathrm{d}^{2n+1}}{ {\mathrm{d} v_i^{(\mathcal{A})}}^{2n+1} } \, E_{\mathrm{kin}}^{(\mathcal{B})}
   \;\;=\;\; \frac{\mathrm{d}^{2n-1}}{ {\mathrm{d} v_i^{(\mathcal{A})}}^{2n-1} } \,
   \!\!\!\!\!
   \underbrace{\frac{\mathrm{d}^2 E_{\mathrm{kin}}^{(\mathcal{B})}}{ {\mathrm{d} v_i^{(\mathcal{A})}}^2 }}_{\stackrel{(\ref{Abschnitt -- SRT Dynamics - calorimeter model - phys connection dun units - d^2 E_B / d v_A^2})(\ref{Abschnitt -- SRT Dynamics - calorimeter model - Quantification energy-momentum - generic particle})}{=}\: \frac{1}{c^2} \cdot E_{\mathrm{kin}}^{(\mathcal{B}) } + 1
   }
   \!\!\!\!\!
   \;\;=\;\;
   \frac{1}{c^{2n}} \cdot \frac{\mathrm{d} }{ {\mathrm{d} v_i^{(\mathcal{A})}} } \, E_{\mathrm{kin}}^{(\mathcal{B}) }  \;
   \nn
\ee
we neglect higher order terms in $v_i^{(\mathcal{A})}=1 \ll c$. $\mathcal{B}$ob also measures a deceleration impulse
\bea\label{Abschnitt -- SRT Dynamics - calorimeter model - phys connection dun units - p_B (p_A)}
   & & \!\!\!\!\!\!\!\!\!\!\!\!\!\!\!\!\!\!\!\!\!\!
   \mathbf{p}^{(\mathcal{B})} \left[ \circMunit_{\: v_i^{(\mathcal{B})}} \Rightarrow \circMunit_{\:v_{\mathcal{A}}^{(\mathcal{B})}} \right]
   \;\; = \;\; \mathbf{p}^{(\mathcal{B})} \left[ \circMunit_{\: v_i^{(\mathcal{B})}} \right] \;\; - \;\; \mathbf{p}^{(\mathcal{B})} \left[ \circMunit_{\: v_{\mathcal{A}}^{(\mathcal{B})}} \right]   \nn \\
   & & \!\!\!\!\!\!\!\!\!\!\!\!\!\!\!\!\!\!\!\!\!
   =  \frac{\mathrm{d}}{\mathrm{d} v_i^{(\mathcal{A})} } \, \mathbf{p}^{(\mathcal{B})}   \cdot \underbrace{{v_i^{(\mathcal{A})}}}_{=\:1} +
   \frac{1}{2} \cdot \frac{\mathrm{d}^2}{ {\mathrm{d} v_i^{(\mathcal{A})}}^2 } \, \mathbf{p}^{(\mathcal{B})}  \cdot \underbrace{{v_i^{(\mathcal{A})}}^2}_{=\:1} + \mathcal{O} \left( \frac{1}{c^2} \right)
   \stackrel{(\ref{Abschnitt -- SRT Dynamics - calorimeter model - phys connection dun units - d p_B / d v_A})(\ref{Abschnitt -- SRT Dynamics - calorimeter model - phys connection dun units - d^2 p_B / d v_A^2})}{\simeq}  \gamma  + \frac{1}{2} \cdot \frac{1}{c^2} \cdot \gamma \cdot v_{\mathcal{A}}^{(\mathcal{B})}
\eea
where in the analogous series expansion with
\be\label{Abschnitt -- SRT Dynamics - calorimeter model - phys connection dun units - d p_B / d v_A}
   \frac{\mathrm{d}}{\mathrm{d} v_i^{(\mathcal{A})} } \, \mathbf{p}^{(\mathcal{B})} \left[ \circMunit_{\: v_i^{(\mathcal{B})}} \right] \; = \;
   \frac{\mathrm{d} \mathbf{p}^{(\mathcal{B})} }{\mathrm{d} v_i^{(\mathcal{B})} } \cdot
   \frac{\mathrm{d} v_i^{(\mathcal{B})} }{\mathrm{d} v_i^{(\mathcal{A})} }
   \stackrel{(\ref{Abschnitt -- SRT Dynamics - calorimeter model - Quantification energy-momentum - generic particle})(\ref{Abschnitt -- SRT Dynamics - calorimeter model - phys connection dun units - velocity derivative})}{=}
   \frac{\mathrm{d}}{\mathrm{d} v } \left( m \!\cdot \gamma \cdot v  \right)\,
   \cdot \left( 1 - \frac{ v^2 \,}{c^2}  \right)
   \stackrel{(\ref{Abschnitt -- SRT Dynamics - calorimeter model - reservoir balance for absorption - Ableitung-standard})}{=} \gamma
\ee
\be\label{Abschnitt -- SRT Dynamics - calorimeter model - phys connection dun units - d^2 p_B / d v_A^2}
   \frac{\mathrm{d}^2}{ {\mathrm{d} v_i^{(\mathcal{A})}}^2 } \, \mathbf{p}^{(\mathcal{B})} \; \stackrel{(\ref{Abschnitt -- SRT Dynamics - calorimeter model - phys connection dun units - d p_B / d v_A})}{=} \:
   \frac{\mathrm{d}}{\mathrm{d} v_i^{(\mathcal{B})} }\,  \gamma   \cdot \frac{\mathrm{d} v_i^{(\mathcal{B})} }{\mathrm{d} v_i^{(\mathcal{A})} } \, \stackrel{(\ref{Abschnitt -- SRT Dynamics - calorimeter model - reservoir balance for absorption - Ableitung-standard})(\ref{Abschnitt -- SRT Dynamics - calorimeter model - phys connection dun units - velocity derivative})}{=}  \frac{1}{c^2} \cdot \gamma^3 \cdot v \cdot \left( 1 - \frac{ v^2 \,}{c^2}  \right)
   = \frac{1}{c^2} \cdot \gamma \cdot v_{\mathcal{A}}^{(\mathcal{B})}
\ee
and by induction all higher order terms $n\geq 3$ are suppressed by an extra factor $1/c^n$
\be
   \frac{\mathrm{d}^{2n}}{ {\mathrm{d} v_i^{(\mathcal{A})}}^{2n} } \, \mathbf{p}^{(\mathcal{B})}
   \;\;=\;\; \frac{\mathrm{d}^{2n-2}}{ {\mathrm{d} v_i^{(\mathcal{A})}}^{2n-2} } \, \!\!\!\!\!
   \underbrace{\frac{\mathrm{d}^2 \mathbf{p}^{(\mathcal{B})}}{ {\mathrm{d} v_i^{(\mathcal{A})}}^2 }}_{\stackrel{(\ref{Abschnitt -- SRT Dynamics - calorimeter model - phys connection dun units - d^2 p_B / d v_A^2})(\ref{Abschnitt -- SRT Dynamics - calorimeter model - Quantification energy-momentum - generic particle})}{=}\: \frac{1}{c^2} \cdot \mathbf{p}^{(\mathcal{B}) } }
   \!\!\!
   \;\;=\;\; \frac{1}{c^{2n}} \cdot \, \mathbf{p}^{(\mathcal{B}) }  \nn \;\; .
\ee

Similarly $\mathcal{B}$ob absorbs $\mathcal{A}$lice (boosted) standard spring $\mathcal{S}\big|_{\mathbf{v}_{\mathcal{A}}}$ (resp. the effect of a standard particle pair with velocity $\pm \mathbf{v}_{S^{(\mathcal{A})}}$) with an absorption energy and momentum
\bea\label{Abschnitt -- SRT Dynamics - calorimeter model - phys connection dun units - E_B (S_1^A)}
   E^{(\mathcal{B})} \left[ \mathcal{S} \big|_{\,\mathbf{0}^{(\mathcal{A})}} \right] \: \stackrel{(\mathrm{Leibniz})}{=} \: E^{(\mathcal{B})} \left[ \circMunit_{-\mathbf{v}_{S^{(\mathcal{A})}} } \Rightarrow \circMunit_{\:\mathbf{0}^{(\mathcal{A})}} \right] \;+\; E^{(\mathcal{B})} \left[ \circMunit_{\:\mathbf{v}_{S^{(\mathcal{A})}} } \Rightarrow \circMunit_{\:\mathbf{0}^{(\mathcal{A})}} \right]
   \;\;\;\;\;\;\;\;\;\;\;\;\;\;\; & &
   \\
   \;\; \stackrel{(\ref{Abschnitt -- SRT Dynamics - calorimeter model - phys connection dun units - E_B (p_A)})}{=} \;
   \gamma \cdot v \cdot \underbrace{\mathbf{v}_i^{(\mathcal{A})}}_{=-1}  \:+\: \frac{1}{2} \cdot \gamma \cdot (\underbrace{\mathbf{v}_i^{(\mathcal{A})}}_{=-1}\,)^2
   \:+\:
   \gamma \cdot v \cdot \underbrace{\mathbf{v}_i^{(\mathcal{A})}}_{=\,1}  \:+\: \frac{1}{2} \cdot \gamma \cdot (\underbrace{\mathbf{v}_i^{(\mathcal{A})}}_{=\,1}\,)^2  \:+\: \mathcal{O} \left( \frac{1}{c^2} \right)  \;\simeq\; \gamma  & & \nn
\eea
\bea\label{Abschnitt -- SRT Dynamics - calorimeter model - phys connection dun units - p_B (S_1^A)}
   & & \!\!\!\!\!\!\!\!\!\!
   \mathbf{p}^{(\mathcal{B})} \left[ \mathcal{S} \big|_{\,\mathbf{0}^{(\mathcal{A})}} \right] \; = \; \mathbf{p}^{(\mathcal{B})} \left[ \circMunit_{-\mathbf{v}_{\mathbf{1}^{(\mathcal{A})}} } \Rightarrow \circMunit_{\:\mathbf{0}^{(\mathcal{A})}} \right] \;+\; \mathbf{p}^{(\mathcal{B})} \left[ \circMunit_{\:\mathbf{v}_{\mathbf{1}^{(\mathcal{A})}} } \Rightarrow \circMunit_{\:\mathbf{0}^{(\mathcal{A})}} \right]
    \\
   & & \!\!\!\!\!\!\!\!\!\!
   \stackrel{(\ref{Abschnitt -- SRT Dynamics - calorimeter model - phys connection dun units - p_B (p_A)})}{=}
   \gamma \cdot \underbrace{v_i^{(\mathcal{A})}}_{=-1}  \:+\: \frac{1}{2} \cdot \gamma \cdot v \cdot (\underbrace{\mathbf{v}_i^{(\mathcal{A})}}_{=-1}\,)^2
   \:+\:
   \gamma \cdot \underbrace{v_i^{(\mathcal{A})}}_{=\,1}  \:+\: \frac{1}{2} \cdot \frac{1}{c^2} \cdot \gamma \cdot v \cdot (\underbrace{v_i^{(\mathcal{A})}}_{=\,1}\,)^2  \:+\: \mathcal{O} \left( \frac{1}{c^2} \right) \: \simeq \: \frac{1}{c^2} \cdot \gamma \cdot v \; . \nn
\eea

From $\mathcal{B}$ob's energy-momentum measurements for absorbing one single boosted energy and momentum unit from $\mathcal{A}$lice\footnote{$\mathcal{A}$lice standard energy source $\mathcal{S}\big|_{0\cdot\mathbf{v}_{S^{(\mathcal{A})}}}$ and impulse carrier $\circMunit_{\:\mathbf{v}_{S^{(\mathcal{A})}} }$ represents a \emph{mixed basis} for energy-momentum measurements. In a calorimeter model built from refined standard actions $w_{\epsilon}:  \mathcal{S}_{\epsilon}\big|_{\mathbf{0}} , \circMunit_{\:\mathbf{0}} , \circMunit_{\:\mathbf{0}} \Rightarrow \circMunit_{\:\epsilon\cdot\mathbf{v}_S} , \circMunit_{-\epsilon\cdot\mathbf{v}_S}$ one can construct a \emph{normal basis} $ \lim_{\epsilon\rightarrow 0} \left( \epsilon^{-2}  \cdot \mathcal{S}_{\epsilon}\big|_{\mathbf{0}} \,,\: \epsilon^{-1} \cdot \circMunit_{\epsilon\cdot\mathbf{v}_S} \right)$ with energy sources without momentum and with momentum carriers without energy, i.e. $(E,\mathbf{p})$-''eigenvectors''.}
\be\label{Abschnitt -- SRT Dynamics - calorimeter model - phys connection reference units}
\begin{array}{l}
   \!\!\!\!\left(\!\!
     \begin{array}{c}
       E \\
       \mathbf{p} \\
     \end{array}
   \!\!\right)  \left[ \mathcal{S}\big|_{\mathbf{0}^{(\mathcal{A})}}  \right]
   \;=\;
   \left(\!
     \begin{array}{c}
       1 \cdot E [ \mathcal{S}\big|_{\mathbf{0}^{(\mathcal{A})}}  ] \\
       \\
       0 \cdot \mathbf{p} [ \circMunit_{\:\mathbf{v}_{S^{(\mathcal{A})}}} ] \\
     \end{array}
   \!\right)
   \stackrel{(\ref{Abschnitt -- SRT Dynamics - calorimeter model - phys connection dun units - E_B (S_1^A)})(\ref{Abschnitt -- SRT Dynamics - calorimeter model - phys connection dun units - p_B (S_1^A)})}{=}
   \left(\!
     \begin{array}{c}
       \gamma \cdot E [ \mathcal{S}\big|_{\mathbf{0}^{(\mathcal{B})}}  ] \\
       \\
       \gamma \cdot \frac{\mathbf{v}}{c^2} \cdot \mathbf{p} [ \circMunit_{\:\mathbf{v}_{S^{(\mathcal{B})}}} ] \\
     \end{array}
   \!\right)
   \\
   \\
   \!\!\!\!\left(\!\!
     \begin{array}{c}
       E \\
       \mathbf{p} \\
     \end{array}
   \!\!\right)  \left[ \circMunit_{\:\mathbf{v}_{S^{(\mathcal{A})}} } \right]
   =
   \left(\!
     \begin{array}{c}
       \frac{1}{2}  \cdot E [ \mathcal{S}\big|_{\mathbf{0}^{(\mathcal{A})}}  ] \\
       \\
       1 \cdot \mathbf{p} [ \circMunit_{\:\mathbf{v}_{S^{(\mathcal{A})}}} ] \\
     \end{array}
   \!\right)
   \stackrel{(\ref{Abschnitt -- SRT Dynamics - calorimeter model - phys connection dun units - E_B (p_A)})(\ref{Abschnitt -- SRT Dynamics - calorimeter model - phys connection dun units - p_B (p_A)})}{=}
   \left(\!\!
     \begin{array}{c}
       \left( \gamma \cdot \frac{1}{2} + \gamma \cdot \mathbf{v} \right) \cdot E [ \mathcal{S}\big|_{\mathbf{0}^{(\mathcal{B})}}  ] \\
       \\
       \left( \gamma \cdot \frac{\mathbf{v}}{c^2} \cdot \frac{1}{2} + \gamma \right) \cdot \mathbf{p} [ \circMunit_{\:\mathbf{v}_{S^{(\mathcal{B})}}} ] \\
     \end{array}
   \!\!\right)\!\!\!\!\!\!\!\!
\end{array}
\ee
and by their additivity in a generic calorimeter measurement we find the linear transformation between their numerical energy-momentum values (\ref{Abschnitt -- SRT Dynamics - calorimeter model - phys connection dun units - E,p_B (E,p_A)}) for the same process.
\qed

\subsection{Inertia of energy sources}\label{Kap - SRT Dynamics - On the inertia of energy sources}

The absorption of a boosted energy unit is associated with an impulse transfer; the same holds for every deceleration. The energy stored in a bound body effects the behavior in a head-on collision; hence the basic observable ''inertial mass'', which according to Galilei we define by this elemental comparison (see Definition \ref{Def - vortheor Ordnungsrelastion - inertial mass}).
\begin{co}
The absorption of internal (binding) energy $\Delta E_{\mathrm{int}}^{(\mathcal{A})} \cdot E \left[ \mathcal{S}\big|_{\mathbf{v}_{\mathcal{A}}} \right]$ increases the inertial mass $\Delta m_{\mathcal{A}}$ of the absorbing body $\circMA$
\be\label{Cor - SRT Dynamics - calorimeter model - inertia of E sources - Delta m (Delta E)}
   \Delta E_{\mathrm{int}}^{(\mathcal{A})} \;\; = \;\; \Delta m_{\mathcal{A}} \cdot c^2 \;\; .
\ee
\end{co}
\textbf{Proof:}
Let us couple one (resting) energy unit $\mathcal{S}\big|_{\mathbf{v}_{\mathcal{A}}}$ into a co-moving absorber $\circMA$. We determine the inertial mass $m[\circMA\cup \mathcal{S} ]$ of the absorber by a \emph{physical criterion}. For simplicity we suppress the carrier $\circMA$. The energy source $\mathcal{S} \sim_{m} \circMm$ has the same inertia as a composite $\circMm := \circMunit\ast\ldots\ast\circMunit$ of $m$ standard elements if in an elastic head-on collision test with same initial velocity $v$ (of arbitrary value) each rebounds antiparallel with same velocity
\[
   \mathcal{S}\big|_{\mathbf{v}} \,,\: \circMm_{-\mathbf{v}}  \;\; \Rightarrow \;\;
   \mathcal{S}\big|_{-\mathbf{v}} \,,\: \circMm_{\:\mathbf{v}}  \;\; .
\]
Let $\mathcal{B}$ob observe the process from a perspective co-moving with the test composite $\circMm$
\be
   \mathcal{S}\big|_{v^{(\mathcal{B})}_{\mathcal{A}}} \,,\: \circMm_{0^{(\mathcal{B})}}  \;\; \Rightarrow \;\;
   \mathcal{S}\big|_{0^{(\mathcal{B})}} \,,\: \circMm_{v^{(\mathcal{B})}_{\mathcal{A}}}  \;\; . \nn
\ee
The energy source $\mathcal{S}$ with initial velocity $v^{(\mathcal{B})}_{\mathcal{A}}$ appears to come to rest and kicks the test particle $\circMm$ into the same velocity $v^{(\mathcal{B})}_{\mathcal{A}}$.

$\mathcal{B}$ob can generate the same result by first absorbing the boosted energy source $\mathcal{S}\big|_{v^{(\mathcal{B})}_{\mathcal{A}}}$ in his calorimeter and then by kicking out the test particle $\circMm_{v^{(\mathcal{B})}_{\mathcal{A}}}$ into the same velocity $\mathbf{v}_{\mathcal{A}}$ and providing a resting energy unit $\mathcal{S}\big|_{0\cdot \mathbf{v}_{S^{(\mathcal{B})}}}$. The energy-momentum of the separate steps
\be\label{Abschnitt -- SRT Dynamics - calorimeter model - inertia of E sources - E p vor und nach collision test}
   \left(\!\!
     \begin{array}{c}
       E \\
       \mathbf{p} \\
     \end{array}
   \!\!\right)^{\!\!(\mathcal{B})} \! \left[ \mathcal{S}\big|_{\mathbf{v}_{\mathcal{A}}} \right]
   \stackrel{(\ref{Abschnitt -- SRT Dynamics - calorimeter model - phys connection reference units})}{=}
   \gamma \cdot
   \left(\!\!
     \begin{array}{c}
       1 \\
       \frac{\mathbf{v}}{c^2} \\
     \end{array}
   \!\!\right)
   \: \stackrel{!}{=} \:
   \left(\!\!
     \begin{array}{c}
       E \\
       \mathbf{p} \\
     \end{array}
   \!\!\right)^{\!\!(\mathcal{B})} \! \left[ \circMm_{\mathbf{v}_{\mathcal{A}}} , \mathcal{S}\big|_{0\cdot \mathbf{v}_{S^{(\mathcal{B})}}} \right]
   \stackrel{(\ref{Abschnitt -- SRT Dynamics - calorimeter model - Quantification energy-momentum - generic particle})}{=}
   \left(\!
     \begin{array}{c}
       \!m \cdot c^2 \cdot ( \gamma - 1 )\! \\
       m \cdot \gamma \cdot \mathbf{v} \\
     \end{array}
   \!\right)
   +
   \left(\!\!
     \begin{array}{c}
       1 \\
       \mathbf{0} \\
     \end{array}
   \!\!\right)
\ee
must match; otherwise, from an imbalance in the number of extracted and expended standard energy-momentum carriers Bob could couple them into a circular process and steer a perpetuum mobile (see Lemma \ref{Lem - kin quant Absorptions Wirkung - Reservoirbilanz - p conserved}).

With one (co-moving) energy unit $\mathcal{S}\big|_{\mathbf{v}_{\mathcal{A}}}$ the inertial mass of the absorber $m\left[\circMA\right]$ increases by $\Delta m \left[ \mathcal{S} \right] \stackrel{(\ref{Abschnitt -- SRT Dynamics - calorimeter model - inertia of E sources - E p vor und nach collision test})}{=} \frac{1}{c^2}$.
We can absorb multiple $\Delta E^{(\mathcal{A})} \cdot \mathcal{S}\big|_{\mathbf{v}_{\mathcal{A}}}$ energy units one by one. The total inertial mass increase of the absorber is proportional to their number $\Delta E^{(\mathcal{A})}$
\[
   \Delta m \left[ \Delta E^{(\mathcal{A})} \cdot \mathcal{S} \right]
   \;\; = \;\;
   \Delta E^{(\mathcal{A})} \cdot \frac{1}{c^2}  \;\; .
\]
\qed
\begin{de}
With the emission of internal energy the inertial mass of the emitter decreases. Suppose the mass could vanish entirely, according to Einstein we call the maximum energy gain \underline{rest energy}
\be\label{Abschnitt -- SRT Dynamics - calorimeter model - inertia of E sources - E_0 rest energy}
   E_{0} \;\; \stackrel{(\ref{Cor - SRT Dynamics - calorimeter model - inertia of E sources - Delta m (Delta E)})}{=} \;\;
   m\cdot c^2   \;\; .
\ee
\end{de}
\begin{co}
A free body $\circMm_{v}$ with mass $m$ and velocity $v$ has a \underline{total energy} from the deceleration and emission of internal binding energy (so its inertia would vanish entirely)
\be\label{Abschnitt -- SRT Dynamics - calorimeter model - inertia of E sources - E_tot total energy}
   E_{\mathrm{tot}} \left[  \circMm_{v}  \right] \;\;:=\;\; E_{\mathrm{kin}} \left[  \circMm_{v} \Rightarrow \circMm_{0} \right] \;+\; E_{0} \left[  \circMm \right] \;\;\stackrel{(\ref{Abschnitt -- SRT Dynamics - calorimeter model - Quantification energy-momentum - generic particle})(\ref{Abschnitt -- SRT Dynamics - calorimeter model - inertia of E sources - E_0 rest energy})}{=}\;\;
   \gamma \cdot m \cdot c^2   \;\; .
\ee
\end{co}
We uncover new connections between mass and energy. Though in view of \emph{one} simple equation $E_0 = m \cdot c^2$ one cannot take as absolute, that ''Mass and energy are therefore essentially alike; they are only different expressions of the same thing'' \cite{Einstein-Grundlagen der ART}. Galilei and Leibniz define the \emph{basic observables} independently by a physical criterion. The ''capability to work'' and the ''inertia'' are well-distinguished attributes for any carrier system:
\begin{itemize}
  \item The \emph{inertial mass} is uniquely determined by Galilei's head-on collision test.
  \item With the motion of a free body and its internal binding state (without needing to specify of what nature its bound constituents are) we associate two forms of energy:
      \begin{itemize}
        \item[(i)] \emph{Kinetic energy} is associated with changes of the (collective) motion of a composite body (provided the internal binding state is unchanged).
        \item[(ii)] \emph{Potential energy} is associated with changes in the internal state (provided the body remains at rest). We measure the transformation of (electromagnetic, gravitational, chemical, radioactive etc.) potential energy of the system into kinetic energy of its elements. Leibniz equipollence principle (to measure an energy source by its effect against our reference calorimeter; energy conservation \{\ref{Kap - KM Dynamics - Basic Dynamical Measures - potential Energy}\}) extends to processes, where the nature or number of the particles are not conserved!
      \end{itemize}
\end{itemize}

Einstein's concern, that ''with a different definition of force and acceleration we should naturally obtain other values for the masses'', indicates that commonly mass is introduced formally \{\ref{Kap - SRT Dynamics - Pre-theoretic elements}\} and not seen as a basic observable \cite{Adler - Does mass really depend on velocity dad?}. We quantify the energy and mass directly by counting intrinsically well-defined reference units. Instead we derive the \emph{fundamental equations} between independently measurable quantities from undisputed physical and practical principles \{\ref{Kap - KM Dynamics - Discussion - Principles}\}. The particular relation depends on the concrete situation:
\begin{enumerate}
  \item In the calorimeter model we measure the relation between the impact velocity and the kinetic energy of an incident body
  \[
     E_{\mathrm{kin}} \left[  \circMm_{v}  \right]
     \;\;\stackrel{(\ref{Abschnitt -- SRT Dynamics - calorimeter model - Quantification energy-momentum - generic particle})}{=}\;\;
     m \cdot c^2 \cdot \left( \gamma - 1 \right)  \;\; .
  \]
  \item By continued calorimetric measurements we derive the relation between emitted internal energy $\Delta E_{\mathrm{int}}$ and the inertial mass decrease $\Delta m$ of the emitter
  \[
     \Delta E_{\mathrm{int}}
     \;\;\stackrel{(\ref{Cor - SRT Dynamics - calorimeter model - inertia of E sources - Delta m (Delta E)})}{=}\;\;
     \Delta m \cdot c^2  \;\; .
  \]
  \item We understand ''rest energy'' as potentially expendable energy (limited by the inertial mass of the carrier)
  \[
     E_{0}
     \;\;\stackrel{(\ref{Cor - SRT Dynamics - calorimeter model - inertia of E sources - Delta m (Delta E)})}{=}\;\;
     m \cdot c^2
  \]
  \item and the total energy of a free body from the deceleration and the (hypothetically) complete emission of internal energy
  \[
     E_{\mathrm{tot}} \left[  \circMm_{v}  \right]
     \;\;:=\;\; E_{\mathrm{kin}} + E_{0} \;\;=\;\;
     \gamma \cdot m \cdot c^2    \;\; .
  \]
\end{enumerate}
For example one can assign total energy and momentum to a photon $(E,\mathbf{p})_{\mathrm{photon}} $ based on calorimetric measurements. In Corollary \ref{Cor - SRT Dynamics - calorimeter model - phys connection dun units - E,p_B (E,p_A)} one can compare, according to Leibniz, the kinetic effect of absorbing a photon with the effect of our standard energy source $\mathcal{S}\big|_{\mathbf{v}=0}$. One can test, according to Galilei, whether in a head-on collision the photon has more impact than our standard impulse carrier $\circMunit_{\:\mathbf{v}_S}$. But in no physically meaningful way one can attribute mass to it. A ''photon'' always propagates at the speed of light $c\neq \mathbf{v}_{\mathbf{1}}$ and thus violates the measurement condition of the underlying head-on collision test, that both parties (photon and composite of material standard elements $\circMm := \circMunit\ast\ldots\ast\circMunit)$ have \emph{equal} initial velocity of arbitrary value. Hence a photon is associated with energy-momentum but not mass.\footnote{One can argue whether to regard a ''photon'' (\emph{propagation} of electromagnetic interactions) as particle since one can only determine its potential energy-momentum (Feynman attributes the latter strictly to the emitting material source \cite{Feynman lectures II}) upon absorption, i.e. after its destruction in the absorbing body.}

\subsection{Four-momentum formulation}

By a formal rescaling of the physical energy values $\frac{1}{c^2} \cdot E^{(i)}$ and substitution $\mathbf{v}_{\mathcal{A}}^{(\mathcal{B})}=- \mathbf{v}_{\mathcal{B}}^{(\mathcal{A})}$
%
%
\begin{diagram}
\;\;\;
\left(\!\!
     \begin{array}{c}
       E \\
       \mathbf{p} \\
     \end{array}
   \!\!\right)^{\!\!(\mathcal{B})}   &
   \;\:
   \stackrel{(\ref{Abschnitt -- SRT Dynamics - calorimeter model - phys connection dun units - E,p_B (E,p_A)})}{=} \;\;\;\;\;\;
   \gamma \cdot
   \left(
     \begin{array}{ccc}
       1 & \!\!& \mathbf{v}_{\mathcal{A}}^{(\mathcal{B})} \\
       & \!\!& \\
       \mathbf{v} / c^2 & \!\!& 1 \\
     \end{array}
   \right)
     & \;\;\;\;\;\;\;\;
   \left(\!\!
     \begin{array}{c}
       E \\
       \mathbf{p} \\
     \end{array}
   \!\!\right)^{\!\!(\mathcal{A})} \;\;\;\;\;   \\
\dTo^{ \left(
         \begin{array}{cc}
           1/c^2 & 0 \\
           0 & 1 \\
         \end{array}
       \right)
\;\;}   &   &
   \uTo_{\;\;\; \left(
                 \begin{array}{cc}
                   c^2 & 0 \\
                   0 & 1 \\
                 \end{array}
               \right)
   }   \\
\;\;\;\;\;\;\;
\left(\!\!
     \begin{array}{c}
       \frac{1}{c^2}E \\
       \mathbf{p} \\
     \end{array}
   \!\!\right)^{\!\!(\mathcal{B})} \;\;  &
   \textcolor{green}{
   = \;\;\;
   \gamma \cdot
   \left(
     \begin{array}{ccc}
       1 & \!\!& - \frac{1}{c^2} \cdot\mathbf{v}_{\mathcal{B}}^{(\mathcal{A})} \\
       & \!\!& \\
       -\mathbf{v} & \!\!& 1 \\
     \end{array}
   \right)
   }
   \!\!\!\!\!\!\!
   & \;\;\; \left(\!\!
     \begin{array}{c}
       \frac{1}{c^2}E \\
       \mathbf{p} \\
     \end{array}
   \!\!\right)^{\!\!(\mathcal{A})}
\end{diagram}
the induced relation between $\mathcal{A}$lice and $\mathcal{B}$ob's \emph{rescaled} energy-momentum values coincides with the familiar Lorentz transformation.

The Lorentz transformation from a proper time interval of object $\mathcal{O}$
\be\label{Abschnitt -- SRT Dynamics - calorimeter model - Four-momentum formalism - kin Lorentz trafo}
   \left(\!\!
     \begin{array}{c}
       t \\
       \mathbf{s} \\
     \end{array}
   \!\!\right)^{\!\!(\mathcal{A})} \;\; = \;\;
   \gamma \cdot
   \left(
     \begin{array}{cc}
       1 &  - 1/c^2 \cdot\mathbf{v} \\
       -\mathbf{v} &  1 \\
     \end{array}
   \right)
   \left(\!\!
     \begin{array}{c}
       \tau \\
       \mathbf{0} \\
     \end{array}
   \!\!\right)^{\!\!(\mathcal{O})}
\ee
gives the duration value measured by moving observer $\mathcal{A}$lice $\mathrm{d}t^{(\mathcal{A})}  \stackrel{(\ref{Abschnitt -- SRT Dynamics - calorimeter model - Four-momentum formalism - kin Lorentz trafo})}{=} \gamma \cdot \mathrm{d} \tau^{(\mathcal{O})}$.
\begin{co}
$\mathcal{A}$lice rescaled total energy-momentum values for free body $\circMm_{\:\mathbf{v}}$
\be
   \left(\!\!
     \begin{array}{c}
       \frac{1}{c^2} E_{\mathrm{tot}} \\
       \mathbf{p}
     \end{array}
   \!\!\right)^{\!\!(\mathcal{A})}
   \!\left[ \circMm_{\:\mathbf{v}} \right]
   \stackrel{(\ref{Abschnitt -- SRT Dynamics - calorimeter model - inertia of E sources - E_tot total energy})(\ref{Abschnitt -- SRT Dynamics - calorimeter model - Quantification energy-momentum - generic particle})}{=}
   m \cdot \left(  \gamma \:,\: \gamma \cdot \frac{\mathrm{d}\mathbf{s}^{(\mathcal{A})}}{\mathrm{d}t^{(\mathcal{A})}}  \right)
   \stackrel{(\ref{Abschnitt -- SRT Dynamics - calorimeter model - Four-momentum formalism - kin Lorentz trafo})}{=}
   m \cdot \left(  \frac{\mathrm{d}t^{(\mathcal{A})}}{\mathrm{d}\tau^{(\mathcal{O})}} \,,\: \frac{\mathrm{d}\mathbf{s}^{(\mathcal{A})}}{\mathrm{d}\tau^{(\mathcal{O})}}  \right)  \: =: \: m \cdot \frac{\mathrm{d}x^{(\mathcal{A})}}{\mathrm{d}\tau}
   \nn
\ee
are given in terms of its mass $m$ and the so called ''four-velocity'' $u^{\mu}:= \frac{\mathrm{d}x^{(\mathcal{A})}}{\mathrm{d}\tau}$.
\end{co}
Though after substitution the new values do not refer to the directly countable energy sources $\mathcal{S}\big|_{\mathbf{v}=0}$ and impulse standards $\circMunit_{\:\mathbf{v}_S}$ and the dimensions are not physically connected by the vivid reference process $\mathcal{S}\big|_{\mathbf{0}^{(\mathcal{A})}} , \circMunit_{\:\mathbf{0}^{(\mathcal{A})}} , \circMunit_{\:\mathbf{0}^{(\mathcal{A})}} \stackrel{w_{S^{(\mathcal{A})}}}{\Rightarrow} \circMunit_{-\mathbf{v}_{S^{(\mathcal{A})}}} ,  \circMunit_{\:\mathbf{v}_{S^{(\mathcal{A})}}}$. Despite the appeal of a simpler calculus, without the operationalization the \emph{physical} interpretation is concealed.


\chapter{Discussion}\label{Kap - KM Dynamics - Discussion}

One usually explains physics axiomatically. One derives a formalism from a manageable system of initial propositions which are logically independent from one another. That is good for purposes like formal mathematical elegance, if one wants to teach a concise formalism \cite{Janich Das Mass der Dinge}. Though this approach already begins in the abstract. It obscures the physical meaning and vivid understanding (Anschaulichkeit). Students find it hard to relate the mathematical structure to the physical phenomena.

%
The axiomatic explanation of mechanics assumes mathematical terms (mass $m$, force $\mathbf{F}$, impulse $\mathbf{p}$) as known elements of its description, postulates properties ($m=\mathrm{const}$, $\mathbf{F}/\mathrm{mod} \: \mathbf{v}$) and basic equations ($\mathbf{F}=m\cdot \mathbf{a}$, $\mathbf{p}=m\cdot \mathbf{v}$). The mathematical theory defines derived quantities in terms of basic quantities and proves the equations from a system of postulated basic propositions. This logical mathematical reasoning though starts from \emph{undefined} basic elements and from \emph{unproven} postulates. Helmholtz \cite{Helmholtz - Zaehlen und Messen} demands a further justification and derivation. He does not accept the axioms (of geometry, arithmetics) as unprovable and not proof requiring propositions. After all physics is a measuring natural science and as such distinguished from mathematics.\footnote{Ruben stresses ''A pure mathematical relation (alone) says nothing about a particular sensual concrete domain, thus has no \emph{mechanical meaning} at all. The latter \emph{comes only about if} symbols (for measures and operations), which are connected in a mathematical relation, symbolize properties which are tangible in the experimental conduct'' \cite{Peter '67 - zum Streit um das wahre Mass der Kraft}. Aside from developing the quantitative notions one must also specify the physical and methodical prerequisites. }

One can also look into what actually happens in measurement practice. The main objective for Mach's account of the historic evolution of mechanical principles \cite{Mach - Mechanik in ihrer Entwicklung} was ''to dispel metaphysics out of physics... because we are used to call those notions metaphysical, from which we have forgotten how we arrived at them.''\footnote{According to H\"orz, Wollgast \cite{Helmholtz - Zaehlen und Messen} explaining known laws in scientific theories also requires the description of methods by means of which they were found. New concepts are not given naturally. The development is always connected with particular operations, which determine the actual meaning. The measurement theoretical foundation of physics plays a big role for the ongoing relativization of absolute physical notions even today (e.g. Einstein's revision of the concept of simultaneity). To prevent the continuation of prejudices we always connect the conceptual critique with the inspection of operations which lead to the concepts.} It was also motivated \emph{pedagogically} \cite{Mach - Raum und Geometrie}: ''An insight is grasped best by the learner through the same way along which it was found.'' The primary requirement in education, Plank \cite{Planck - Wege zur physikalischen Erkenntnis} explains, ''is not so much the amount of material but rather the way of its treatment... A single mathematical proposition, which is truly understood by students, has more value for them than ten memorized formulas... School is not supposed to convey technical routines but consequent methodical thinking.''

\section{Basic measurement operations}\label{Kap - KM Dynamics - Discussion - Equivalence Relations}

Our novel approach to the foundation of physics starts from the underlying measurement practice. We are long familiar with basic measurements, as in a length measurement by placing copies of a ruler side by side. Wallot \cite{Wallot - Groessengleichungen Einheiten und Dimensionen} defines ''physical measures are never tangible things, but always \emph{attributes of things}, properties, which we can notice on the things of our experience''. Helmholtz regards attributes of objects, which in a comparison allow the difference of larger, equal or smaller. The tangible procedure for finding ''how many times'' more is the measurement.

Helmholtz \cite{Helmholtz - Zaehlen und Messen} starts from counting same objects. He begins with basic questions:
\begin{enumerate}
  \item ''What is the physical meaning if we declare two objects as \emph{equal} in a certain relation?''
  \item ''Which character must the physical concatenation of two objects have, that we may consider comparable attributes thereof as connected \emph{additively}?''
\end{enumerate}
In this way Helmholtz elucidates familiar examples like the weight $m_{\mathcal{O}}$ of a material $\mathcal{O}$bject, the length $s_{\overline{\mathcal{A}\mathcal{B}}}$ of a straight line $\overline{\mathcal{A}\mathcal{B}}$, the duration $t_{\mathcal{P}}$ of a physical $\mathcal{P}$rocess etc. Thereby we notice, that the way of concatenation generally depends on the kind of measure. ''We add e.g. weights, by simply placing them on the same weighing-pan. We add time periods, by letting the second begin at exactly the moment, where the first stops; we add lengths, by placing them next to each other in a certain way, namely in a straight line etc.'' For a basic measurement we specify a method of comparison for the energy (of directly observable phenomena) and a method for their physical concatenation.

In everyday work experience one becomes aware of what is meant by ''length'', ''duration'', ''capability to work'' (Wirkungsverm\"ogen) and ''impact (in a collision)'' (Wucht). We use the familiar notions for the relative motion of nearby rigid bodies and the behavior in work actions to prevent a premature anticipation of the mathematical formalism. Our foundation is circularity free. We neither presuppose equations of motion nor formal proportionalities, conservation laws, symmetries etc. Every new basic measure has to be explained in words or by examples because definition-\emph{equations} for basic quantities do not exist \cite{Wallot - Groessengleichungen Einheiten und Dimensionen}.\footnote{One defines derived quantities by equations. Helmholtz \cite{Helmholtz - Zaehlen und Messen} calls them more accurately coefficients. ''\emph{Basic quantities} cannot be deduced by equations onto other already explained quantities.''} We define the basic observables from practical comparison ''longer than'', ''heavier than'', ''more impact than'' etc. Without a word of mathematics one can assess
\begin{itemize}
\item   $>_{l}$ \;\;\; if two extended objects lie on top of each other: one will \emph{cover} the other
\item   $>_{t}$ \;\;\; if two processes begin simultaneously: one will \emph{outlast} the other
\item   $>_{E}$ \;\; if against the same system $\{\mathcal{G}_I\}$: the effect of one source \emph{exceeds} the other
\item   $>_{\mathbf{p}}$ \;\;\; in a head-on collision: one body \emph{overruns} the other.
\end{itemize}
Each comparison method is universally reproducible in an observer independent way. In German one uses simple denominations (\"uberdecken, \"uberdauern, \"ubersteigen, \"uberrennen) with a direct colloquial meaning.\footnote{Ruben zeigte, da{\ss} die Bedeutung von Ausdr\"ucken f\"ur elementare logische Operation zur\"uckf\"uhrt auf ''Bezeichnungen von Handlungen, die samt und sonders aus der Sprache des r\"omischen Bauern stammen'' \cite{Peter '76 - Praedikationstheorie}. Zur Klarstellung der Bedeutung von physikalischen Ausdr\"ucken empfiehlt Ruben: Halte Dich nahe dran an der Muttersprache und achte genau darauf, was Du damit eigentlich sagst.}

\section{Principles}\label{Kap - KM Dynamics - Discussion - Principles}

We make serious with presupposing the tangible operations and a colloquial description for the foundation of the physical theory. For reproducible measurement procedures one manufactures equally long meter sticks, standard springs, impulse carriers etc. and concatenates them in suitable ways. We assemble elementary standard kicks to a \emph{functioning} calorimeter configuration. The pre-theoretic building blocks $w_{\mathbf{1}}$ \{\ref{Kap - KM Dynamics - Basic Dynamical Measures - Quantification - Dynamical Unit}\} are well-defined. We construct a basic measurement device following one simple task: whatever gets absorbed in an external reservoir must generate only standard energy and momentum carriers. The construction of the calorimeter model \{\ref{Kap - KM Dynamics - Basic Dynamical Measures - Measurement Means - Kinematic Quantification Calorimeter Action}\} relies on undisputed \emph{physical} principles:
\begin{itemize}
  \item   \emph{causality}: Under certain conditions something certain happens.
  \item   \emph{inertia} \{\ref{Kap - SRT Massbestimmung - classical metric - principle of inertia}\}
  \item   \emph{impossibility of a perpetuum mobile}: After a circular process one can neither extract ''capability to work'' from the unaltered system nor additional ''impact for a collision''.
  \item   \emph{sufficient reason}: A reasonable external cause is responsible for the change in the state of motion of an object.
  \item   \emph{relativity}: Internal processes are intrinsically equivalent under a relative boosts of the system or observer; spacetime intervals transform active resp. passive covariant.
  \item   \emph{superposition}: The intrinsic actions $w$ and the external steering resp. measurement actions $w_{\mathbf{1}}$ and $W_{\mathrm{cal}}$ are compatible.
\end{itemize}
and on \emph{methodical} principles for the constructor:
\begin{itemize}
  \item   \emph{basic measurement}: One doubles the physical measures. The act of a basic measurement is a pair comparison between a measurement object and a material model.
  \item   \emph{congruence}: For a reliable quantification one counts congruent reference units.
  \item   \emph{equipollence}: One measures the cause of an action by its effect \{\ref{Kap - KM Dynamics - Basic Dynamical Measures - Energy - quantification scheme}\}.
\end{itemize}
which include the test procedures for manufacturing sufficiently constant reference devices (straight rulers, uniform running clocks, equally loaded springs etc.) and for a reproducible way of concatenation (locally regular grid of light clocks, standardized calorimeter process). Further we assume a \emph{social} condition: Physicists must cooperate to create material models. A team of assistants has to know, when, where and how to couple initially resting reservoir elements $\circMunit_{\:\mathbf{v}=0}$ into the deceleration process, to steer a useful absorption. The cooperation is community-building. Team work is crucial for the conduct of basic measurements. Basic physical quantities are a joint product and not generated individually.

We start with the \emph{basic observables}: the ''capability to work'' and ''impact'' of an incident object $\circMO_{\:\mathbf{v}}$. We count how many standard springs $\mathcal{S}_{\mathbf{1}}\big|_{\mathbf{v}=0}$ it can overcome before it stops and thus find ''how many times larger'' its kinetic energy is than the reference energy of one standard spring. We transfer its impact onto a certain number of standard impulse carriers $\circMunit_{\:\mathbf{v}_{\mathbf{1}}}$ and thus quantify its momentum. We obtain a \emph{quantified basic observable}, a certain physical quantity (number of reference units) times the dimension of the reference device. True physical quantities originate from counting congruent reference units.

Usually
%
%
one ''calculates'' the value of basic dynamical observables as a derived quantity $F=m\cdot a$, $p=m\cdot v$ etc. Scope and limitations of the postulated equations remain unclear. Our program presents a novel method for defining basic measures. The method is of a physical nature (no formal preassumptions): (i) The objects have to be of physical nature. (ii) Their properties have to be of physical nature. (iii) The measurability of these properties is achieved by the fact that the reference standards ''$\mathcal{S}$'' are producible and reproducible in a physical way and that one realizes the concatenation ''$\!\ast$'' of measurement units by tangible operations. Only if a standard energy source is defined, one can count how many copies the calorimeter model generates; only then therefore comes the \emph{counting}! We investigate the foundation of a non-mathematical science which however uses mathematics \{\ref{Kap - KM Dynamics short Review - On physical objects and mathematical objects}\}. We first determine the physical operations really in a strictly physical way. The mathematical bookkeeping comes into being at the moment we introduce reference units.

\section{Genesis of algebra}\label{Kap - KM Dynamics - Discussion - Basic Arithmetic Relations}

''Calculating is an indirect or mediated way of counting'', explains Mach \cite{Mach - Zahl und Mass}. Every arithmetic law establishes an abbreviating relation between two ways - the laborious direct and the elegant indirect - of counting. The result can always be deduced in the more complicated direct way from the definition of elementary counting. In a \emph{mathematical proof} one essentially reduces complex relations between derived terms (sums, products, integrals etc.) to the relation $1=1$ of basic elements (number symbols).

Our \emph{physical proof} relies on physical principles and constructing measurement models. We couple the energy-momentum of moving bodies $\circMa_{\:\mathbf{v}_a}$\,, $\circMb_{\:\mathbf{v}_b}$ by absorbing them in the same calorimeter reservoir $\{\circMunit_{\:\mathbf{v}=0}\}$. We count the total number of extractable energy-momentum units from the (consecutive or simultaneous) absorption. We introduce a meaningful addition of (extensive) energy-momentum quantities $(E, \mathbf{p})_a+(E,\mathbf{p})_b$ from direct counting.

We can perceive the validity of the elemental comparisons ''more capability to work than'' and ''more impact than'' $>_{E,\mathbf{p}}$ in practice and conduct the coupling of standard actions $w_{\mathbf{1}}\ast w_{\mathbf{1}}$ manually. ''In Gedanken'' we can measure energy and momentum directly. From matching the layout and form of the building blocks in our machinery we can count the coinciding numbers of activated standard springs, standard bodies, velocity units. We independently measure the physical quantities (of reference units) and derive their fundamental equations. When we build the model in Galilei kinematics we prove, that the measured values for kinetic energy $\frac{E \left[ \circMO_{\:\mathbf{v}} \right]}{E \left[ \mathcal{S}_{\mathbf{1}}\big|_{\mathbf{0}} \right]} := \sharp \left\{ \mathcal{S}_{\mathbf{1}}|_{\mathbf{0}} \right\}$, momentum $\frac{\mathbf{p} \left[ \circMO_{\:\mathbf{v}} \right]}{\mathbf{p} \left[ \circMunit_{\:\mathbf{v}_{\mathbf{1}}} \right]} := \sharp \left\{ \circMunit_{\:\mathbf{v}_{\mathbf{1}}} \right\}$, inertial mass $\frac{m\left[ \circMO \,\right]}{m\left[ \circMunit \,\right]} =: \sharp \left\{ \circMunit \,\right\}$ and velocity $\frac{v_{\mathcal{O}}}{v_{\mathbf{1}}} := n$ satisfy the equations $E_{\mathrm{kin}} = \frac{m}{2}\cdot \mathbf{v}^{2}\:$, $\mathbf{p}=m\cdot \mathbf{v}\:$; and for same procedure in Poincare kinematics we derive $E_{\mathrm{kin}}=m\cdot c^2 \cdot (\frac{1}{\sqrt{1-\frac{v^2}{c^2}}}-1) \, ,$ $\mathbf{p}=m\cdot \frac{\mathbf{v}}{\sqrt{1-\frac{v^2}{c^2}}}\,$. We derive the basic equations from physical and methodical principles instead of postulating a formula.
For the judgement of this methodology we can remark with Planck \cite{Planck - Wege zur physikalischen Erkenntnis}
\begin{quote}
''Activities from the \emph{Axiomatiker} are useful and necessary but therein also hides the dubious danger of one-sidedness, that the physical world view loses its meaning and degenerates into empty formalism. Because if connection with reality is detached then a \underline{physical law} \emph{appears} - not anymore as relation between quantities which can all be measured independently from one another but - \emph{as definition}, by means of which one of those quantities is reduced to the others. Such \emph{reinterpretation} is particularly tempting, because a physical quantity can be defined much more exactly by an equation than by a \emph{measurement}; however that fundamentally represents \emph{abandonment of its true meaning}.''\footnote{Poincar\'es \cite{Szabo Geschichte der Mechanischen Prinzipien}
proclaims ''What science can grasp are not the things themselves but the relations between things''. Planck \cite{Planck - Wege zur physikalischen Erkenntnis} emphasizes: ''The two sentences - There is a real outside world. The real outside world is not directly accessible - are the crux of the matter of entire natural science physics.'' He distinguishes (i) the natural world - we actually inhabit and of which we are natural part of - (ii) our sensual perceptions and (iii) our physical conception of the world (physikalisches Weltbild). For Planck \emph{physicists} are workers, who develop the physical world view. He classifies three groups depending on their main interest and method how they treat the theory: A \emph{Metaphysician} stresses its relation to the real outside world. A \emph{Positivist} stresses its relation to the sensual world. The \emph{Axiomatiker} focus attention neither on its relation to the real nor on those to the sensual world but rather on the inner coherence and logical structure of the physical theory.}
\end{quote}

One can couple local calorimeters into e.g. gravitational or electromagnetic systems. Provided also a swarm of light clocks one can measure the relation between geodesic deviation and energy-momentum sources intrinsically. When in an elementary particle collision new particle generations develop, one can capture a jet with the calorimeter and measure the energy and momentum of separate decay products. We presuppose all interactions solely as a completed process with unknown inner structure. We pick the compression of a standard spring as an elementary building block for our calorimeter model because they are congruent and reproducible. Physicists can couple these units in a controlled way and count; and then measure the other interactions with this reference interaction $w_{\mathbf{1}}$.

By consecutive calorimeter interventions $\mathrm{RB}^{(i)}$ one can steer the course of an intrinsic action $w$ and analyze the corresponding steering effort. We analyze the extractable energy-momentum from infinitesimal configuration changes. From the basic quantities of length, duration and energy, momentum we define ''force'' as a derived quantity and ''displacement work'' in steered actions. We analyze the properties and ultimately derive \emph{Newton's equations of motion} for a conservative system.\footnote{Newton's force used to be the main and basic concept of mechanics. In more recent development of physics - Planck \cite{Planck - Wege zur physikalischen Erkenntnis} states - ''Newton force has lost its fundamental importance for theoretical physics. In the modern layout of mechanics it only appears as a secondary quantity, replaced by another higher and broader notion of work and potential energy.'' Just what measurement theoretical analysis yields directly!} Finally we examine the physical process
behind basic variation maneuvers, the coupling of external steering actions $\mathrm{RB}^{(1)}\ast w_1 \ast \mathrm{RB}^{(2)}\ast w_2 \ast \ldots$ into the course of an intrinsic action $w$. For the variational analysis (of the steering effort) we compare the steered process and the free evolving process. For every local Hamilton type variation $\delta\gamma^{(\mathrm{Ham})}$ of the free evolution $\gamma$ of intrinsic action $w$ the variation of Hamilton's quantity of action is positive definite
$ 0  <  \delta S_{\mathrm{Ham}} \left[ \gamma \right]  \;\;\; \forall \; \delta\gamma^{(\mathrm{Ham})}$.
Steered actions are associated with extra steering effort.
\\

\begin{figure}    
  \begin{center}           
  \includegraphics[height=20cm]{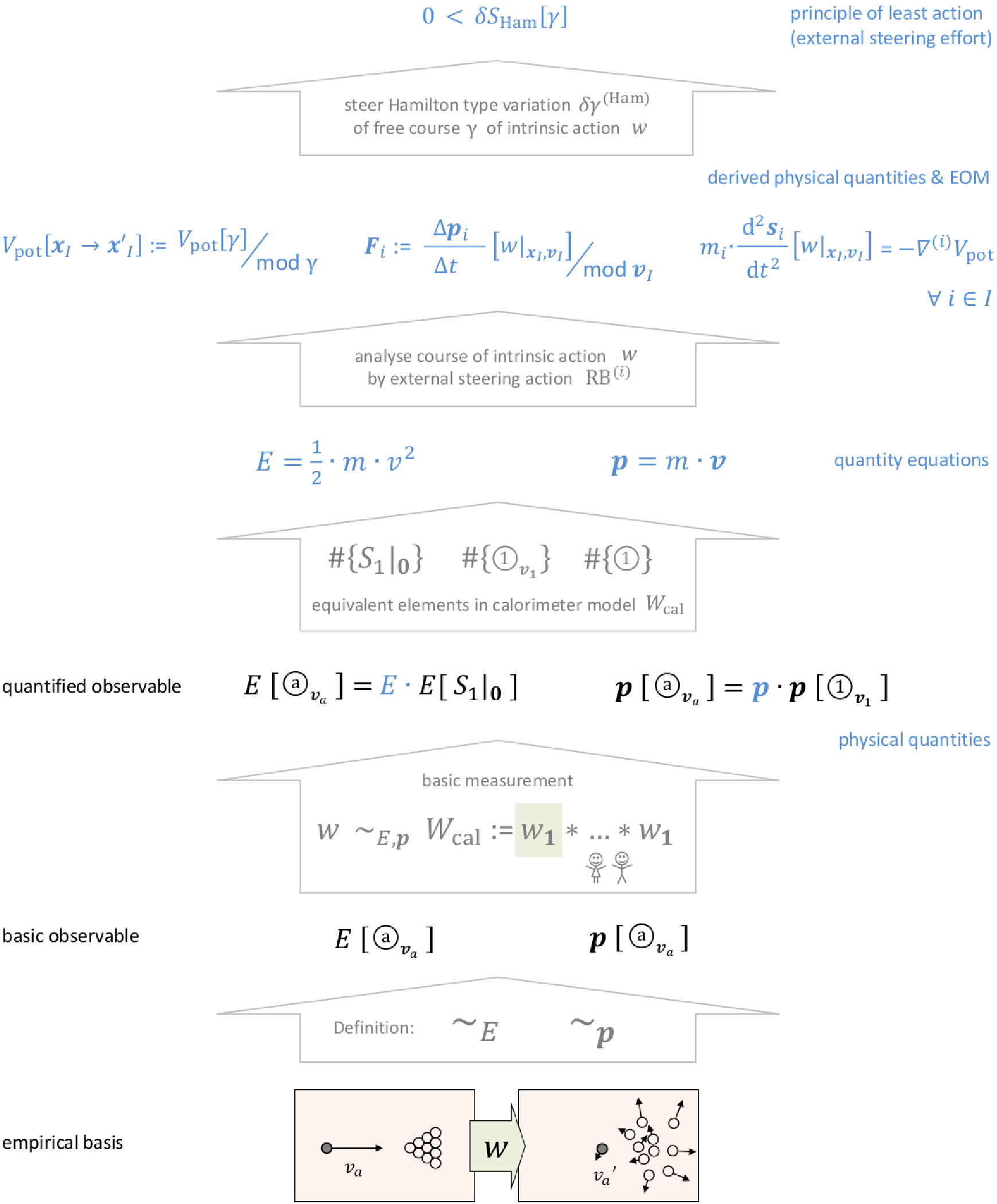}  
  \end{center}
  \vspace{-0cm}
  \caption{\label{pic_Begriffsbildung_als_Arbeitsprozess} practical and theoretical steps for the physical determination
    }
  \end{figure}

We start from a physical interaction and we arrive at a mathematical action (functional). We begin with generic actions (billiard-collision $w$, gravitational process, standard kick $w_{\mathbf{1}}$); in the end we have determined them by physical measures. We develop all physical terminology from empirical grounds.\footnote{Despite the Nobel - Laughlin \cite{Laughlin reinventing physics from the buttom down} ''focussed on the question whether physics was a logical creation of the mind or a synthesis built on observations... Seeing our understanding of nature as a mathematical construction has fundamentally different implications from seeing it as an empirical synthesis... (from a) world view that mathematics grows out of experimental observation, not the other way around. The world we actually inhabit, as opposed to the happy idealization of modern scientific mythology, is filled with wonderful and important things we have not yet seen because we have not looked, or have not been able to look at due to technical limitations.'' ''It is not the case that in natural science, as in any other science, we start from fixed basic concepts and search for their realization in our surrounding world - explains Planck \cite{Planck - Wege zur physikalischen
Erkenntnis} - rather it is quite the reverse. By birth we humans are all simply placed into the midst of life without prior preparation even without being informed and in order to cope with this imposed life... we seek to \emph{establish} certain \emph{concepts which are suitable for usage} to past and future (work) experiences \emph{in real life}.''}
In figure \ref{pic_Begriffsbildung_als_Arbeitsprozess}
%
we demonstrate every step as a \emph{production process}.\footnote{Ruben \cite{Ruben - Arbeitskonzept} examines the reflection of practical actions in the conduct of consciousness: ''Products of consciousness are linked with practical actions. For Hegel the natural scientist is a producer of knowledge, not an owner of knowledge which exists independently of the work of scientists per se. Hegel assumes the theoretical conduct itself as work. 'Thinking as work' - that is the great basic idea of Hegel's philosophy (Wissenschaft als Arbeit). Under the expression 'work' Hegel understands both that an active subject finds an object, which he reshapes for his purpose (thus converts into something different to what it has been before) and also that in this conduct the subject realizes his own ability in the other object.''}
The \emph{tangible operations} of the physicists lead to the \emph{mathematical formulation}, not the other way around.  In measurement practice we uncover the origin of basic observables \{\ref{Kap - KM Dynamics - Discussion - Equivalence Relations}\}, quantification \{\ref{Kap - KM Dynamics - Discussion - Principles}\} and the genesis of the fundamental equations. The physical determination of an action demonstrates how from a pre-theoretic definition of basic observables \{\ref{Kap - KM Dynamics - Physical Measurement - Pre-theoretical Ordering Relation}\} the quantified observables in the four-vector formulation \{\ref{Kap - SRT Dynamics - quantitative equations}\} arise.

The existence of basic physical quantities for dynamics \{\ref{Kap - KM Dynamics - Basic Dynamical Measures}\} and the associated fundamental equations \{\ref{Kap - KM Dynamics - Differentiated Analysis}\} \{\ref{Kap - KM Dynamics - Principle of Least Action}\} is tied to conditions. Provided the building blocks $w_{\mathbf{1}}$ of the model satisfy all physical conditions and the concatenation procedure ''$\!\ast$'' is realizable, then by counting the standard elements in our model one arrives at well-defined physical quantities of energy and momentum. From the geometric structure of the model (impulse reversion figure \ref{pic_elast_collision_2-5}, absorption figure \ref{pic_calorimeter_model}) we derive the fundamental equations. One can check whether the underlying operations are feasible in the gravitational and quantum mechanical domain and to what extent one can use or alter the corresponding mathematical formalism. The operationalization reveals the scope and limitations of the familiar approach but also from what grounds a mathematical formulation faithful to the basic measurement operations under quantum mechanical and gravitational conditions can arise.

We have successfully completed Hertz program \{\ref{Kap - KM Dynamics short Review - Hertz program}\} for an operationalization of energy-momentum as a \emph{comprehensible} basis for elementary dynamics \{\ref{Kap - Mechanics}\}, \{\ref{Kap - Analytical mechanics}\}, \{\ref{Kap - Relativistic energy-momentum}\} and also for kinematics \{\ref{Kap - Kinematics}\}. Next one can extend it to more nebolous fields. We open up the possibility for a complementary operational clarification of conceptual problems in the theoretical description of gravitational and electromagnetic interactions. One can proceed from expedient measurement instructions via the definition of basic measures to the formulation as a principle of least ''action'' and present its mathematical form of appearance in union with its physical content. As next step one can develop the field description of electromagnetic and gravitational interactions, scope and limitations - without presupposing the Maxwell or Einstein equations resp. Yang-Mills or Einstein-Hilbert formalisms - from basic methodical principles (of intrinsic measurement practice \{\ref{Kap - SRT Massbestimmung}\}, \{\ref{Kap - KM Dynamics - Basic Dynamical Measures}\}) and simple physical principles (retarded Coulomb principle \cite{Feynman lectures II} \cite{Rosser - classical EM via relativity}, Newton's principle of universal gravitation \cite{Baez - Meaning of GR}).
\\

When we retrospect what we actually have done, we notice with Ruben \cite{Peter '69 - Dissertation} that mathematics is not only applied in physics. In reverse, physics also appears as the mother of \emph{its} mathematics in empirical practice. We begin from pre-theoretic ordering relations ''more capability to work than'' and ''more impact (in a collision) than'' and Helmholtz program for basic measurements. In a physical model, built by steering congruent unit actions $w_{\mathbf{1}}$, we can quantify the basic observables. The \emph{basic quantities} (as opposed to a coordinate parametrization) have a precise physical meaning; under arbitrary reparametrization of the fundamental equations the measurement interpretation will be concealed. Fundamental equations can (and for didactic purposes should) be \emph{derived} without postulating dubious base equations. From physical and methodical principles we arrive at the fundamental equations and conservation laws and ultimately at the principle of least action. In retrospect pure mathematics appears like something half  \cite{Burton First Principles}. We have completed physics with the practical. Now we know more than before. This work is a contribution to understanding the active role of physicists, their interventions in basic measurements. We reconcile these operations (from the standardization of conducting experiments) with their words in the theory.

\backmatter

\chapter*{Selbst\"andigkeitserkl\"arung}

Ich erkl\"are, dass ich die vorliegende Arbeit selbst\"andig und nur unter Verwendung der angegebenen Literatur und Hilfsmittel angefertigt habe. Ich habe mich anderweitig nicht um einen Doktorgrad beworben und besitze einen entsprechenden Doktorgrad nicht. Ich erkl\"are die Kenntnisnahme der dem Verfahren zugrunde liegenden Promotionsordnung der Mathematisch-Naturwissenschaftlichen Fakult\"at der Humboldt-Universit\"at zu Berlin.

\vspace{2\baselineskip}
\noindent Berlin, den 29.04.2014\hfill Bruno Hartmann

\end{document}